\documentclass[article]{elsart}
\usepackage{epsfig}
\usepackage{graphicx}
\usepackage{fancybox}

\newcommand{\vas}{v_{\rm {as}}}
\newcommand{\ds}{{\rm d}}

\begin{document}

\begin{frontmatter}
\title{Front propagation into unstable states}
\author{Wim van Saarloos} 
\address{Instituut--Lorentz, Universiteit Leiden, Postbus 9506,
  2300 RA Leiden, \\The Netherlands}
\maketitle
\begin{abstract}
This paper is an introductory review of the problem of front
propagation into unstable states. Our presentation is centered around
the concept of the asymptotic linear spreading velocity $v^*$, the
asymptotic rate with which initially localized perturbations spread
into an unstable state according to the linear dynamical equations obtained
by linearizing the fully  nonlinear equations about the unstable state. 
This allows us to give a precise definition of pulled fronts,
nonlinear fronts whose asymptotic propagation speed equals $v^*$,
and pushed fronts, nonlinear fronts whose asymptotic speed $v^\dagger$ is larger
than $v^*$. In addition, this approach allows us to clarify many aspects of the 
front selection problem, the question whether for a given dynamical
equation the front is pulled or pushed. It also is the basis for the
universal expressions for the power law rate of approach of the
transient velocity $v(t)$ of a pulled front as it converges toward its
asymptotic value $v^*$.  Almost half  
of the paper is devoted to reviewing many experimental and
theoretical examples of front propagation into unstable states from
this unified perspective.  The
paper also includes   short sections on the derivation of the
universal power law relaxation behavior of $v(t)$, on the absence of a moving boundary
approximation for pulled fronts, on the relation between so-called global
modes and front propagation, and on stochastic fronts.
\end{abstract}
\end{frontmatter}

\tableofcontents
\newpage

\section{Introduction} 
\subsection{Scope and aim of the article}\label{scopeaim}
The aim of this article is  to introduce, discuss and review the
main aspects of front propagation into an unstable state.  By this we
mean that we will consider situations in spatio-temporally extended
systems where the  (transient) dynamics is dominated by  a well-defined front
which invades a domain in which the system is in an unstable
state. With the statement that the system in the domain into which the
front propagates is in an unstable state, we mean that the state of
the system in the region far ahead of
the front is {\em linearly unstable}. In the prototypical case
in which  this unstable state is a stationary {\em homogeneous}
state of the system, this  simply means that if one takes an arbitrarily
large domain of the system in this state and analyzes its linear
stability  in terms of Fourier modes,  a
continuous set of these modes is  unstable, i.e., grows in time.

At first sight, the subject of {\em front propagation into unstable states}
might seem to be an esoteric one. After all, one might think that examples of
such behavior would hardly ever occur cleanly in nature, as they appear to
require that the system is first prepared carefully  in an unstable
state, either by
using special initial conditions in a numerical simulation or by
 preparing an  experimental system in a state it does not
naturally stay in.  In reality, however, the subject is not at all of
purely academic
interest, as there are many examples where either front
propagation into an unstable state is an essential element of the
dynamics, or where it plays an important role in the transient
behavior. There are several  reasons for this. First of all, there are
important experimental examples where the system is essentially
quenched rapidly into an unstable state. Secondly, fronts naturally
arise in convectively unstable systems,  systems in which a
state is unstable, but where in the relevant frame of reference 
perturbations  are convected away faster than they grow out --- it is
as if in such systems the unstable state is actually dynamically
produced since the convective effects naturally sweep the system
clean. Even if this is the case in an infinite system, fronts do play
an important role when the system is finite. For example, noise or a 
perturbation or special boundary condition near a fixed inlet can then
create patterns dominated by 
fronts. Moreover, important changes in the dynamics usually occur when
the strength of the instability increases, and the 
analysis of the point where the instability  changes over from convectively
unstable to absolutely unstable (in which case perturbations in the
relevant frame do grow faster than they are convected away) 
 is intimately connected with the theory of front
propagation into unstable states.  Thirdly, as we shall explain in
more detail later, close
to an instability threshold front propagation always wins over the
growth of bulk modes.

The general goal of our discussion of front propagation into unstable
states is to investigate the following  {\em front
propagation problem:}

\begin{tabular}{p{0.1cm}|p{13.2cm}}
& {\em If initially a spatially extended system is in an unstable
state everywhere except in some  spatially localized region, what will
be the large-time dynamical properties and speed  of the nonlinear front which
will  propagate into the unstable state? Are there classes of initial
conditions for which the front dynamics converges to some  unique asymptotic
front state? If so, what characterizes these initial conditions, and
what can we say about the asymptotic front properties and the
convergence to them? }
\end{tabular}

Additional questions that may arise concern the sensitivity of the
fronts to noise or a fixed perturbation modeling an 
experimental boundary condition or an inlet, or the question under what conditions the fronts can
be mapped onto an effective interface model when they are weakly
curved.

Our approach to introducing and reviewing front propagation into unstable states
will be based on the insight that  there is a single unifying
concept that allows one to approach essentially all these questions for
a large variety of fronts. This concept is actually very simple
and intuitively appealing, and allows one 
to understand the majority of examples  one
encounters  with just a few related theoretical tools. Its essence 
 can actually be stated in  one single sentence: 

\begin{tabular}{p{0.1cm}|p{13.2cm}}
& {\em Associated with  any given
unstable state is a well-defined and easily calculated so-called
``linear'' spreading velocity $v^*$, the velocity with 
which arbitrarily small linear perturbations about the unstable state grow out and
spread according to the dynamical equations obtained by linearizing
the full model about the unstable state;  nonlinear fronts can either have their asymptotic speed
$\vas$ equal to $v^*$ (a so-called pulled front) or larger
than $v^*$ (a pushed front). }
\end{tabular}

 The name pulled front stems from the
fact that such a front is,  as it were, being ``pulled along'' by the
{\em leading edge} of the front, the region where the dynamics of the front is to good 
approximation governed by the equations obtained by linearizing about the unstable state. 
The natural propagation speed of the leading edge is hence the  asymptotic linear 
spreading speed
$v^*$. In
this way of thinking, a pushed front is being pushed from behind by the nonlinear
growth in the nonlinear front region itself
\cite{paquette1,paquette2,stokes}.

\begin{figure}[t]
\begin{center}
\epsfig{figure=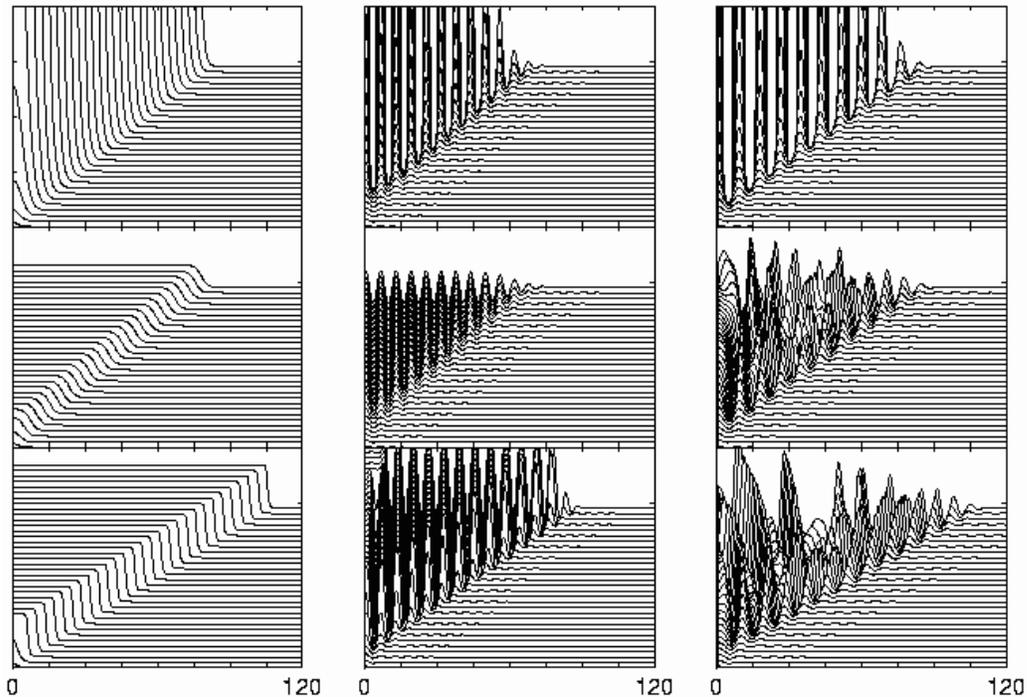,angle=-90,width=0.98\linewidth} 
\end{center}
\caption[]{Graphical summary of one of the major themes of this paper.
From top to bottom:  {\em linear spreading}, {\em pulled fronts } and
{\em pushed fronts}. From left to right: {\em uniformly translating
fronts, coherent pattern forming fronts and incoherent pattern forming
fronts}. The plots are based on  numerical simulations of
three different types of dynamical equations discussed in this paper. In all cases, the initial 
condition was a Gaussian of height 0.1, and the state to the right is linearly unstable.  
To make the dynamics visible in these space-time plots, successive traces 
of the fronts have been moved upward. Thicks along the vertical axes are placed
a distance 2.5 apart.  Left column: F-KPP equation (\ref{fkpp})
with a pulled front with $f(u)=u-u^3$ (middle) and a pushed one
for $f(u)=u + 2\sqrt{3}u^3-u^5$ in the lower row, for 
times up to 42. Middle column: the  Swift-Hohenberg equation of section
\ref{sectionsh} (middle) and an extension of it as in
Fig.~\ref{figsh}({\em (b)} (bottom). Right column: Kuramoto-Sivashinsky equation discussed
in section \ref{sectionks} (middle) and an extension of it, as  in
Fig.~\ref{figks} ,  but with $c=0.17$ (bottom).
   }\label{F1}
\end{figure}

The fact that the linear spreading velocity is the organizing
principle for the problem of front propagation is illustrated in
Fig.~\ref{F1} for all three classes of fronts,  simple {\em uniformly translating fronts},
 and {\em coherent} and {\em incoherent pattern forming fronts}.
 In the upper panels, we show 
simulations of the spreading of an initial  perturbation into
the unstable state according to the linear equations, obtained by linearizing the model
equation about the unstable state. This illustrates the linear
spreading problem associated with  the linear dynamics. The asymptotic
linear spreading speed  $v^*$ can be calculated explicitly for any given dynamical equation.
Note that since the dynamical equations have been linearized, there is
no saturation: The dynamical fields
in the upper panels continue to grow and grow (in the  plots in the middle and on the right, 
the field values also grow to negative
values, but this is masked in such a hidden-line plot).  The middle panels show
examples of  pulled fronts: These are seen to advance asymptotically with the {\em same}  speed 
$v^*$ as
the linear spreading problem of the upper panel. The lower panels illustrate pushed fronts, whose
asymptotic speed is larger than the linear spreading speed $v^*$.  The fronts in the left
column are {\em uniformly translating}, those in the middle column are
{\em coherent pattern forming fronts}, and
those in the right {\em incoherent pattern forming fronts}.  We
will define these front classifications more precisely later in section \ref{sectiontypesoffronts} --- for
now it suffices to become aware of the remarkable fact that in spite of the
difference in appearance and structure of these fronts, it is useful  to divide fronts
into two classes, those which propagate with asymptotic speed $v^*$
and those whose asymptotic speed $v^\dagger$ is larger. 
Explaining and exploring the origin and ramifications of this basic fact is one
of the main goals of this article.

In line with our philosophy to convey 
the power of  this simple concept, we will first only present the essential
ingredients that we think a typical non-expert reader should know, and
then  discuss a large variety  of experimental and theoretical
examples of front propagation that can indeed be understood to a large
extent with the  amount of theoretical baggage that we equip the reader
 with in chapter \ref{secoverview}. Only then will we turn to a more
detailed exposition of some of the more technical issues underlying
the presentation of chapter \ref{secoverview}, and to a number of
advanced topics. Nevertheless, throughout the
paper our philosophy will be to focus on the essential ideas and to
refer for the details to the literature --- we will try not to mask
the common and unifying features with too many details and special
cases,  even though making some caveats along the lines will be
unavoidable.  In fact,  even in these later chapters, 
 we will see that the above simple insight is the
main idea that also brings together various important recent theoretical developments:
  the derivation of an exact results for the
universal power law convergence of pulled fronts to their asymptotic
speed,  the realization that many of these results  extend  without
major modification to fronts in difference equations or fronts with
temporal or spatial kernels, the realization that curved pulled fronts in more than one
dimension can not be described by a moving boundary approximation or
effective interface description, as well as 
the effects of a particle cutoff on fronts, and the effects of
fluctuations.

A word about referencing: when referring to several papers in one
citation, we will do so in the numerical order imposed by the alphabetic
reference list, not in order of importance of the references. If we
want to distinguish papers, we will reference them separately.

\subsection{Motivation: a personal historical perspective}\label{perspective}

My choice to present the theory this way is admittedly very personal and
unconventional,  but is made deliberately. The theory of front
propagation has had a long and twisted but interesting history, with essential
contributions coming from different directions. I feel it is time
to take stock.  The field started essentially some 65 years ago\footnote{As mentioned by
Murray \cite{murray} on
  page 277, the Fisher equation was apparently already considered in 1906 by
  Luther, who obtained the same analytical form as Fisher for the
  wave front.}
with the work of Fisher \cite{fisher} and Kolmogoroff, Petrovsky, and
Piscounoff \cite{kpp}  on fronts in nonlinear diffusion type equations motivated by
population dynamics issues. The subject  seems to
have remained mainly in mathematics initially, culminating in the 
classic work of Aronson and Weinberger \cite{aw1,aw2} which contains a
rather complete set of results for the nonlinear diffusion equation (a
diffusion equation for a single variable with a nonlinear growth
term, Eq.~(\ref{fkpp}) below). The special feature of the nonlinear diffusion equation that
makes most of the rigorous work on this equation possible
is the existence of a so-called comparison theorem, which allows one
to {\em bound} the actual solution of the nonlinear diffusion equation by 
suitably chosen simpler ones with known properties. Such an approach
is mathematically  powerful, but  is essentially limited by its nature to  the nonlinear
diffusion equation and its extensions:  A  comparison theorem basically only holds
for the nonlinear diffusion equation or variants thereof, not for the
typical types  of equations that we encounter in practice and that
exhibit front  propagation into an unstable state in a pattern forming
system.

In the early eighties of the last century, the problem of front
propagation was brought to the attention of physicists by Langer and
coworkers \cite{bj,dl,langermk}, who noted that
there are some similarities between what we will call the regime of
pulled front propagation and the  (then popular) conjecture that the
natural operating point of dendritic growth was the  ``marginally
stable'' front solution \cite{langerrmp,langermk}, i.e., the
particular front solution for which the least stable stability
eigenmode changes from stable to unstable (for dendrites, this
conjecture was later abandoned). In addition, they re-interpreted the
two modes of operation\footnote{Their  ``case I'' and ``case II'' \cite{bj} are examples of what we
refer to here as pulled and pushed front solutions.} of front dynamics in terms of the stability of
front solutions \cite{bj}. This point of
view also brought to the foreground the idea that front propagation
into unstable states should be thought of as an example of {\em pattern
selection}: since there  generally exists a continuum of front
solutions,  the question becomes which one of these is 
``selected'' dynamically for a large class of initial conditions.
For this reason, much of the work in the physics community  following 
this observation was focused on understanding this apparent
connection between the stability of front profiles and the dynamical
selection mechanism 
\cite{paquette3,paquette1,paquette2,powell,shraiman,vs1,vs2}. 
Also in my own work along these lines
\cite{vs1,vs2} I pushed  various of the arguments for the connection between
stability and selection. This line of approach showed indeed that  the two 
regimes of front propagation that were already apparent from the work
on the nonlinear diffusion equation do in fact have their counterparts for
pattern forming fronts, fronts which leave a well-defined
finite-wavelength pattern behind. In addition, it showed that  the
power law convergence to the asymptotic speed
known for the nonlinear diffusion equation \cite{bramson} is just a
specific example of a generic property of fronts in  the  
``linear marginal stability'' \cite{vs1,vs2} regime  --- the ``pulled'' regime as we
will call it here.
Nevertheless, although some of these arguments have
actually made it into a review \cite{ch}  and into textbooks
\cite{gc,nishiura}, they remain at best 
a plausible set of arguments, not a real  theoretical framework; this is
illustrated by the fact that  the term ``marginal stability
conjecture'' is still often used in the literature, especially when
the author seems to want to  underline
its somewhat mysterious character.

Quite naturally, the starting point of the above line of research was
the {\em nonlinear}  evolution of fronts solutions. From this perspective 
it is understandable that many researchers were intrigued but 
also surprised to see that in the pulled (or linear
marginal stability) front regime  almost all the essential properties of the
fronts are determined by the dispersion relation of the {\em
  linearized } dynamics of arbitrarily small
perturbations about the unstable state. Perhaps this also explains, on 
hindsight, why for over 30 years there was a {\em virtually independent}
line of research that originated in plasma physics and fluid
dynamics. In these fields, it is very common that even though a system 
is linearly unstable (in other words,   that when linearized about a homogeneous
state, there is a continuous range of unstable Fourier modes), it is
only {\em convectively
unstable}. As mentioned before, this means that in the relevant frame
a localized perturbation is convected away faster than it is growing
out. To determine whether a system is either convectively unstable or
absolutely unstable mathematically translates into studying the
long-time asymptotics of the Green's function of the dynamical
equations, 
linearized about the unstable state.\footnote{Some readers may be
  amused to note that there are traces of such arguments in the
  original paper of Kolmogorov {\em et al.} \cite{kpp}: although
hidden, the Green's
  function of the diffusion equation plays an important role in their
  convergence proofs. This makes me believe that it is likely that
  it will be possible to prove results concerning front propagation
  into unstable states for more complicated equations like higher
  order partial differential equations, by putting bounds on the
  Green's function.} The technique to do so was
developed in the 1950-ies \cite{briggs} and  is even treated in one of the
volumes of the Landau and Lifshitz course on theoretical physics
\cite{ll}, but appears to have gone unnoticed in the mathematics literature.
It usually goes by the name of ``pinch point
analysis'' \cite{bers,huerre1,huerre2}. As we will discuss, for simple systems it amounts to a
saddle point analysis of the asymptotics of the Green's function. 
In 1989 I pointed out \cite{vs2} that the equations for the linear
spreading velocity of   perturbations, according to this analysis,  the velocity we will
refer to as $v^*$, are actually the {\em same} as the expressions for
the speed in the ``linear marginal stability'' regime of nonlinear
front propagation \cite{bj,dl,vs2}. Clearly, this can not be a
coincidence,  but the general implications of this observation appear not to have
been explored for several more years. One immediate simple but powerful consequence of this
connection is that it shows that the concept of
the linear spreading velocity $v^*$ applies equally well to   difference equations in
space and time --- after all  in Fourier language, in which the asymptotic
analysis of the Green's  function analysis is most easily done,
putting a system on a lattice just means that the Fourier integrals
are restricted to a finite range (a physicist would say: restricted to
the Brillouin  zone). The concept of linear spreading velocity also
allows one to connect front propagation with  work in 
recent years on the concept of global modes in weakly inhomogeneous
unstable systems \cite{couairon1,couairon2}.
 
Most of the work summarized above was confined to fronts in one
dimension. The natural  approach to analyze nontrivial patterns in
more dimensions on 
scales much larger than the typical front width is, of course, to view
the front on the large pattern scale as a sharp moving interface ---
in technical terms, this means that one would like to apply singular
perturbation theory to derive a moving
boundary approximation or an effective interface approximation (much
like what is often done for the so-called phase-field models that have
recently become popular \cite{batesfife,caginalp,karma}). When this
was attempted for discharge patterns whose dynamics is governed by
``pulled'' fronts \cite{streamers1,streamers2}, the standard
derivation of a moving boundary approximation was found to break
down. Mathematically, this traced back to the fact that
for pulled fronts the dynamically important region is {\em ahead} of
the nonlinear transition zone which one normally associates with the
front itself. This was another important sign that {\em one really has to
take the dynamics in the region ahead of the front seriously!}

My view that  the linear spreading velocity is the proper starting point for
understanding  the two regimes of front propagation into unstable
states, and  for tying 
together the various theoretical developments and experiments --- and  hence that 
an  introductory review  should be  organized around it ---  is
colored by   the
developments sketched above and in 
particular  the fact that  Ebert and I have recently been able
to derive from it important  {\em new and exact}  results for the power law
convergence of the velocity and shape of a pulled front to its
asymptotic value \cite{evs1,evs2}. The fact that starting from this
concept one can set up a fully explicit calculational scheme to study
the long time power law convergence or relaxation and that this
yields new universal terms which are exact (and which even
for the nonlinear diffusion equation \cite{aw1,aw2}  go beyond
those which were  previously known \cite{bramson}),
shows more than anything else that we have moved from the stage of
speculations and  intuitive concepts  to 
what has essentially become a well-defined and powerful theoretical
framework. My whole presentation builds on the picture coming out of
this work \cite{evs1,evs2,evs3,evsp,esvs}.

As mentioned before, the subject of front propagation also has a long
history in the mathematics literature; moreover,  especially in the last ten
to fifteen years a lot of work has been done on coherent structure
type solutions like traveling fronts, pulses etc. With such a diverse
field, spread throughout many disciplines, one can not hope to do
justice to all these developments. My choice to approach the subject
from the point of view of a physicist just reflects  that I only feel
competent to review the developments in this part of the field, and that I do want
to open up the many advances
that have been made recently  to
researchers with different backgrounds who typically will not scan the
physics literature. I will  try to give a 
fair assessment of some of the more mathematical developments  but
there is absolutely no claim to be exhaustive in that regard. Luckily, authoritative
reviews of the more recent mathematical literature are available
\cite{freidlin,freidlin3,volpert,xin}.
The  second reason for my choice is indeed that most of the mathematics
literature is focused on equations that admit uniformly translating
front solutions. For
many pattern forming systems, the relevant front solutions are not of this
type, they are either coherent or incoherent pattern forming fronts of
the type we already encountered in Fig.~\ref{F1}
(these concepts are defined precisely in section
\ref{sectiontypesoffronts}).
Even though not much is known rigorously about these more
general pattern forming fronts, our presentation will allow us to
approach all types of fronts in a unified way that  illuminates what
is and is not known. We hope this will also stimulate  the more
mathematically inclined reader to take up the challenge of entering
an area where we do know most answers but lack almost any proof. I am
convinced a gold mine is waiting for those who dare.

As explained above, we will first  introduce in chapter
\ref{secoverview}   the key ingredients of front propagation into unstable states
that provide the basic  working knowledge for the non-expert
physicist. The introduction along this line  also allows us to
identify most clearly the open problems. 
We then turn right away to a discussion of a large number
of  examples of  
front propagation. After this, we will give  a more detailed discussion
of the slow convergence of pulled fronts to their asymptotic velocity
and shape. We are then in a position to discuss  what
patterns, whose dynamics is dominated by  fronts propagating into an
unstable state, can be analyzed in terms of a moving boundary
approximation, in the limit that the front is thin compared to the
pattern scale.  This is followed by a discussion of  the relation 
with the existence of ``global modes'' and of some of the
issues  related to stochastic fronts.

\section{Front Propagation into Unstable States: the basics}\label{secoverview}

The central theme of this paper is to study fronts  in  spatio-temporally
extended systems  which propagate into a linearly unstable
state. The special character and, to a large extent, simplicity of
such fronts arises from the fact that  arbitrarily small perturbations about the unstable
state already grow and spread by themselves, and thus have a
nontrivial --- and, as we shall see,  rather universal --- dynamics by
themselves.  The ``linear spreading speed'' $v^*$ with which small
perturbations spread out is then automatically an important reference
point. This is different from  fronts which separate two 
linearly stable states --- in that case the perturbations about each
individual stable state just damp out and there is not much to be
gained from studying precisely how this happens; instead, the motion of such
fronts is inherently nonlinear.\footnote{Technically, determining the
asymptotic fronts speed then usually amounts to a nonlinear eigenvalue
problem. The spreading of the precursors of such fronts is studied in \cite{kessler}.}

It will often be instructive to illustrate our analysis and arguments
with a simple explicit  example; to this end we will use the famous nonlinear
diffusion equation with which  the field started,
\begin{equation}
\partial_t u (x,t) = \partial_x^2 u(x,t) + f(u)
,\hspace*{0.8cm}\mbox{with}~ \begin{array}{l} f(0)=0, ~f(1)=0, \\ f^\prime
(0)=1, ~f^\prime(1) <0. \end{array}\label{fkpp}
\end{equation}
This is the equation studied by Fisher \cite{fisher} and Kolmogorov,
Petrovsky and Piscounov
 \cite{kpp} back in 1937, and we shall therefore follow the
convention to refer to it as the F-KPP equation. As we mentioned already in the introduction,
this equation and its extensions have been the main focus of
(rigorous)  mathematical studies of front propagation into unstable
states, but these are not the main focus of this review ---  rather, we
will use the F-KPP equation only as the simplest equation to
{\em illustrate} the points which are generic to the front propagation
problem, and will not rely on  comparison-type methods or bounds
which are special to this equation.\footnote{A physicist's
introduction to comparison type arguments can be found in the appendix
of \cite{bj}; for recent work on bounds on the asymptotic speed of
fronts in the F-KPP equation see  in particular
\cite{depassier0,depassier4,depassier1,depassier3,depassier2}. A new look at the  
the global variational  structure was recently given in \cite{muratov}. A
recent example of a convergence proof in coupled equations can be
found in \cite{matano}.   } 
At this point it simply suffices to note
that  the  state $u =0$ of the real field $u$ is 
 an unstable state: when $u$ is positive but small, $f(u)\approx
 f^\prime(0) u= u $, 
so  the second term on the right hand side of the F-KPP equation drives $u$  away 
from zero. The front propagation problem we are interested in was
already illustrated in Fig.~\ref{F1}: We want  to determine the long time
asymptotic behavior of the front which propagates to the right into the unstable state $u=0$, given 
initial conditions for which $u(x\to \infty,t=0) = 0$.  A simple
analysis based on constructing the uniformly translating front
solutions $u(x-vt)$ does not suffice, as there is a continuous family
of such front solutions. Since the argument can be found at many places
in the literature \cite{aw1,bj,ch,evs2,langerma,logan,vs2,vsalt}, we will not repeat it here.

\subsection{The linear dynamics: the linear spreading speed $v^*$}\label{sectionv*}
Our approach to the problem via the introduction of the linear
spreading speed $v^*$ is a slight reformulation of the analysis
developed over 40 years ago in plasma physics
\cite{bers,briggs,ll}. We first formulate the essential concept
having in mind a simple partial differential equation or a set of
partial differential equations, and then briefly discuss the
minor complications that more general classes of dynamical equations
entail. We postpone the discussion of fronts in higher dimensions to
section \ref{sectionmba}, so we limit the discussion here to fronts in one dimension.

Suppose we have a dynamical problem for some field, which we will
generically denote by $\phi(x,t)$, whose
dynamical equation is translation invariant and has a homogeneous
stationary state $\phi=0$ which is {\em linearly 
unstable}.  With this we mean that if we linearize the dynamical
equation in $\phi$  about the unstable state,  
 then Fourier
modes grow  for some
range of spatial wavenumbers $k$.  More concretely, if we take a
spatial Fourier transform and write
\begin{equation}
\tilde{\phi}(k,t) = \int_{-\infty}^{\infty} \ds x \,\phi(x,t) e^{-ikx} ,
\end{equation}
substitution of the  Ansatz
\begin{equation}
\tilde{\phi}(k,t) = \bar{\phi}(k) e^{-i\omega(k)t}  \label{phikevolution}
\end{equation}
yields the {\em dispersion relation} $\omega(k) $ of Fourier modes of the linearized equation. We will
discuss the situation in which the dispersion relation has more than
one branch of solutions later,  and for now 
assume that it  just has a single branch. Then the statement that the
state $\phi=0$ is linearly unstable simply means that 
\begin{equation}
\phi=0 ~\mbox{\em linearly unstable:} \hspace*{0.2cm}
\mbox{Im} \,\omega (k) > 0 \hspace*{0.2cm} \mbox{for some range of } ~k\mbox{-values}.\label{definitionunstable}
\end{equation}
At this stage, the particular equation we are studying is simply  
encoded in the dispersion relation  $ \omega(k)$.\footnote{The reader
who may prefer to see an example of a dispersion relation is invited
to check the dispersion relation (\ref{fourthorderdispersion}) for
Eq.~(\ref{fourthorderlinear}).}  This dispersion
relation can be quite general --- we will come back to the conditions
on $\omega(k)$ in section \ref{sectionmoregeneral}  below, and for now will simply assume
that $\omega(k)$ is an analytic
function of $k$ in the complex $k$-plane.

\begin{figure}[t]
\begin{center}
 \epsfig{figure=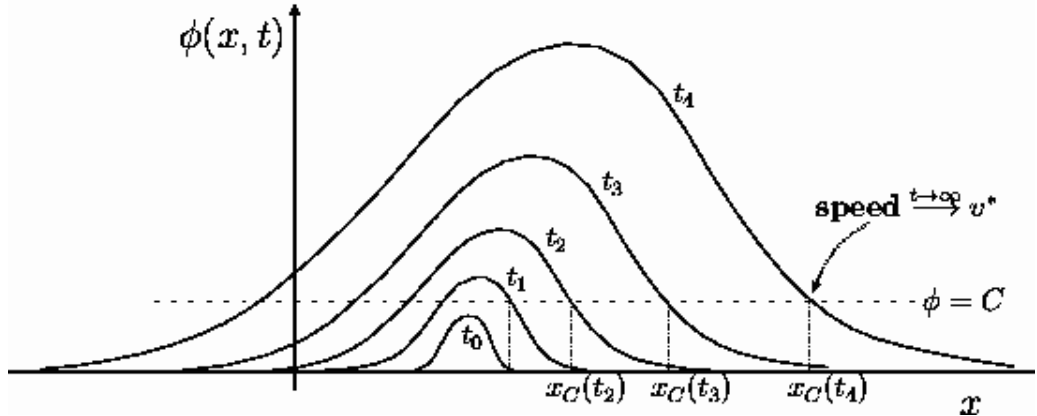,width=0.98\linewidth} 
\end{center}
\caption[]{Qualitative sketch of the growth and spreading of the field
$\phi(x,t)$ according to the dynamical evolution equation linearized
about the unstable state
 $\phi=0$. The successive curves illustrate the initial condition
$\phi(x,t_0)$ and the field $\phi(x,t)$ at successive times
$t_1<t_2<t_3<t_4$. Note that  there is obviously no saturation of the
field in the linearized dynamics: The asymptotic spreading velocity
$v^*$ to the right is the asymptotic speed of the positions $x_C(t)$ where
$\phi(x,t)$ reaches the level line $\phi=C$: $\phi(x_C(t),t)=C$. The
asymptotic spreading velocity to the left is defined analogously.
}\label{figspreading}
\end{figure}
We are interested in studying the long-time dynamics emerging from
some generic initial condition which is sufficiently localized in
space (we will  make the term ``sufficiently
localized''  more precise in section \ref{sectiontransients} below).
Because there is a range of unstable modes which grow exponentially in time
as $e^{{\rm Im}\omega (k) t}$, a typical localized initial condition will grow out and spread in time
under the linear
dynamics  as sketched in
Fig.~\ref{figspreading}.  If we now trace the level curve $x_C(t)$
where $\phi(x_C(t),t) =C$ in space-time, as indicated in the figure, the
{\em linear spreading speed} $v^*$ is  defined 
as the {\em asymptotic} speed  of the point $x_C(t)$:
\begin{equation}
v^* \equiv  \lim_{t\rightarrow \infty} \frac{\ds x_C(t)}{\ds t}.
\end{equation}
Note that $v^*$ is independent of the value of $C$ because of the
linearity of the evolution equation. However, for systems whose dynamical
equations are not reflection symmetric, as happens quite often in
fluid dynamics and plasmas, one does have to distinguish between a spreading speed to
the left and one to the right. In order not to overburden our
notation, {\em we will in this paper by convention always focus on the spreading velocity
 of the right flank} of $\phi$; this velocity is counted positive if this
flank spreads to the right, and negative when it recedes to the left. 

Given  $\omega(k)$ and $\bar{\phi}(k)$, which according to
(\ref{phikevolution}) is just the Fourier
transform of the initial condition $\phi(x,t=0)$, one can write
 $\phi(x,t)$ for $t>0$ simply as the inverse Fourier transform
\begin{equation}
\phi(x,t) = \frac{1}{2\pi} \int_{-\infty}^{\infty} \ds k \, \bar{\phi}(k)\,
e^{ikx -i\omega(k) t }. \label{finversex}
\end{equation}
The more general Green's function formulation will be discussed later
in section \ref{sectionmoregeneral}. 
Our definition of the linear spreading speed $v^*$ to the right is
illustrated in Fig.~\ref{figspreading}. We will work under the
assumption that the asymptotic spreading speed $v^*$ is finite;
whether this is true can always be verified self-consistently at the
end of the calculation. The existence of a finite $v^*$ implies that  if we look in  frame 
\begin{equation}
\xi = x - v^* t \label{xiframe}
\end{equation} moving with this speed, we neither see the right flank grow nor decay
exponentially. To determine $v^*$, we therefore first write the inverse
Fourier formula  (\ref{finversex})  for $\phi$ in this  frame,
\begin{eqnarray}
\phi(\xi,t) &= &\frac{1}{2\pi} \int_{-\infty}^{\infty} \ds k \, \bar{\phi}(k)\,
e^{ik(x-v^*t) -i[\omega(k) -v^* k]t} , \nonumber \\
&= &\frac{1}{2\pi} \int_{-\infty}^{\infty} \ds k \, \bar{\phi}(k)\,
e^{ik \xi  -i[\omega(k) -v^* k] t} ,
\label{finverse}
\end{eqnarray}
and then determine $v^*$ self-consistently by analyzing when this
expression neither leads to exponential growth nor to decay {\em in the
limit $\xi$ finite, $t \to \infty$}.  We can not simply 
evaluate the integral by closing the contour in the upper half of the $k$-plane, since the large-$k$ 
behavior of the exponent is normally dominated by the large-$k$ behavior of $\omega(k)$. 
However, the large-time  limit clearly calls for a
saddle-point approximation \cite{bender} (also known as stationary
phase or steepest descent approximation):  Since
$t$ becomes arbitrarily large, we  deform the $k$-contour to go through the
 point in the complex $k$ plane where the
term between square brackets varies least with $k$, and the integral is then dominated by the
 contribution  from the region near this  point. This so-called  saddle
point $k^*$ is given by
\begin{equation}
\left. \frac{\ds [\omega(k)-v^*k ] }{\ds k}\right|_{k^*} = 0
\hspace*{0.7cm} \Longrightarrow \hspace*{0.7cm}  v^* =
\left. \frac{\ds \omega(k)}{\ds k}\right|_{k^*}. \label{saddle1}
\end{equation}
These saddle point equations will in general have solutions in both the 
upper and the lower half of the complex $k$-plane; the  ones in the
upper half 
correspond to the  asymptotic decay towards large $x$ in the Fourier
decomposition (\ref{finverse}) associated with the right flank of the perturbation
sketched in Fig.~\ref{figspreading}, and those in the lower half ty to an exponentially growing
solution for increasing $x$ and thus to the behavior on the left
flank. By convention, we will focus on the right flank, which may invade
the unstable state to the right.
If we deform the $k$-contour into the complex plane to go through the
saddle point in the upper half plane, {\em and assume for the moment that $\bar{\phi}(k)$}, 
the Fourier transform of the initial condition, {\em  is an entire function} (one that is analytic
in every finite region of   the complex $k$-plane), the dominant term to the integral is nothing but the 
exponential factor  in (\ref{finverse})  evaluated  at the
saddle-point, i.e., $e^{i[\omega(k^*)-v^* k^*] t}$. The
self-consistency requirement that this term neither grows nor decays  exponentially
thus simply leads to
\begin{equation}
{\rm Im}\, \omega  (k^*) - v^* {\rm Im}\, k^* = 0  
\hspace*{0.5cm} \Longrightarrow \hspace*{0.5cm}  v^* =
\frac{{\rm Im}\, \omega   (k^* ) }{\,{\rm Im}\, k^*} =
\frac{\omega_{\rm i} }{k_{\rm i}}. \label{saddle2}
\end{equation}
The notation $\omega_{\rm i}$ which we have
introduced here for the imaginary part of $\omega$ will be used
interchangeably from now on with ${\rm Im }\,\omega$; likewise, we
will introduce the subindex r to denote the real part of a complex quantity.
Upon  expanding  the factor in the exponent in   (\ref{finverse}) around the saddle point
value given by Eqs.~(\ref{saddle1}) and (\ref{saddle2}),  we then get from the resulting Gaussian
integral
\begin{eqnarray}
\phi (\xi,t) &\simeq &\frac{1}{2\pi}  \int_{-\infty}^{\infty} \ds k \, \bar{\phi}(k)\,
e^{-i\omega_{\rm r}^*t + i(k^*+ \Delta k) \xi   - {\mathcal D}  t (\Delta k)^2 }  , \hspace*{1.6cm} (\Delta k=k-k^*), \nonumber\\
 &\simeq &\frac{1}{2\pi} \, e^{ik^*\xi -i\omega^*_{\rm r}t } \,
\int_{-\infty}^{\infty} \ds k \, \bar{\phi}(k)\,
e^{   - {\mathcal D}  t [ \Delta k - i\xi /2{\mathcal D}t ]^2 - \xi^2/4{\mathcal D}t}  , 
\nonumber    \\
 & \simeq &
 \frac{1}{\sqrt{4\pi {\mathcal D} t}}\, e^{ik^* \xi -i(\omega^*_{\rm r}-k_{\rm r}^*v^*)t }\, e^{-\xi^2/4{\mathcal D}t} \, \bar{\phi}(k^*)  , \hspace*{1.5cm} ( { \xi} ~ { \mbox{fixed}, t\to \infty }) ,\label{phibasic}
\end{eqnarray}
where all parameters are determined by the dispersion relation through the saddle point values,
 \begin{equation} 
\left. \frac{\ds \omega(k)}{\ds  k}\right|_{k^*} = \frac{\omega_{\rm i}  (k^*
) }{k_{\rm i}^*} ,
\hspace*{0.9cm}
 v^* =   \frac{\omega_{\rm i}  (k^* ) }{k_{\rm i}^*} , \hspace*{0.9cm}
{\mathcal D} = \left. \frac{i}{2} \frac {\ds ^2\omega(k)}
{\ds k^2}\right|_{k^*}  .\label{saddlepoint}
\end{equation} 
Since $\omega $ and $k$ are in general complex, the first equation can actually be thought of as
two  equations for the real and imaginary parts, which can be used to
solve for $k^*$. The second and third equation then give 
$v^*$ and ${\mathcal D}$.

The exponential factor $e^{ik^*\xi}$ gives the dominant spatial
behavior of $\phi$ in the co-moving frame $\xi$ on the right flank in
Fig.~\ref{figspreading}: if we define the 
asymptotic spatial decay rate $\lambda^*$ and the effective {\em diffusion coefficient}\footnote{We 
stress that $D$ is the effective diffusion coefficient associated with the saddle point governing
the linear spreading behavior of the {\em deterministic} equation.  In
section \ref{sectionstochasticfronts}  we will also encounter  a front
diffusion coefficient $D_{\rm front}$ which 
is a measure for  the stochastic front wandering, but this
is an entirely different quantity.}  $D$ by
\begin{equation}
\lambda^* \equiv {\rm Im}\, k^* , \label{lambda*} \hspace*{1.8cm} \frac{1}{D } \equiv 
{\rm Re}\, \frac{1}{\mathcal D}, 
\end{equation}
then we see that the modulus of $\phi$  falls off as
\begin{equation}
\left| \phi (\xi,t)\right|   \sim
 \frac{1}{\sqrt{t} }\, e^{- \lambda^* \xi}\,
e^{-\xi^2 /4D t}  , \hspace*{2.5cm} ( { \xi}
 ~ { \mbox{fixed}, t\to \infty }),  \label{phibasic2}
\end{equation}
i.e., essentially as $ e^{-\lambda^*\xi}$ with a Gaussian
 correction that is reminiscent of diffusion-like behavior. 

We will prefer not to  name  the point $k^*$ after the way it
arises mathematically
(e.g., saddle point or  ``pinch point'',  following  the formulation discussed  in section
\ref{sectionmoregeneral}). Instead, we will usually refer to $k^*$ as the {\em linear
spreading point}; likewise, the expressions (\ref{phibasic}) and
(\ref{phibasic2}) for $\phi$ will be referred to as the linear spreading profiles.  

For an ordinary diffusion process to be stable, the diffusion
coefficient has to be positive. Thus we expect that in the present
case $D$ should be positive. Indeed, the requirement that the linear spreading
point corresponds to a maximum of the exponential term in
(\ref{finverse}) translates into the condition,  $\mbox{Re}\,
{\mathcal D}>0$, and this entails $D>0$. We will come back to this and
other conditions  in section \ref{sectionmoregeneral} below.

In spite of the simplicity of their derivation and form, equations
(\ref{phibasic}) and (\ref{saddlepoint}) express  the crucial result 
that as we shall see permeates the field of front propagation into unstable states:

\begin{tabular}{p{0.1cm}|p{13.2cm}}
&  {\em  associated with any linearly unstable state is a
  linear spreading speed $v^*$ given by (\ref{saddlepoint}); this is
  the natural asymptotic  spreading speed with which small ``sufficiently localized''
  perturbations  spread into a domain of the unstable state according
  to the linearized dynamics.}
\end{tabular}

Before turning to the implications for front propagation, we will in
the next sections
discuss various aspects and generalizations of these concepts,
including  the precise condition under which 
``sufficiently localized'' initial conditions do lead to an asymptotic
spreading velocity $v^*$ (the so-called steep initial conditions given
in (\ref{steepincond}) below).

\subsubsection*{$\Box$ Example: application to the linear F-KPP equation}
Let us  close this section by applying the above formalism to the F-KPP equation (\ref{fkpp}).
Upon linearizing the equation in $u$, we obtain
\begin{equation}
\mbox{linearized F-KPP:}\hspace*{0.5cm} \partial_t u(x,t) =
\partial_x^2 u (x,t) + u. \label{linearfkpp}
\end{equation}
Substitution of a Fourier mode $e^{-i \omega t + ikx}$ gives the
dispersion relation
\begin{equation}
\mbox{ F-KPP:}\hspace*{0.5cm}\omega(k) = i(1 -k^2),\label{dispersionfkpp}
\end{equation}
and from this we immediately obtain from (\ref{saddlepoint}) and (\ref{lambda*})
\begin{equation}
\mbox{ F-KPP:}\hspace*{0.5cm} v_{\rm FKPP}^*=2, \hspace*{0.8cm} \lambda^*=1,
\hspace*{0.8cm} \mbox{Re}\,k^* =0, 
\hspace*{0.8cm} D = {\mathcal D}=1. \label{fkppvalues}
\end{equation}
The special simplicity of the F-KPP equation lies in the fact that $\omega(k)$ is quadratic in
$k$. This not only implies that  the effective diffusion coefficient
$D$ is nothing but the diffusion coefficient entering the F-KPP
equation, but also that the exponent in (\ref{finverse}) is in fact a Gaussian form without higher order
corrections. Thus, the above evaluation of the integral is actually exact in this case. Another 
instructive way to see this is to note that the transformation $u= e^t n$ transforms 
 the linearized F-KPP equation (\ref{linearfkpp}) into the diffusion
equation $\partial_t n= \partial_x^2 n$ for $n$. The fundamental
solution corresponding to delta-function initial condition is the
well-known Gaussian;  in terms of  $u$ this yields
\begin{equation}
\mbox{ F-KPP:}\hspace*{0.5cm} u (x,t) = \frac{1}{\sqrt{4\pi t}}
e^{t-x^2/4t} \hspace*{6mm}\mbox{(delta function initial cond.)}. \label{linearfkppsol}
\end{equation}

\subsection{The linear dynamics: characterization of exponential solutions}\label{sectionexponential}

In the above analysis, we focused immediately on the importance of
the linear spreading point $k^*$ of the dispersion relation $\omega(k)$ in determining
the spreading velocity $v^*$. Let us now pay more attention to the
precise initial conditions for which this concept is important.

In the derivation of the linear spreading velocity $v^*$, we took the
Fourier transform of the initial conditions to be an entire function,
i.e., a function which is analytic in any finite region of the complex
plane. Thus, the analysis applies to the case in which $\phi(x,t=0)$
is a delta function ($\bar{\phi}(k)$ is then
$k$-independent\footnote{Most of the original literature
  \cite{bers,briggs,huerre1,ll} in which the asymptotic large-time
  spreading behavior of a perturbation is obtained through a similar
  analysis or the more general ``pinch point'' analysis, is implicitly
  focused on this case of delta-function initial conditions, since
  the analysis is based on a large-time asymptotic analysis of the
  Greens function of the dynamical equations. Note in this connection
  that (\ref{linearfkppsol}) is indeed the Green's function solution
  of the linearized F-KPP equation. }), has compact support (meaning
that $\phi(x,t=0)=0 $ outside some finite interval of $x$), or falls
off faster than any exponential for large enough $x$ (like, e.g., a
Gaussian).

For all practical purposes, the only really relevant case in which
$\bar{\phi}(k)$ is not an entire function is when it has poles off the
real axis in the complex plane.\footnote{Of course, one may consider
  other examples of non-analytic behavior, such as power law
  singularities at $k=0$. This would correspond to a power law initial
  conditions $\phi(x,t=0) \sim x^{-\delta}$ as $x\to \infty$. Such
  initial conditions are so slowly decaying that they give an infinite
  spreading speed, as $\phi(x,t)\sim e^{{\rm
      Im}\,\omega(0)t}x^{-\delta}$. Also the full nonlinear front
  solutions have a divergent speed in this case \cite{needham3}. }
This corresponds to an initial condition $\phi( x,t=0)$ which falls
off exponentially for large $x$. To be concrete, let us consider the
case in which $\bar{\phi}(k)$ has a pole in the upper half plane at
$k=k^{\prime}$.  If we deform the $k$-integral to also go around this
pole, $\phi(x,t)$ also picks up a contribution whose modulus is
proportional to\footnote{We are admittedly somewhat cavalier here;
a more precise analysis of the crossover between the various
contributions is given in the next subsection  below Eq.~(\ref{phibasicnew}).}
\begin{equation}
\left| e^{-i\omega(k^{\prime}) t + ik^{\prime} x } \right|  = e^{-\lambda (x-
v(k^{\prime}) t) } ,  \hspace*{0.6cm}\mbox{with} \hspace*{0.6cm} \lambda \equiv {\rm Im}\, k^{\prime},   \label{phienv}
\end{equation}
and whose {\em envelope velocity} $v(k^{\prime}) $ is given by
\begin{equation}
v(k^{\prime}) = \frac{{\rm Im}\,\omega(k^{\prime})}{{\rm Im}\,k^{\prime}} .
\end{equation}
We first characterize these solutions in some detail, and then
investigate their relevance for the full dynamics.

Following \cite{evs2}, we will refer to the exponential decay rate
$\lambda$ of our dynamical field as the {\em steepness}.  For a {\em
  given} steepness $\lambda$, $\omega(k^\prime)$ of course still
depends on the real part of $k^\prime$. We choose to
introduce a {\em unique}  envelope velocity $v_{\rm env}(\lambda)$ by taking
for ${\rm Re}\,k^{\prime}$ the value that maximizes ${\rm Im}\,\omega$
and hence $v(k^{\prime})$,
\begin{equation}
v_{\rm env}(\lambda\equiv k^{\prime}_{\rm i} ) = \left. \frac{\omega_{\rm i} (k) }{k_{\rm i}
}\right|_{k=k^{\prime}}, \hspace*{0.6cm}\mbox{with} \hspace*{0.6cm} \left. \frac { \partial  \omega_{\rm i} (k)}{\partial k_{\rm r}}\right|_{k=k^{\prime}} 
= \left. {\rm Im} \frac{\ds \omega}{\ds  k}\right|_{k=k^{\prime}} 
  = 0, \label{venv}
\end{equation}
where the second condition determines $k_{\rm r}$ implicitly as a
function of $\lambda= k^{\prime}_{\rm i}$.  The rationale to focus on this
particular velocity as a function of $\lambda$ is twofold: First of
all, if we consider for the fully linear problem under investigation
here an initial condition whose modulus falls of as $e^{-\lambda x}$
but in whose spectral decomposition a whole range of values of $k_{\rm
  r}$ are present, this maximal growth value will dominate the large
time dynamics. Secondly, in line  with this, when we consider
nonlinear front solutions corresponding to different values of
$k_{\rm r}$, the one not corresponding to the maximum of $\omega_{\rm
i}$ are unstable and therefore dynamically irrelevant --- see section
\ref{stabilitygeneral}. Thus, for all practical purposes the branch of
velocities $v_{\rm env}(\lambda)$ is the real important one. 

\begin{figure}[t]
\begin{center}
  \epsfig{figure=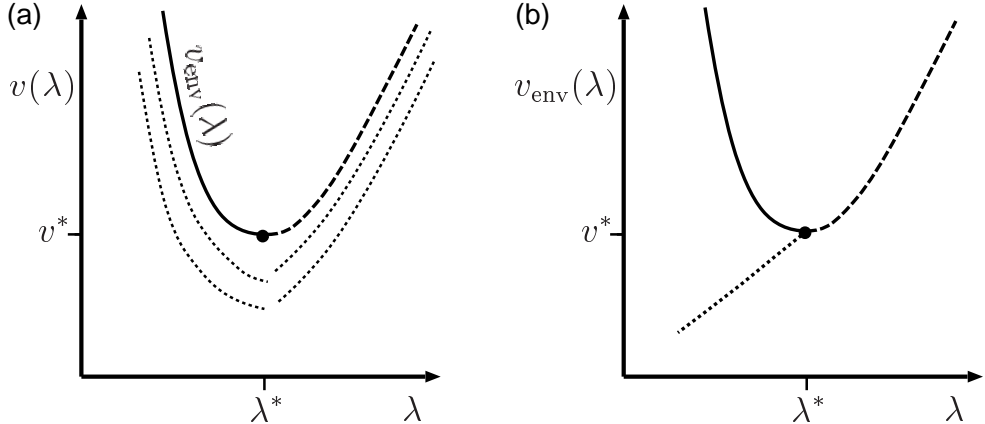,width=0.98\linewidth}
\end{center}
\caption[]{(a) Generic behavior of the  velocity $v(\lambda) $ as a function of
  the spatial decay rate $\lambda$. The thick full line and the thick
dashed line indicate the envelope velocity defined in (\ref{venv}):
for a given $\lambda$ this corresponds to the largest value of
$\omega_{\rm i}$ and hence to the largest velocity on these branches.
The minimum of  $v_{\rm env}$ 
  is equal to the linear spreading speed $v^*$. (b) The situation in
the special case of uniformly translating solutions which obey $\omega/k = v$. The dotted line marks
  the branch of solutions with velocity less than $v^*$ given in (\ref{velocityexpansion}). 
  }\label{figvversuslambda}
\end{figure}

The generic behavior of $v_{\rm env}(\lambda)$ as a function of
$\lambda$ is sketched in Fig.~\ref{figvversuslambda}{\em (a)}.  In
this figure, the dotted lines indicate branches not corresponding to
 the envelope velocity given by (\ref{venv}): For a given value
of $\lambda$, the other branches correspond to a smaller value of
$\omega_{\rm i}$ and hence to a smaller value of
$v(\lambda)$. Furthermore, since we are
considering the spreading and propagation dynamics at a linearly
unstable state, the maximal growth rate $\omega_{\rm i}(\lambda) >0$
as $\lambda \downarrow 0$. Hence $v_{\rm env}(\lambda) $ diverges as
$1/\lambda$ for $\lambda\to 0$. When we follow this branch for
increasing values of $\lambda$, at some point this branch of solutions
will have a minimum. This minimum is nothing but the value $v^*$:
Since along this branch of solutions $\partial \omega_{\rm i}/\partial
k_{\rm r}=0$, we simply have
\begin{equation}
\frac{\ds  v_{\rm env}}{\ds \lambda} =   \frac{1}{\lambda} \left(
  \frac{\partial \omega_{\rm i}}{\partial \lambda } + \frac{\partial
    \omega_i}{\partial k_r} \frac{\ds  k_r}{\ds  \lambda } -
  \frac{\omega_{\rm i}}{\lambda} \right)  = 
 \frac{1}{\lambda} \left(
  \frac{\partial \omega_{\rm i}}{\partial \lambda }  -
  \frac{\omega_{\rm i}}{\lambda} \right)  ,
\end{equation}
and so at the linear spreading point $k^*$
\begin{equation}
\left. \frac{\ds  v_{\rm env}}{\ds \lambda}\right|_{k^*}  = \frac{1}{\lambda^*} \left(
  \left. \frac{\partial \omega_{\rm i}}{\partial \lambda
      }\right|_{k^*} -
  \frac{\omega^*_{\rm i}}{\lambda^*} \right) =0, 
\end{equation}
since at the point $k^*$ the term between brackets precisely vanishes,
see Eq.~(\ref{saddlepoint}).  By differentiating once more, we see
that the curvature of $v_{\rm env}(\lambda)$ at the minimum can be related 
 to  the effective diffusion coefficient\footnote{Aside for the reader
familiar with amplitude equations   \cite{ch,fauve,gc}: The relation
between the curvature of $v_{\rm env} (\lambda)$ at the
minimum and the diffusion coefficient $D$ bears some intriguing
similarities to the   relation between the  curvature
of the growth rate as a function of $k$ of the pattern forming mode
near the bifurcation to a finite-wavelength pattern and the parameters
in the amplitude equation \cite{ch,gc,hakim,newell2,nishiura}.  That curvature
is also essentially the diffusion constant that enters the amplitude
equation. Nevertheless, one should keep in mind that the minimum of
$v_{\rm env}(\lambda)$ is associated with the saddle point of
an invasion mode which falls off in space, {\em not} with the maximum
growth rate of a Fourier mode. Moreover, while the amplitude equations
only describe pattern formation near the instability threshold, the
pulled front propagation mechanism can be operative far above an
instability threshold as well as in pattern forming problems which
have no obvious threshold.}
$D$ 
introduced in (\ref{lambda*}),
\begin{eqnarray}
\left. \frac{\ds^2 v_{\rm env}(\lambda)} {\ds \lambda^2}   \right|_{\lambda^*}  & = &  
\frac{1}{\lambda^*}    \left[
\left. \frac{\partial^2 \omega_{\rm i}}   {\partial \lambda^2}  \right|_{k^*}  +  
2  \left.    \frac{\partial^2 \omega_{\rm i}}{\partial \lambda \partial k_{\rm 
    r}  }\right|_{k^*} 
       \left. \frac{\ds  k_{\rm r}} {\ds  \lambda} \right|_{\rm k^*}  
 + \left. \frac{\partial^2 \omega_{\rm i}} {\partial  k^2_{\rm   r}  }\right|_{k^*} 
\left(  \frac{\ds  k_{\rm r}} {\ds  \lambda} \right)^2_{\rm k^*}  \right]
\nonumber\\
 & = & \frac{2}{\lambda^*} \left[ {\mathcal D}_{\rm r} + 2 {\mathcal
     D}_{\rm i}\left( \frac{{\mathcal D}_{\rm i}} {{\mathcal D}_{\rm r}} \right) -
     {\mathcal D}_{\rm r} \left(
 \frac{{\mathcal D}_{\rm i}} {{\mathcal D}_{\rm r}} \right)^2 \right] 
\nonumber = \frac{2}{\lambda^*} \left[ {\mathcal D}_{\rm r} +  \frac{{\mathcal D}_{\rm i}^2}{{\mathcal D}_{\rm r}} \right] ,
\\
& = & 
 \frac{2 D}{\lambda^*} ,   \label{Dandveq}
\end{eqnarray} 
where ${\mathcal D}$ was defined in (\ref{saddlepoint}) and where we used the fact that according
to the definition (\ref{lambda*}) of $D$, we can write $D= {\mathcal D}_{\rm r}+ {\mathcal D}_{\rm i}^2 / {\mathcal D}_{\rm r}$.
Furthermore, in deriving these results, we have repeatedly used the Cauchy-Riemann
relations  for complex analytic functions that relate the various derivatives of
the real and imaginary part, and  the fact that along the branch
of solutions $v_{\rm env}$, the relation $\partial \omega_{\rm
i}/\partial k_{\rm r}=0$ implies ${\mathcal D}_{\rm i } - {\mathcal
D}_{\rm r} (\ds k_{\rm r}/\ds \lambda)= 0$.

If we investigate  a dynamical equation  which admits a uniformly translating front
solution of the form $\phi(x-vt)$, the  previous results need to be consistent which this
special type of asymptotic behavior.  Now, the exponential
leading edge 
behavior  $e^{ikx-i\omega t}$ we found above only corresponds to uniformly translating
behavior provided 
\begin{equation}
\mbox{uniformly translating solutions:}~~~v(\lambda) =
\frac{\omega(k)}{k} , ~~~~~(\lambda=k_{\rm i} ). \label{uniform1}
\end{equation}
The real part of this equation is consistent with the earlier condition $v=\omega_{\rm i}/k_{\rm i}$
that holds for all fronts, but for uniformly translating fronts it implies that
 in addition  $\mbox{Im}\, (\omega/k)=0$.

Hence, the above discussion is only self-consistent for uniformly
translating solutions if the branch $v_{\rm env}(\lambda)$ 
where the growth rate $\omega_{\rm i}$ is maximal for a given
$\lambda$
coincides with the condition  (\ref{uniform1}). In all the cases that I know of,\footnote{As
we shall see in section \ref{sectionefk}, the EFK equation illustrates that when the
linear spreading point ceases to obey (\ref{uniform2}), the pulled
fronts change from uniformly translating to coherent pattern forming solutions.}
the branch of 
envelope solutions for $v> v^*$ coincides with uniformly translating solutions
because the  dispersion relation is 
such that  the growth rate
$\omega_{\rm i}$ is maximized for $k_{\rm r}=0$: 
\begin{equation}
\mbox{uniformly translating solutions with $v>v^*$:}~~~ k_{\rm r} = \omega_{\rm r}
=0, ~~~{\mathcal D}_{\rm i}=0. \label{uniform2}
\end{equation}
Obviously, in this case the branch $v_{\rm env}(\lambda)$
 corresponds to the simple exponential behavior $\exp({-\lambda x + \omega_{\rm i} t})$
which is neither temporally nor spatially oscillatory.\footnote{For
  uniformly translating fronts, it would be more appropriate to use in
  the case of uniformly translating fronts the usual Laplace transform
  variables $s=-i\omega$ and $\lambda=-ik$ as these then take
  real values.  We will refrain from doing so since most of the
  literature on the asymptotic analysis of the Green's function on
  which the distinction between convectively and absolutely unstable
  states is built, employs the $\omega$-$k$ convention.}

We had already seen that  there generally are also
solutions with velocity $v<v^*$, as the branches with velocity
$v_{\rm env}>v^*$ shown in Fig.~\ref{figvversuslambda}{\em (a)} are only those
corresponding to the maximum growth condition $\partial \omega_{\rm
  i}/\partial k_{\rm r}=0$, see Eq.~(\ref{venv}).  It is important to
realize that if an equation admits uniformly translating solutions,
there is in general also a branch of uniformly translating solutions
with $v<v^*$. Indeed, by expanding
the curve $v_{\rm env}(\lambda )$ around the minimum $v^*$ and looking for
solutions with $v<v^*$, one finds that these are given
by\footnote{Note that the formula given on page 53 of \cite{evs2} is
  slightly in error.}
\begin{equation}
\lambda-\lambda^* \approx \frac{ v^{\prime \prime \prime}}{3 (v^{\prime\prime})^2} (v-v^*),~~~ ~~k_{\rm r}-k^*_{\rm r} \approx \sqrt{2 
|v-v^*|/ 
v^{\prime\prime}} ~~~ (v<v^*). \label{velocityexpansion}
\end{equation} 
The situation in the special case of uniformly translating solutions
is sketched in Fig.~\ref{figvversuslambda}{\em (b)}; in this figure,
the dotted line shows the branch of solutions with $v< v^*$. 
  Since solutions for $v<v^*$ are always
spatially oscillatory ($k_{\rm r}\neq 0$), they are sometimes disregarded in the analysis
of fronts for which the dynamical variable, e.g. a particle density,
is by definition non-negative. It is important to keep in mind,
however, that they do actually have relevance as intermediate
asymptotic solutions during the transient dynamics: as we shall see
in section \ref{sectionunirelsimple}, the asymptotic velocity $v^*$ is always
approached slowly from below, and as a result the transient dynamics
follows front solutions with $v<v^*$ adiabatically. Secondly, this
branch of solutions also pops up in the analysis of fronts in the case
that there is a small cutoff in the growth function --- see section \ref{finiteparticles}.

The importance of this simple connection between the minimum of the
curve $v_{\rm env}(\lambda)$ and the linear spreading speed $v^*$ can
hardly be overemphasized:

\begin{tabular}{p{0.1cm}|p{13.2cm}}
& {\em For equations of F-KPP type, the special
significance of the minimum of the $v_{\rm env}(\lambda) $ branch as
the selected asymptotic velocity in the pulled regime is well known,
and it plays a crucial role in more rigorous comparison-type arguments
for front selection in such types of equations. The line of argument
that we follow here emphasizes that $v^*$ is the asymptotic speed that
naturally arises from the linearized dynamical problem, and that this
is the proper starting point both to understand the selection problem,
and to analyze the rate of convergence to $v^*$ quantitatively.}
\end{tabular}

\subsubsection*{$\Box$ Example: application to the linear F-KPP equation}
We already gave the dispersion relation of the F-KPP equation in  (\ref{dispersionfkpp}); 
using this in Eq.~(\ref{venv}) immediately gives for the upper
branches with $v_{\rm env} \ge v^*=2$
\begin{equation}
\mbox{ F-KPP:}\hspace*{0.5cm}  \lambda = \frac{v_{\rm env}\pm
\sqrt{v_{\rm env}^2-4}}{2} ~~~~\Longleftrightarrow~~~~ v_{\rm env}= \lambda+
\lambda^{-1} , \label{vforfkpp}
\end{equation}
and for the branches below $v^*$
\begin{equation}
\mbox{ F-KPP:}\hspace*{0.5cm}  \lambda = v/2,~~~~ k_{\rm r} = \pm  \frac{1}{2}
\sqrt{4- v^2} ~~~~(v  < v^*=2),
\end{equation}
in agreement with the above discussion and with (\ref{velocityexpansion}).

\subsection{The linear dynamics: importance of initial conditions and  transients}\label{sectiontransients}

 We now study the dependence on initial conditions and the transient
behavior.  This question is obviously relevant: The discussion in the
previous section shows that simple 
exponentially decaying solutions can propagate faster than
$v^*$ --- at first sight, one might wonder how a  profile spreading with
velocity $v^*$ can ever
emerge from the dynamics if solutions exist which tend to propagate
faster.  Moreover, as we shall see, initial conditions which fall with
an exponential decay rate $\lambda< \lambda^*$ do give rise to
a propagation  speed $v_{\rm env} (\lambda)$ which is larger than $v^*$.

If the initial condition is a delta function, or, more generally,
if the initial condition has compact support (i.e. vanishes
identically outside some finite range of $x$), then the Fourier
transform $\bar{\phi}(k)$ is an entire function. This means that
$\bar{\phi}(k)$ is analytic everywhere in the complex $k$-plane. The
earlier analysis shows that whatever the precise initial conditions,
the asymptotic spreading speed is always simply $v^*$, determined by
the saddle point of the exponential term.

The only relevant initial conditions which can give rise to spreading
with a constant finite speed are the exponential initial conditions
already discussed in some detail in the previous subsection. Let us assume that
$\bar{\phi}(k)$ has a pole in the complex $k$-plane at $k^\prime$,
with spatial decay rate $k^\prime_{\rm i} = \lambda$.
In our first round of the discussion, we analyzed the limit $\xi $
fixed, $t \to \infty$, but it is important to keep in mind that  the limits $\xi $ fixed, $t \to \infty$ and $t $
fixed, $\xi \to \infty$ do not commute. Indeed, it follows directly from the inverse
Fourier formula that the spatial asymptotic behavior as $x \to
\infty$ is  the same as that of the initial conditions,\footnote{For 
  the F-KPP equation, this is discussed in more detail in section 2.5
  of \cite{evs2}, where this property is referred to as ``conservation
  of steepness''. }
\begin{equation}
\phi(x\to \infty ,t=0) \sim e^{-\lambda x}
~~~~\Longrightarrow ~~~~~\phi(x \to \infty,t) \sim e^{-\lambda x} .
\end{equation}
In order to understand the competition and crossover between
 such exponential parts
and the contribution from  the saddle point, let us return to
the intermediate expression (\ref{phibasic}) that arises in analyzing the large-time
asymptotics,
\begin{equation}
\phi (\xi,t) \simeq \frac{1}{2\pi} \, e^{ik^*\xi -i \omega_{\rm r}^*t}
\, \int_{-\infty}^{\infty} \ds k\,  \bar{\phi}(k)\,
e^{   - {\mathcal D}  t [ \Delta k - i\xi /2{\mathcal D}t ]^2 - \xi^2/4{\mathcal D}t}  ,
\label{phibasicnew}
\end{equation}
and analyze this integral more carefully in a case in which  $\bar{\phi}(k)$ has a 
pole whose strength is small.
The term $-i\xi/2{\mathcal D}t$ in the above expression 
gives a  shift in the value of the $k$ where the quadratic
term vanishes. For fixed $\xi$, this shift is very
small for large  $t$, and the Gaussian integration yields the
asymptotic result (\ref{phibasic}). However,  when $\xi$ is large enough that the
point where the growth rate is maximal moves close to the pole, the
saddle point approximation to the integral breaks down. This
clearly happens when the term between brackets in the exponential in
(\ref{phibasicnew}) is small at the pole, i.e., at the $c$ross$o$ver
point $\xi_{\rm co} $
for which
\begin{equation}
\xi_{\rm co}\, {\rm Re} \left(\frac{1}{2{\mathcal D} t }\right) 
\sim  (\lambda
  -\lambda^*) ~~~~\Longrightarrow ~~~~ \xi_{\rm co} \sim  2D   (\lambda
  -\lambda^*)t, \label{crossover}
\end{equation}
where we used the effective diffusion coefficient $D$ defined in
(\ref{lambda*}).  This rough argument  relates the velocity and
direction of motion of
  the crossover point to the difference in steepness $\lambda$ of the
initial condition and the steepness $\lambda^*$, and gives insight into how
the contributions from the initial condition  and the saddle point  dominate in different regions.
Before we will
discuss this, it is instructive to give a more direct derivation of a 
  formula for the
velocity of the crossover region by matching the expressions for the
field $\phi$ in the two regions. Indeed, the expression for $\phi$ in
the region dominated by the saddle point is the one given in
(\ref{phibasic2}),
\begin{equation}
|\phi (\xi,t) | \simeq 
 \frac{1}{\sqrt{4\pi D t}}\, e^{-\lambda^* \xi}\, e^{-\xi^2/4 Dt} \, |
 \bar{\phi}(k^*)|  ,   \label{phi*profile}
\end{equation}
while in the large $\xi$ region the profile is simply exponential: in
the frame  $\xi$ moving with the linear spreading speed $v^*$
the profile is according to (\ref{phienv}) 
\begin{equation}
|\phi (\xi,t) | \simeq 
  A   e^{-\lambda [\xi- (v_{\rm env}(\lambda)-v^*) t]}, \label{expprofile}
\end{equation}
where $A$ is the pole strength of the initial condition. The crossover
point is simply the point where the two above expression match; by
equating the two exponential factors and writing  $\xi_{\rm
  co}= v_{\rm co}t$,  we obtain from the dominant 
terms  linear in $t$
\begin{equation}
-\lambda^* v_{\rm co} - v^2_{\rm co}/{4D}= -\lambda
v_{\rm co} + \lambda [ v_{\rm env} (\lambda) - v^*],
 \end{equation}
and hence 
\begin{equation}
v_{\rm co} = 2D (\lambda - \lambda^*) \pm 2D
\sqrt{(\lambda-\lambda^*)^2 - \lambda [v_{\rm env}(\lambda)-v^*]/D} .\label{crossoverfull}
\end{equation}
It is easy to check that for for equations where $\omega(k)$ is
quadratic in $k$, the F-KPP equation as well as the Complex Ginzburg
Landau equation discussed in section \ref{sectioncglcubic},  the
square root vanishes in view of the relation 
(\ref{Dandveq}) between $D$ and the curvature of $v_{\rm
env}(\lambda)$ at the minimum. Hence,  
(\ref{crossoverfull}) then reduces to (\ref{crossover}).  This is simply
because when $\omega(k)$ is quadratic, the Gaussian integral in the
first argument is actually exact.  Since the square root term in (\ref{crossoverfull}) is always smaller
than the first term in  the expression, we see that the sign of
$v_{\rm co}$, the velocity of the crossover point, is the same as the sign
of $\lambda-\lambda^*$. Thus,  the upshot of  the analysis is that the crossover point to a tail
with steepness $\lambda$ {\em  larger}  than $\lambda^*$ moves to the
{\em right}, and the crossover point
to a tail which is {\em less steep}, to the {\em left}.\footnote{ Note that when the velocity is
expanded in the term under the
square root sign, the terms of order $(\lambda-\lambda^*)^2$ always
cancel in view of Eq.~(\ref{Dandveq}). Thus the argument of the the square root term generally
grows  as $(\lambda-\lambda^*)^3$, and 
depending on  $v_{\rm env}^{\prime \prime \prime}(\lambda^*)$ 
 the roots of Eq.~(\ref{crossoverfull}) are complex either for $\lambda >
 \lambda^*$ or for $\lambda< \lambda^*$.  This indicates that the detailed 
 matching in the regime where the roots of (\ref{crossoverfull}) are
 complex is more complicated than we have assumed in the analysis, but 
 the general conclusion that the direction of the motion of the
 crossover point is determined by the sign of $\lambda-\lambda^*$ is unaffected.}

\begin{figure}[t]
\begin{center}
 \epsfig{figure=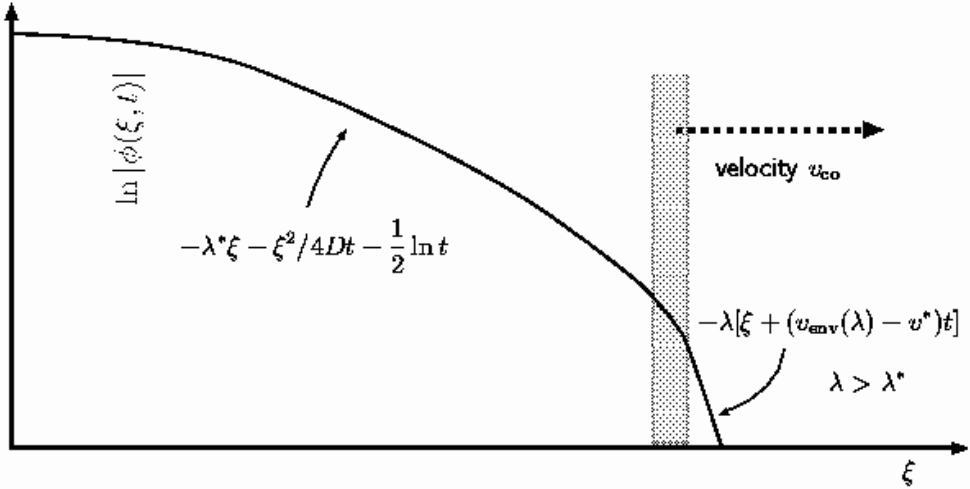,width=0.93\linewidth} 
\end{center}
\caption[]{Illustration of the crossover in the case of an initial
  condition which falls of exponentially with steepness $\lambda >
  \lambda^*$, viewed in the frame $\xi=x-v^*t$ moving with  the
  asymptotic spreading speed. Along the vertical axis we plot the logarithm of the
  amplitude of the transient profile. The dashed region marks the
  crossover region between the region where the linear spreading point
  contribution dominates and which spreads  asymptotically with
  speed $v^*$ in the lab frame, and the exponential tail which moves
  with a speed $v_{\rm env}> v^*$. As indicated, the crossover region
  moves to the right, so the steep fast-moving exponential tail
  disappears from the scene. The speed of the crossover region is
  obtained by matching the two regions, and is given by (\ref{crossoverfull}).
}\label{figlinearcrossover1}
\end{figure}

\begin{figure}[t]
\begin{center}
 \epsfig{figure=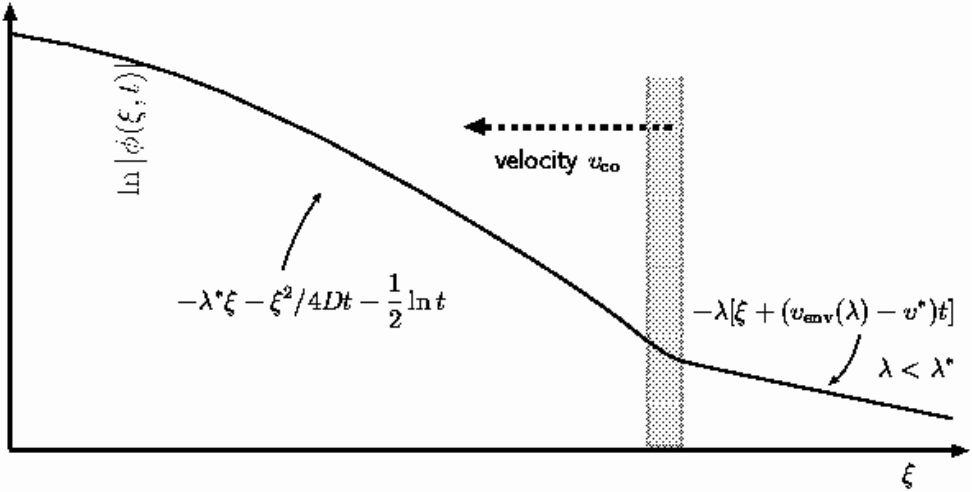,width=0.93\linewidth} 
\end{center}
\caption[]{As Fig.~\ref{figlinearcrossover1} but now for the case of an initial
  condition which falls of exponentially with steepness $\lambda <
  \lambda^*$. In this case, the dashed crossover region moves to the
  left, so the slowly decaying exponential tail gradually overtakes
  the region spreading with velocity $v^*$ in the lab frame. In other
  words, the asymptotic rate of propagation for initial conditions
  which decay slower than $\exp(-\lambda^* x)$ is $v_{\rm env} > v^*$.
}\label{figlinearcrossover2}
\end{figure}

 The picture that  emerges from this analysis is
illustrated in Figs.~\ref{figlinearcrossover1} and
\ref{figlinearcrossover2}. When $\lambda> \lambda^*$, i.e. for initial 
conditions which are steeper than the asymptotic linear spreading
profile, to the right for large enough $\xi$ the profile always
falls of fast, with the steepness of the initial conditions. However, 
as illustrated in Fig.~\ref{figlinearcrossover1}
the crossover region between this exponential tail and the region
spreading with velocity $v^*$ moves to the right in the frame moving
with $v^*$, i.e. moves out of sight! Thus, as time increases larger and
larger portions of the profile spread with $v^*$.\footnote{There is an
amusing analogy with crystal growth: the shape of a growing faceted
crystal is dominated by the slowing growing facets, as the fast ones
eliminate themselves \cite{vs1}.}

Just the opposite happens when the steepness $\lambda$ of the initial conditions 
is less than $\lambda^*$. In this case $v_{\rm co}<0$, so as
Fig.~\ref{figlinearcrossover2} shows, in this case the exponential tail
expands into the region spreading with velocity $v^*$. In this case,
therefore, as time goes on, larger and larger portions of the profile
are given by the exponential profile (\ref{expprofile}) which moves
with a velocity larger than $v^*$.

Because of the importance of  initial conditions whose
steepness $\lambda$ is larger than $\lambda^*$, we will henceforth
refer to these as {\em steep initial conditions}:
\begin{equation}
\mbox{{\em steep} initial conditions:} ~~~~ \lim_{x\to\infty} \phi(x,0)
e^{\lambda^*x} = 0, \label{steepincond}
\end{equation}
We will specify  the term  ``localized initial conditions'' more precisely when we
will discuss the nonlinear front problem in section
\ref{sectionleadingedgedominated}.

In conclusion, in this section we have seen that

\begin{tabular}{p{0.1cm}|p{13.2cm}}
& {\em According to the linear dynamics, initial conditions whose
  exponential decay rate (``steepness'') $\lambda$ is larger than
  $\lambda^*$ lead to profiles which asymptotically spread with the
  linear spreading velocity $v^*$. Initial conditions which are less
  steep than $\lambda^*$ evolve into profiles that advance with a
  velocity $v_{\rm env} > v^*$. }
\end{tabular}

As we shall see, these simple observations also have strong implications for the
nonlinear behavior: according to the linear dynamics, the fast-moving
exponential
tail moves out of sight. Thus, with steep  initial
conditions we can only get fronts which move faster than $v^*$ {\em if this
exponential tail matches up with a nonlinear front}, i.e. if there are
nonlinear front solutions whose asymptotic spatial decay rate
$\lambda > \lambda^*$. These will turn out to be the pushed front solutions.

\subsubsection*{$\Box$ Example: crossover in the linear F-KPP equation}
The above general analysis can be nicely illustrated by the initial value problem
$u(x,0)= \theta(x)e^{-\lambda x}$ for the linearized F-KPP equation
(\ref{linearfkpp}),  taken from section 2.5.1 of \cite{evs2}. 
Here $\theta $ is the unit step function. The
solution of the linear problem is
\begin{equation}
u(x,t)=  \exp[-\lambda x - v_{\rm env}(\lambda) t] \, \frac{1 +
\mbox{erf} [(x-2\lambda t)/\sqrt{4t}]}{2}, \label{fkppcrossover}
\end{equation}
where $\mbox{erf}(x) = 2 \pi^{-1/2} \int_0^x e^{-t^2}$ is the error
function and where $v_{\rm env}(\lambda )$ is given in
(\ref{vforfkpp}). The position of the crossover region is clearly $x
\approx 2 \lambda t$, which corresponds to a speed $2
(\lambda-\lambda^*)$ in the $\xi =x-2t$ frame, in accord with 
(\ref{crossover}) and  (\ref{crossoverfull}) with $D=1$, $\lambda^*=1$ 
and $v^*=2$ [Cf. (\ref{fkppvalues})]. Moreover, this 
crossover region separates the two regions where the asymptotic
behavior is given by 
\begin{eqnarray} \nonumber
u(x,t) & \approx & \exp[-\lambda [x-v_{\rm
      env}(\lambda)t]] ,\\ & = &  \exp[-\lambda [\xi-(v_{\rm
      env}(\lambda)-v^*) t]]  , \hspace*{1cm}  \mbox{for}~ \xi \gg 2(v_{\rm env}
    -2)t ,
\end{eqnarray} 
and
\begin{eqnarray}
u(x,t) & \approx  & \nonumber
\frac{1}  { \sqrt{4\pi t}\,
  \lambda(1-x/(2\lambda t)}  \exp [ -(x-2t) -(x-2t)^2/4t] , \\
&  \approx & \frac{1}{ \sqrt{4\pi t}\,
  \lambda }  \exp [ -\xi -\xi^2/4t] ,  \hspace*{1.7cm}  \mbox{for} ~\xi~ \ll   2(v_{\rm env}
    -2)t,
\end{eqnarray}
in full agreement  with the general expressions (\ref{expprofile})
and (\ref{phi*profile}). Finally, note that according to
(\ref{fkppcrossover}) the width of the crossover region grows
diffusively, as $\sqrt{t} $.  We expect this width $\sim \sqrt{t}$
behavior of the crossover region to hold more generally.

\subsection{The linear dynamics: generalization to more complicated
  types of equations} \label{sectionmoregeneral}

So far, we have had in the back of our minds the simple case of a
partial differential equation whose dispersion relation $\omega(k)$ is
a unique function of $k$. We now briefly discuss the generalization of
our results to more general classes of dynamical equations, following \cite{evs2}.

First, consider difference equations. The only difference with the
previous analysis is that in this case the $k$ -space that we introduce
in writing a Fourier transform, is periodic --- in the language of a
physicist, the $k$ space can be limited to a finite Brillouin zone.
Within this zone, $k$ is a continuous variable and $\omega(k)$ has the
same meaning as before. So,  if $\omega(k)$ has a saddle point in
the first Brillouin zone, this saddle point is given by the {\em same}
saddle point equations (\ref{saddlepoint}) as before, and the
asymptotic expression (\ref{phibasic2}) for the dynamical field $\phi$
is then valid as well!\footnote{When $\omega(k)$ is periodic in
  $k$ space, there will generally also be saddle points  at the
  boundary of the Brillouin zone. These will usually not correspond to
  the unstable modes --- they correspond to a an oscillatory
  dependence of the dynamical field (like in antiferromagnetism) --- but
  there is no problem in principle with the linear spreading being determined
  by a saddle point associated with the edges of the Brillouin zone.}

In passing, we note that although the above conclusion is simple but
compelling, one may at first sight be surprised by it. For, many
coherent solutions like fronts and kinks are susceptible to
``locking''  to the underlying lattice when one passes from a
partial differential equation to a difference equation
\cite{fath,keener}. Mathematically this is because perturbations to
solutions which on both sides approach a stable state are usually
governed by a a solvability condition. The linear spreading dynamics
into an unstable state, on the other hand, is simply governed by the
balance of spreading and growth, and this is virtually independent of
the details of the underlying dynamics.

The concept of linear spreading into an unstable state can be
generalized to sets of equations whose linear dynamics about the
unstable state can, after spatial Fourier transformation and temporal
Laplace transform, be written in the form
\begin{equation}
\sum_{m=1}^N  \hat{S}_{nm} (k,\omega) \hat{\phi}_m(k,\omega) =
\sum_{m=1}^N \hat{H}_{nm}(k,\omega)
\tilde{\phi}_m(k,t=0),~~~n=1,\cdots,N. \label{generalform}
\end{equation}
Here $n$ is an index which labels the fields. The above formulation is
the one appropriate when we use a temporal Laplace transform,
\begin{equation}
\hat{\phi}_n(k,\omega) =  \int_0^\infty \ds t \int_{-\infty}^\infty
\ds x\,
\phi_n(x,t) e^{-ikx+i\omega t}. \label{lftransform}
\end{equation}
In the Laplace transform language, terms on the right hand side arise
from the partial integration of temporal derivative terms
$\partial^k_t \phi_m(x,t)$ in the dynamical equation;  the coefficients
$H_{nm}$ therefore have no poles in the complex $\omega$
plane but poles in the $k$ plane can arise from exponentially decaying
initial conditions.

It is important to realize that the class of equations where the
linearized dynamics about the unstable state can be brought to the
form (\ref{generalform}) is extremely wide: in includes sets of
partial differential equations, difference equations, equations with a
spatial and temporal kernels of the form $\int \ds x^\prime \int^t
\ds t^\prime K(x-x^\prime, t-t^\prime) \phi(x^\prime,t^\prime)$, as well
as equations with a mixture of such terms.\footnote{I have have the impression
 that  population dynamicists \cite{odo,metz0} realized most
clearly first that the front speed of what we refer to as pulled fronts can be
calculated explicitly also for equations with a memory kernel. Within
the physics community, this was realized of course from the start by the plasma
physicists when they developed the ``pinch point formulation''
discussed below. Quite surprisingly, it appears that many of these early  developments have
never become standard knowledge in  the mathematics literature. The ``new method'' proposed in
\cite{fokas} is essentially  a reinvention of parts  of the work half a century before in plasma
physics and fluid dynamics referenced below, 
and this paper contains no references to these earlier developments. Even 
in this paper, the analysis is presented as a method that applies to  partial differential equations only. }  
 In addition, we conjecture
that much of the analysis  in this section can quite straightforwardly
be extended to front propagation into
periodic media (see section \ref{sectioncombustion}). We will give a
few simple examples based on extensions of he F-KPP equation below.
 
The Green's function $\underline{\underline{\hat{G}}}$ associated with
the equations is the inverse of the matrix
$\underline{\underline{\hat{S}}}$,
\begin{equation}
\underline{\underline{\hat{G}}} (k,\omega) \equiv
\underline{\underline{\hat{S}}}^{-1} (k,\omega) . \label{gsequation}
\end{equation} 
and the formal solution of (\ref{generalform}) can be written simply in
terms of $\underline{\underline{\hat{G}}}$ as
\begin{equation}
\underline{\hat{\phi}} (k,\omega) =
\underline{\underline{\hat{G}}}(k,\omega)\cdot 
\underline{\underline{\hat{H}}}(k,\omega)\cdot 
\underline{\bar{\phi}}(k,t=0). \label{inverseeq}
\end{equation}

When we invert the Fourier-Laplace transform, the term on the right
hand side has, in view of (\ref{gsequation}), poles at the points where 
the determinant $| \underline{\underline{\hat{S}}}|$ of
$\underline{\underline{\hat{S}}}$ vanishes. There may generally be various
branches of solutions of the equation   $| \underline{\underline{\hat S}}|=0$.  
In discussing the large-time behavior, one first assumes that the
initial conditions have compact support, so that their spatial Fourier 
transform is again an entire function of $k$. The analysis then
amounts to extracting the long-time behavior of the Green's function
$\underline{\underline{G}}$. 

The poles given by the zeroes of $|\underline{\underline{\hat{S}}}|$
 determine the dispersion relations $\omega_{\alpha}(k)$ of
the various branches $\alpha$.  The branches on which all modes are
damped do not play any significant role for the long-time asymptotics. For each of the branches on
which some of the modes are unstable, the analysis of the previous
sections applies, and for the linear problem {\em the} linear
spreading velocity $v^*$ is simply the largest of the linear spreading
speeds $v^*_\alpha$ of these branches. 

In fact, the long time asymptotics of $\phi_n(x,t)$ can be extracted in two
ways from (\ref{inverseeq}), depending on whether one first evaluates
the $\omega$-integral or the $k$-integral. The first method
essentially reproduces the formulation of the previous sections, the
second one leads to the so-called pinch-point formulation \cite{bers,briggs,huerre1,ll}
developed in plasma physics in the 1950-ies.
  We discuss their differences,  as well as their
advantages and disadvantages in  
appendix \ref{appendixconnection}, and proceed here keeping in mind
that the two methods invariably give the same expressions for the linear
spreading velocity $v^*$ and associated parameters. 

In order to keep our notation simple, we will from now on drop the
branch index $\alpha$, assuming that the right linear spreading point
has been selected if there is more than one, and we will usually also
drop the index $n$ or the vector notation for the dynamical field
$\phi$. 

\subsubsection*{$\Box$ Example: finite difference version of the F-KPP
equation}

As a simple example of the implications of the above discussion, imagine we integrate
the F-KPP equation with a cubic nonlinearity with a simple Euler
scheme.\footnote{The equivalent expressions  for the second-order implicit
(``Crank-Nicholson'') integration scheme are given in section 5.6.4 of
\cite{evs2}.} This amounts to
replacing the F-KPP equation  by the following finite difference
approximation:
\begin{equation}
\frac{u_j(t+\Delta t) -  u_j(t)}{\Delta t} =
\frac{u_{j+1}(t) -2u_j(t)+u_{j-1}(t)}{(\Delta x)^2} + u_j(t)-
u^3_j(t).\label{finitedifferencefkpp}
\end{equation} 
If we linearize  the equation by ignoring the last term and substitute
a linear mode $u_j\sim \exp(s t-\lambda j\Delta x)$ (this amounts
to writing $\omega=i s$ with  $s$ real) we obtain the
dispersion relation
\begin{equation}
\frac{\exp[s \Delta t] -1}{\Delta t} = 1 + \left( \frac{\sinh \half
\lambda \Delta x}{\half \Delta x} \right)^2.
\end{equation}
The saddle point equations or, what amounts to the same, the minimum
of the curve $v_{\rm env}(\lambda) = s/\lambda$ is easy to determine
 numerically. For small $\Delta t$ and $\Delta x$ one can also solve
the equation analytically by expanding about the values for the
continuum case given in (\ref{fkppvalues}), and one finds  \cite{evs2}
\begin{equation} \mbox{Euler approximation to F-KPP} ~~\left\{
\begin{array}{l}
v^*=  2 - 2 \Delta t + \frac{1}{12} (\Delta x)^2+
\cdots, \\
\lambda^* =  1+\Delta t - \frac{1}{8} (\Delta x)^2+
\cdots,\\
D=   1 - 4 \Delta t + \frac{1}{2} (\Delta x)^2+
\cdots, \end{array} \right.
\end{equation}
Although  these expressions look simply like  error estimates  for the
finite difference approximation of the F-KPP partial differential
equation, they are actually more than that:  they give the {\em exact}
parameters $v^*, \lambda^*$ and $D$  of the finite difference approximation. So
when the precise values of these parameters are not so important, e.g.,
if one want to study  the emergence of patterns or the power law
relaxation discussed below in section \ref{sectionunirelsimple}, one can take advantage of
this by  doing numerical simulations  with
relatively large values of $\Delta t$ and $\Delta x$ using the above
properties  as the reference values, rather than those obtained in the
continuum limit $\Delta t, \Delta x \to 0$.

\subsubsection*{$\Box$ Example:  F-KPP equation with a memory kernel}

\begin{figure}[t]
\begin{center}
\epsfig{figure=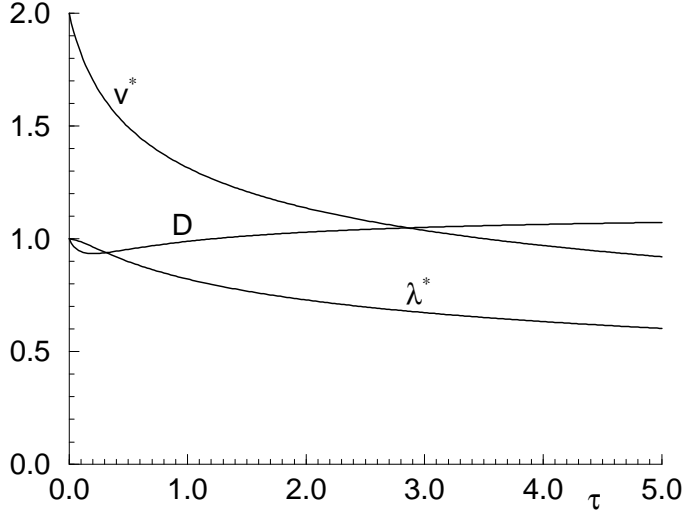,width=0.65\linewidth,bbllx=94pt,bblly=84pt,bburx=442pt,bbury=344pt}
\end{center}
\caption[]{Plot of $v^*$, $\lambda^*$ and $D$ as a function of $\tau$ for the extension
(\ref{delayedfkppeq1}) of the F-KPP equation with   memory kernel
(\ref{delayedfkppeq2}). From \cite{evs2}.
}\label{figdelayedfkpp}
\end{figure}

The extension of the F-KPP equation in which the linear growth term is
replaced by a term with a memory kernel,
\begin{equation}
\partial_t u(x,t)= \partial^2_x u + \int_0^t \ds t^\prime \,
K(t-t^\prime) \, u(x,t^\prime) - u^k(x,t), \label{delayedfkppeq1} ~~~~~(k>1),
\end{equation}
is an example of a dynamical equation which can still be treated along
the lines laid out above, as its Fourier transform is of the form
(\ref{generalform}). If we take for instance \cite{evs2}
\begin{equation}
K(t-t^\prime) = \frac{1}{\sqrt{\pi} \, \tau} \exp \left[
\frac{-(t-t^\prime)^2}{4 \tau^2}\right], \label{delayedfkppeq2}
\end{equation}
the implicit equation for $s(\lambda)=\omega_{\rm i}(\lambda)$ becomes
\begin{equation}
\lambda^2 - s +\exp[\tau^2s^2]\, \mbox{erfc}(\tau s) ,
\end{equation}
where erfc is the complementary error function. The result for
$v^*$, $\lambda^*$ and $D$ obtained by solving numerically for  the minimum of
$v_{\rm env}= s(\lambda)/\lambda$, are shown in Fig.~\ref{figdelayedfkpp}.

Note that when $\tau \ll1$ we can to a good approximation expand
$u(x,t^\prime)$ in  the memory term in
(\ref{delayedfkppeq1})  around $u(x,t)$ to second order. In this
approximation,  we then arrive at  a second order version of the F-KPP
equation. Such extensions have often been used as a simple way to
model delay effects, as will be discussed briefly in section
\ref{sectionbiologicalinvasion}.  From the point of view of
determining the linear spreading speed, however, there is no real advantage in
using a second order equation rather than an equation with a kernel.

\subsection{The linear dynamics: convective versus absolute instability}\label{sectionconvversusabs}

\begin{figure}[t]
\begin{center}
 \epsfig{figure=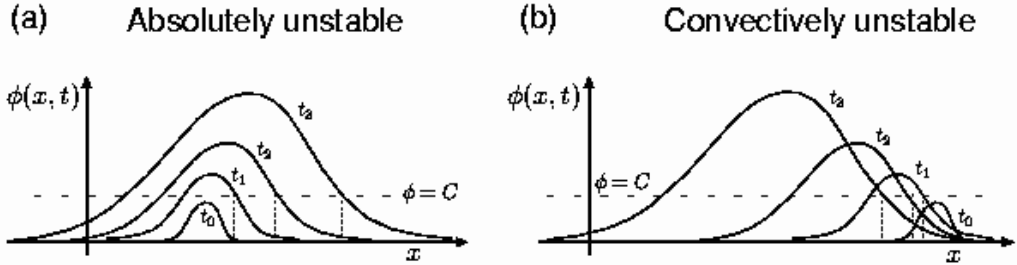,width=0.98\linewidth} 
\end{center}
\caption[]{Illustration of the distinction between an absolute
instability and  a convective instability, according to
the linearized dynamics. In the first case, sketched on the left in { (a)}, the
perturbation about the unstable state grows for sufficiently long
times at any position $x$. In the convectively unstable case sketched
on the right in {(b)}, the
perturbation grows but is at the same time advected away so fast that
it eventually dies at any fixed position $x$. Note that as a result of
it, the point on the right where $\phi$ reaches the level line $\phi=C$
retracts to the left. The transition from  convective to absolute
instability occurs when this point does not move for long times,
i.e. when the left spreading velocity $v^*_{\rm L}$ equals zero
\cite{bers,briggs,huerre1,huerre2,ll}. Note that the distinction  between absolutely and
convectively unstable depends on the frame of reference.  The
case sketched on the left which is absolutely unstable in the $x$
frame is convectively unstable in a frame moving with a
sufficiently large speed to the right,  while the case sketched on the
right in the $x$-frame becomes   absolutely unstable in a frame moving
with speed larger than $v^*$ to the left.
}\label{figabsconv}
\end{figure}

The case that we will typically have in mind  is the one in which  the
growth rate of the unstable modes is so strong that the amplitude of a generic localized
perturbation grows for long times at any fixed position, as sketched 
in Fig.~\ref{figabsconv}{\em  (a)}. It thus 
spreads into the unstable state on both flanks of the perturbation.

However, even when a state is linearly unstable, so that according
to (\ref{definitionunstable}) a range of modes has a positive growth 
rate $\omega_{\rm i}$, 
if there are symmetry breaking convective terms in the dispersion
relation a localized perturbation may be convected away faster than it
grows out. Figure \ref{figabsconv}{\em (b)} illustrates how in this
case the amplitude of the perturbation for any fixed position on the
right actually decreases in time, even though the overall amplitude
grows. Even for any position on the left of the figure, if we wait
sufficiently long   the amplitude of the perturbation eventually
decays. This regime is usually referred to as the {\em convectively
unstable regime}, while the other regime is referred to as the {\em
absolutely unstable regime} \cite{bers,briggs,huerre1,huerre2,ll}.

Clearly, these concepts  are intimately connected with the
 linear spreading speed discussed above. Indeed, given our convention to
focus on the right flank of a perturbation, the two regimes are
distinguished according to whether $v^*$ is positive or negative:
\begin{equation}
\begin{array}{ll} v^* > 0: & \mbox{linearly absolutely unstable regime,} \\
 v^* < 0: & \mbox{linearly convectively unstable regime.} \end{array}
\end{equation}

It is important to keep in mind that the two regimes are 
defined in relation to a particular frame of reference: if $v^*>0$ in
the lab frame, so that the instability is absolute in that frame, the
instability becomes convectively unstable for an observer moving to
the right with velocity larger than $v^*$.  Furthermore, a
convectively unstable system ideally remains in or will return to the unstable
state according to the linear deterministic dynamics. Nevertheless, small
perturbations or fluctuations are amplified while they are advected
away. This makes convectively unstable systems particularly sensitive
to fluctuations and  to small but fixed perturbations at a particular
point. The latter could arise in real experiments due to
imperfections in experimental setup or due to local perturbations at
an inlet in systems with a throughflow. We will encounter  a nice  experimental illustration of the
sensitivity to noise near the convective to absolute transition in
section \ref{secfrontsnoise1}.

We finally note that we have followed here the standard practice to distinguish the two regimes
according to the fully linear dynamics. The generalization of these
concepts to the nonlinear regime will be given later in section
\ref{sectionnonlinearconv2abs}.

\subsubsection*{$\Box$ Example:  the F-KPP equation with a convective
term}
Consider the F-KPP equation with a convective term,
\begin{equation}
\partial_t u (x,t) + s \partial_x u (x,t) = \partial_x^2 u (x,t) + f(u)
,\label{fkppwithv}
\end{equation}
where, as in the standard form (\ref{fkpp}), $f(0)=0$ and $f^\prime(0)=1$. Since this
equation simply becomes  the standard F-KPP equation  (\ref{fkpp})  upon
transforming to the moving frame $x^\prime= x-st$,  the linear spreading
velocity associated with (\ref{fkppwithv}) is $v^*=s+ v^*_{\rm FKPP}=
s+2$. Thus the linear instability changes from absolute to convective
at $s=-2$. Whether nonlinearly the transition also happens at this
point depends on the function $f(u)$: when $f$ is such that fronts
are pulled, the nonlinear transition remains at $s=-2$, but when the
 fronts become pushed,  the transition shifts to more negative values
of $s$ --- see section \ref{sectionnonlinearconv2abs}. In the
Taylor-Couette experiments reviewed in section \ref{secfrontsnoise1}
the velocity $s$ in the closely related amplitude equation is controlled with a flow through  the cell.

\subsection{The two-fold way of front propagation into linearly unstable
states: pulled and pushed  fronts}\label{sectiontwofold}

In the previous sections, we have analyzed the {\em  linear} spreading
dynamics into a linearly unstable state. We concluded that starting
from sufficiently localized initial conditions --- the precise
condition being given in 
(\ref{steepincond}) ---  the perturbations spread into the unstable
state with asymptotic speed $v^*$. This asymptotic spreading speed and
associated parameters are determined through the dispersion relation
$\omega(k)$ via Eqs.~(\ref{saddlepoint}).  

We now turn to the  genuine {\em   nonlinear  front propagation problem} already stated in the
introduction,  questions like: 
 {\em If initially a spatially extended system is in a linearly  unstable
state everywhere except in some  spatially localized region, what will
be the large-time dynamical properties and speed  of the nonlinear front which
will  propagate into the unstable state? Are there classes of initial
conditions for which the front dynamics converges to some  unique asymptotic
front state? If so, what characterizes these initial conditions, and
what can we say about the asymptotic front properties and the
convergence to them? }

As before, our discussion is aimed to be as general as is
possible. The only restriction  we will make, barring
pathological cases, is that the nonlinear dynamical equations of the
system under investigation have a sufficient  degree of ``locality'': they can
involve partial derivative terms and nonlocal terms with a kernel
whose range is essentially finite, but the dynamics of the dynamical
variables at a given position should not depend crucially on the nonlinear state
of the system infinitely  far away.\footnote{This requirement is
comparable to the fact that in statistical physics the standard
universality classes of equilibrium phase transitions only apply to
Hamiltonians in which the interactions are sufficiently
short-ranged. This does include Hamiltonians where the interactions
decay as a power $r^{-\sigma}$ provided $\sigma$ is large
enough.}\footnote{A simple example that illustrates this is the
following extension of the F-KPP equation: $\partial_t u =
\partial_x^2 u + u(1 - m \tanh[ \int_{-\infty}^{\infty} \ds x\,u^2])
-u^3$. For $m<1$ fronts in this equation are still pulled, albeit
with a spreading speed renormalized down from the value obtained by
linearizing the equation in $u$ about $u=0$. For $m>1$ there is no
finite asymptotic front speed.  This example illustrates that although
it is difficult to specify the general conditions under which our analysis
applies precisely,  in practice common sense gets one  quite far
 for any given problem. } 

A characteristic feature of front propagation into unstable states
which carries over from the linear dynamics that we discussed above,
is that initial conditions which decay very slowly spatially lead to
a large front speed whose value depends on the spatial decay rate. We
will come back to this later. Hence, in order to make precise 
 statements  we need to limit the  class of initial conditions
we consider. Now, for any given set of deterministic dynamical
evolution equations  with an unstable state, the 
linear spreading speed $v^*$ and associated steepness $\lambda^*$ are
explicitly and uniquely calculable according to the discussion of the
previous sections. This motivates us to distinguish the front
propagation properties of dynamical equations according to the long-time
evolution starting from steep initial conditions:

\begin{tabular}{p{0.1cm}|p{13.2cm}} &
{\em Whenever we   associate   a particular 
 front propagation mechanism with a  given
dynamical equation, this is a statement about the dynamical evolution
of fronts that evolve from localized  initial
conditions. These  include all initial conditions which are ``steep'',
i.e., which as defined in (\ref{steepincond})  fall off faster than
$\exp[-\lambda^* x]$.}\footnote{We will later in section
\ref{sectionleadingedgedominated}  enlarge
the class of initial conditions which lead to pushed front solutions.}
\end{tabular}

Let $v_{\rm front}(t)$ be some suitably defined instantaneous front velocity. 
 The  crucial insight on which our presentation
will be based is the following  simple insight:  {\em For front
propagation into a linearly unstable state, there are only two possibilities  if
we start from steep initial conditions,}

\begin{tabular}{p{0.1cm}|p{1cm} p{6cm}p{1cm}p{5cm}} & 
I.  & $v_{\rm as} \equiv \lim_{t \to \infty} v_{\rm front} (t) = v^*$  &
$ \Longleftrightarrow $ & {\em ``pulled''} front,\\ &  
II. & $ v_{\rm as} \equiv \lim_{t \to \infty} v_{\rm front} (t) = v^\dagger > v^*$ &
$\Longleftrightarrow$  & {\em ``pushed''} front.
\end{tabular}

This statement amounts to the claim that nonlinear fronts propagating
into a linearly unstable state can asymptotically not propagate
with speed slower than $v^*$ for equations of the type we consider, in
which the dynamics is ``local'' in the sense that  the linear dynamics is
not suppressed by nonlinear behavior arbitrarily far away. To see this, suppose we start with
a front solution which propagates with speed less than $v^*$. Any sufficiently small perturbation ahead
of it will then grow out and spread, asymptotically with the linear spreading speed $v^*$. Eventually
these perturbations will grow large enough that nonlinear behavior kicks in, {\em but the crossover
region where this happens  must
 asymptotically advance at least with the  speed} $v^*$. In other
words, since the nonlinearities
can not suppress the linear growth arbitrarily far ahead of the front, they can not prevent the
linear spreading to asymptotically move ahead with asymptotic speed $v^*$ and thus create a front
with at least this speed. Consequently, nonlinearities can only
drive front speeds up, i.e., to $v^\dagger > v^*$.

Admittedly, while the above conclusion that fronts can asymptotically
not move slower than $v^*$ may be dynamically ``obvious'',  I know of
no general mathematical proof of it. For some of the fourth order
equations reviewed in section \ref{section4thorder}, this conclusion is 
either implicit or explicit in a formulation in which one proves that
in any frame moving with velocity less than $v^*$ one sees
perturbations grow \cite{collet3,evs2}. 

At this point, the names {\em ``pulled''} and {\em ``pushed''} front
 fall fully into place.\footnote{The
names {\em pulled} and {\em pushed} were introduced back in 1976 by
Stokes \cite{stokes} and revived in the physics literature
in the mid nineties by Paquette {\em et al.}  \cite{paquette1,paquette2}. Ben-Jacob {\em
et al.} \cite{bj} referred to them as {\em Case I} and {\em Case II}
marginal stability and I initially used the words {\em linear} and
{\em nonlinear marginal stability} \cite{vs2}. The great advantage
of  the nomenclature of Stokes is that the notions of ``pulled'' and
``pushed'' 
tie in nicely with the general concept of linear spreading velocity,
and hence that they can be defined independently of whether or not a uniformly
translating or coherent front solution exists.} With a
pulled front we literally mean one which  
is pulled along by the linear spreading of small perturbations into
the linearly  unstable state. Any front which asymptotically moves with speed
$v^\dagger$ faster than
$v^*$ is somehow ``pushed'' into the unstable state by the nonlinear
behavior in the front region itself or the region behind  it --- if there
were no nonlinear behavior, one would find spreading with velocity
$v^*$. We will thus refer to $v^\dagger$ as the pushed front speed.

The present line of argument, where one takes the classical linear
spreading of perturbations into an unstable state as the starting
point, has several important advantages and ramifications. First of
all, it allows one to make statements irrespective of the nature of
the nonlinear state behind the front: Quite literally the picture that 
underlies it and that both the analytical and numerical results
presented later confirm, is that a pulled front spreads in the way the 
linear spreading point conditions (\ref{saddlepoint}) {\em force it to
do}. In other words, the nonlinear dynamics in the region behind the
leading edge just has to adapt to whatever is forced by the linear
spreading. Depending on the existence, stability and nature of
nonlinear front solutions, this behavior can be coherent or incoherent, 
but this by itself does not really feed back onto the linear
spreading.\footnote{Actually, we can make this quite precise: 
As we shall see, the  first consequence of the   feedback of
the nonlinear behavior onto front propagation is the
change of the prefactor of the $1/t$ relaxation term by a factor
3. See section \ref{sectionunirelsimple}.} This simple idea can be made explicit and
quantitative: it lies at the basis of the exact results for the
universal relaxation behavior of pulled fronts discussed in sections
\ref{sectionunirelsimple} and \ref{sectionuniversalrel}.

The second advantage is that the present line of argument focuses on
the fact that the most clear and relevant  issue is to understand the
mechanism through which we can get fronts to propagate at speeds
larger than $v^*$, i.e., to be pushed.

We finally note that the concept of a pulled front is only
well-defined for propagation into a linearly unstable state:  A small perturbation
around a linearly stable state dies out and does not spread.  As a
result, the propagation of a front into a linearly stable state is
always driven by the nonlinearities and the nonlinear or dynamical competition
between this state and a different one. From this perspective,
especially near subcritical bifurcation points, it is sometimes useful
to think of a front
which propagates into a stable state as being ``pushed''. 

\subsection{Front selection for uniformly translating fronts and
  coherent and incoherent   pattern forming
  fronts}\label{sectiontypesoffronts} 

In order to  discuss the selection of a particular type of
front and its propagation mechanism, it is useful to  distinguish three
different classes of nonlinear front dynamics:\footnote{With our classification we have in mind
fronts which propagate in a homogeneous
background medium. If the medium itself is periodic, fronts which are uniformly
translating in a homogeneous medium become automatically periodic; likewise if the parameters
of the dynamical equation are randomly varying the fronts will always be incoherent. Strictly speaking,
  fronts in a periodic medium, which  are
sometimes referred to as  ``periodically pulsating fronts'', are of the type 
(\ref{phiperiodic}) below  but as stated,
we will always have in mind that the medium itself is homogeneous, and when considering
periodic front solutions concentrate on
fronts which then generate a nontrivial pattern. For the linear dynamics of small perturbations
about the unstable state of a homogeneous system, 
the dispersion relation can be obtained from a simple Fourier transformation. As we shall
discuss below in   section
\ref{sectioncombustion}, the analysis of pulled fronts can be extended to the case of periodic media by
recognizing that  one simply has to do a Floquet analysis to determine $\omega(k)$. }
\begin{itemize}
\item
 {\em Uniformly translating fronts} are nonlinear front solutions for the dynamical
variable $\phi$ of the form
\begin{equation}
\phi (x,t )  = \Phi_v(\zeta) , \label{utsol}
\end{equation}
where $\zeta $ is the co-moving coordinate
\begin{equation}
\zeta \equiv x-vt ,
\end{equation}
which we will use for general velocity $v$, to distinguish it from the
 coordinate  $\xi=x-v^*t$ moving with velocity $v^*$. We remind the
reader that, unless noted otherwise, we do not distinguish notationally between a
 single dynamical variable or a set (vector) of them. Equation
 (\ref{utsol}) expresses that in a frame moving with velocity $v$, the 
 front solution is stationary, i.e., invariant in time. Uniformly
 translating solutions are normally only appropriate solutions for
 problems in which the front leaves behind a homogeneous state, i.e.,
 does not generate a pattern. The overwhelming majority of asymptotic
 front solutions which have been analyzed in the mathematics
 literature are of this type. An example of a pushed front in the
F-KPP equation is shown in the lower left panel of Fig.~\ref{F1}.
\item
{\em Coherent pattern forming fronts} are the
generalization of uniformly translating solutions to systems where the 
front leaves behind a pattern with a well-defined wavenumber. As the
 middle panel of Fig.~\ref{F1} and the
examples discussed in section \ref{section4thorder} and chapter
\ref{sectionexamples}  illustrate, usually
the pattern behind the front is stationary or  moves
with a velocity different from  the front speed. In other words,  when viewed in the frame
moving with the front, the pattern behind the front is {\em not}
stationary. Hence the front solution can not be of the form (\ref{utsol}). However,
 we may generalize the concept by introducing a {\em coherent pattern
   forming front solution}  as a front solution which is periodic in the frame $\zeta=x-vt$
moving with speed $v$, i.e.,
\begin{equation}
\phi(x,t) = \Phi(\zeta, t), ~~~~~\mbox{with}~~\Phi(\zeta,t+T)= \Phi(\zeta,t) . \label{phiperiodic}
\end{equation}
We can equivalently write this as
\begin{equation}
\phi(x,t) = \sum_{n=0,\pm 1,\cdots}  e^{-i n \Omega t }\, \Phi_v^n(\zeta) , ~~~~~~~~~(\Omega=2\pi/T),\label{cfsol}
\end{equation}
where for real dynamical fields
$\phi$ the complex functions $\Phi_v^n$ obey the symmetry
\begin{equation}
\Phi_v^{-n}(\zeta) = \Phi_v^n(\zeta) .
\end{equation}
Clearly, Eq.~(\ref{phiperiodic}) expresses that in the co-moving frame the
front solution is periodic.\footnote{We prefer to use the general name
``coherent pattern forming front solution'' instead of periodic front
solutions for two reasons. First of all, this name is consistent with
the name ``coherent structure solutions'' for special type of
solutions of the CGL equation (see  Eq.~(\ref{coherentfront}) in
section \ref{sectioncglquintic}). Secondly, we wish to stress 
the fact that we aim at pattern forming solutions and so want to
distinguish them from the simpler periodic or ``pulsating'' types of
fronts that arise in equations like the F-KPP equation when the medium
itself is periodic --- see section \ref{sectioncombustion}. }
Hence,  in the representation (\ref{cfsol}) the functions 
$\Phi_v^n(\zeta)$ can at every position be viewed as the Fourier coefficients of the
time-periodic function $\Phi(\zeta, t) $.

These {\em time-periodic coherent front solutions} of
   the form (\ref{phiperiodic}) or (\ref{cfsol})  are relevant for e.g. the Swift-Hohenberg equation
   discussed in section \ref{sectionsh} and were to my knowledge for
the first time  introduced
   for the analysis of fronts in this equation by Collet and Eckmann
   \cite{collet}. Moreover, they can also be viewed  
   as an extension of  the ``coherent structure solutions''
   \cite{vsh} introduced in the
context of the Complex Ginzburg Landau equations of sections
\ref{sectioncglcubic} and \ref{sectioncglquintic} --- because of the
fact that the  Complex Ginzburg Landau equation arises as the lowest
order amplitude equation for traveling wave patterns, only one term in 
the sum (\ref{cfsol}) is nonzero in this case.  As we shall
see,  both pulled and pushed coherent fronts solutions are found
in certain parameter ranges. 
\item
 {\em Incoherent pattern forming fronts} is the name we will 
 use for all the incoherent fronts like those of the right column of
Fig.~\ref{F1} which do not fit into  one of
 the two previous classes. We will encounter other  pulled and pushed  examples of them
 later, but they are the least understood of all. 
\end{itemize}
We now proceed  to discuss the dynamical
mechanism that  distinguishes 
pushed and pulled fronts for these various classes of equations. The
essential ingredient  will be to maximally exploit the constraints imposed
by the linear dynamics about the unstable state.

\subsubsection{Uniformly translating front
solutions}\label{selectionutfs}
When a dynamical equation also admits  a homogeneous stable stationary
state $\phi_{\rm ss}=const.$  in addition to the unstable state $\phi=0$, then  usually this
dynamical equation also admits  uniformly translating front
solutions of the form (\ref{utsol}), $\phi(x,t)= \Phi_v(\zeta)$ with
$\zeta=x-vt$. To investigate whether  it does, one substitutes this
Ansatz into the dynamical equation and analyzes its behavior. The
simplicity of uniformly translating fronts lies in the fact that the
function $\Phi_v(\zeta)$ depend on the single variable $\zeta$ only,
so that the function $\Phi_v$ then obeys an ordinary differential equation
rather than a partial differential equation. The existence of front
solutions can from there on be analyzed using standard methods
\cite{arnold1,arnold2,guckenheimer}: the 
homogeneous stationary states $\phi=0$ and $\phi=\phi_{\rm ss}$ are
fixed points of this ordinary differential equation. By linearizing
about these fixed points and studying the dimensions of the stable and
unstable manifolds  (i.e. the dimensions of the manifolds flowing into
and out of each fixed point), one can then study  the multiplicity of
front solutions (see e.g. \cite{vsh} for an extensive use of such
``counting arguments'' in the context of CGL equations). For the F-KPP
equation the analysis, which is summarized below,  is relatively trivial; one finds that there is
a one-parameter family of front solutions that connect the stable
homogeneous state with the unstable state $\phi=0$. For the F-KPP
equation,  one can easily go beyond a simple
counting argument based on the analysis near the fixed points to
prove more rigorously  when there is such a family of solutions.  These front
solutions can be parametrized by their
velocity $v$.  We stress that for an arbitrary dynamical
equation that admits  a stable homogeneous state and a homogeneous
unstable state, there is not necessarily always a one-parameter family of front solutions
connecting the two. However,  this {\em is}  what is typically found
and what is intuitively most reasonable based on the fact that  a
special feature of the linear dynamics about an unstable state is that
any exponential tail with steepness $\lambda$ less than $\lambda^*$
can propagate into the unstable state with speed $v_{\rm
env}(\lambda)$ --- see section \ref{sectionexponential}.\footnote{An explicit
counting argument for partial differential equations that are symmetric
under spatial reflection confirms that for such equations one
typically expects a one-parameter family of front solutions  (see appendix A of \cite{vs1}).
There are two  intimately related ways to understand why this is so often the case:
 {\em (i)} Suppose we first change the dynamical equation so as to make the
unstable state stable. One then expects there to be at least one  front solution (or
maybe  a discrete set of them) connecting the two stable states. Now
when we change the equation back to its original form, at the moment
when the $\phi=0$ state becomes unstable again, the dimension of the
stable manifold flowing into the $\phi=0$ fixed point increases by
one, because a new root appears which is related to the left branch in
Fig.~\ref{figvversuslambda};  {\em (ii)}
According to the
linear dynamics  discussed in section \ref{sectionexponential}, any
any exponential tail with $\lambda
<\lambda^*$ can propagate into the
unstable state with envelope velocity $v_{\rm env} (\lambda)$. When
the dynamical equation admits a stable homogeneous state as well, it
is not unreasonable that every  exponential tail propagating with some
velocity $v$ in the leading edge can match up smoothly with a
saturating behavior behind the front. }

Let us first proceed by assuming the equations do admit a
one-parameter family of front solutions, parametrized by their
velocity. Then this family generically will include  
a uniformly translating front solution $\Phi_{v^*}(\xi)$
moving with velocity $v^*$, to which  the front will
asymptotically converge in the pulled regime, i.e., starting from steep
initial conditions. 

The behavior near the fixed points of the ordinary differential
equation for $\Phi_v(\zeta)$, obtained by linearizing the flow
equations  around the fixed point, corresponds to an exponential
 $\zeta$-dependence. Thus, 
for an arbitrary velocity $v$, the solutions $\Phi_v(\zeta)$ will
decay to zero exponentially for large $\zeta$,
\begin{equation}
\Phi_v(\zeta) \approx  a_1 e^{-\lambda_1 \zeta} + a_2 e^{-\lambda_2
\zeta} + \cdots ~~~~~(\zeta \to \infty), \label{decayformula}
\end{equation}
where all roots $\lambda_i$ are positive and where the real coefficients
$a_1$, $a_2$, etcetera can only be determined by solving the equations
for    the fully nonlinear front solution.\footnote{In
(\ref{decayformula}) we want to bring out that there are contributions
from the various roots, or, in more technical language, from the
various directions that span the stable many-fold of the fixed point
corresponding to $\phi=$. If $\lambda_1 < \lambda_2/2$ then of course
terms $\exp[-2\lambda_1\zeta]$ dominate over terms
$\exp[-\lambda_2\zeta]$, but this is all hidden in the dots.}  We take these roots ordered,
$0< \lambda_1 < \lambda_2  \cdots $.  The relation between the
velocity $v$ and the two smallest roots in the generic
case\footnote{We assume here for simplicity that there are no other
branches at even smaller $\lambda$; this is usually  the case, as
the branch which in Fig.~\ref{figvversuslambda} diverges as
$\lambda\downarrow 0$ is associated with the instability, the fact
that $\omega_i(\lambda=0)>0$. We will briefly come back to the case
where other branches intervene in section \ref{sectiontwospreadingpoints}.}  is 
shown in Fig.~\ref{figvversuslambda}: For $v >v^* $ the root
$\lambda_1 < \lambda^*$ while $\lambda_2 > \lambda^*$. In passing we
note that  at $v=v^*$ two roots coincide; general results
\cite{arnold1,arnold2,guckenheimer} for the
flow behavior in the presence of degenerate eigenvalues then imply
\begin{equation}
\Phi_{v^*} (\xi) \approx  (a_1\xi + a_2) e^{-\lambda^*\xi} + a_3
e^{-\lambda_3\xi} + \cdots. \label{Phi*eq}
\end{equation}

\begin{figure}[t]
\begin{center}
 \epsfig{figure=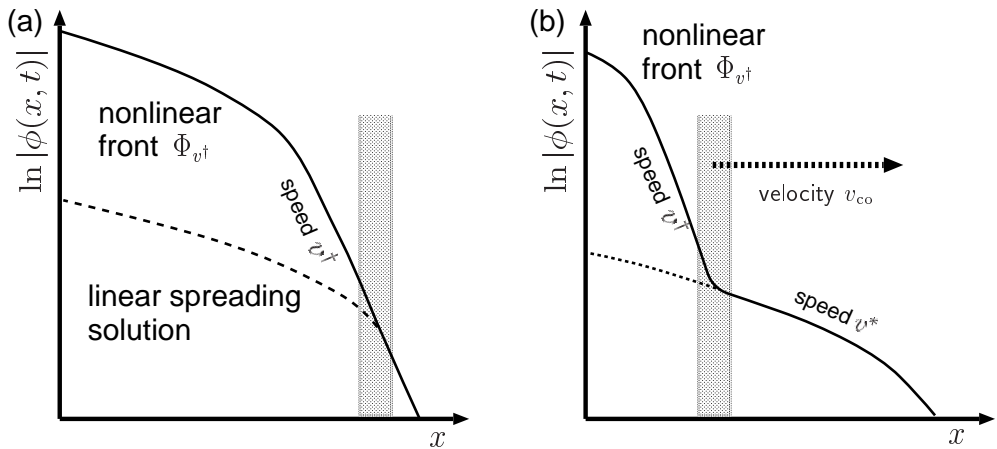,width=0.98\linewidth} 
\end{center}
\caption[]{Illustration of the fact that when a nonlinear front solution 
$\Phi_{v^\dagger}$ exists, whose asymptotic steepness  is according
to (\ref{pushedexpr}) larger 
than $\lambda^*$, then this front will generically emerge: it is the
selected pushed front solution. Figure  (a) should be compared to
Fig.~\ref{figlinearcrossover1}: If the equation would be fully linear,
the steep tail on the right, which moves at velocity larger than
$v^*$, would cross over in the dashed region to the dashed profile
moving with asymptotic speed $v^*$. However, this does not happen.
When the steep profile 
matches up arbitrarily well with a fully nonlinear front profile
$\Phi_{v^\dagger}$ this steep region does not disappear from the
scene: Instead a fully nonlinear profile with speed $v^\dagger$
emerges. While  (a) shows how only the pushed front solution
$\Phi_{v^\dagger}$ can asymptotically emerge, (b) illustrates how a
pushed front solution invades a region where the profile is close to
that given by the linear spreading analysis. The dashed line indicates
the continuation of the profile as given by the linear spreading
analysis of previous sections, but in the dashed region the profile
crosses over to the steep tail of the nonlinear pushed profile. The
fact that the crossover region moves to the right with the speed
$v_{\rm co}$ determined earlier, confirms that the pushed front
solution invades the leading edge.
}\label{figpushedscenario}
\end{figure}

When do we expect the equation to be pushed, i.e., when do we expect
that a pushed front solutions with velocity $v^\dagger > v^*$ will
emerge starting from steep initial conditions?  The answer lies hidden
in Fig.~\ref{figlinearcrossover1}: {\em according to the fully linear dynamics}
 an exponential tail which is
steeper than $\lambda^*$ and which corresponds to the dashed branch
$\lambda_2$ in Fig.~\ref{figvversuslambda}, {\em does run faster than $v^*$
but disappears from the scene}  because the crossover point to the
linear spreading profile moves to the right.  According to our
``locality'' assumption for the dynamics, the same holds arbitrarily
far into the leading edge of a front, {\em unless the nonlinearities
in the dynamical equation
allow the fast exponential tail to match up perfectly with a nonlinear front
solution!} This is illustrated in Fig.~\ref{figpushedscenario}: if a uniformly translating
front solution with $v= v^\dagger > v^*$ exists for which $a_1=0$ so
that its asymptotic
behavior is given by the root $\lambda_2$, 
\begin{equation}
\Phi_{v^\dagger} (\zeta) \approx a_2 e^{-\lambda_2 \zeta}  + a_3
e^{-\lambda_3\zeta}  + \cdots
~~~~~ (\lambda_2 > \lambda^*,~~\zeta=x-v^\dagger t \to \infty  ),\label{pushedexpr}
\end{equation}
then this front solution {\em can and will overtake} any transient dynamical
tail in the leading edge. In other words, such a solution is the
asymptotic pushed front solution sought for, and {\em if it exists it is
the dynamically relevant front solution emerging from steep initial
conditions. }

As we noted before, front solutions $\Phi_v(\zeta)$ obey an ordinary
differential equation or a set thereof; such a differential equations can be formulated
as a flow in phase space \cite{arnold1,arnold2,guckenheimer}. In such
an interpretation, a front solution corresponds to a so-called
heteroclinic  orbit, an orbit which goes from one
point (the asymptotic state behind the front) to another (the state
$\phi=0$). For general $v$, the orbit approaches  the fixed point
corresponding to the state $\phi=0$ asymptotically along the
eigendirection whose rate of attraction, given by $\lambda_1$, is slowest. The
pushed front solution $\Phi_{v^\dagger}(\zeta)$, however, approaches
it along the  slowest but one eigendirection , the one with eigenvalue $\lambda_2$. For this reason, a
uniformly translating pushed front solution is sometimes referred to as
a {\em strongly heteroclinic orbit} \cite{powell}.

\begin{figure}[t]
\begin{center}
\epsfig{figure=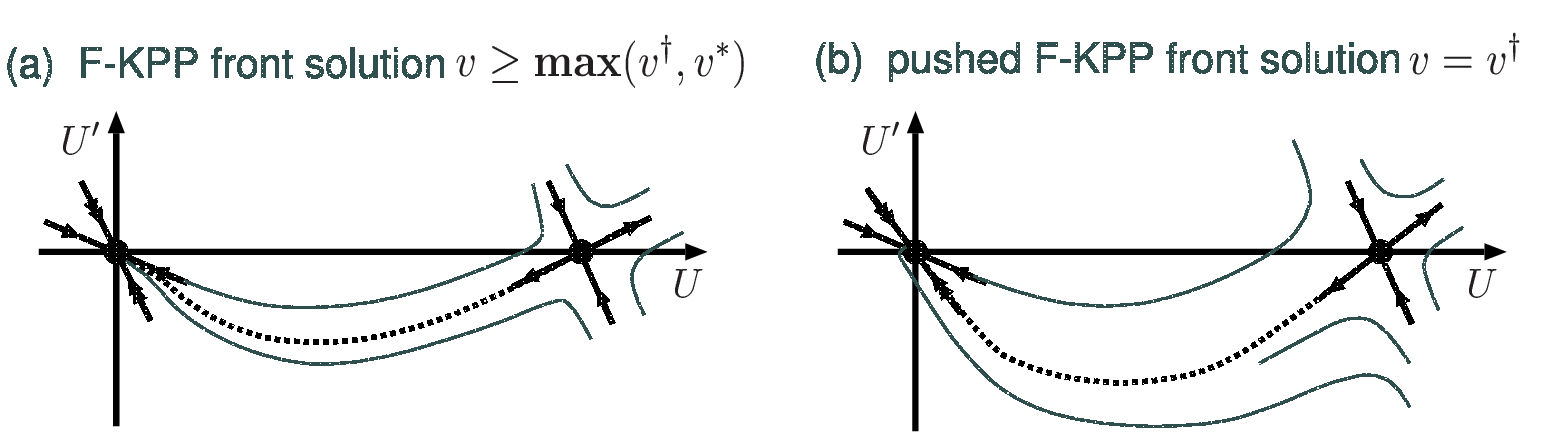,width=1.0\linewidth}
\end{center}
\caption[]{Sketch of the flow in the phase space $(U,U^\prime)$ of the
flow equations (\ref{floweq1}) and (\ref{floweq2}) that govern the
uniformly translating solutions of the F-KPP equation. The solid lines
at the two fixed points indicate the shape of the stable and unstable
manifolds near each fixed point, obtained by linearizing the flow
around each fixed point. The arrows indicate the direction of the flow,
and the double arrow at the $(0,0)$ fixed point indicates that the
contraction along this eigendirection is stronger than along the other
one $(\lambda_2 > \lambda_1)$. The dashed trajectory is the
heteroclinic orbit connecting the two fixed points; it corresponds to
the front solution connecting the stable and the unstable state.  (a) Flow in the
case $v\ge \mbox{max}(v^\dagger, v^*)$. The trajectory approaches the
$(0,0)$ fixed point along the slowest contracting direction. (b) If
for some velocity $v=v^\dagger$ the dashed trajectory becomes a
``strongly heteroclinic orbit'' which approaches the 
$(0,0)$ fixed point along the more strongly contracting direction,
the equation admits a pushed front solution.
}\label{figphasespaceflow}
\end{figure}

Let us illustrate the above considerations briefly for two equations,
the F-KPP equation  and an extension of it, the EFK equation.
 Uniformly translating solutions $U(x-vt)=U(\zeta)$ of the
F-KPP equation (\ref{fkpp}) obey the ordinary differential equation
\begin{equation}
-v \frac{dU}{d\zeta} = \frac{d^2U }{d\zeta^2} + f(U). \label{fkppode2}
\end{equation}
It is convenient to use the standard trick of writing this second
order equation as a set of first order equation: by introducing the
variable $U^{\prime} = dU/d\zeta$ we can write (\ref{fkppode2}) as
\begin{eqnarray}
\frac{dU}{d\zeta} & = &  U^{\prime} , \label{floweq1}\\
\frac{dU^{\prime}}{d\zeta} & = & - v\, U^{\prime} - f(U) .\label{floweq2}
\end{eqnarray}
These equations describe the flow in the two-dimensional phase space
$(U, U^\prime )$, with $\zeta$ playing the role of time. Because
$f(0)=f(1)=0$ according to (\ref{fkpp}), the points $(0,0)$ and
$(1,0)$ are fixed points of these flow equations; the first one
corresponds to the stable state and the second one to the unstable
state. With our convention that fronts move into the unstable state on
the right, a uniformly translating front solution 
corresponds to a trajectory flowing from the $(1,0)$ fixed point to
the $(0,0)$ fixed point. Such a ``heteroclinic orbit'' is sketched for
large arbitrary $v$ in Fig.~\ref{figphasespaceflow}{\em
(a)}.\footnote{For $v<v^*$ the eigenvalues at the fixed point $(0,0)$
are complex so trajectories spiral into this fixed point. }  The solid
lines near the two fixed points  in this figure denote the directions of
the stable and unstable manifolds flowing into and out of each fixed
point. These are easily determined by linearizing equations
(\ref{floweq1}), (\ref{floweq2}) about the fixed point solutions and
solving for the eigenvalues of the linearized flow. The arrows in
Fig.~\ref{figphasespaceflow} indicate the directions of the flow for
increasing $\zeta$ (``time''). As is indicated in the figure,  there
is one stable and one unstable stable direction at the $(1,0)$ fixed
point. For any fixed velocity $v$, there is hence a unique trajectory
coming out of this fixed point in the direction of decreasing $u$. At
the $(0,0)$ fixed point, however, both eigendirections are attracting;  we
have indicated the direction along which the  contraction is largest
with a double arrow.  Now, because 
there is a two-dimensional manifold flowing into the $(0,0)$ fixed
point, the unique dashed trajectory that flows out of the $(1,0)$ fixed point
will flow into the $(0,0)$ fixed point, and it will asymptotically
flow in along the slowest contracting eigendirection --- these
observations correspond to the statements that there is a front
solution for a any $v$  and that  the asymptotic large-$\zeta$
behavior  in (\ref{decayformula}) is dominated by the smallest
eigenvalue $\lambda_1$.  

Depending on the form of the nonlinearity $f(u)$, the situation as
sketched in Fig.~\ref{figphasespaceflow} may occur:  
for a particular value $v^\dagger$ of the velocity, it may happen that
the unique
dashed trajectory that flows out of the $(1,0)$ fixed point flows into
the stable $(0,0)$ fixed point along the most rapidly contracting
direction which is indicated in the figure with the double arrow. In
other words, for this trajectory the asymptotic behavior of $U(\zeta)$
goes as $e^{-\lambda_2\zeta}$, i.e., has $a_1=0$ in
(\ref{decayformula}). This ``strongly heteroclinic orbit'' thus
corresponds to a  pushed front solution --- if it exists, the selected front in the
F-KPP equation is a pushed front. Clearly, {\em whether such a pushed
front solution exists depends on the global nonlinear properties of
the equation}, the full nonlinear behavior of $f(u)$ in this case ---
determining or proving whether for a given equation the selected front
is  pushed,   requires a  fully nonlinear global analysis of the
flow. All the details of the nonlinear behavior count. In the example at the end of this section
we will show that for certain classes of nonlinearities $f(u)$, the
pushed front solution of the F-KPP equation can be 
obtained analytically.

To give an idea of the complications that one immediately encounters
when one goes beyond the F-KPP equation, let us briefly consider the
equation
\begin{equation}
\partial_t u = \partial^2_x u - \gamma\partial^4_xu + f(u), \label{generalizedefk}
\end{equation}
which can be thought of as an extension of the F-KPP equation. Indeed,
with $f(u)=u-u^3$ this equation is the EFK (``Extended
Fisher-Kolmogorov'')  equation which we will discuss in more detail in
section \ref{sectionefk}. As we shall see there, the pulled fronts in
this equation exhibit a transition from uniformly translating fronts
to coherent pattern forming fronts at $\gamma=1/12$, but we focus for
now on the uniformly translating fronts for $\gamma<1/12$. If we
substitute the Ansatz $=U(\zeta)$ into the equation and write the
resulting ordinary differential equation as a set of first order
equations, we get in analogy with (\ref{floweq1}) and (\ref{floweq2})
\begin{eqnarray}
\frac{dU}{d\zeta} & = &  U^{\prime} , \label{efkfloweq1}\\
\frac{dU^\prime}{d\zeta} & = &  U^{\prime \prime} , \label{efkfloweq2}\\
\frac{dU^{\prime\prime} }{d\zeta} & = &  U^{\prime \prime \prime} , \label{efkfloweq3}\\
\frac{dU^{\prime\prime \prime}}{d\zeta} & = & \gamma^{-1} \left[
U^{\prime\prime} + v\, U^{\prime} +
f(U)\right] .\label{efkfloweq4}
\end{eqnarray}
\begin{figure}[t]
\begin{center}
\hspace*{-4mm}
{\tt (a)} 
\hspace*{-2mm} 
\epsfig{figure=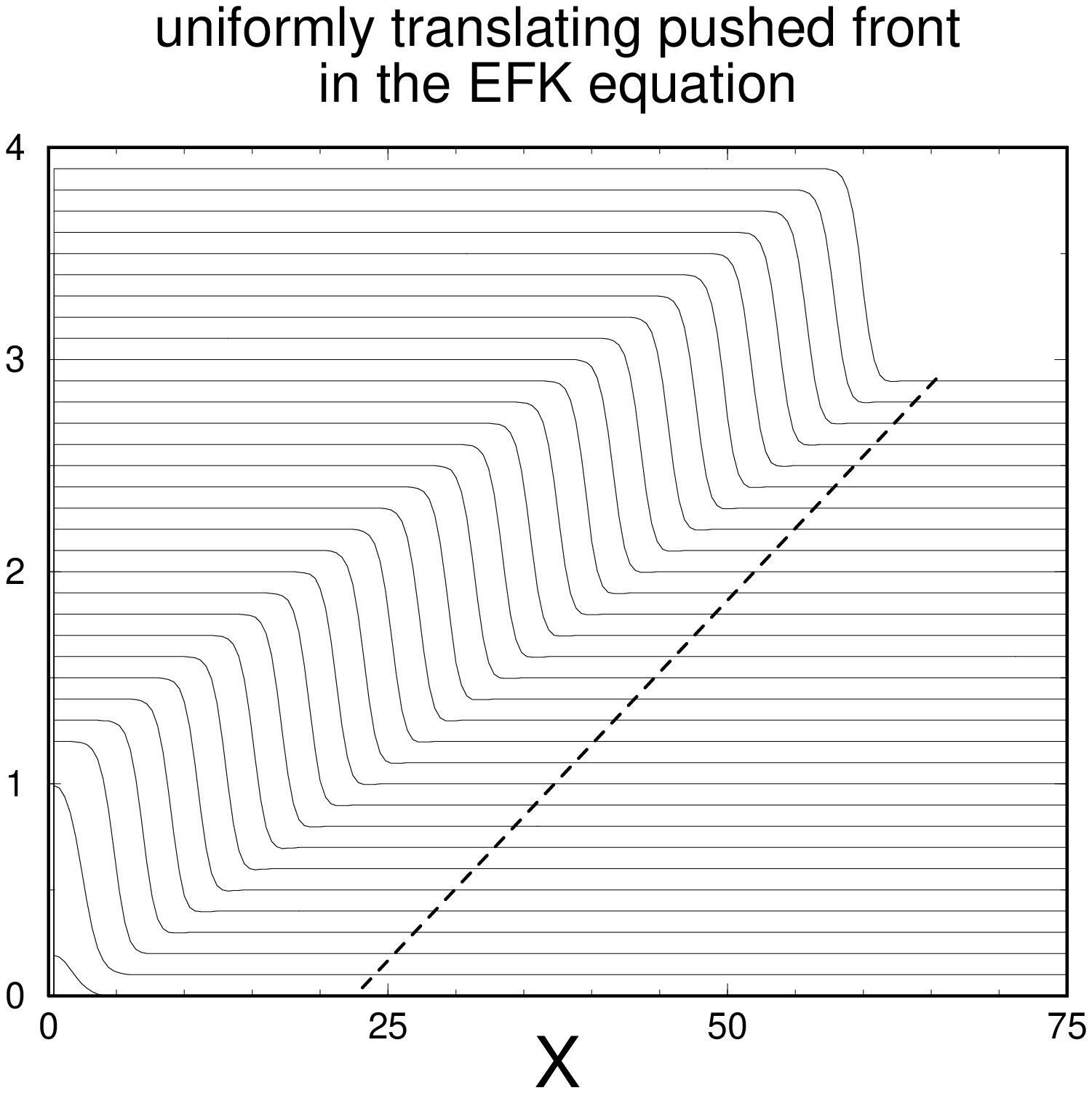,width=0.33\linewidth}
\hspace*{0.3cm}
{\tt (b)} \hspace*{-1mm}
\epsfig{figure=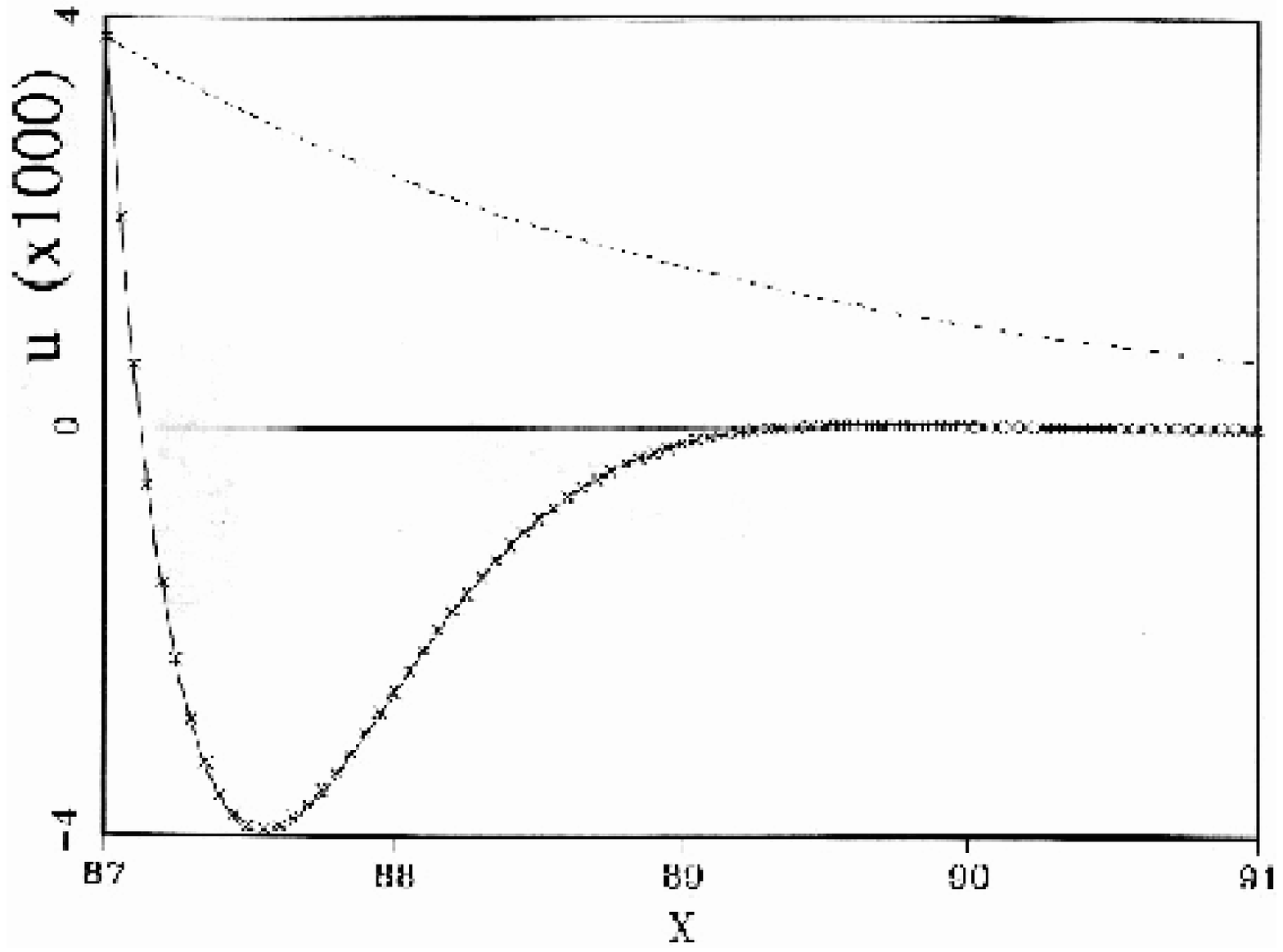,width=0.46\linewidth}
\end{center}
\caption[]{(a) Space time of a numerical solution of  equation (\ref{generalizedefk}) with nonlinearity
$f(u) = u+9u^2-10u^3$. This is an extension of the EFK equation
discussed in section \ref{sectionefk}. For this nonlinearity, the
selected front is a pushed front propagating 
with speed $v^\dagger=2.751 $\cite{vs2}. The time increment between successive time-slices
is 0.75; the initial condition  is a small Gaussian peak centered at the origin, and the total
simulation time is 22.5. The dashed line indicates a position moving with the speed $v^*=1.8934$. 
Although the front might appear to be  monotonically decaying to the right,
close inspection of this  pushed front  in the tip shows that it
is  actually non-monotonic;   the enlargement in panel { (b)}
shows this more clearly. (b) Blow up of the asymptotic front profile in the leading
edge. The symbols denote the actual front   
values at the grid points in the simulations, while the full line is the profile in the leading 
edge given by the three terms in (\ref{pushedexpr}). The dashed line denotes
the mode $\exp(-\lambda_1 \zeta)$. 
Clearly, for the pushed front solution, $a_1$ =0 indeed, in accordance
with the mechanism of pushed front propagation set forth  in the main
text. From \cite{vs2}. }\label{figpushedefk}
\end{figure}
Since the flow is now in a four-dimensional phase space, it is clear that the question
of existence of such uniformly translating front solutions is much more subtle than
for the F-KPP equation: all the simplifications  of flow in a
two-dimensional plane, special to the F-KPP equation, are lost.\footnote{If one again linearizes the flow
near the two fixed points $(0,0,0,0)$ and $(1,0,0,0)$, the
multiplicity of the stable and unstable manifolds is such that one
would indeed expect the existence of a one-parameter family of front
solutions \cite{vs2}. A more rigorous study of such front solutions and
of the flow in this four-dimensional phase space has been taken up
only recently in the mathematics literature. See in particular
\cite{rottschafer2} for a proof of the existence of front solutions of
the above flow equations and \cite{bert,vandenberg} for other types of
solutions.}  Let us here simply
illustrate that one important simplifying property of the F-KPP
equation is immediately lost. From the phase space arguments sketched
above for the F-KPP equation, it immediately follows that the selected
front is the front solution with the smallest speed for which the
front solution is  monotonically decaying with $\zeta$. Therefore,
this idea has sometimes been proposed in the literature as a general
principle for front selection. However, it is simply a property of the
F-KPP equation and a limited class of extensions of it --- it  is a property that does not really have
anything to do with front selection and just
does not hold generally. The simulations of
Fig.~\ref{figpushedefk} illustrate this for the above equation (\ref{generalizedefk})
with $f(u)=u+9u^2-10u^3$. As the left panel illustrates, the selected
front is a uniformly translating pushed front in this case. Although
this is hardly visible in the space-time plot on the left, 
 this pushed front solution is not monotonically decaying towardss the
right. The enlargement of  the
leading edge in the right panel shows how $U$ first goes through zero
and then has a local maximum at negative $U$. This plot
also confirms fully the mechanism for pushed fronts that
we have identified above:  the selected profile fits the
form (\ref{decayformula}) with $a_1=0$.  

Let us return to our general formulation of the mechanism through
which pushed fronts arise for uniformly translating fronts.
While this scenario  is arrived at simply by exploiting our insight into
the leading edge where the dynamics is essentially given by the linear
dynamics, it is in complete agreement  with {\em (i)}  what is known rigorously for the
second order F-KPP equation, {\em (ii)}  the (marginal) stability arguments
for front selection (section \ref{stabilityandothermechanisms})
which show that if a strongly heteroclinic front
solution $\Phi_{v^\dagger}$  exists, all front solutions with lower
speed are unstable to invasion by this front, and {\em (iii)}  all numerical
results known to me.  Furthermore, the extension of the argument to coherent pattern
forming fronts  is fully consistent with the 
analytical and numerical results for the quintic CGL equation
discussed in section
\ref{sectioncglquintic}).

We finally return to the question what happens if the front solutions
do not admit a one-parameter family of front solutions. If this
happens, then we generically expect that the equations will not admit
a uniformly
translating pulled front solution moving with velocity $v^*$. A
pulled front can then not be uniformly translating --- the pulled
front solution must then show nontrivial dynamics in the front
region. Presumably, the dynamics is then either that of a coherent or
incoherent pattern forming front, even if the state it finally leaves
behind is structureless. Furthermore, the absence of a one-parameter
family of uniformly translating front solutions makes it also very
unlikely that there will be uniformly translating pushed front
solutions, as these have to obey one additional constraint $a_1=0$.

\subsubsection*{$\Box$ Examples of pushed front solutions of the F-KPP
equation and the reduction of order method }

For the F-KPP equation, for which the dispersion relation $\omega(k)$
is quadratic, a general uniformly translating front solution
$u(x-vt)=U(\zeta)$ falls off to $u=0$ with two exponentials --- in
other words, for a front solution with arbitrary velocity $v$ there
are two terms in the expression (\ref{decayformula})  for the large
$\zeta$ behavior. However, as we already discussed above, a pushed
front solution of the F-KPP equation  approaches $u=0$
with a single exponent, as $a_1=0$. A good strategy  to look for exact
pushed front solutions of the F-KPP equation is therefore \cite{vs2} to investigate
when front solutions of the first order equation 
\begin{equation}
\frac{\ds U_v}{\ds  \zeta} = h(U_v)\label{ansatz1}
\end{equation}
are also front solutions of the full ordinary differential equation
for front solutions $U_v(\zeta)$ of the F-KPP equation. This is the
case if they satisfy Eq.~(\ref{fkppode2}), 
\begin{equation}
-v \frac{\ds U_v(\zeta)}{\ds \zeta} = \frac{\ds ^2 U_v(\zeta)}{\ds \zeta^2} +
f(U_v(\zeta))\label{fkppode}.
\end{equation}
Searching for solutions of the type (\ref{ansatz1}) is very easy; to see this, note
that upon
substitution of (\ref{ansatz1}) into (\ref{fkppode}) we get
\begin{equation}
-v h(U_v) = h^\prime (U_v) \,h(U_v) + f(U_v), \label{hfeq}
\end{equation}
so that we just need to look for combinations of functions $h$ and $f$
which obey this last equation (for any given $h$ one can trivially find a function $f$
but the converse is not true). This is especially simple for
polynomial functions. Indeed, it is straightforward to check that the
functions \cite{evs2}
\begin{equation}
h(u) =   -\lambda u(1-u^n), \hspace*{1cm} f(u) =  \tilde{\varepsilon}
u+u^{n+1} -(1+\tilde{\varepsilon})u^{2n+1} \label{fandhnonlinear}
\end{equation} solve (\ref{hfeq}) 
provided we take
\begin{equation}
\lambda^2 =  \frac{\tilde{\varepsilon}+1}{n+1}, \label{lambdaverg} \hspace*{1cm}
v =  (n+2) \lambda -1/\lambda.
\end{equation}
In order that the front solution of these equations corresponds to a pushed
front solution of the F-KPP equation, the decay rate  $\lambda$ given by (\ref{lambdaverg})
needs to be bigger than $\lambda^*= \sqrt{\tilde{\varepsilon}}$; this
is easily found to be the case for $\tilde{\varepsilon}< 1/n$.  Thus,
the F-KPP equation with a nonlinearity $f(u)$ given by\footnote{The
prefactor of the term $u^{n+1}$ looks special, but  the analysis
covers all cases with $f(u)$ of the form \ref{fandhnonlinear}) since
all other cases can be brought to this form by a proper scaling of
space, time and $u$ --- see \cite{powelltabor} and appendix C of \cite{evs2}.}
(\ref{fandhnonlinear})     has the pushed to pulled transition at
$\tilde{\varepsilon}=1/n$, and for $ \tilde{\varepsilon}<1/n$ the pushed
front velocity $v^{\dagger} $ is given by (\ref{lambdaverg}). 

For further discussion of these results for the most common cases $n=1$ ($f$ a cubic polynomial
in $u$) and $n=2$ ($f$ a function of  $u$, $u^3$ and $u^5$), we refer to \cite{bj,vs2}. Other
examples  are discussed  in
\cite{depassier4,evs2,hereman,kaliappan,otwinowski,powelltabor,wang,yang2} and section
\ref{sectionsmectic}, where we 
will discuss a nontrivial case relevant to liquid crystals. Quite surprisingly this ``reduction
of order method'', as it is sometimes called, also allows one to find
the pushed front solutions of  the quintic CGL equation (see
section \ref{sectioncglquintic}) and  to construct
other types of exact solutions of the CGL equation  \cite{conte1,musette,nozakibekki,vsh}.
Deep down, the  method is  related to the construction of exact solutions of nonlinear 
equations using Painlev\'e analysis 
 \cite{conte1,conte2,ramani}. Also symmetry reduction
methods have been used to search for exact solutions of the nonlinear
diffusion equation \cite{clarkson,oron}.

\subsubsection{Coherent pattern forming front solutions}\label{selectioncpfs}
Let us now turn to coherent pattern forming front solutions of the
form (\ref{cfsol}). Every function
$\Phi^n_v(\zeta) $ in this expression also depends only on the
co-moving coordinate $\zeta$. Hence if we substitute the expansion
(\ref{cfsol}) into the dynamical equation under investigation, the
functions $\Phi^n_v(\zeta)$ will in general obey a set of
coupled ordinary differential equations. In general, this infinite set
of ordinary differential equations is quite cumbersome; however, for
fronts we need to have $\Phi_v^n(\zeta) \to 0$ as $\zeta \to \infty$,
and to linear order all the $\Phi^m_v$ obey essentially the same
equation obtained by linearizing the dynamical equation about the
unstable state.  We  then have the freedom to take the term with $n=1$
to correspond to asymptotic decay $\exp[-\lambda_1 \zeta]$
associated with the smallest root $\lambda_1$; in analogy with
(\ref{decayformula}) we then have
\begin{equation}
\Phi_v^1 (\zeta) \approx  A_1 e^{ik_{{\rm r},1}\zeta - \lambda_1\zeta } + A_2
e^{ik_{{\rm r}2}\zeta -\lambda_2\zeta} + \cdots , \label{Phicoherent}
\end{equation}
where now the coefficients $A_1$, $A_2$ etcetera are complex. Again it
follows from the general arguments underlying
Fig.~\ref{figvversuslambda} that for any $v> v^*$, we generally have
$\lambda_1< \lambda^*$, $\lambda_2>\lambda^*$. 

One may, like in the case of uniformly translating fronts, wonder
about the multiplicity of coherent pattern forming fronts. There are
even fewer results in this case, but indications are that for front
propagation into unstable states the typical case is that there is a
two-parameter family  of front solutions. That this is the case has
been proved for the Swift-Hohenberg equation \cite{collet} and is
also found  \cite{vsh} for the quintic CGL equation of section
\ref{sectioncglquintic}. Moreover,
counting-type arguments for the pattern-forming regime of the EFK
equation (section \ref{sectionefk})  are consistent with this ---  we expect that the
methods developed by Sandstede and Scheel \cite{sandstede2} will allow
one to establish this more generally and cleanly.
Intuitively, the existence of a  two-parameter family of front
solutions is the natural extension of the existence of a one-parameter
family of uniformly translating front solutions, since the leading
edge of pattern forming
fronts generically is   characterized by a  wavenumber $k_{\rm r}$
in addition to the steepness $\lambda$. In other words,  our
conjecture that the generic situation is that there is a two-parameter
family of coherent pattern forming fronts, means that for every
velocity $v$, there is a one-parameter family of fronts parametrized
e.g. by the wavenumber $k_{\rm r}$ in the leading edge  or by the
wavenumber of the coherent pattern behind the front. The front
solutions whose wavenumber leads to maximal growth in the leading edge
for given steepness ($\partial \omega_{\rm i}/\partial k_{\rm r}=0$),
then correspond to the left branch in Fig.~\ref{figvversuslambda}.

Let us proceed first by assuming that  indeed the equations for the
coherent pattern forming front solutions admit a two-parameter
family of front solutions.  The above considerations then imply that
there will in general exist a coherent pulled front solution, i.e., a solution whose asymptotic behavior to
the right matches the linear spreading point behavior,
\begin{equation}
\Phi_{v^*}^1 (\xi) \sim (A_1 \xi + A_2) e^{-\lambda^*\xi + i(k^*_{\rm r}
\xi - (\omega^*_{\rm r}-k_{\rm r}^*v^*)t)},~~~~~~(\xi=x-v^*t \to \infty) , \label{Phi1*eq}
\end{equation}
where  in analogy with (\ref{Phi*eq}) the term linear in $\xi$
arises because two roots coincide at $v^*$. 
The coherent pulled front solutions determined this way will be the
ones that one will observe in the regime of pulled front propagation. 

The mechanism to  get coherent pushed solutions moving at a speed
$v^{\dagger}>v^*$ is completely analogous to the one we discussed
above for uniformly translating solutions with the aid of
Fig.~\ref{figpushedscenario}: starting from steep initial 
conditions an arbitrary front solution with speed $v>v^*$ can not emerge, since
that would be incompatible with the dynamics in the leading
edge, but any special coherent front solution whose asymptotic spatial decay is
steeper in that
\begin{equation}
\Phi^1_{v^\dagger} (\zeta ) \sim  A_2
e^{ -\lambda_2\zeta +ik_{{\rm r}2}\zeta} + \cdots ,   \label{pushedcoherent}
\end{equation}
{\em can and will} invade the leading edge. In other words, if a
solution for which $A_1=0$ in (\ref{Phicoherent}) exists, {\em this is the
pushed front solution that will be selected by the dynamics}. Note that
since the coefficients $A_1$, $A_2$ etcetera in (\ref{Phicoherent})
are complex coefficients, the condition that $A_1=0$ amounts to {\em two}
conditions. Whether there are front solutions which obey this
condition depends on the equation under investigation, but since we
assume that there is a {\em  two}-parameter family of 
front solutions,  if they exist, pushed front solutions are expected to be isolated
solutions. 

This  selection mechanism for  coherent pattern forming front
solutions is clearly quite  analogous to the one we discussed for
uniformly translating fronts, but for coherent fronts I am not aware
of any rigorous work on the pulled to pushed transition of coherent
fronts. The numerical and analytical work on a modified version of the
Swift-Hohenberg equation  and on
the quintic CGL equation, summarized briefly in sections
\ref{sectionsh} and \ref{sectioncglquintic} are in full accord with
the above scenario.

It is important to keep in mind that our discussion has been based on
the idea that coherent  front solutions  come as  a two-parameter
family  {\em if they exist.} For a particular equation, it is not
guaranteed that they do exist, of course. Indeed, the quintic CGL equation discussed in  section
\ref{sectioncglquintic} illustrates this: In some parameter ranges one
can show that the equation does not admit any coherent pulled front
solution.  In the parameter range where this happens, the pulled
fronts become incoherent --- see Fig.~\ref{figcglcubic}.
Likewise, if the coherent front solutions do not come as a
two-parameter family, then neither coherent pulled front solutions nor
coherent pushed front solutions are 
expected to exist: the dynamics is then expected to be incoherent.

We finally note an important point. We already noted above that
 the condition
$A_1=0$ is equivalent to {\em two}  conditions, and that this implies
that pushed fronts solutions, if they exist, generically come as
isolated (discrete) solutions.  Moreover, this simple counting argument also shows that the 
dominant  spatial decay  rate $\lambda_2$ of this isolated solution
(\ref{pushedcoherent}) {\em generally does not lie } on the thick dashed
branch in Fig.~\ref{figvversuslambda}.\footnote{In \cite{vs2}, we
suggested that the combinations $v^\dagger, \lambda_2$ of pushed
fronts of the Swift-Hohenberg equation should lie on the dashed
branch. This suggestion is wrong.}  By the same token, this
implies that at the pulled to pushed transition, i.e. when $v^\dagger
\downarrow v^*$, $\lambda_2 \neq \lambda^*$ in
(\ref{pushedcoherent}).  So, while 
for uniformly translating profiles
the pulled to pushed transition corresponds to a continuous transition
in the front shape, for coherent pattern forming fronts the transition
is discontinuous in the front shape!  Such discontinuous behavior,
which was
first discovered  for the quintic CGL equation of  section
\ref{sectioncglquintic}, imply that ``structural stability''
conjectures do not apply to pattern forming fronts. We will come back
to this in section \ref{structuralstability}. 

Examples of pushed
coherent fronts will be encountered later in sections
\ref{sectionefk}, \ref{sectionsh}, \ref{sectioncglquintic},
\ref{sectionsupercond}, and \ref{sectionmullinssekerka}; we also
conjecture that pushed coherent fronts can arise in the models of 
\ref{sectionpropagatingrt} and \ref{secpropRayleigh}, for which so far
only pulled fronts have been found.

\subsubsection{Incoherent pattern forming front solutions}\label{sectionincoherentpffs}

Incoherent pattern forming fronts are those fronts which do not leave
behind a coherent pattern. Hence they can not be of the form
(\ref{cfsol}).  From the numerical simulations of the
cubic and quintic CGL equation discussed in sections
\ref{sectioncglcubic} and \ref{sectioncglquintic} and the full-blown
numerical simulations of turbulence fronts discussed in section
\ref{sectionparallelshear} we know that both pulled and pushed
incoherent fronts can  exist. Of course, as always we can calculate
the speed of incoherent pulled fronts, but to my knowledge there is
essentially no good understanding  of what drives an incoherent front to be
pushed.  Thus I can not give a precise mathematical formulation,
analogous to (\ref{pushedexpr}) or (\ref{pushedcoherent}),  that
identifies what property will give rise to a 
pushed incoherent front. My conjecture is that in analogy with what we
found for uniformly translating and coherent pattern forming fronts,
incoherent pulled fronts are solutions which in some average sense do
fall off with a well-defined steepness larger than $\lambda^*$. The simulations of
turbulence fronts discussed in section \ref{sectionparallelshear}
confirm this idea, but I  do not know
how to give {\em predictive power}  to this statement. Unfortunately, the
problem of the transition from pulled to pushed incoherent fronts
appears to be as hard as the spatio-temporal chaos and turbulence
problem itself!

Examples of pushed incoherent fronts will be encountered in
\ref{sectionks}, \ref{sectioncglquintic}, \ref{sectionparallelshear}
and \ref{sectionerrorpropagation}.

\subsubsection{Effects of the stability of the state generated by the front}\label{sectioneffectsofstab}

In our discussion of how pushed fronts can arise in the various modes
of front propagation,  a central role is played by the  {\em existence} of the special
type of uniformly translating or coherent pattern forming solutions
which for $\zeta \to \infty$ decay with steepness larger than
$\lambda^*$ --- these become the pushed front solutions. To what extent
does the stability of fronts play a role? As we will discuss in
section \ref{stabilityandothermechanisms} below, one can relate the
above selection mechanism of 
pushed fronts to the  stability properties of front solutions with
$v<v^{\dagger}$, but it is more appropriate  to think of this  as a
{\em consequence}
rather than a {\em cause} of front selection.

The stability or instability of the state generated by the front does
have important consequences for the dynamics, however. The examples
discussed in section \ref{section4thorder} will illustrate this most clearly: it is
quite  possible that the state which emerges behind the front is
itself unstable. When the selected state is nonlinearly convectively unstable in
the frame moving with the front, the
state behind the first front is  invaded by a second front which moves
slower than the first. If, however,  the state generated by the first
front is absolutely unstable in the frame moving with the first front,
then the second one will catch up with the first and alter its
properties. In the examples we will discuss the second front is
usually  an incoherent front, and when the one with which it catches
up  is a coherent front, this induces a transition from coherent to
incoherent front dynamics in the leading front.  In practice,
therefore, what type of front dynamics one will get is
{\em determined both by the existence of front solutions and by the
stability of the state these solutions generate.}

\subsubsection{When to expect pushed fronts?}\label{sectionwhenpushed}
The requirement for existence of a pushed front solution,  $a_1=0$ according to Eq.~(\ref{pushedexpr}) 
 for
a uniformly translating front solution to 
exist, or $A_1=0$ according to Eq.~(\ref{pushedcoherent}) 
for a coherent pattern forming solution, is a
condition  on the global properties of the solution of a nonlinear
ordinary differential equation. Hence whether a pushed front solution
exists depends on the full nonlinear properties of the dynamical
equation we wish to investigate. There does not appear to be a general mathematical
framework that allows us to predict from the appearance or the  global structure of an
equation whether a strongly heteroclinic front solution does exist, and hence whether
fronts will be pulled or pushed.
For the F-KPP equation one can derive
general conditions  on the nonlinear function $f(u)$ such that the
selected fronts
in the equation are pulled \cite{depassier1,depassier2}. One of the
simple results is  that when $f(u)/u \le f^\prime (0)$ the fronts are
pulled, which confirms our  intuitive understanding that for a front to
become pushed one needs  the 
nonlinearities to enhance the growth. For (sets of) equations
that admit uniformly translating fronts one typically finds the same
trend, namely  that the enhancement of the growth by  the nonlinear terms in the dynamical equations
tends to favor the occurrence of pushed fronts.  

It is important to realize, however, that for coherent and incoherent pattern forming
fronts, these intuitive ideas do not necessarily apply. Several of the examples to
be discussed in section \ref{section4thorder} will illustrate this: in
the quintic CGL equation the cubic term enhances the growth, but there
are large regions of parameter space where nonlinear dispersion
effects completely suppress the occurrence of pushed fronts. Likewise,
fronts in the Kuramoto-Sivashinsky equation of section \ref{sectionks} are pulled, but if one
adds a linear term to the equation, a pulled to pushed
transition is found.  However, in dynamical pattern forming
equations which derive from a Lyapunov functional  and for which one
has an reasonable  understanding of whether the nonlinearities enhance
or suppress the growth about the unstable state,  one's intuition of
what to expect is usually correct. Our discussion of front dynamics in
the  Swift-Hohenberg and Cahn-Hilliard equation will illustrate this.

As the example discussed at the end of section \ref{selectionutfs} illustrated,
for the  second order partial F-KPP equation, the pushed front
solutions can in a number of cases be explicitly constructed 
analytically. The reason is the following. While for an arbitrary $v$
the large-$\zeta$ asymptotics (\ref{decayformula}) is characterized by
two 
different exponentials, the pushed front solutions (\ref{pushedexpr})
of the F-KPP equation are characterized by one root. It is therefore
possible  (but not necessary) that   the pushed solutions  obey a
first order equation. The ``reduction of order method'' briefly
reviewed in section \ref{selectionutfs}
 is based on substituting a
first order Ansatz into the second order ordinary differential
equation for the uniformly translating solutions. The most remarkable
success of this method is that the pushed solutions of the quintic CGL
equation have been found this way --- see section \ref{sectioncglquintic}.

\subsubsection{Precise determination of localized initial conditions
which give rise to pulled and pushed fronts, and leading edge dominated
  dynamics for non-localized initial conditions}\label{sectionleadingedgedominated}

So far, we have focused on the front dynamics
emerging from  initial conditions which are ``steep'', in the sense
that they fall off faster than $\exp[-\lambda^* x]$. This is because
only for such initial conditions does the front selection problem have
a sharp and unique answer --- whether fronts are pushed or pulled is an {\em  inherent}
property of the dynamical equations. Indeed,  the values of the 
corresponding velocities   $v^*$ and $v^\dagger$ are determined
completely  by the equation itself. 

What happens if the initial conditions are not steep, i.e., fall off
slower than $\exp[-\lambda^* x]$? Given our assumption of locality of
our dynamical equations (no influence of from points arbitrarily far
away)  the answer lies hidden again in our analysis of the linear
dynamics illustrated in Fig.~\ref{figlinearcrossover2}. If initial
conditions fall off with  steepness $\lambda< \lambda^*$, the spatially
slowly decaying leading edge on the right, which moves with velocity $v_{\rm
env}(\lambda)$, expands in time. In other words, in the frame moving
along with this leading edge, the crossover region  to the slower part of the
profile  recedes to the left  and larger and larger parts of the
profile move with asymptotic velocity $v_{\rm env}$. Whatever the
nonlinear dynamics is, it  must be compatible with this dynamical
constraint in the leading edge of the profile. Indeed, such initial
conditions which are not steep necessarily lead to a front moving with speed
$v_{\rm env}(\lambda)$ with $\lambda<\lambda^*$, unless a pushed front
solution exists whose speed $v^{\dagger}$ is larger than $v_{\rm
env}$. For, if such a pushed front solution with $v^\dagger > v_{\rm
env}(\lambda)$ exists, this solution will invade the leading edge
according the the mechanism sketched in
Fig.~\ref{figpushedscenario}{\em (b)}. Thus we conclude:

\begin{tabular}{p{0.1cm}|p{13.2cm}} &
{\em For an equation whose fronts are pulled, all initial
conditions with steepness $\lambda< \lambda^*$ lead to fronts moving
with speed $v_{\rm env}(\lambda) > v^* $. In other words, in the
pulled regime we can identify 
``localized'' initial conditions  which lead to pulled fronts with
initial conditions which are steep, i.e., which fall off faster than
$\exp[-\lambda^* x]$.} 
\end{tabular}
\begin{tabular}{p{0.1cm}|p{13.2cm}} &
{\em For an equation whose fronts are pushed, only initial conditions with
steepness $\lambda<\lambda_1$ lead to fronts moving with speed $v_{\rm
env}(\lambda) > v^\dagger$, where $\lambda_1$ is the smallest root for
which $v_{\rm env}(\lambda_1) = v^{\dagger}$. In other words,
localized initial conditions which lead to pushed fronts are initial
conditions which fall off faster than $\exp [-\lambda_1 x]$, where
$\lambda_1$ is determined implicitly by the pushed front solution
through the requirement $v_{\rm env}(\lambda_1) = v^\dagger$.  }
\end{tabular}

Intuitively the mechanism through which non-localized initial
conditions lead to fronts that move faster than the naturally selected
pulled or pushed speed, is very much like ``pulling along'' the
nonlinear front. However, in order to distinguish them from the pulled
front solutions which naturally emerge from all localized initial
conditions, we  refer to this type of dynamics more generally as ``leading edge
dominated'' dynamics \cite{evs2}. 

\subsubsection{Complications when there is more than one linear spreading point} \label{sectiontwospreadingpoints}

In our discussion of the pushed fronts, we have so far assumed that
the linear dispersion is such that the $v_{\rm env}(\lambda)$ versus $\lambda$
diagram is of the type sketched in Fig.~\ref{figvversuslambda}, i.e.,
that there is only one branch which in the limit $\lambda \downarrow
0$ corresponds to a positive growth rate $\omega_{\rm i}$ and hence to
a divergent $v_{\rm env}(\lambda)$. This is the normal situation for
problems where there is essentially one branch of linear modes which
is unstable.  We now briefly discuss the  subtleties associated with
having more than one unstable branch of the dispersion relation.

\begin{figure}[t]
\begin{center}
 \epsfig{figure=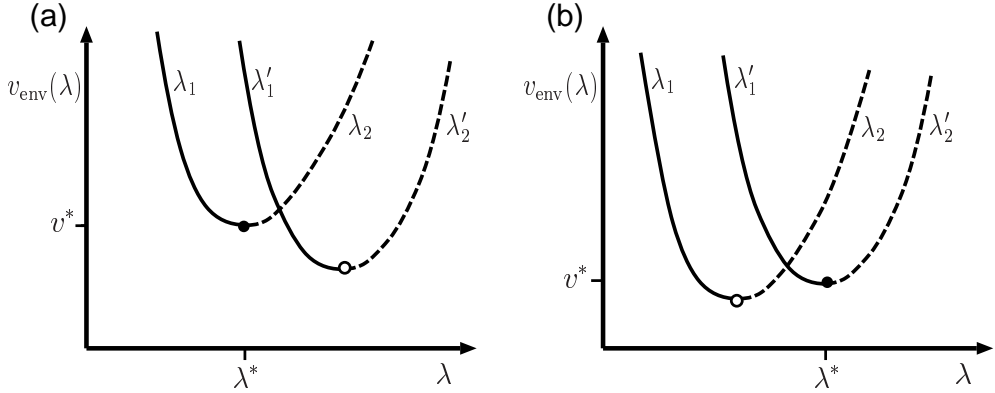,width=0.98\linewidth} 
\end{center}
\caption[]{Illustration of  possible $v_{\rm env}(\lambda)$ versus
$\lambda$ diagrams  in the case in which
there is more than one unstable branch of modes. The relevant linear
spreading velocity in each case is indicated with a filled circle. As
explained in the
text, the asymptotic steepness of a pushed front has to be associated
with one of the two dashed branches. }\label{figtwospreadingpoints}
\end{figure}

Two
examples of possible   $v_{\rm env}(\lambda)$ 
diagrams for  the  case of two unstable branches  are sketched qualitatively in
Fig.~\ref{figtwospreadingpoints}.
As we discussed before, according to the fully linear spreading
analysis, the relevant linear spreading velocity $v^*$ is the largest
spreading velocity 
of the two branches. The crucial point to understand for the proper
extension of the concept of a pushed front in such cases is that the
linear dynamics associated with the two individual branches is
completely independent --- we can repeat the earlier discussion of how
pushed fronts can emerge for each branch individually, and this leads
to the conclusion that a pushed front is again a front whose
asymptotic exponential decay is faster than
$e^{-\lambda^*\zeta}$. However, 
 in addition the exponential steepness $\lambda^{\dagger}$ of the
pushed front
should be larger than the value $\lambda^*_1$ or $\lambda^*_2$
associated with the minimum of the
branch to which it belongs \cite{chomaz}.\footnote{Another way to arrive at the same
conclusion, stressed in particular by Chomaz and Couairon
\cite{chomaz}, is to note that asymptotic behavior with steepness
given by one of the two left branches $\lambda_1$ or $\lambda_2$ can
only emerge dynamically from non-localized  initial conditions  falling off
with this steepness:  Chomaz and Couairon \cite{chomaz} call these branches
non-causal. In the language of \cite{vs2} one can understand this from
the fact that on these branches the group velocity is less than the
envelope velocity --- perturbations can not work their way up towards
the far tip.}.  In other words, the leading steepness of a pushed
front should correspond to\footnote{In the case of coherent pattern forming front solutions,
``correspond to'' should be read as ``analytically connected to'', since as we explained in section
\ref{selectioncpfs}  the asymptotic decay of a pushed front does not correspond to a value on the dashed
branch of $v_{\rm env}$. } one of the two dashed branches in the
figure. In the case of 
Fig.~\ref{figtwospreadingpoints}{\em (a)}, a uniformly translating pushed front solution is
a solutions whose 
 asymptotic decay  is as $e^{-\lambda_2 \zeta}$, not as
$e^{-\lambda^{\prime}_1\zeta}$. Likewise, for the case sketched in
Fig.~\ref{figtwospreadingpoints}{\em (b)}  a
transition from the pulled front with  velocity $v^*$ to a pushed one,
whose asymptotic behavior is as $e^{-\lambda_2\zeta}$, could 
occur.  

Like before, the possibility of such situations to occur is
intimately related with counting arguments for the multiplicity of
front solutions. E.g., for uniformly translating fronts with a
dispersion relation consistent with one of the cases sketched in
Fig.~\ref{figtwospreadingpoints}, one in general expects there to be a
two-parameter family of front solutions. For the case of
Fig.~\ref{figtwospreadingpoints}{\em (a)} pushed front solutions with
the required asymptotic behavior are then expected to come as isolated
front solutions, just as before.

\subsection{Relation with existence and stability of front stability and relation with previously
proposed  selection mechanisms}  \label{stabilityandothermechanisms} 

In this section, we briefly discuss the relation between front
selection and stability, and the relation with some of the older
proposed selection mechanisms.  

\subsubsection{Stability versus selection}\label{stabvsel}

The main difference in  perspective with the  ``marginal
stability'' selection mechanism of the 1980ies \cite{bj,dl,vs1,vs2}
is that we  here emphasize that by starting from the classical linear spreading analysis
\cite{bers,briggs,ll} the concept of a linear spreading velocity $v^*$
naturally arises. This concept holds  irrespective of whether or
not there
are well-defined front solutions, and applies equally well to partial
differential equations, difference equations and  integro-differential
equations. From this perspective, the possibility of having pulled
fronts arises most naturally and independently of whether they are
coherent or incoherent. In addition, the saddle point
integration immediately shows that the intermediate asymptotic
dynamics giving the approach to the asymptotic spreading speed is
that of a diffusion equation, even if the equation itself is not at
all a simple diffusion equation. This latter fact lies at the basis of
the universal relaxation behavior discussed in the next section.
Moreover, by exploring the constraints 
imposed by the linear dynamics, the essential properties of pulled
fronts follow. 

The drawback of the  ``marginal stability'' formulation of \cite{vs1,vs2}
is that when one wants to relate selection with the stability
properties of fronts, one can at best only understand the selection
once one is already close to an asymptotic front solution.\footnote{
Booty {\em et al.} \cite{booty} have also analyzed the dynamics in the
leading edge of the F-KPP equation. Conceptually, their approach has
much in common with that of \cite{vs1,vs2}.  } Why the
intermediate asymptotics bring one  there is less clear --- even
though the two are quite consistent 
the attempt in the marginal stability approach to treat selection and
linear spreading in  one fell swoop makes the problem unnecessarily
cumbersome and masks the generality of the linear spreading
concept. Likewise, the universal relaxation behavior of pulled
fronts that we will discuss in the next section is virtually
impossible to get from  a stability analysis (for uniformly
translating or coherent pulled fronts, the spectrum of the stability
operator is continuous with arbitrarily small eigenvalues), whereas it
naturally emerges from the linear spreading concept.

Our simple observation that {\em whenever a solution (\ref{pushedexpr}) or
(\ref{pushedcoherent}) with $v^\dagger > v^*$ exists whose 
asymptotic steepness is larger than} $\lambda^*$, this solution {\em can and will}
invade the leading edge and lead to pushed front propagation, is
actually a  general formulation which encompasses  the ``nonlinear marginal
stability'' scenario  that when one considers fronts as a function of
there velocity, front solutions with speed $v< v^\dagger$  are
unstable \cite{bj,vs2}. One can  see this as follows (\cite{evs2}, appendix I).  For
dynamical  equations which are translation invariant, the translation
mode is always a mode with zero eigenvalue of  the stability
operator. At $v=v^\dagger$ the front solution is a strongly
heteroclinic orbit, and hence the zero eigenmode of the stability
operator corresponds to a strongly heteroclinic orbit.  Continuity and
counting arguments then imply that for $v$ close to but different from
$v^\dagger$, there is a strongly heteroclinic solution of the
stability operator whose eigenvalue crosses zero at $v=
v^\dagger$. In other words, when the stability of the front solutions
is studied for varying velocity $v$, the stability changes at
$v^\dagger$ --- front solutions with velocity $v< v^\dagger$ are
unstable due to the invasion of the pushed front solution into the
leading edge.

Another reason for  separating front stability from front selection
is that the simulations discussed in section \ref{section4thorder}
will show that the state generated by the front becomes absolutely
unstable in the frame moving with the front,  this entails a
transition from a coherent pattern forming front to an incoherent
pattern forming front, not necessarily a change from pulled to pushed
dynamics.

\subsubsection{Relation between the multiplicity of front solutions
and their stability spectrum} \label{stabilitygeneral}
The above discussion of the implication of the existence of a pushed
front solution for the stability of fronts  also illustrates that there generically is an intimate
connection between the multiplicity of uniformly translating and
coherent front solutions and the
stability properties.  We can illustrate this for the  other stability
modes
as follows. As we discussed in 
the previous section, a uniformly translating or coherent front
solution corresponds to the an orbit in the phase space of the
ordinary differential equations that govern these solutions, and the
multiplicity of these solutions is determined by ``counting
arguments'' for the dimension of the manifolds that flow into and out of
the fixed points that correspond to the asymptotic states.  Consider
now the case in which the dynamical equations admit a one-parameter
family of front solutions, parametrized by their speed $v$  --- we argued
that this is the usual case for 
front propagation into unstable states. If we now pick an arbitrary
front solution from this family and write down the stability operator
for perturbations about the front solution, then the counting arguments for the
stability modes is essentially unchanged because the linearization
about the asymptotic states before and after the front is
unchanged. Hence the existence of a one-parameter family of fronts,
parametrized by $v$, generically implies that if we fix $v$ and
consider the stability modes, then we expect there to be in general a
one-parameter family of stability modes, parametrized by their
growth/decay rate. Together with the fact that translation invariance
of the dynamical equation implies that there is a zero mode of the
stability operator, this also implies that there generically is a continuous
spectrum of the stability operator near zero.  A general analysis of the
asymptotic behavior of these modes about the unstable state shows that
 the modes from this spectrum which decay faster than the front
solution are stable, and that those which decay less fast than the
front solution are unstable \cite{vs2}. This line of analysis in
combination with the one above for the possible presence of a localized stability mode
if a pushed front solution exists, gives a quite complete generic
picture of the stability of the front solutions in the generic
case. In short: if no pushed front solution exists, so that the selected
fronts are pulled, then the front  solutions with $v>v^*$ are stable to
perturbations which decay faster than the front itself, and unstable
to those which spatially decay less fast than the front itself. If a
pushed front solution exists, then the generic picture is that 
front solutions with $v< v^\dagger$ are unstable to the localized
mode, and those with speed $v> v^\dagger$ are stable to this mode and
to perturbations whose spatial decay to the right is faster than that
of the front solution itself.

For coherent pattern forming fronts, similar arguments apply. Let us
again focus on the generic case (see section \ref{selectioncpfs}) that
these front solutions come into a two parameter family. If we consider
a particular front solution at a fixed velocity $v$, then the generic
scenario that results from similar continuity arguments is that any
arbitrary front solution which in the tail does not match up with the
maximal growth rate $\omega_{\rm i}$ in the leading edge, or, in other
words, whose asymptotic behavior does not correspond with the left
branch drawn with a full line in Fig.~\ref{figvversuslambda}{\em (a)}, is
unstable. This is the reason we focused on the analysis of $v_{\rm
env}$  defined in 
 Eq.~(\ref{venv}) in section \ref{sectionexponential}: only that
branch matches up with asymptotic coherent front solutions which are
stable to perturbation that decay spatially faster to the right than the front
solutions themselves. The discussion of stability of a possible
localized stability mode, associated with the existence of a pushed
front solution, is analogous to the one given above. 

The explicit stability calculation for the uniformly translating 
fronts in the F-KPP equation or extensions of it can be found in a number of papers
\cite{bricmont,stability1,evs2,stability6,stability7,stability5,stability2,stability4,rottschafer2,stability3}
and will not be repeated here. The results are completely in accord  with the
above general discussion.

\subsubsection{Structural stability}\label{structuralstability}

We finally comment briefly on the proposal to connect
propagating front selection  with   ``structural stability''
ideas. According to this conjecture \cite{paquette3,paquette2}, the pulled front is the natural
front speed as it is the only front solution which is ``structurally
stable'' to small changes in the dynamical equations (like those which
would suppress the instability or make the state $\phi=0$ even
linearly stable). It is easy to convince oneself that uniformly
translating pulled solutions do have this property, as the dynamically
relevant front solutions are characterized by a {\em real} spatial
decay rate only, but that as we discussed at the end of section
\ref{selectioncpfs} coherent pattern forming front solutions which
are characterized by a decay rate and wavenumber in the leading edge,
generically do {\em not}   have this property. Indeed, the quintic CGl
equation provides an explicit counterexample to the ``structural
stability'' postulate: As we shall see under
{\em (iii)} in 
section \ref{sectioncglquintic}, for the quintic CGL equation the selected wavenumber can
jump at the pulled to pushed transition. 

\subsubsection{Other observations and conjectures} \label{sectionotherconjectures}

The issue of front selection issue has intrigued many authors, so various other observations
and conjectures  have been made. 
In appendix \ref{otherconjectures} we briefly discuss some of these: a
(wrong) conjecture  concerning the
analytic structure of pushed front solutions, an observation
about obtaining the selected uniformly translating
front by studying the front solutions on a finite interval, and the connection with Hamilton-Jacobi
theory and renormalization group ideas.

\subsection{Universal power law relaxation of pulled
fronts}\label{sectionunirelsimple}

Up to now, we have focused on the asymptotic front velocity. Let us  now
assume we study an equation whose fronts are pulled, and ask how the
asymptotic front velocity  is approached. To do so, we first have to
state how we define a time-dependent velocity of a front during the
transient regime when it approaches its asymptotic value. We will be
quite pragmatic in our discussion:  the
asymptotic convergence to $v^*$ will turn out to be very slow, so slow
that the differences between various conventions do not really
matter. We will therefore focus simply on the most natural definition.

In this section, we will simply state the results for the convergence
to $v^*$; the essential ingredients of the derivation are reviewed in
section \ref{sectionuniversalrel}. 

It is good to stress that the universal relaxation only holds for
pulled fronts. The discussion of section \ref{stabilityandothermechanisms} implies that the stability
spectrum of a pushed front solution is gapped, and hence that a pushed
front relaxes exponentially fast to its asymptotic velocity and
shape \cite{evs2,kessler,vs2}.
 This is in line with the intuitive idea illustrated in Fig.~\ref{figpushedscenario}
that a pushed front invades the region ahead of it with a finite speed.

\subsubsection{Universal relaxation towards a uniformly translating
pulled front}\label{relaxtoutpulled}

If a
pulled front is asymptotically uniformly translating, we can simply
follow the same idea as in our discussion of the linear spreading
problem in section \ref{sectionv*}: In essence our convention will
always be to  determine a front speed $v_C(t)$ by 
tracking the position $x_C(t)$ of the level line where the dynamical
field $\phi$ reaches a level $C$; the velocity is then the speed of
this point
\begin{equation}
 \phi(x_C,t)=C  \hspace*{1.0cm}\Longleftrightarrow \hspace*{1.0cm}  v_C(t) = \frac{dx_C}{dt} .
\end{equation}
Of course, for dynamical fields with more than one component, the
transient velocity could in principle depend on which component we
track, but we will not distinguish this possibility notationally, as
we will focus on those aspects which are independent of those details.

 The  analysis \cite{evs2} of the convergence of the
nonlinear pulled front speed to $v^*$  is based on matching the
behavior in the leading edge, where corrections to the asymptotic
exponential behavior $e^{-\lambda^* \xi }$ are  governed by an
equation which in dominant order is a diffusion-type equation, to the
behavior in the fully nonlinear region. In this step, the fact that
the asymptotic front solution $\Phi_{v^*}(\xi)$ has according to (\ref{Phi*eq})
a $\xi e^{-\lambda^* \xi} $ behavior plays a crucial role --- see
section \ref{sectionuniversalrel}. The final result of the matching
analysis \cite{evs2} is the following {\em exact}  expression for the velocity
$v(t)$  of an asymptotically uniformly translating front,
\begin{equation}
v(t) = v^* - \frac{3}{2\lambda^*t} + \frac{3 \sqrt{\pi}}{2 \sqrt{D}
(\lambda^*)^2 t^{3/2} } + {\mathcal O}\left( \frac{1}{t^2}\right) , \label{v(t)relaxation}
\end{equation}
which holds provided one starts from steep initial
conditions which fall off faster than $e^{-\lambda^*x}$.\footnote{To be
more precise, the result holds provided the initial 
conditions fall off exponentially faster than $\exp[-\lambda^*x]$, i.e., provided
there is some $\delta>0$ such that
$\phi(x,t=0)e^{\lambda^*x} < e^{-\delta x}$ as $x\to \infty$.  The
special case in which $\phi(x,t=0)\sim x^\nu  e^{-\lambda^*x}$ is
discussed  in \cite{bramson,evs2,needham2,needham3,needham1}.}\label{footnotelabel1}
This slow power-law like relaxation of the velocity to its asymptotic
value $v^*$ entails a slow relaxation of the front profile to its
asymptotic shape. Indeed, if we define $X(t)$ as the shift of the
front position in the frame moving with speed $v^*$, 
\begin{equation}
X(t) = \int^t dt^\prime \, (v(t^\prime) -v^*) 
\hspace*{0.7cm}\Longleftrightarrow \hspace*{0.7cm} X(t) \simeq
-\frac{3}{2\lambda^* } \ln t + {\mathcal O} (t^{-1/2}),
\end{equation}
then one finds  for the relaxation of the
front profile to its asymptotic shape
\begin{equation}
\phi(x,t) \simeq \Phi_{v(t)} (\xi_X)   + {\mathcal O}\left(
\frac{1}{t^2}\right) ,\hspace*{1cm} \xi_X \ll \sqrt{Dt}, \label{phirelaxshape}
\end{equation}
where $\xi_X$ is the frame
\begin{equation}
\xi_X \equiv \xi - X(t) = x -v^* t - X(t)
\end{equation}
which includes the logarithmically increasing shift $X(t)$.

\begin{figure}[t]
\begin{center} \hspace*{-1.5cm}
\epsfig{figure=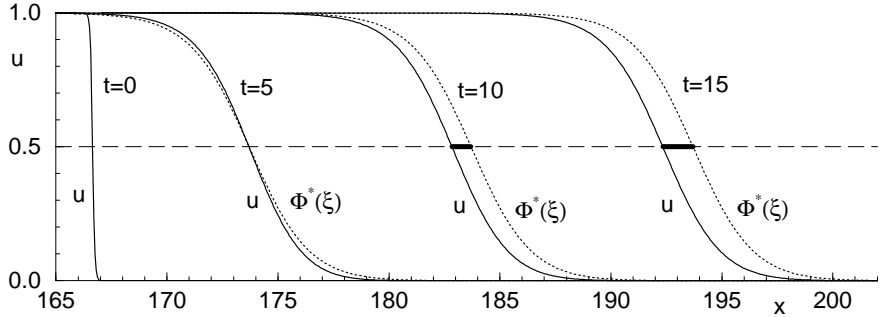,width=0.75\linewidth,bbllx=103pt,bblly=85pt,bburx=610pt,bbury=281pt}
\end{center}
\caption[]{Illustration of the logarithmic shift of the transient
profile relative to the asymptotic profile $\Phi_{v^*} $ moving with
constant speed $v^*$, for the case of the F-KPP equation with
$f(u)=u-u^3$. The solid line  shows the actual shape and position of the
front emerging from an initial condition at time $t=0$ at three
successive times. The dotted line shows the uniformly translating
solution $\Phi_{v^*} $ (labeled $\Phi^*$ in the figure) at three
successive times; at $t=5$ this solution was placed such that it
intersects the actual solution at $u=0.5$. The thick solid line
indicates that the transient solution and the asymptotic solution
separate more an more as time goes on, even though the shape of the
transient front is alway close to the asymptotic one. From \cite{evs2}. }
\label{figfkpprelaxation}
\end{figure}
There are many points to note about these universal relaxation results
for pulled fronts and the physical
picture underlying their derivation:
\begin{enumerate}
\item While the above results are exact for all pulled fronts which asymptotically are uniformly
translating, for the special case of the F-KPP equation, the $1/t$
term was known for over 20 years \cite{delft,delft2,mckean} and was derived rigorously
in 1983  by Bramson \cite{bramson}.  Since then it  has been re-derived 
 by various methods (including a matching analysis of \cite{derrida} similar
to ours), but to my knowledge even for the
F-KPP equation the universal $1/t^{3/2}$ term was not known before the matching analysis
\cite{evs2} described in more detail in section \ref{sectionuniversalrel}. 
\item
The dominant logarithmic shift in $X(t)$  which is driven by   the
diffusive dynamics in the leading edge implies that when viewed in the
frame moving with the asymptotic velocity $v^*$, the front moves back
 while it is relaxing. This is illustrated in
Fig.~\ref{figfkpprelaxation} for a simulation of the F-KPP equation
starting from a localized initial condition. The plot shows the actual
front profile at times $t=5$, 10 and 15 (full line), and compares
these with the uniformly translating asymptotic profile $\Phi_{v^*}$
(dotted line)
which has been placed so that it intersects the transient profile at
$\phi=1/2$ at time $t=5$. The increase of the length of the thick line
with time illustrates that the actual transient profile and the
asymptotic profile $\Phi_{v^*}$ separate more and more in time. 
\item 
The results which are illustrated in Fig.~\ref{figfkpprelaxation} show
quite clearly that any method based on linearizing about the
asymptotic front solution $\Phi_{v^*}$ will not work: the differences
between the actual front solution and this  asymptotic one grow
arbitrarily large! This is why the insight to use the logarithmically
shifted frame $\xi_X$ is crucial for any theoretical analysis.
\item
Nevertheless, the figure does confirm   that already quite soon  the
transient front {\em shape}  is close to the asymptotic shape
$\Phi_{v^*}$.  This fact is expressed in
precise mathematical terms by 
 (\ref{phirelaxshape}): to order $t^{-2}$ {\em the  front shape is
given by the expression for the uniformly translating front solution
$\Phi_v$, provided we use  the instantaneous value $v(t)$ given by
(\ref{v(t)relaxation}) for the velocity, and put the front at the
appropriately shifted position $\xi_X$.}  In other words, to
${\mathcal O}(t^{-2})$ the transient front shape  follows the uniformly translating solutions
adiabatically.
This fact was first noted
empirically in simulations by Powell {\em et al.}  \cite{powell}.
\item 
The above results are valid for any pulled front which asymptotically is
uniformly translating: {\em it is independent on the precise
nonlinearities and applies to all pulled fronts  in equations for
which the linear spreading speed $v^*$ can be determined   according to
our discussion of section \ref{sectionmoregeneral}. } The fact that these results apply
equally well to difference equations  like  the finite difference
version (\ref{finitedifferencefkpp}) of the F-KPP equation or the
difference-differential equation discussed in section \ref{sectionvanzon}, is that the
saddle point expression for $v^*$ ensures that the corrections to the
asymptotic exponential $e^{-\lambda^* x}$ behavior are governed by a
diffusion-type equation. In other words, as  these corrections become
arbitrarily smooth for long times,  the lattice effects only give
higher order corrections \cite{evs2,evsp}. See section
\ref{sectionuniversalrel} for further details.
\item
Since the above results are independent of the precise form of the nonlinearities in the
dynamical equation as long as the fronts are pulled, one may wonder
where the nonlinearities are hidden. Comparison with the result
for the convergence to $v^*$ according to the
fully linear dynamics discussed in section
\ref{twofeatures} shows that the prefactor 3/2 of the $1/t$
term and  the subdominant $t^{-3/2}$ term both reflect  the nonlinear
behavior.\footnote{As we will show in section \ref{twofeatures},
Eq.~(\ref{xClinear}),
according to the fully {\em linear} dynamics a level line asymptotically recedes as $-(\ln
t)/2 \lambda^*$ in the frame moving with $v^*$, rather than as the $-3(\ln
t)/2\lambda^*$ term of the nonlinear front profile. Hence if we would
also draw in Fig.~\ref{figfkpprelaxation} the profile as it evolves according to the
linearized equation, it would for large times   and not too large
$u$ lie  in between the
asymptotic front profile $\Phi_{v^*}$ and the actual transient
profile. }
\item
In line with the earlier conclusion  of section
\ref{sectiontransients} that the limits
$t\to\infty$, $\xi$ fixed and $\xi\to \infty$, $t $ fixed do not
commute, we stress that the expression (\ref{phirelaxshape}) is the correct
asymptotic expression for $t\to \infty$, $\xi_X $ fixed. At
$\xi_X\simeq \sqrt{Dt}$, there is a crossover to a different
asymptotics that governs the large $\xi_X$ limit \cite{evs2}.
\item
Note that the above result for the universal velocity relaxation holds
independent of the initial conditions
  and of the level line which is used to track the
position of the front, provided one is not at the pulled to pushed
transition.\footnote{At the pulled to pushed transition the
prefactors of the $1/t$ and $1/t^{3/2}$ term are different --- see
appendix G of \cite{evs2}.}\label{footnotelabel2}  In fact, the correction to the velocity
relaxation associated with the shape relaxation is according to
(\ref{phirelaxshape}) of order $[\delta
\Phi_v(\zeta) /\delta v] dv(t)/dt \sim t^{-2}$.  
\item
The terms displayed in  (\ref{v(t)relaxation})  for $v(t)$ and in
(\ref{phirelaxshape}) for the
relaxation of the front profile are also the {\em only} universal terms. This can easily be seen as follows.
Suppose we compare the velocity formula for two cases, one starting from some steep 
initial condition at time $t=0$ and the other one by viewing the
dynamical state at time $t=\Delta t$ as the  initial condition.   For
large times the $1/t$ terms for the two cases differ of order $\Delta
t/ t^2$. Thus the term of order $1/t^2$ depends on the initial
conditions, i.e., is non-universal.
\item
The formula (\ref{v(t)relaxation}) shows that the asymptotic velocity
is always approached {\em from below}. This explains why in many
finite time simulations of pulled fronts the published velocity data
are slightly below $v^*$ (we will encounter several examples of this
later). This fact together with Eq.~(\ref{phirelaxshape}) for the
shape relaxation of the front imply that for the transient dynamics
the front solutions  with speed $v<v^*$  are important. This is the
reason for our cautionary note in section \ref{sectionexponential} about using the phrase
``the minimum velocity''. 
\item
Extensive numerical investigations and illustrations of the universal relaxation
behavior of uniformly pulled fronts can be found in
\cite{evs2,evsp}. An example of such tests will be discussed in
section \ref{sectionmatching1}.

\end{enumerate}

\subsubsection{Universal relaxation towards a coherent pattern forming
pulled front}\label{relaxtodpfpulled}

How to define  the instantaneous front velocity for a 
 coherent pattern forming fronts is   subtle issue. 
 If one traces the position of the foremost point where a
dynamical field which develops oscillations reaches a given value,
this position will make finite jumps when a new oscillatory part in
the leading edge grows large enough that it reaches the predetermined
level $C$. Since the time between successive jumps will be finite,
averaging over some finite time then already gives a crude idea of the slow
long-time convergence. A better way  is to determine numerically an envelope of the front
profile from traces of  e.g. the maxima of the oscillations during one
period, and to then determine the velocity from the positions of a level line of this
empirical envelope of the front \cite{storm1,willem1}. The advantage of this method is that
it also works well when one wants to trace level lines in the range where the nonlinearities
in the dynamical equation are clearly important. If, on the other hand, one decides to track 
the front velocity in the leading edge only, then a good method is to fit a decaying oscillatory 
exponential to the front profile, and to determine the position of a
given level from that fit. We will not dwell on the advantages and
disadvantages of the these methods further, as they are of little
relevance when one wants to probe the universal slow long-time relaxation.

Under the same assumptions as before for the uniformly translating
fronts (see footnotes \ref{footnotelabel1} and \ref{footnotelabel2}),
the front velocity of a coherent pulled front relaxes to its
asymptotic value $v^*$ as \cite{esvs,willem1,storm1} 
\begin{equation}
v(t) \equiv v^*+\dot{X}(t) = v^* - \frac{3}{2\lambda^*t} +\, \frac{3 \sqrt{\pi}}{2
(\lambda^*)^2 t^{3/2} }\, {\rm Re}\frac{1}{\sqrt{\mathcal D}} + {\mathcal O}\left( \frac{1}{t^2}\right) , \label{v(t)2relaxation}
\end{equation}
which reduces to the result (\ref{v(t)relaxation}) for uniformly
translating fronts when ${\rm Im}\,{\mathcal D}=0$. The relaxation of
the front profile to its asymptotic behavior (\ref{cfsol})  is characterized by
the relaxation of the velocity 
and a global phase $\Gamma(t)$: it is found that in this case one can
write for the long-time asymptotics
\begin{equation}
\phi(x,t) \approx \sum_{n=0,\pm 1,\cdots}  e^{-i n [\Omega t +
\Gamma(t)]}\, \Phi_{v(t)}^n(\xi_X)  + {\mathcal O}\left(\frac{1}{t^2}\right) ,\label{cfsolrelax}
\end{equation}
where $\Gamma(t)$ is given by \cite{esvs,willem1,storm1} 
\begin{equation}
\dot{\Gamma}(t)  = -k^*_{\rm r} \dot{X}(t) 
- \, \frac{3 \sqrt{\pi}}{2
\lambda^*  t^{3/2} }\, {\rm Im}\frac{1}{\sqrt{\mathcal D}} + {\mathcal
O}\left( \frac{1}{t^2}\right) .
\label{gamma(t)}
\end{equation}
These equations simply express that  to  ${\mathcal O}(t^{-2})$ also
the  pattern forming fronts   follow adiabatically a family of coherent  
front solutions.

Essentially all the remarks made about the slow relaxation behavior of
uniformly translating fronts apply equally well to the relaxation
behavior of the pattern forming fronts --- the only additional feature
is the analogous slow convergence of the frequency $\dot{\Gamma}(t)$:
it implies a slow convergence of the wavelength of the pattern
generated by the front. To our knowledge, this behavior has been
verified both qualitatively as well as quantitatively only for the
Swift-Hohenberg equation --- see section \ref{sectionsh} and
Fig.~\ref{figcglquintic2} below. It does not appear to have been
studied  experimentally in a systematic way, although the slow
velocity relaxation of pattern forming fronts does appear to play a
role in some pattern forming experiments --- see section
\ref{sectiontcrb}. 

\subsubsection{Universal relaxation towards an  incoherent pattern forming
pulled front}\label{relaxtodincohpfpulled}

Quite remarkably, though we can not make a prediction for the front
shape relaxation for an incoherent pulled front, we we will argue in
section \ref{sectionuniversalrel} that the {\em same } velocity relaxation
formula (\ref{v(t)2relaxation}) applies to an incoherent pattern forming pulled
front. The reason for this is that even when a pulled front is
nonlinearly incoherent, the behavior in the leading edge is still very
smooth and coherent, as it is governed by the {\em same} linear dynamical equation
for the expansion about the linear spreading point. This latter
observation also implies that the velocity of an incoherent front is
in practice most easily measured in the leading edge. Even the phase
correction $\Gamma(t)$ is well-defined there and follows (\ref{gamma(t)}) --- the
results for the quintic CGL equation shown in
Fig.~\ref{figcglquintic2} provide evidence in support of this claim.

\subsection{Nonlinear generalization of convective and absolute
instability  on the basis of the results for front propagation } \label{sectionnonlinearconv2abs}

In section \ref{sectionconvversusabs}  we discussed 
the difference between convective and absolute instabilities
in the case that the evolution of a perturbation around the unstable state evolves 
{\em according to the linearized equations}. As Fig.~\ref{figabsconv}
illustrated, if we focus on the right flank of the local perturbation,
the linear instability is absolute if $v^*>0$ and convective if $v^*<0$ ---
in 
the latter case the perturbation ``retreats'' so that it eventually dies out at any fixed position.  
If perturbations have enough time to grow that the nonlinearities
become important and that a nonlinear front develops, it is straightforward to extend the
distinction between the two regimes to the nonlinear case \cite{chomaz3,couairon2,vanhecke8}. 
Clearly,  in the pulled regime the criterion is the same since the
nonlinearities do not affect the asymptotic front speed, so 
the instability is nonlinearly convectively unstable
for $v^*<0$ and nonlinearly absolutely unstable for $v^*>0$. When the fronts are pushed
in the dynamical equation 
under consideration, the criterion becomes simply that the instability is nonlinearly convectively
unstable if $v^{\dagger}<0$ and nonlinearly absolutely unstable when
$v^\dagger >0$. That's all there is to it. A recent nice experiment on the
Kelvin-Helmholtz instability in which the
transition from a nonlinearly convective to a nonlinearly absolute
instability was observed, can be found in \cite{gondret}.

\subsection{Uniformly translating fronts and coherent and incoherent pattern forming fronts in
fourth order    equations and  CGL amplitude  equations}\label{section4thorder}

In this section we   illustrate  many of the basic issues
of front propagation into unstable states by reviewing the  diversity of such fronts in   
Complex Ginzburg Landau (CGL) equations and  in a
number of well known 
fourth order partial differential equations which have been introduced in the literature
as model problems for a variety of physical phenomena.\footnote{We focus in the section on rather well known
model equations. From a historical perspective, it is interesting to note that pulled propagating fronts
were also found back in 1983 in a fourth order caricature model aimed at describing sidebranching
in dendrites \cite{langermk}. The dispersion relation for the sidebranch instability in this model is
of the same form as the general form we consider here, Eq.~(\ref{fourthorderdispersion}) below.}
    In fact, we encounter   all three 
different modes  of front propagation dynamics  both for pushed and for pulled fronts
  in  these equations, so taken together the examples discussed below give a good idea of the  richness
of front propagation into unstable state as well as of the power of
the concepts of a pulled and pushed fronts.

We will limit  our discussion
completely to the dynamics of fronts which propagate into an unstable state, so
it is important to keep in mind that such behavior constitutes just a small fraction of the wide
range of dynamical
behavior  that is found in these equations.  The books by Greenside and Cross \cite{gc} and by
Nishiura \cite{nishiura} provide nice complementary introductions to these model equations.

While  the nonlinear front dynamics will be found to be different for all of the
 fourth order equations we will discuss, the linear
dynamics obtained by linearizing the equations about the unstable
state will always be   a special case of
\begin{equation}
\partial_t \phi = a  \phi - b \partial_x^2 \phi -
\partial^4_x \phi. \label{fourthorderlinear}
\end{equation}
The negative prefactor of the fourth order derivative term is required
to ensure stable behavior of the equations at the short  length scales.
Indeed, the dispersion relation corresponding to this equation is
\begin{equation}
\omega = i (a+ bk^2 -k^4), \label{fourthorderdispersion}
\end{equation}
so for any value of $a $ and $b$ the short wavelength modes at large
$k$ are always damped ($\omega_{\rm i}<0$).
The requirement that the state about which we linearize is unstable
for a range of modes implies  that the coefficients should obey $a+ b^2/4> 0$.
It is simply a matter of some straightforward algebra to work out the
linear spreading point equations (\ref{saddlepoint}) and 
(\ref{lambda*})  explicitly. We find
that there are two regimes: for $a> 0$ and $b< -\sqrt{12 \,a}$ we get
\begin{eqnarray}
\nonumber
v^*_{ {\rm 4}^{th}{\rm ord}}(a,b)  & = &  \frac{2}{3 \sqrt{6}} \left( -2b +
\sqrt{b^2-12a }\right)  \left(- b - \sqrt{b^2-12 a} \right)^{1/2}   ,
 \\ \nonumber
\lambda^*_{ {\rm 4}^{th}{\rm ord}}(a,b) & = & \frac{1}{\sqrt{6}} \left( -b - \sqrt{b^2-12a}
\right)^{1/2} , \\
k^*_{{\rm r},{ {\rm 4}^{th}{\rm ord}}}(a,b)  & = &  0 , \label{v*efk} \\
D_{ {\rm 4}^{th}{\rm ord}}(a,b) &=& \sqrt{b^2-12a} ,  \nonumber
\end{eqnarray}
while for $a>0$ and $b> -\sqrt{12\, a} $ as well as for $a<0$ but $b> 2 \sqrt{-a}$
\begin{eqnarray} \nonumber
v^*_{ {\rm 4}^{th}{\rm ord}}(a,b)  & = &  \frac{2}{3\sqrt{6} } \left( 2b +
\sqrt{7 b^2+24a}\right)  \left(- b +\sqrt{7b^2+24a} \right)^{1/2}   ,
 \\ \nonumber
\lambda^*_{ {\rm 4}^{th}{\rm ord}}(a,b) & = & \frac{1}{2 \sqrt{6}} \left( -b + \sqrt{7b^2+24a}
\right)^{1/2} , \\
k^*_{{\rm r},{ {\rm 4}^{th}{\rm ord}}}(a,b)  & = & \pm  \frac{{1}}{2\sqrt{2}} \left(
3 b+ \sqrt{7b^2+24a} \right),  \label{v*sh} \\
D_{ {\rm 4}^{th}{\rm ord}}(a,b) &=& 2 \sqrt{7b^2+ 24 a} ,
\nonumber
\end{eqnarray}
As we shall see below, in the first regime  $a>0$, $b< -\sqrt{12\,a}$,
fronts are
uniformly translating, while in the other regime they are pattern
forming.

\subsubsection{The Extended Fisher-Kolmogorov equation}\label{sectionefk}
\begin{figure}[t]
\begin{center}
\epsfig{figure=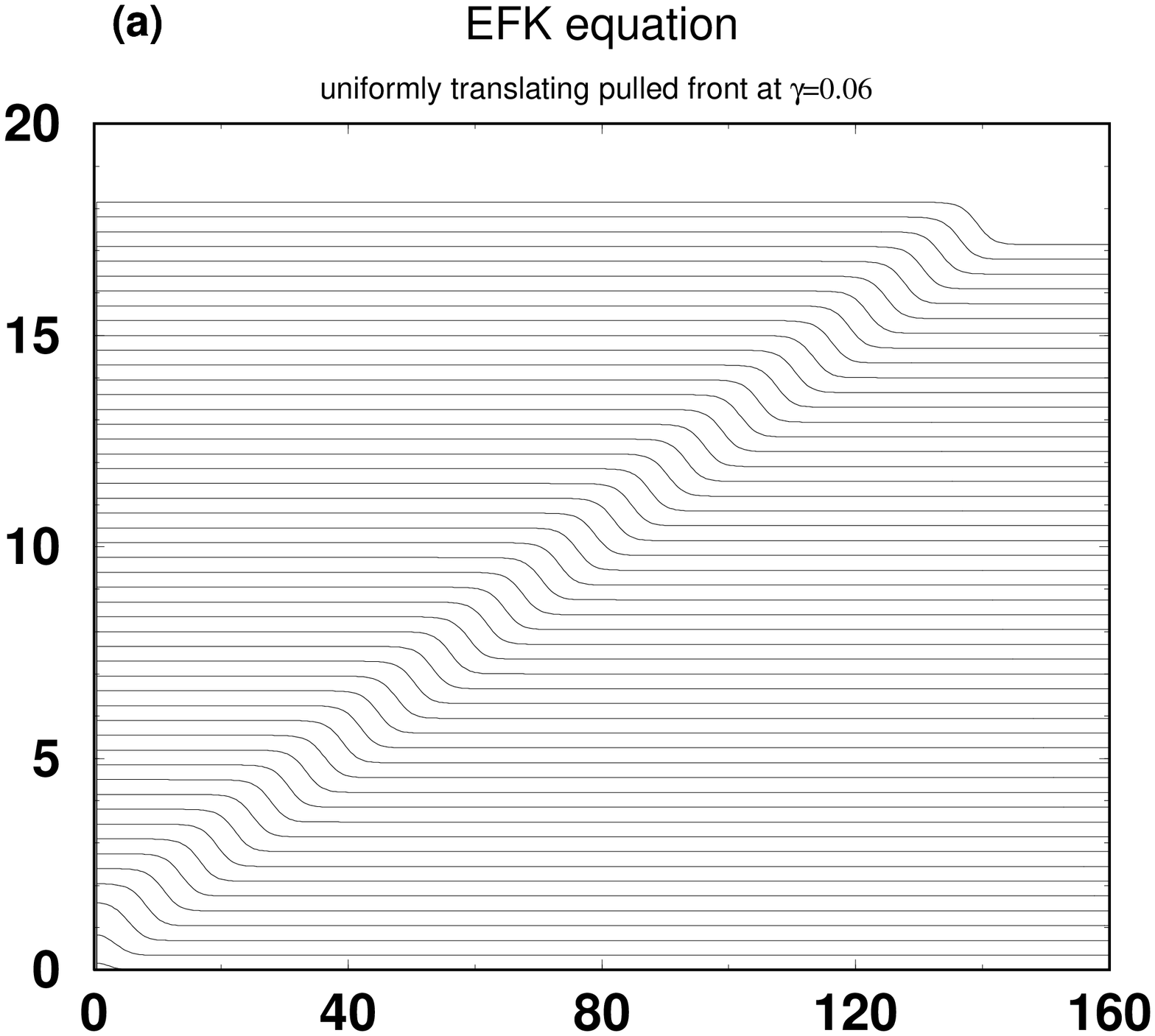,width=0.432\linewidth,angle=-90} 
\hspace*{2mm} \epsfig{figure=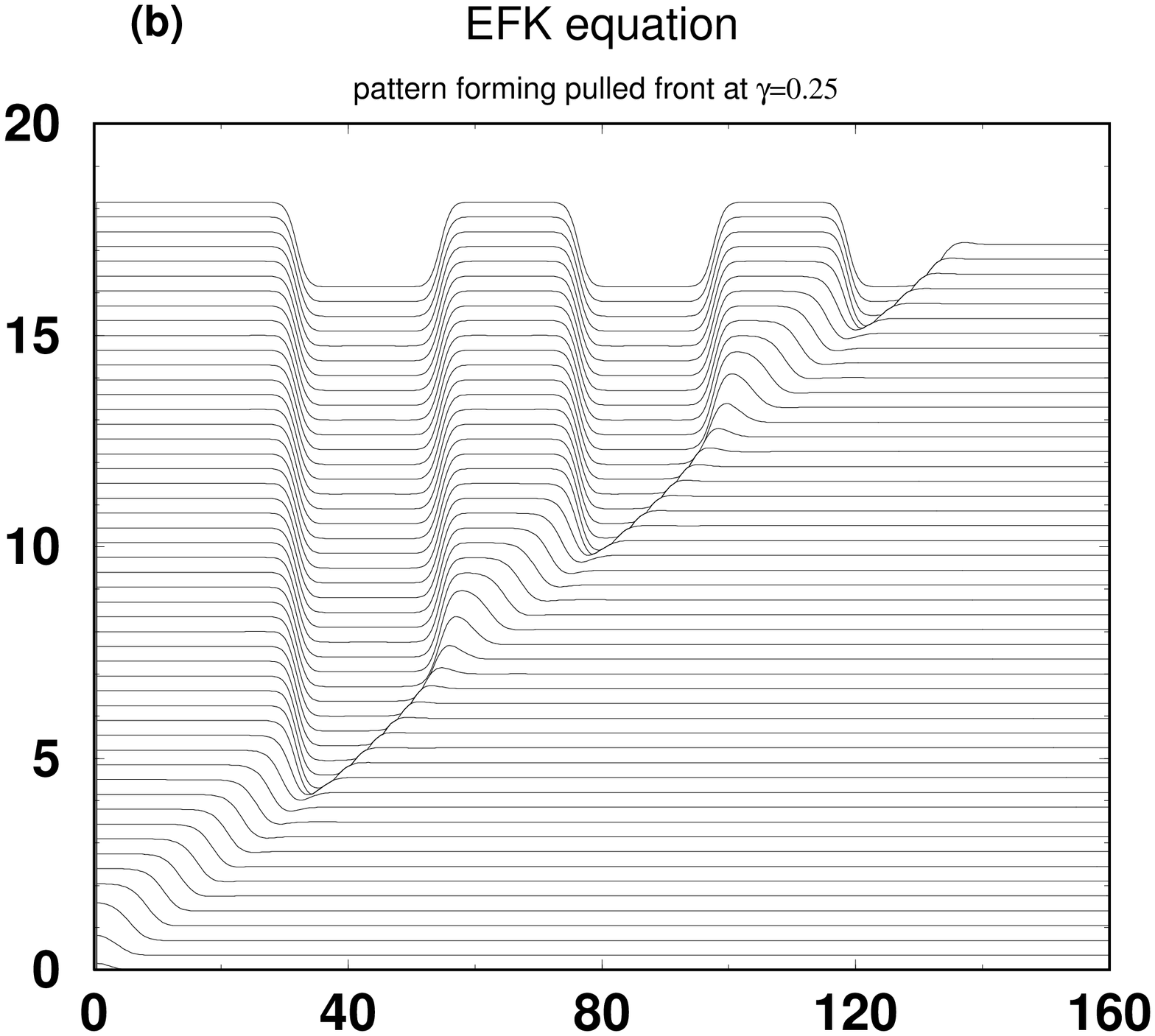,width=0.432\linewidth,angle=-90}
\end{center}
\caption[]{ Space-time plot of simulations results for pulled fronts in 
the EFK equation (\ref{efkeq}). The  lines denote the front
position at successive time intervals of 1.5, and are shifted upward relative to
each-other for clarity. The initial condition is a Gaussian of amplitude
0.1 centered at the origin and the total simulation time is 75. (a) Illustration of the front
dynamics in the regime $\gamma< 1/12$. The plot, made for $\gamma=0.06$ illustrates that
the fronts converge to uniformly translating solutions propagating with  velocity $v^*_{\rm EFK}$. 
 (b) The simulation for $\gamma=0.25$ illustrates that in the regime
$\gamma>1/12$ the pulled fronts generate patterns: they leave behind
an array of kinks. As can be seen from the graph, nodes are conserved
under the dynamics, and as a result the wavelength of the pattern is
given by the node conservation formula (\ref{lstateq}). Note also the
smooth uniform propagation of the leading edge, in spite of the
complicated nonlinear dynamics in the front region itself.}\label{figefk}
\end{figure}

While the relevant asymptotic front solutions of the F-KPP equation
are uniformly translating 
and  monotonically decaying in space to the right, if one includes
a fourth order derivative term in the equation,  the front dynamics
becomes much more rich. If we  consider the case in which the
nonlinear term is a simple cubic nonlinearity so as to study pulled fronts, we arrive at what is
usually referred to as the Extended Fisher-Kolmogorov or EFK equation \cite{deevs,vs2}
\begin{equation}
\partial_t u = \partial^2_x u - \gamma \, \partial^4_x u  + u
-u ^3, ~~~~~~(\gamma>0 ).  \label{efkeq}
\end{equation}
This form of the equation is the most suitable for studying the
behavior in the F-KPP limit $\gamma\to 0$. For any nonzero $\gamma$,
however, upon transforming to the scaled coordinate $x^\prime = x
/\gamma^{1/4}$ the linear terms in this equation are precisely of the
form (\ref{fourthorderlinear}) with $a=1 $ and $b= -1/
\sqrt{\gamma}$. Thus we can use the above expressions provided we
rescale all lengths by a factor $\gamma^{1/4}$; e.g., the linear
spreading speed of the EFK equation is simply 
\begin{equation}
v^*_{\rm EFK}= \gamma^{1/4 } \, v^*_{ {\rm 4}^{th}{\rm ord}} (1,-1/\sqrt{\gamma}).
\end{equation}
The most interesting aspect of the EFK equation as far as front
propagation is concerned is the bifurcation that occurs in the pulled
front behavior at $\gamma =1/12$: For $\gamma < 1/12$, we are in the
regime where the pulled front  parameters are given by (\ref{v*efk})
as $b<-\sqrt{12}$, and hence where we have  $k^*_{\rm r}=0$. This means that
the asymptotic front profile falls off  monotonically, just like in
the F-KPP equation. Fig.~\ref{figefk}{\em (a)} shows an example of the
front dynamics in this regime. For $\gamma>1/12$, however, $k^*_{\rm
r} \neq 0$, and the leading edge of the pulled front falls off in an
oscillatory manner. The dynamics that results from this oscillatory
behavior is illustrated in Fig.~\ref{figefk}{\em (b)}: the 
oscillations in the leading edge periodically grow in size and ``peel
off'' from the leading edge to generate an array of kinks between the
two stable states $u=\pm 1$ \cite{deevs}. 

In  the numerical simulation  one empirically observes that, in this process, nodes
in the profile where $u$ goes through zero, never disappear nor are
 generated spontaneously: They are formed in the far leading edge, and
then just gradually drift through the leading edge until they detach
and come to rest as a kink. To our knowledge, there is no real theory
which shows under what conditions or for what type of equations nodes
are ``conserved'' under the dynamics. However,  given that we empirically
observe it, we can  calculate the wavelength of the stationary
nonlinear kink pattern behind the front \cite{dl,deevs}! In the frame $\xi$ moving
with the asymptotic speed $v^*$, the oscillatory part of the profile
is for long times given by $\exp[-i(\omega_{\rm r}^* - k^*_{\rm r}
v^*)t]$, so the angular frequency with which nodes pass any fixed position in
the leading edge is $|\omega_{\rm r}^*-k^*_{\rm r}v^*|$. Far behind the
front, the kinks come to rest in the $x$ frame. If the wavenumber  of
the kink profile is $q_{\rm stat}$, then the  angular frequency with
which  nodes  pass a fixed position $\xi$ far behind the front in the moving
frame is simply $v^* q_{\rm stat} $. Node conservation together with
the expressions (\ref{v*sh}) therefore implies that 
\begin{equation}
q_{\rm stat} =  \left| \omega_{\rm
r}^*/v^*- k^*_{\rm r}\right| =   
\frac{3 }{8\sqrt{2} } \,
\frac{\left(3b+\sqrt{7b^2+24a}\right)^{3/2}}{2b+ \sqrt{7b^2+24a} }  . \label{lstateq}
\end{equation}
The expression for the front speed in both regimes, as well as this
one for the pattern wavenumber in the regime $\gamma > 1/12$, was found
to be fully consistent with the results of numerical
simulations.\footnote{Inspection of the data points for the velocity
in \cite{deevs} shows that they fall slightly below the predictions
for $v^*_{\rm EFK}$. This is due to the fact that in a finite time
simulation one always observes a velocity slightly less than $v^*$ due
to the power law convergence to $v^*$ from below discussed in section
\ref{sectionunirelsimple}. Similar effects can be observed in virtually all published data of
studies of pulled fronts, some of which are reviewed in  section
\ref{sectionexamples}. }
In fact, an ideal periodic kink pattern of the type generated behind
the front is not  necessarily stable, so the emerging near-periodic state
 is usually only an intermediate asymptotic state: For $1/12  < \gamma < 1/8$,
successive kinks and antikinks attract each-other. This attraction is
very weak for widely spaced kinks, and so the bunching instability
that it gives rise to is virtually unnoticeable in the simulations. For
$\gamma > 1/8$, the spatially oscillatory tail of a single kink gives
rise to an oscillatory interaction between kinks \cite{deevs,bert}. This is just one
example of the rich type of behavior that the EFK displays --- see
\cite{bert,vandenberg} for further details or \cite{rottschafer,rottschafer2} for
work on the existence and stability of the front solutions.

Just like the F-KPP equation exhibits a transition to pushed fronts if the cubic nonlinearity
$u^3$  is changed into a nonlinear function $f(u)$ whose growth is enhanced over
the linear term at intermediate values (e.g. by a quadratic nonlinearity $u^2$), so does the
EFK equation \cite{vs2} --- we already illustrated this in
Fig.~\ref{figpushedefk}. Even for $\gamma< 1/12$, when pulled fronts in the EFK equation
are spatially monotonic, pushed fronts in this regime can be
non-monotonic even though they remain uniformly translating, see
Fig.~\ref{figpushedefk} and   \cite{vs2}. So, as we already stressed
before, unlike
 fronts in the F-KPP equation, both pulled and pushed fronts in the EFK equation can be 
non-monotonic --- monotonicity has nothing to do with front selection.

What else does the EFK equation teach us? Well, this equation 
 with its bifurcation from uniformly translating fronts for $\gamma< 1/12$ to
pattern forming fronts for $\gamma>12$ is probably the simplest equation that illustrates
the essence of pulled dynamics: the dynamics in the leading edge just does what the linear
spreading point conditions impose, while the rest of the front where the nonlinear saturation
term becomes important just can not do anything but adjust 
to the dynamics enforced  by the leading edge dynamics. In the
present case, where essentially  the only basic coherent structures admitted by the equation are kinks,
the state behind the front is in first approximation an array of such kinks or domain walls. This  state 
itself may be unstable or have other nontrivial dynamics, but this does not really affect the propagation
of the pulled front.

As we discussed in section \ref{sectiontypesoffronts}, the coherent
pattern forming front solutions (\ref{cfsol}) are coherent in that they are
temporally periodic in the frame $\xi$ moving with velocity $v^*$. The
node-conservation argument is one immediate implication of this. Hence
the empirical observation of node-conservation in the EFK equation for
$\gamma>0$ strongly suggests that the pattern forming solutions are 
 coherent, i.e., of the form (\ref{cfsol}). Indeed one can convince oneself
\cite{wvsunpub} term by term
in the expansion that the EFK equation for $\gamma>1/12$ should admit
a two-parameter family of front solutions, parametrized by their
velocity $v$ and the wavelength of the kink pattern behind the front.

We finally note that while the EFK equation has been introduced mainly
as a model equation to illustrate the transition from uniformly
translating fronts to coherent pattern forming fronts as the spreading
points parameters change when $\gamma$ is increased beyond
1/12. Zimmermann \cite{zimmermann} has shown that the equation can
arise as a type of amplitude equation in electro-hydrodynamic
convection in liquid crystals. In this interpretation the point
$\gamma=1/12$ can be viewed as a dynamical Lifshitz point.\footnote{A
Lifshitz point is a point in parameter space where a phase transition
from a homogeneous to a modulated state occurs.}

\subsubsection{The Swift-Hohenberg equation}\label{sectionsh}
The Swift-Hohenberg equation 
\begin{equation}
\partial_t u  = \varepsilon u  - (\partial^2_x+1)^2 u  -u^3 =
(\varepsilon-1) u  - 2 \partial^2_x u  -\partial^4_x u - u^3 \label{swifthohenberg}
\end{equation}
was introduced as a simple model equation for  the dynamics just
above a supercritical finite-wavelength instability \cite{sh}.\footnote{It also arises in the context of
laser physics \cite{lega} or as an amplitude equation near a particular type of
 co-dimension 2  bifurcation point \cite{vivi}.} In the
Swift-Hohenberg equation,
the $u =0$ state is  linearly unstable for
$\varepsilon>0$; for $0 < \varepsilon \ll 1$  the equation gives rise
to patterns of wavenumber about 1 whose dynamics can be analyzed in
terms of an amplitude equation. The dispersion relation for small
perturbations about the unstable state is given by
(\ref{fourthorderdispersion}) with $a=\varepsilon-1 $ and $b=2$.

\begin{figure}[t]
\begin{center}
\epsfig{figure=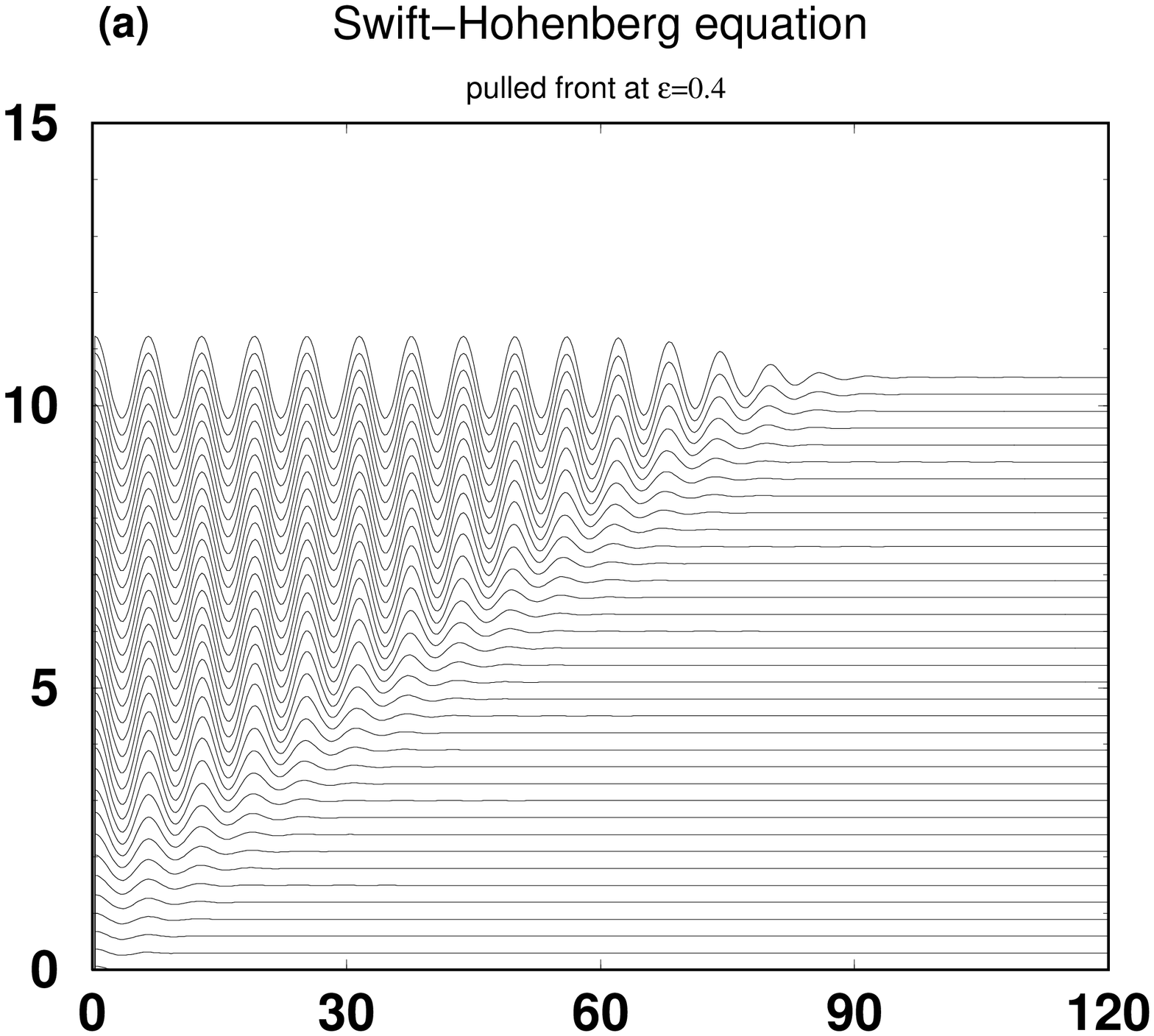,width=0.432\linewidth,angle=-90} 
\hspace*{2mm} \epsfig{figure=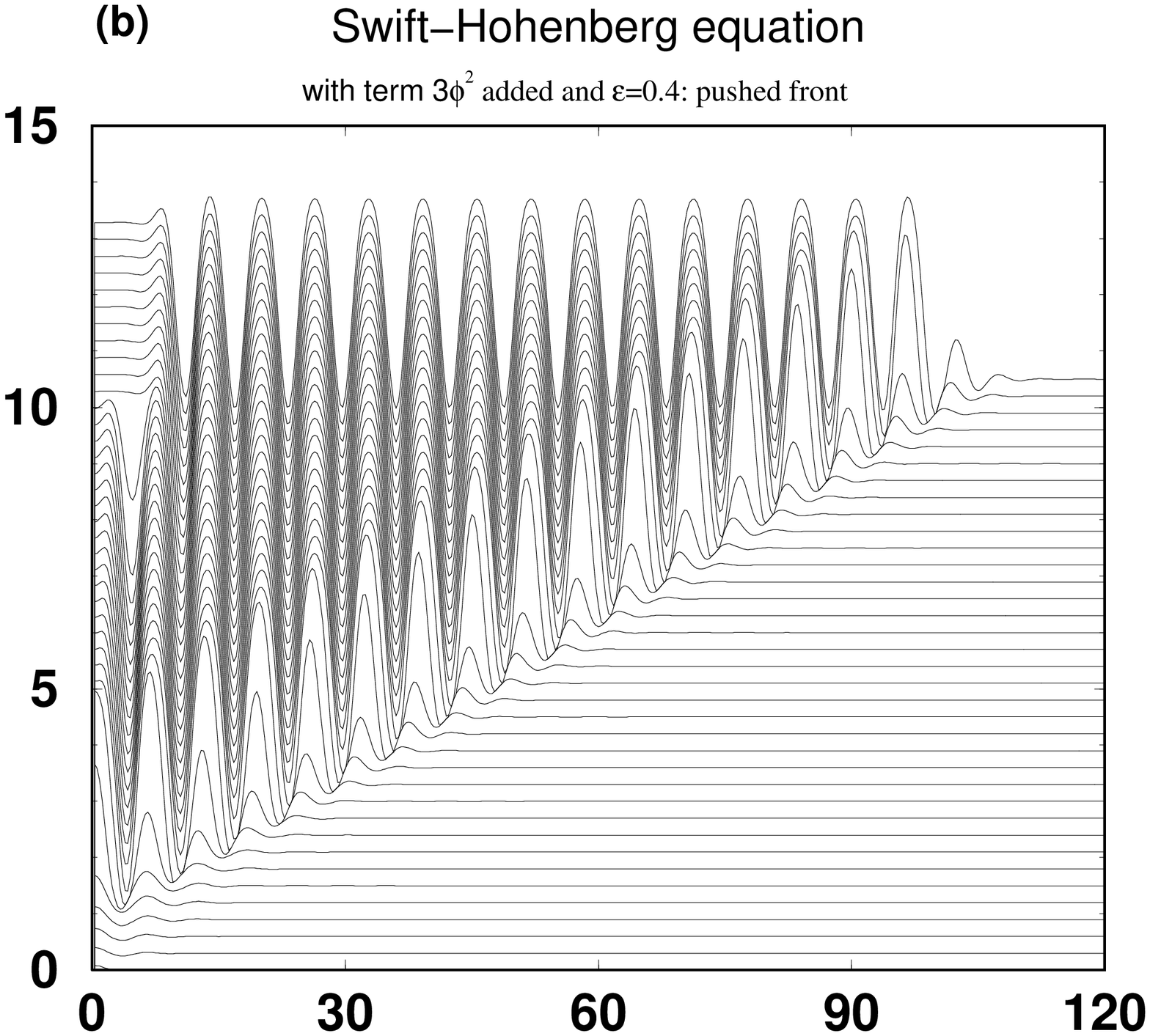,width=0.432\linewidth,angle=-90}
\end{center}
\caption[]{ Space-time plot similar to Figure \ref{figefk}, but now
for the Swift-Hohenberg equation. Total run time is 35, and the time
difference between successive lines is 1; the simulations started
from a Gaussian initial condition. The plots in this figure are more
detailed version of those in the middle column of Fig.~\ref{F1}.  (a) A pulled front  at
$\varepsilon=0.4$. (b) Simulation of the Swift-Hohenberg equation with
a symmetry breaking term $b u^2$ added to the right hand side. For
$b>b_{\rm c}(\varepsilon)$ fronts in this equation are pushed, where
$b_{\rm c}(\varepsilon \downarrow 0) = \sqrt{27/38}$
\cite{vs2}. The simulation shown is for $b=3$ and $\varepsilon=0.4$,
while $b_{\rm c}(0.4) \approx 1.85$ \cite{vs2}. Note that after a
while, a second front develops; at later times this  moves into the periodic state generated
by the first front. This front generates a stable homogeneous state $u
=const.$  }\label{figsh}
\end{figure}

For $\varepsilon>0$ the Swift-Hohenberg equation admits  a band of
stable periodic states, and this has made the Swift-Hohenberg equation
an attractive model equation for front propagation in the past, as
fronts propagating into the unstable state $u=0$ are found to
generate one of these regular and stable periodic patterns behind
them --- see Fig.~\ref{figsh}{\em (a)}. 
For this reason  the Swift-Hohenberg equation has 
played an important role both from a historical and from a practical
point of view:
\begin{enumerate}
\item  The first
numerical study of a pulled pattern forming front was done for this
equation and  the expression  (\ref{v*sh}) for $v^*$ 
was confirmed in detail \cite{dl,vs2}.
\item  The node conservation that we discussed
in the previous subsection on the EFK equation was first observed
empirically for this equation by Dee and Langer \cite{dl} and this
lead these authors to derive Eq.~(\ref{lstateq}) with
$a=\varepsilon-1$ and $b=2$ for the wave number
$q_{\rm stat}$ of the stationary periodic profile generated
behind the front. Numerical results were found to be in full accord
with this expression \cite{dl,vs2}, indicating, as for the EFK
equation, that these front solutions appear to be of the coherent
pattern forming type (\ref{cfsol}). 
\item  It was the first pattern forming equation
for which the transition from pushed to pulled  fronts was studied in
detail both  numerically and analytically by including either a
symmetry breaking quadratic term $u^2$ or by changing the sign of
the cubic term and adding a stabilizing quintic term
\cite{vs2}. Simulations of a pushed front in the presence of a
quadratic term are show in Fig.~\ref{figsh}{\em (b)}. The dynamics of
the  pushed fronts generated this way was confirmed 
to be associated with the existence of nonlinear front solutions solutions with
steepness larger than $\lambda^*$, in agreement with the arguments
presented in section \ref{sectiontypesoffronts}. 
\item  The Swift-Hohenberg equation  is to our knowledge
essentially the only pattern forming equation for which the {\em existence}
of pattern forming fronts has been proved rigorously  and for which
the   pulled front selection  has been established with
some rigor \cite{collet2,collet,collet3}. In fact, 
the first formulation of what we refer to as coherent pattern
forming front solutions appears to have been made for this equation by
Collet and Eckmann \cite{collet}.
\item The Swift-Hohenberg equation has recently
also been used in extensive numerical tests of the power law
convergence discussed in  section \ref{sectiontypesoffronts} of  the wavenumber
of the pattern generated behind a pulled front \cite{storm1}. The
results which are reported in Fig.~\ref{figcglquintic2} below are  found
to be in excellent  agreement with the theoretical predictions.
\end{enumerate}

\subsubsection{The Cahn-Hilliard equation}\label{sectionch}
The Cahn-Hilliard equation is a simple model equation for the problem
of phase separation and dynamic coarsening   \cite{bray,gunton},
which we will briefly review in the context of polymer phase
separation in section \ref{secspinodal}. The dynamical equation
is based on the idea that in the phase separation regime the coarse
grained free energy functional 
$ {\mathcal F} = \int dx\, \half (\partial_x u )^2   -  u^2 /2 + u^4/4 $
for the composition field $u $ 
has two minima,  at $u=\pm 1$, and on the idea that this composition field  is a
``conserved order parameter''. This means  that mass is neither created nor destroyed, but that
it  can exchange due to diffusion in
response to a free energy gradient. These ingredients lead to the Cahn-Hilliard equation
\begin{equation}
\partial_t u = \partial^2_x \left( \frac{\delta {\mathcal F}}{\delta u}\right) 
= - \partial^2_x (u -u ^3) - \partial^4_x u  . \label{cahnhilliard}
\end{equation}
Note that this  gradient structure of the equation implies
the conservation of composition in the following form: if we imagine
taking a large system of size $L$ with zero flux boundary conditions,
then  it follows immediately that 
\begin{equation}
\frac{d}{dt} \int_0^L dx \, u (x,t) =0. \label{conservation}
\end{equation}
Thus the spatially averaged composition is conserved under the
dynamics.\footnote{Moreover, the form (\ref{cahnhilliard}) implies
that ${\mathcal F}$ is also a Lyapunov function for the dynamics:
$d{\mathcal F}/dt = - \int dx\, [\partial (\delta {\mathcal F}/\delta
u )/\partial x]^2 \le 0$.}
The long-time dynamical behavior implied  by (\ref{cahnhilliard}) is that the
system is driven towards a phase consisting of domains where $u$ is
close to one of the two minimal values $\pm 1$, separated by domain
walls or kinks. In higher dimensions the dynamics of these domains is
driven by the curvature of the domains (``droplets''), while in one
dimension the coarsening is driven by the 
interaction between the domain walls. This interaction is
exponentially small for large separations, and the coarsening dynamics in one dimension is
therefore quite slow.

Consider now a homogeneous state $u=u_{\rm c}=constant$. If we
linearize the Cahn-Hilliard equation about $u_{\rm c}$, the
resulting equation for $u-u_{\rm c}$ is of the form
(\ref{fourthorderlinear}) with $a=0$ and $b= 1-3 u_{\rm c}^2$. For
$u_{\rm c}< 1/\sqrt{3}$ the homogeneous state is thus unstable and
one can consider the following 
front propagation problem \cite{liu2,ball}: we consider an initial condition where
$u \approx u_c < 1/\sqrt{3}$ everywhere except in a small localized region near
the left edge, where we impose the boundary conditions $u=1$,
$\partial_x u =0$ and investigate the front which propagates into the
linearly unstable state, whose linear spreading point values are given
by (\ref{v*sh}) with $a=0$ and $b=1-3u_{\rm c}^2$. 

An interesting aspect of the Cahn-Hilliard equation (and of similar
equations with conserved dynamics) is that one can immediately see
that even if uniformly translating front solutions
of the type $u (x-vt)$ obeying  the boundary conditions  exist,
they can not be relevant for the dynamics. To see this, suppose such a solution exists; for such
a solution, we would then have
\begin{equation}
\frac{d}{dt} \int_0^L dx\, u(x-vt) = - v \int_0^L dx \, \partial_x u
(x-vt) = -v 
(u_{\rm c} -1)  \neq 0,
\end{equation}
in contradiction with (\ref{conservation}).  A more intuitive way to
understand this is as  follows: because of mass conservation, the
phase separation in the front region can only occur through the
formation of domains of the two preferred phases. An isolated wall
between two domains has no intrinsic motion. So,  the region behind
the front is a modulated phase with only a slow coarsening dynamics
associated with  the slow motion of domain walls in either direction, and a
coherent solution moving with the front speed is dynamically
impossible.\footnote{Such arguments apply of course more generally to
conserved equations. As an example of this, the complex dynamics of
bacterial flagella traces back partially to the conserved nature of
the underlying dynamical equations \cite{coombs}.}
\begin{figure}[t]
\begin{center}
\epsfig{figure=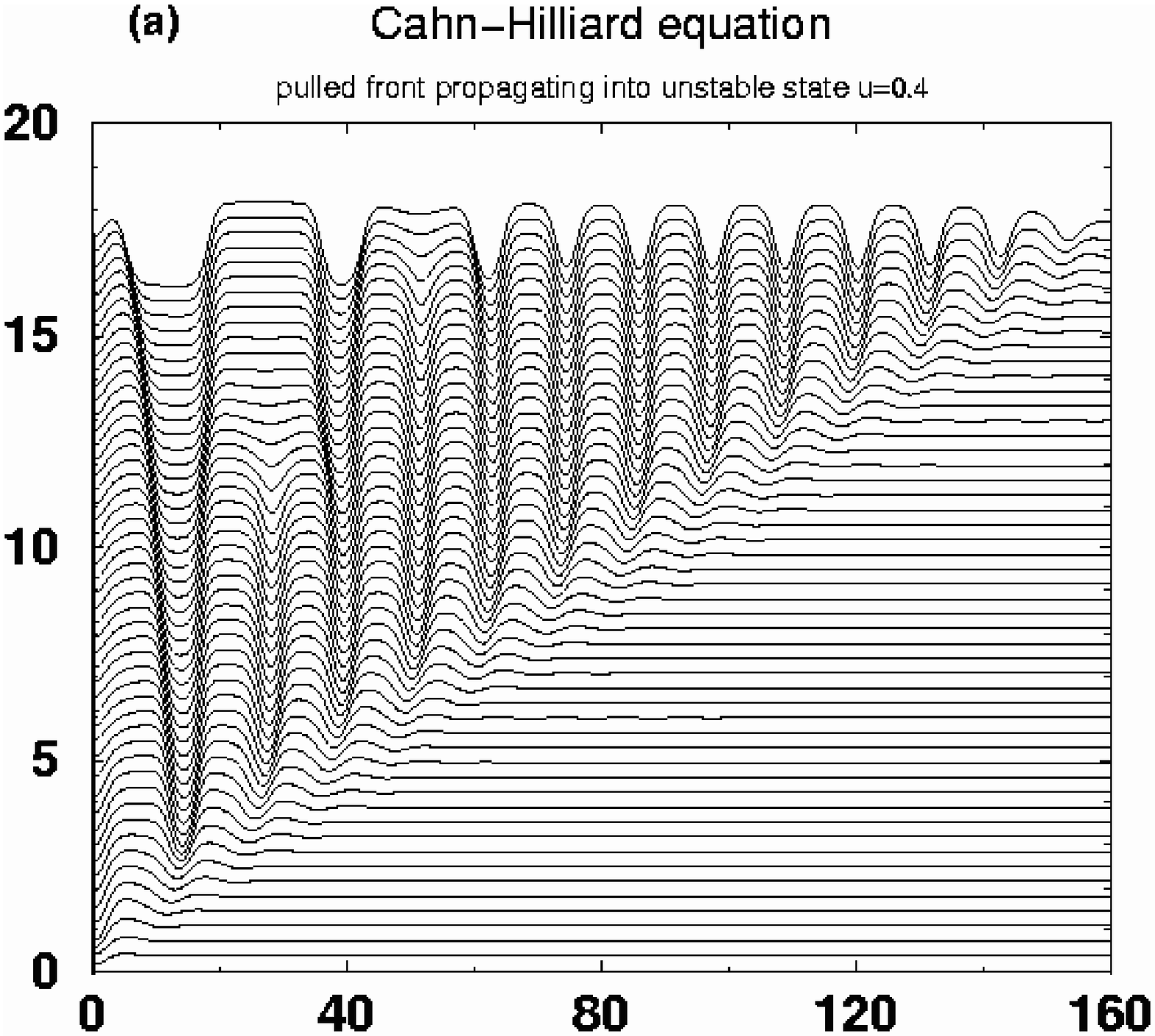,width=0.432\linewidth,angle=-90} 
\hspace*{2mm} \epsfig{figure=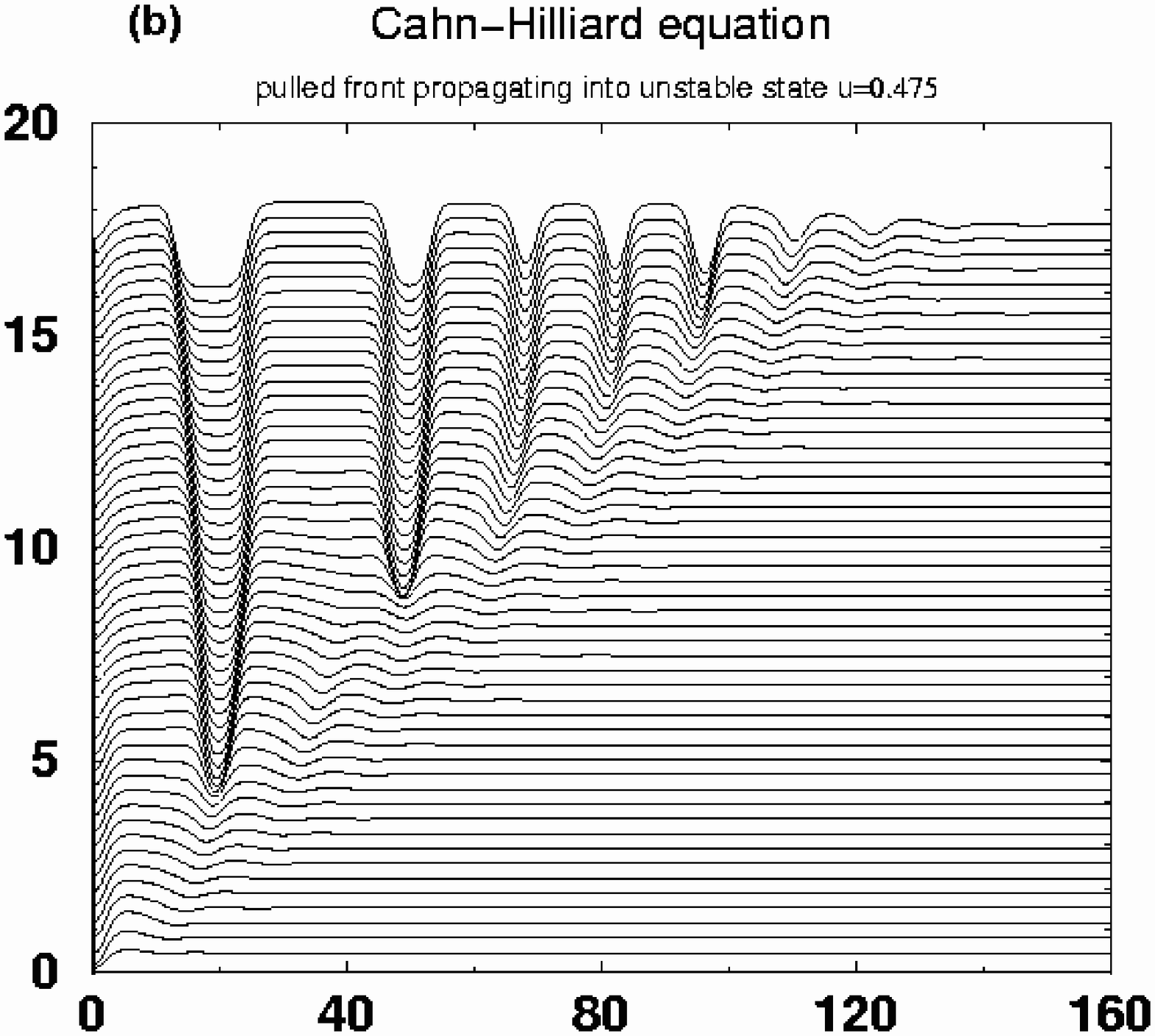,width=0.432\linewidth,angle=-90}
\end{center}
\caption[]{ Space-time plot similar to Figure \ref{figefk}, but now
for the Cahn-Hilliard equation, starting 
from a Gaussian initial condition.  (a) A pulled front  propagating
into the unstable state $u=0.4$. Total run time is 300, and the
time difference between successive lines is 6. Note the slow
coarsening (domain merging)
behind the profile at later times.
(b)  The same for an homogeneous initial state $u=0.475$.  The total
run time is 500, time difference between successive lines is 10. Note
that the front propagates slower than in the case shown in (a), due to
the fact that the homogeneous state on the right is less unstable. }\label{figch}
\end{figure}

The simulations of the Cahn-Hilliard equation of \cite{ball,liu} and
those shown in
Fig.~\ref{figch} confirm  that fronts in the Cahn-Hilliard
equation are pulled in agreement with intuitive notion that the
nonlinearities slow down the growth.  Beyond this, not much is known
with certainty, as the fronts in this equation do not appear to have
been investigated in any detail.  The most likely scenario, it appears
to us, is that the relevant front solutions are incoherent ones, since
the coarsening of the pattern sets in as soon as it is
formed. However, it is not excluded that the equation does admit
coherent pattern forming front solutions, and that the state generated
by the front is unstable to coarsening. Indeed, since  during the
coarsening process domains can merge, a simple node counting argument
can not be applied to predict the final state pattern. Nevertheless, it does
appear from the plots that such an analysis 
might give a reasonable estimate of the transient  domain size in the
regime where the front  propagation speed is rather high.\footnote{If
so, the picture is somewhat similar to the one shown in
Fig.~\ref{figcglcubic}{\em (a)} for the cubic CGL equation: the first
pulled front generates a state which is unstable, and which hence is
invaded by a second front. } Whether there might possibly be   a transition from
coherent to incoherent pattern forming fronts as $u_c$ is increased,
has to my knowledge not been investigated.

Thus, while the
linear spreading dynamics of the Cahn-Hillard equation is just a
special case of that of the Swift-Hohenberg equation, the dynamical
nonlinear behavior of the patterns that the front gives rise to is
very different. This confirms the picture that a pulled front just
propagates as is dictated in essence by the linear spreading point
conditions, while the nonlinear dynamics behind it just follows in
whatever way is allowed or imposed by the dynamics in the nonlinear region. 

\subsubsection{The Kuramoto-Sivashinsky equation}\label{sectionks}

The Kuramoto-Sivashinsky equation \cite{kuramoto2,sivashinsky}
\begin{equation}
\partial_t u = - \partial_x^2 u - \partial_x^4 u + u\partial_x u \label{kseq}
\end{equation}
has played an important role in the study of nonlinear chaotic dynamics. Even though the equation
admits periodic solutions, it has nearby weakly  turbulent states. Even though the latter states may
be transient, they usually  dominate the 
dynamics, as their survival time grows exponentially fast with the
size of the system \cite{shraiman2,zaleski}.  
The scaling properties of these turbulent states, in particular the question whether they are in the
so-called KPZ universality class, has also been a subject of intense
research \cite{bohr,lvov,rost,yakhot}.

Clearly, the state $u=0$ is unstable due to the negative prefactor
of the diffusion-like term, and the dispersion relation  is again of the
standard form (\ref{fourthorderdispersion}) with $b=1$ and $a=0$;
substitution of these values into (\ref{v*sh}) gives the linear
spreading parameters for this equation for fronts propagating into the 
state $u=0$. 

\begin{figure}[t]
\begin{center}
\epsfig{figure=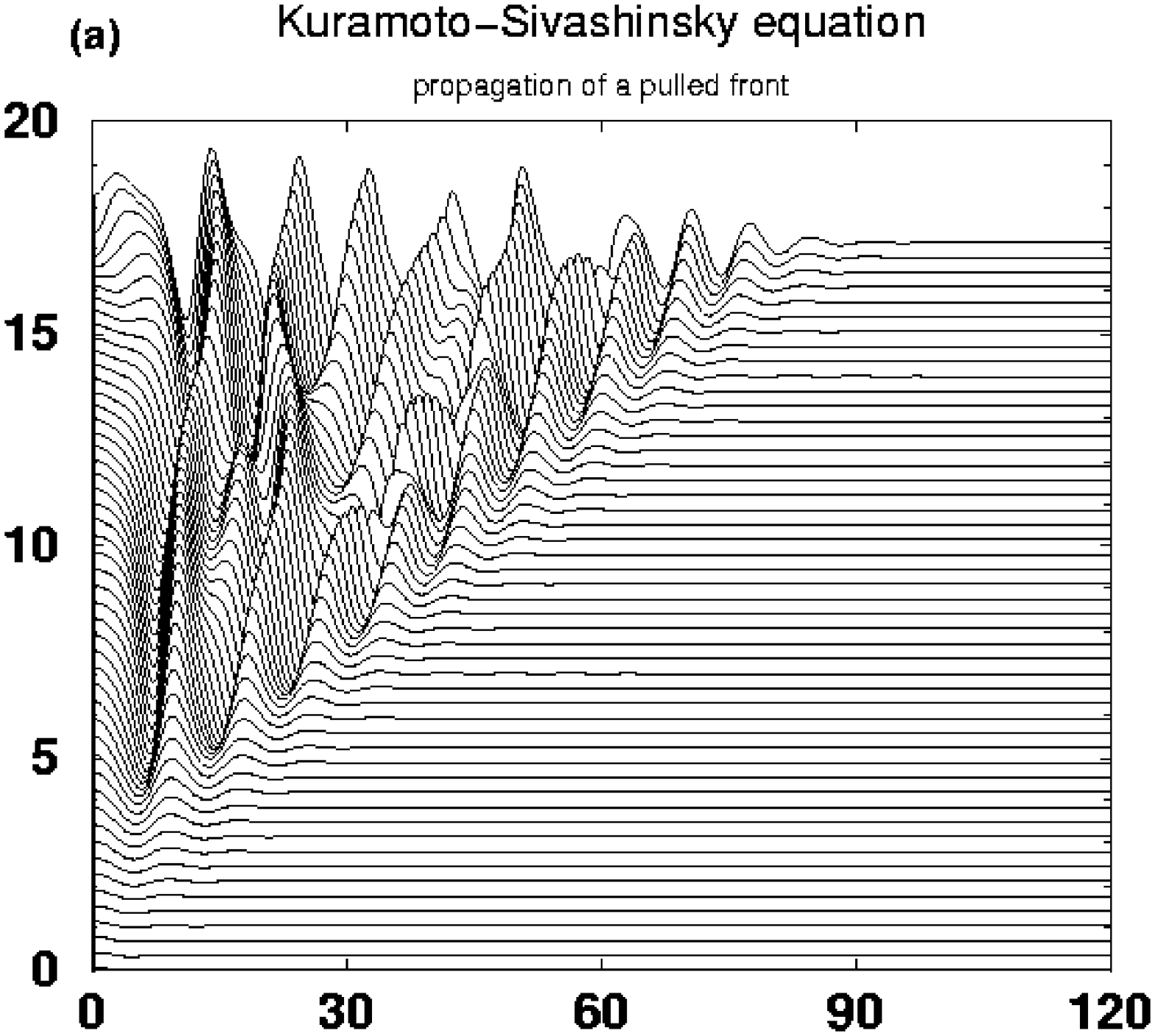,width=0.432\linewidth,angle=-90} 
\hspace*{2mm} \epsfig{figure=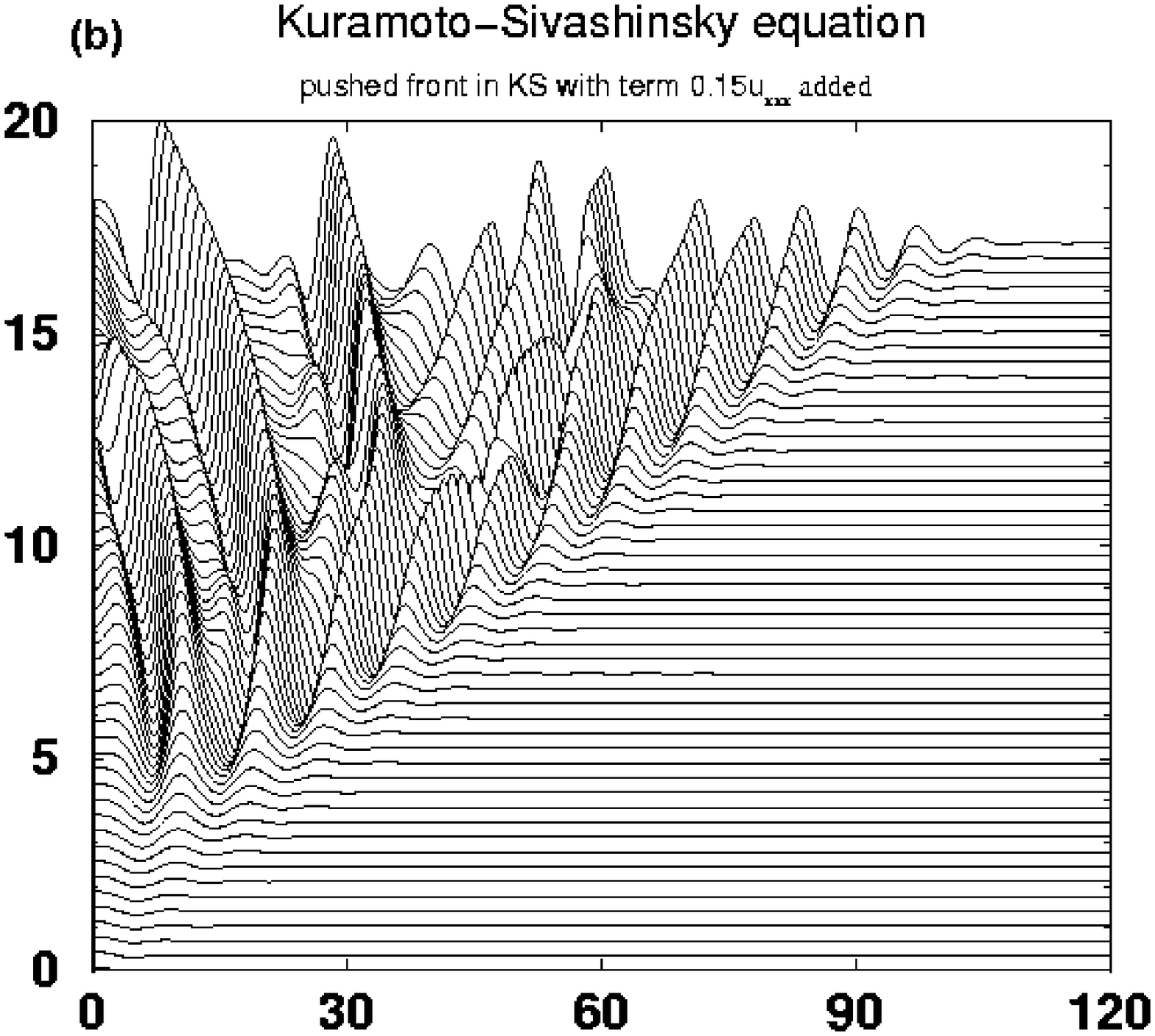,width=0.432\linewidth,angle=-90}
\end{center}
\caption[]{  Space-time plot similar to Figure \ref{figefk}, but now
for the Kuramoto-Sivashinsky equation, starting 
from a Gaussian initial condition.  Total run time is 60, and the
time difference between successive lines is 1.2. (a) A pulled front  propagating
into the unstable state $u=0$.  The state generated by the front is a weakly
chaotic state of the type typically found for this equation.
(b)  The same but now for an extension of the Kuramoto-Sivashinsky equation, obtained
by adding a term $c\partial_x^3 u $ to the right hand side. The results of \cite{chang,hansen}
imply that upon increasing $c$, a transition from pushed to pulled fronts occurs at $c\approx 0.14$.
The simulations shown in the plot  are for  $c=0.15$ as in
\cite{hansen}, i.e.,  just above the transition, which itself 
is not  understood in detail. Note that in this case the transition is tuned
by changing a linear term in the equation, not a nonlinear term!  The
plots in this figure are more detailed versions of those in the
right column of Fig.~\ref{F1}, except that there $c=0.17$ in the lower
panel.}\label{figks}
\end{figure}

Several authors  \cite{conrado} have studied the front propagation
problem  in this equation both numerically and analytically. 
Some representative simulation results   are shown in Fig.~\ref{figks}, where starting from a localized
initial condition one obtains two  pulled fronts which propagate out
with velocity $v^*_{ {\rm 4}^{th}{\rm ord}}(0,1)$ in
opposite directions. Although this has to my knowledge not been
investigated explicitly, it is quite clear from such simulations that these
front solutions are inherently incoherent. Whether the
Kuramoto-Sivashinsky admits (unstable) coherent pattern forming front
solutions is not known to me; in any case this issue is probably not very relevant for the dynamics. 

Another interesting feature is that when  the Kuramoto-Sivashinsky equation
is modified by adding a cubic term $c\partial_x^3 u $ to the right
hand side of (\ref{kseq}), then one of the two fronts  becomes pushed
for a critical value of $c$ of about   $0.14$
\cite{chang,hansen}. The pushed fronts appear to be driven by a pulse-type structure
which propagates  faster than $v^*_{ {\rm 4}^{th}{\rm ord}}$, and which leaves behind a series
of pulse-type structures as well. It is an understatement to say that
these empirical observations are not very well understood 
theoretically. Possibly, in some limit it is more appropriate to think of the pushed
dynamics in terms of the splitting and birth of pulse-type solutions
\cite{arjen,nishiura} than in
terms of a propagating front.

\subsubsection{The cubic Complex Ginzburg-Landau equation}\label{sectioncglcubic}

The cubic Complex Ginzburg-Landau (CGL) equation 
\begin{equation}
\partial_t A= (1+ic_1) \partial_x^2A + \varepsilon A -(1-ic_3)|A|^2A \label{cubiccgl}
\end{equation}
with $\varepsilon >0$
is the appropriate amplitude equation just above the onset to  finite
wavelength traveling wave patterns. The amplitude $A$ is a complex
quantity, and describes the slow space and time modulation of
the finite wavelength mode that goes unstable
\cite{aranson,ch,fauve,gc,newell2,nishiura,vhhvs,vsbeginners,walgraef}.  The phase
diagram of the dynamics behavior of this equation as a function of the
real parameters
$c_1$ and $c_3$ is extremely rich, and includes   chaotic phases as
well as coherent dynamics phases
\cite{aranson,chate1,ch,shraimanpumir,vanhecke6,vanhecke7}. 

An important simple set of solutions of the  CGL equation are the
periodic or {\em phase winding solutions}
\begin{equation}
A=a e^{iqx-i \Omega t}, \label{phasewinding}
\end{equation}
with $a$ a constant.
Upon substitution of this  expression in to the cubic CGL, it is easy
to check that these solutions form a one-parameter family parametrized
by $q$, as we can
express $a$ and $\Omega$ in terms of $q$ as
\begin{equation}
a^2= \varepsilon -q^2, ~~~~~~~~~ \Omega = ( c_1+ c_3) q^2 -
c_3 \,\varepsilon,~~~~~~~~ -\sqrt{\varepsilon} <q< \sqrt{\varepsilon} . \label{phasewindingvalues}
\end{equation}
In an amplitude description, the phase winding solutions represent
periodic traveling patterns whose wavenumber differs by $q$ from the critical
mode at the bifurcation. Because the CGL equation admits a family of
these pattern-type solutions, fronts typically  generate  a pattern
and  hence are pattern forming fronts --- although the CGL equation
looks superficially like the F-KPP equation,  the fact that the
dynamical field is complex makes its dynamical behavior much more
intricate (One might even say  the pulled front propagation mechanism is
virtually the only  element they have in common!).

Not all the phase winding solutions are stable. Those near the edge of
the band are always linearly unstable (the so-called Benjamin-Feir
instability), while the stability of those near the center of the band
depends strongly on $c_1$ and $c_3$. In fact, for $c_1 c_3>1$ none of
the phase winding solutions is stable \cite{ch,malomed1,malomed2,newell}. Chaotic behavior is typically
found in this region of the phase diagram. 

The state $A=0$ in the   cubic CGL (\ref{cubiccgl}) is linearly
unstable; since the dispersion equation is quadratic in $k$, it is
easy to obtain the
spreading point parameters
\begin{equation}
v^*_{\rm CGL} =   2 \sqrt{\varepsilon (1+c^2_1) }, ~\,
\lambda^*  =  \sqrt{\frac{\varepsilon}{1+c^2_1} }, ~\, k^*_{\rm r} =
c_1  \sqrt{\frac{\varepsilon}{1+c^2_1} } , ~\, D= 1+c^2_1. \label{v*cgl}
\end{equation}

CGL equations generally admit an important class of so-called {\em coherent structure
solutions }  of the form
\begin{equation}
A(x,t) = e^{-i\hat{\Omega} t} \hat{A}(x-vt) ,\label{coherentfront}
\end{equation} 
which is a special case of the general form (\ref{cfsol}) of coherent
pattern forming fronts.
Coherent structure solutions of this type --- fronts,
 sources, sinks, pulses \cite{vsh} and  other types of localized solutions \cite{vanhecke6,vanhecke7} ---
turn out to be the crucial building blocks of the dynamics of the CGL,
but we limit our discussion  here exclusively to
fronts. The simplicity of the coherent fronts in the CGL equation lies
in the fact that they are of the coherent pattern forming type (\ref{cfsol}),
but that there is {\em only one } nonzero term in this sum (we can
think of the function $\hat{A}$  as the term $\Phi_1$ in this
expression). Since
$\hat{A}$ depend only on a single co-moving coordinate $\zeta=x-vt$,
$\hat{A}$ obeys an ordinary differential equation. The multiplicity of
coherent structure solutions (do they come in discrete sets, one- or
two parameter families?) can therefore be studied from the number of
stable and unstable manifolds near the fixed points that describe the
asymptotic behavior of these solutions (so-called ``counting
arguments''). For fronts, these  counting arguments show \cite{vsh} that there
generically is a two-parameter family of front solutions which to the
right decay exponentially and on the back side approach a phase
winding solution. Since $k^*$ is a complex quantity, this implies that
there generally will be a unique coherent front solution
(\ref{coherentfront}) with velocity
$v^*$ and which smoothly connects the linear spreading point behavior
$e^{ik^*(x-v^*t)}$ on the right with a phase winding solution like
(\ref{phasewinding}) on the left. 

Since the temporal phase factor is a global ($x$-independent) factor,
one can use an argument reminiscent of the conservation of nodes
argument discussed in subsection \ref{sectionefk} to calculate the 
wavenumber of the coherent pulled front solution. Indeed, if we
compare in the frame moving with velocity $v^*$ two points a fixed
distance apart, this {\em phase  difference is fixed } since $\hat{A}(\xi)$ is
time-independent in this frame while the prefactor $e^{-i{\hat{\Omega}} t}$ is
common to both points. By equating the temporal phase winding in this
frame at  the two points we then get the wavenumber $q$ behind the
front:  In the leading edge to the right, $\hat{\Omega}$ simply needs to be
equal to the temporal phase winding  in the moving frame, 
${\hat{\Omega}} = \omega_{\rm r}^* - v^* k_{\rm r}^* = -c_1$.
Likewise, let us write the asymptotic phase-winding behavior behind
the front for $\xi \to -\infty$ as $\hat{A}(\xi) = ae^{iq\xi}$; the amplitude $a$,
wavenumber $q$ and frequency $\Omega$ are then related by 
$\hat{\Omega} = (c_1+c_3) q^2 - c_3\,\varepsilon + v^* q$,
which is the analog of Eq.~(\ref{phasewindingvalues})
in the moving frame. Equating the two expressions for $\hat{\Omega}$ and solving for the
root $-\sqrt{\varepsilon}<q<\sqrt{\varepsilon}$  simply yields for the
wavenumber $q_{\rm sel}$ selected by 
the coherent pulled front solution \cite{nb}
\begin{equation}
q_{\rm sel} = \frac{\sqrt{1+c^2_1}- \sqrt{1+c_3^2}}{c_1+c_3}\;\sqrt{\varepsilon}   . \label{cglqsel}
\end{equation}

\begin{figure}[t]
\begin{center}
\epsfig{figure=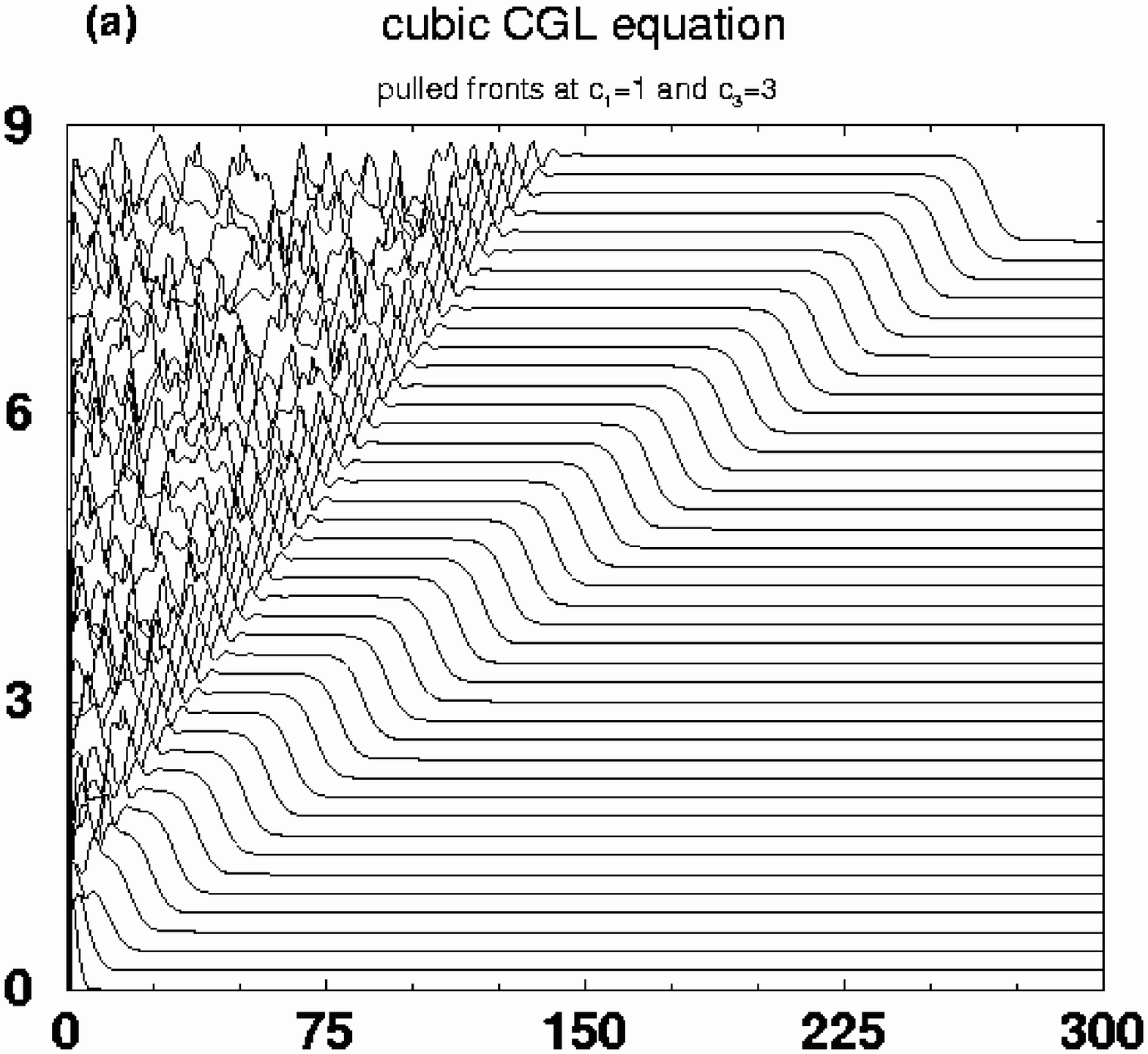,width=0.432\linewidth,angle=-90} 
\hspace*{2mm} \epsfig{figure=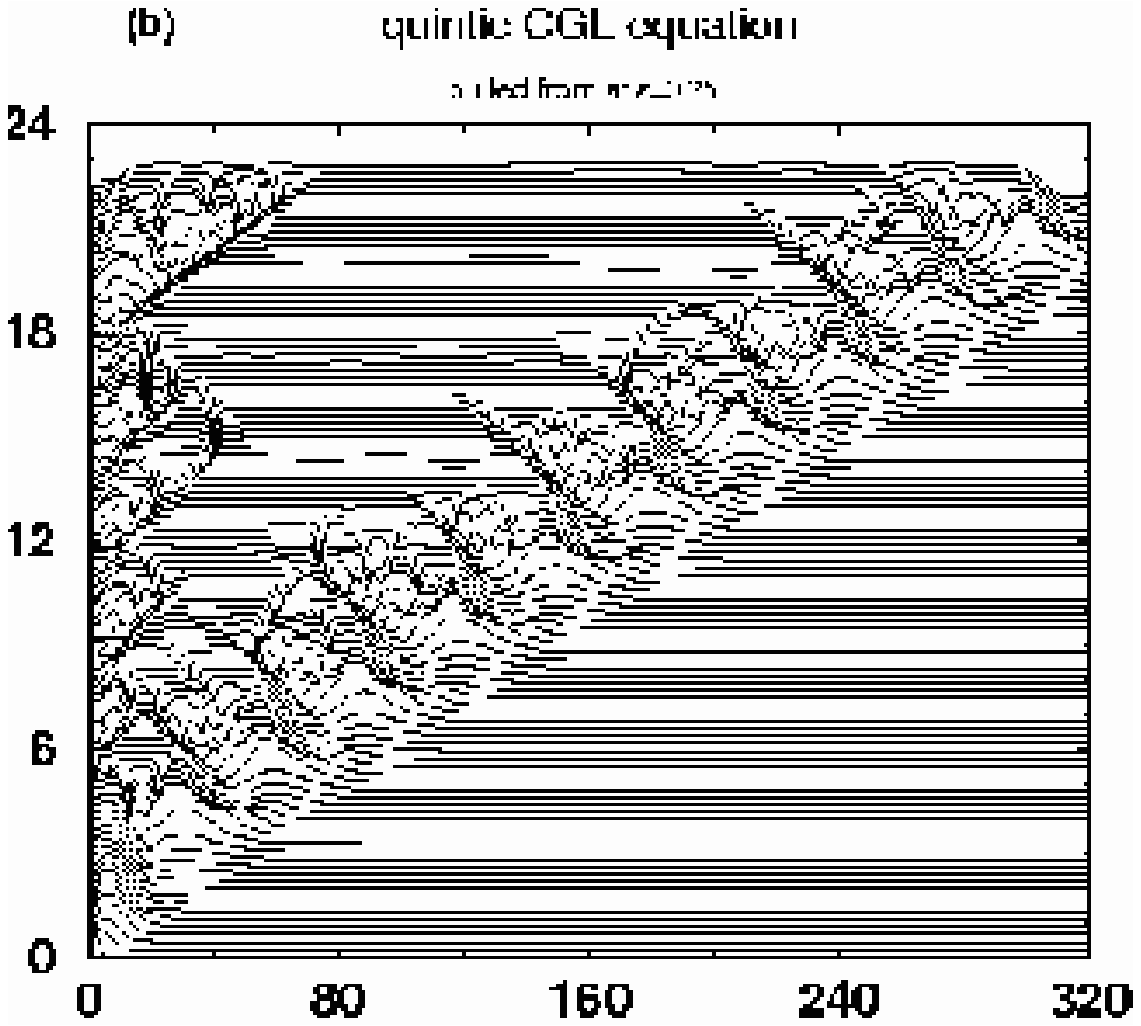,width=0.432\linewidth,angle=-90}
\end{center}
\caption[]{ (a) Space-time plot of the amplitude $A$ in a 
simulation of the cubic GL equation (\ref{cubiccgl}) with
$\varepsilon=1$, $c_1=1$ and $c_3=3$.  Time difference between successive lines is 2
and the total simulation time is 80. The plot illustrates how the
first front is a pulled coherent pattern forming front of the type
(\ref{coherentfront}), but that the state behind the front is
unstable, so that the domain with the state selected by the first
front is invaded by a second front. This second front is also a pulled
one. Note the similarity between the leading part of the second front
and the pulled front found in the Kuramoto-Sivashinsky equation ---
see Fig.~\ref{figks} {\em (a)}. This is no accident, as the
Kuramoto-Sivashinsky equation is the lowest order amplitude equation
just above the Benjamin-Feir instability \cite{ch,gc}. (b) Space-time plot of a
propagating pulled front in the quintic CGL equation with
$\varepsilon=0.25$, $c_1=1$, $c_3= -1$ and $c_5=-1$. For these parameter
values, no coherent nonlinear front solution of the type
(\ref{coherentfront}) exists \cite{vsh}.  The first front is followed
by a region with phase slips (points where $A$ vanishes), before
giving rise to a state close to a phase-winding solution. At the later
stage, this region is invaded by another  front.} \label{figcglcubic}
\end{figure}

Soon after the broader applicability of what we now refer to as the
pulled front concept started to emerge \cite{bj,dl}, Nozaki and Bekki
\cite{nb} studied the front propagation problem in the CGL
numerically. Their results can be understood simply in combination
with the results of the stability analysis of the phase winding
solution $q_{\rm sel}$:
\begin{enumerate}
\item In large parts of the phase diagram, the first
front propagating into the unstable state $A=0$ is a coherent pulled
front of the type (\ref{coherentfront}), with velocity $v^*_{\rm CGL}$
given by (\ref{v*cgl}) and selected wavenumber $q_{\rm sel}$ given by
(\ref{cglqsel}). 
\item
 If the selected mode $\sim \exp({iq_{\rm sel}x})$
is stable, a domain with this state remains behind the
front.\footnote{Depending on the initial conditions, this domain can 
subsequently be invaded by a front connecting to another phase winding
solution.}
\item
 As Fig.~\ref{figcglcubic} illustrates, if the $q_{\rm sel}$ is unstable due to
the Benjamin-Feir instability, the 
domain behind the first coherent front is invaded by a second front. This front is
again a pulled front \cite{dee2,nb}, whose speed $v^*_{\rm BF} $ can be calculated
explicitly in terms of $q_{\rm sel}$ and  $c_1$ and
$c_3$.\footnote{Since the Kuramoto-Sivashinsky equation is the
amplitude equation for the Benjamin-Feir instability in the limit when the
instability is weak  \cite{ch}, the second front is essentially a
pulled Kuramoto-Sivashinsky front.}   Depending on the parameters $c_1$ and $c_3$ this front can 
give rise either to a more stable phase winding solution, or a chaotic
state. An example of the latter case is shown in
Fig.~\ref{figcglcubic} {\em (a)}. Clearly, this regime can only exist
if the first coherent front outruns the second one, hence in the
parameter range  where $v_{\rm CGL}^* > v^*_{\rm BF}$. In the frame
moving with the first front, the state generated by the first front is
nonlinearly convectively unstable to the Benjamin-Feir instability.
\item
 In the parameter range where the Benjamin-Feir instability becomes
nonlinearly absolutely unstable in the frame moving with the first
front, i.e., in the range where $v^*_{\rm
CGL}< v^*_{\rm BF}$ (which is most easily entered by tuning $c_3$ as
$v^*_{\rm CGL}$ does not depend on $c_3$) the coherent pulled front
solution is irrelevant. Instead, one observes a pulled incoherent
front:\footnote{Actually, it is conceivable that in some parameter
ranges the Benjamin-Feir instability of the nonlinear region is so
strong, that the very first front becomes a pushed incoherent
front. To my knowledge this issue has not been studied
systematically.}    the chaotic behavior sets in right at the front region front.
The dynamical behavior in this regime is similar to that shown in
Fig.~\ref{figcglquintic2} {\em (a)} below for the pulled chaotic behavior of
the quintic CGL. 
\end{enumerate}

In conclusion, the cubic CGL equation admits a two-parameter family of
coherent front solutions. This family includes the pulled front solution, for
which the selected wavenumber can be determined explicitly. The
dynamics behind the first front depends on the stability of this
state. When $v^*_{\rm BF} <v^*_{\rm CGL} $ the first front is a 
coherent pulled front, while in the opposite regime an incoherent pulled
front emerges. 

\subsubsection{The quintic Complex Ginzburg-Landau equation}\label{sectioncglquintic}
Unlike the cubic CGL equation, which arises as the lowest order
amplitude equation near a supercritical (forward) bifurcation to
traveling wave patterns, the quintic CGL equation 
\begin{equation}
\partial_t A= (1+ic_1) \partial_x^2A + \varepsilon A +(1+ic_3)|A|^2A
-(1-ic_5) |A|^4 A \label{quinticcgl}
\end{equation} 
is  a model equation for the case in which such a bifurcation
is subcritical (backward). Because of the different sign of the cubic 
term, this term now enhances the growth of amplitude, while saturation 
is only caused by the quintic term. Indeed,  the destabilizing cubic
term implies that the only stable   phase winding solutions
 have a finite amplitude  for any $\varepsilon$. So an expansion based
on assuming that the amplitude is
small is not really justified. In other words, the quintic CGL  is  not a consistent
lowest order amplitude equation. Nevertheless, the equation has played
an important role in identifying the main coherent structures and
dynamical regimes near a subcritical bifurcation.  Note that the terms 
linear in $A$ are the same in the quintic and cubic CGL, so the linear 
spreading point expressions (\ref{v*cgl}) apply to the quintic CGL
equation for $\varepsilon>0$ as well.

In the subcritical range, $\varepsilon<0$, the state $A=0$ is linearly 
stable, and hence a small perturbation does not spread with a finite
speed:   instead  it will die out.  However,  nonlinear states, in
particular phase winding solutions, can perfectly well be stable. In
analogy with 
thermodynamic systems, where one is used to interface-type solutions
between two stable phases separated by a first order transition, 
one  expects there to be front solutions in the subcritical range 
$\varepsilon<0$. When $\varepsilon$ is increased towards small but
positive values, these front solutions would then be  expected to remain
dynamically relevant as they  have   have nonzero speed larger
than the linear spreading speed $v^*_{\rm CGL} \sim
\sqrt{\varepsilon}$ . In other words, one would naively expect that
for small $\varepsilon$ fronts in the quintic CGL are pushed. 

The remarkable feature of the quintic CGL is that the front solutions
corresponding to the pushed fronts can be obtained {\em analytically}
\cite{vsh0,vsh} by a  generalization of the ``reduction of order method''
reviewed  briefly in the example at the end of section \ref{selectionutfs}: When $\hat{A}(\zeta)$ with
$\zeta=x-v^\dagger t$ is written in the form $\hat{A}= a(\zeta) e^{i \phi
  (\zeta) } $ and the Ansatz 
\begin{equation}
  \frac{da}{d\zeta } = \sqrt{ e_1(a^2-a_N^2)},
~~~~~~~~\frac{ d\phi}{d\zeta}  =q_N+e_0(a^2-a_N^2) ,
\end{equation} is substituted into the equations, one arrives at
algebraic equations for the coefficients $e_0$ and $e_1$ and the
wavenumber $q_N$ and amplitude $a_N$ of the phase winding solution
behind this front solution. These coherent pattern forming pushed
front solutions, which are unique for a
fixed set of parameters, only exist in parts of the  $\varepsilon,
c_1, c_3, c_5$ parameter space: Effectively they  exist only  in a band 
around the line $c_3 =  -c_5$ where the nonlinear dispersion terms almost
cancel and especially in the subcritical range $\varepsilon <0$.

\begin{figure}[t]
\begin{center}
\epsfig{figure=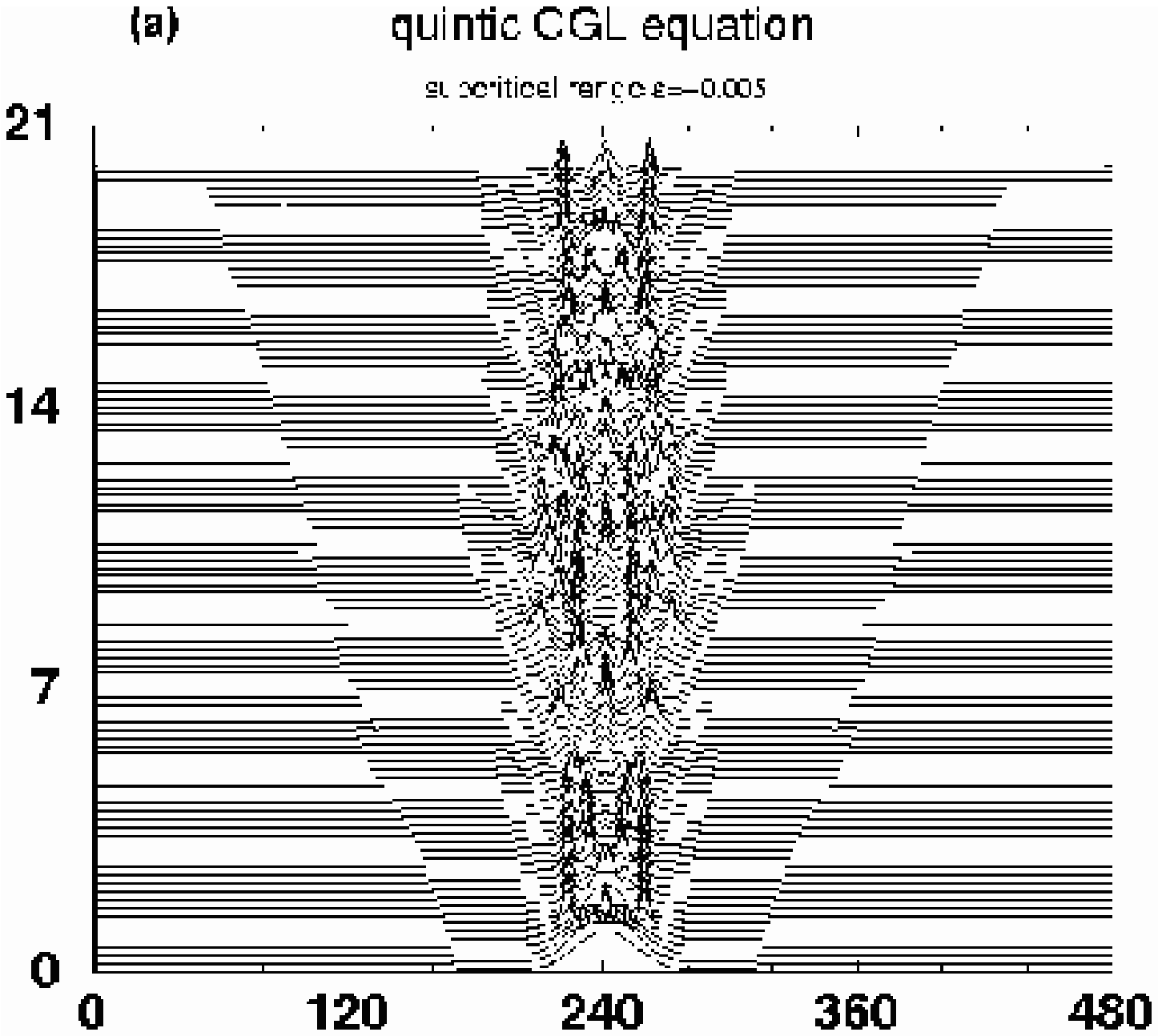,width=0.432\linewidth,angle=-90} 
\hspace*{2mm} \epsfig{figure=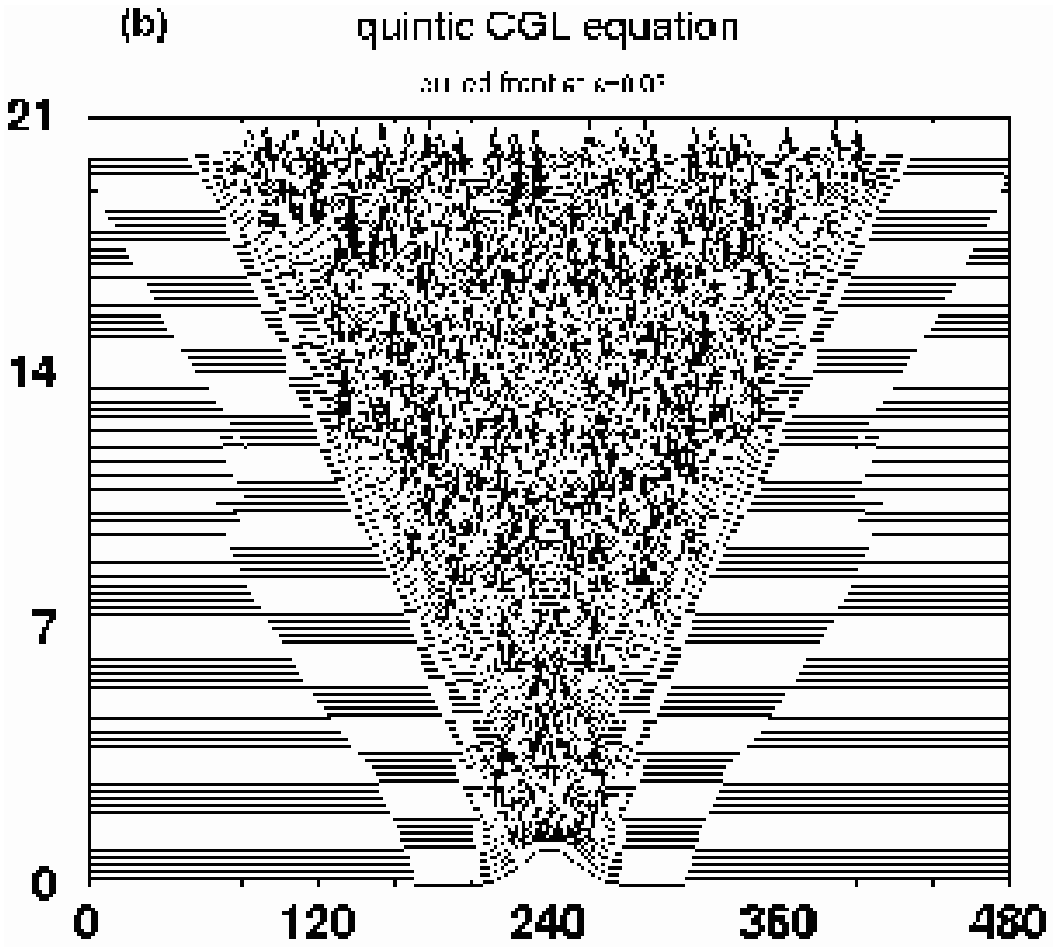,width=0.432\linewidth,angle=-90}
\end{center}
\caption[]{ Space-time plot of $A$ illustrating  the pushed to pulled transition for
incoherently propagating fronts in the quintic CGL equation
(\ref{quinticcgl}) \cite{deissler2}.   In each plot, the total simulation time is 100,
and the time difference between successive lines  is 2. (a) Simulation for $c_1=-2.2$, $ c_3=-0.5$,
$c_5=-2$ in the weakly subcritical range
$\varepsilon=-0.005$. According to figure 15 of \cite{vsh}, in this
range the incoherent domain in the center slowly spreads,
corresponding with a pushed incoherent front with velocity $\approx
0.15$, although this may be a finite-size effect.
 (b) The same for $\varepsilon =0.03$.  } \label{figcglquintic1}
\end{figure}

The main findings of a detailed study of the existence of these solutions
and of their competition with pulled front solutions are:
\begin{enumerate}
\item
 Just like we discussed for the
pulled front solutions of the cubic CGL equation, the phase winding
solution with wavenumber $q_N$ behind the pushed front can be unstable
to the Benjamin-Feir instability. If it is, and if the propagation
velocity $v^*_{\rm BF}$ is smaller than $v^\dagger$, then the size of
the  domain 
behind the first front grows as $(v^\dagger-v^*_{\rm BF} )t$ in
time. If $v^*_{\rm BF} > v^\dagger$, however, an incoherent pattern
forming pushed front results (see figures 13 and 14 of \cite{vsh}).
\item
 Even though the quintic Ginzburg-Landau
equation always admits phase winding solutions in a  subcritical
range $\varepsilon_c<\varepsilon<0$, outside a band where $c_3\approx -c_5$ there are no
pushed front solutions. This means that the ``thermodynamic''
intuition formulated above, according to which one always expects
a pushed to pulled transition at some {\em finite } positive value of
$\varepsilon$, does not apply to the quintic CGL equation (in line
with the fact that there is no standard Lyapunov function for this
equation). In other words, in some parameter ranges, the dynamically
relevant fronts for {\em any} positive $\varepsilon $ are pulled
fronts! This is quite remarkable, as it means that in the parameter
ranges where this happens, from the point of view of the front
propagation problem the dynamical behavior is more like that found near a
supercritical bifurcation.\footnote{For $\varepsilon <0$, a new class of retracting
fronts  ($v<0$) was recently found \cite{coulletkramer}; these solutions determine  much of
the dynamical behavior in the subcritical regime, especially outside
the band where the pushed front solutions exist. } This also
illustrates that while for the F-KPP equation the nonlinear behavior
of the growth function $f(\phi)$ determines whether fronts are pushed
or pulled, the distinction between the two regimes is generally more
subtle ---   in the quintic CGL it is determined by the strength of the
nonlinear dispersion, in the extension of the Kuramoto-Sivashinsky
equation it is tuned  by a symmetry-breaking linear third
order derivative term. 
\item
In equations where the relevant front solutions are uniformly
translating ones, the pushed solutions normally bifurcate off the pulled ones
continuously. In the quintic CGL equation, this is generically not the
case  for pattern forming fronts --- see the discussion in section \ref{stabilityandothermechanisms}. As a
result, the selected wavenumber can {\em  jump}  as one goes from a coherent
pulled front solution to a coherent pushed front solution (see figure
1{\em (c)}  of \cite{vsh0} or 23  of \cite{ch}). Thus, the coherent
pattern forming front solutions found in the quintic CGL provide a
counterexample to the ``structural stability'' postulate
\cite{paquette1,paquette2} that pulled 
front solutions are smoothly connected to the pushed front solutions,
as one tunes one of the parameters in the dynamical equation.
\item
 As we discussed, for the cubic CGL
equation, so-called ``counting arguments'' for the multiplicity of
front solutions indicate that there generally is a two-parameter
family of front solutions. Although this has not been checked
explicitly, we believe that for any value of the parameters there is indeed
a unique pulled front solution, in line with the fact that
(\ref{cglqsel}) fixes $q_{\rm   sel}$ uniquely. For pulled fronts
in the quintic CGL equation, a new feature is encountered: an
extension of the earlier analysis to the quintic  
case shows that in some parameter ranges the equations for $q_{\rm
  sel}$ behind the front do {\em not} admit a  solution. This implies that in these 
parameter ranges no coherent front  solutions (\ref{coherentfront}) can exist
\cite{vsh}. Figure \ref{figcglcubic}{\em (b)} shows an example of the
type of  dynamical behavior that is found in this regime:
in the front region the dynamics is very incoherent and associated
with the occurrence of ``phase slips'' (points in space-time where
$A=0$). The fact that the incoherent front region  is followed by a zone where $A$ is close to a
stable phase-winding solution illustrates that the incoherent front
dynamics in this case is not due to an  instability of
the selected state, but {\em to the absence of a coherent pulled front
solution}. 

\begin{figure}[t]
\begin{center}
 \epsfig{figure=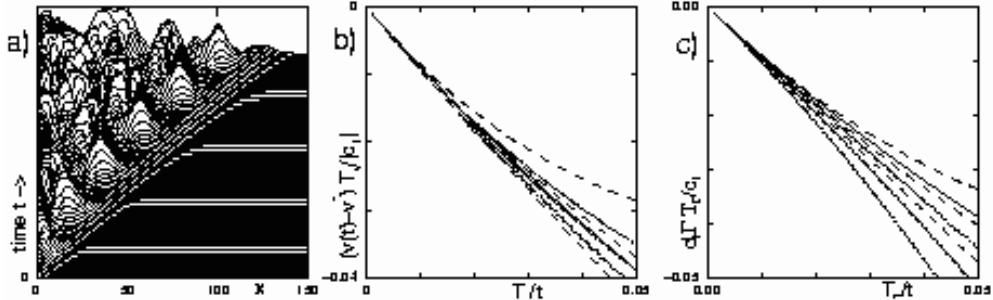,width=0.95\linewidth} 
\end{center}
\caption[]{ (a) Space-time plot of the amplitude $|A|$  of a pulled incoherent
  front in the quintic CGL at $\varepsilon=0.25$, $c_1=1$,
  $c_3=c_5=-3$. Note that in the leading edge, where $|A|$ is small,
the dynamics is actually quite coherent, in agreement with the fact
that the linear spreading dynamics is always coherent. The front only
becomes incoherent once the nonlinear behavior sets in. (b) Scaling plot of the scaled velocity relaxation
  $[v(t)-v^*]T /C_1$ versus $T/t$  for
  simulations of coherent fronts in the Swift-Hohenberg equation and
  the quintic CGL simulations shown in (a). Here $T $ is a properly defined time scale
introduced so that both equations fall on the same asymptotic curve,
and $C_1$ is an appropriate combination of the linear spreading point parameters
\cite{storm1}.  From left to right, the
  first three lines are Swift-Hohenberg data with $\varepsilon=5$, at
  heights $u=\sqrt{\varepsilon}, 0.01 \sqrt{\varepsilon} $ and $0.0001 
  \sqrt{\varepsilon}$, and the next three lines show the CGL data
  traced at amplitudes $0.002$, $0.0002$ and $0.00002$. The solid line 
  is the universal asymptote predicted by the analytic theory.  (c)
  The same for the appropriately scaled $\dot{\Gamma}(t)$ relaxation (\ref{gamma(t)}). From
  \cite{storm1}.} \label{figcglquintic2}
\end{figure}

\item
 Because of the subcritical nature  of the transition,
the quintic CGL equation also has parameter ranges at
$\epsilon<0$ where a 
chaotic phase is statistically stable.  If one puts such a  chaotic
phase next to a domain of the stable  $A=0$ phase, the front between
them can move. Upon increasing $\varepsilon$ one can then observe a
pushed to pulled transition of an incoherent front --- see Fig.~\ref{figcglquintic1}.

\item As we discussed in section \ref{sectionunirelsimple} , the universal 
$1/t$ power law relaxation even applies to the coherent and incoherent 
pattern forming fronts. This was checked numerically for the
simulations of the incoherent fronts of the CGL equation shown in
Fig.~\ref{figcglquintic2}{\em (a)}.  The other panels  
confirm that after scaling  the velocity and time  with the predicted 
values, 
both the velocity [panel {\em (b)}] and wavenumber [panel {\em (c)}]
relaxation data of  the incoherent quintic CGL fronts and the 
coherent fronts in the Swift-Hohenberg equation are  in accord
with the universal predictions.
\end{enumerate}

\subsection{Epilogue}\label{epilogue}

The insight we have attempted to bring across in section
\ref{secoverview} is that the special nature and generality of
front propagation into unstable states lies in the fact that there is
a well-defined linear spreading speed $v^*$ associated with any linear
instability. The linear spreading point is the organizing center for
all subsequent developments:  We can make a large number of general
predictions because the linear dynamics puts a strong constraint on
any  possible nonlinear dynamics. On the one hand, it forces us to
focus on the question ``how can a front faster than $v^*$ emerge?''
and this naturally leads to the concepts of pulled and pushed front. On the other
hand it shows that  the approach to the asymptotic velocity and state
of a pulled front is governed by
an equation which in lowest order is of diffusion-type, and this
allows one to derive the universal asymptotic relaxation of a pulled
front.

For uniformly translating fronts and coherent pattern forming fronts,
the regime of pushed front propagation is  essentially governed by the
{\em existence} of special solutions which we can think of
as ``strongly heteroclinic orbits''. 
For incoherent pushed fronts we are at present unable to
formulate the pushed front propagation mechanism sharply, although
intuitively we expect it to be very similar to that of the other two types
of fronts. 

The various examples in the previous section illustrate that a pulled
front essentially does what it should do, namely propagate with
asymptotic speed  and shape dictated by the linear spreading point,
while the rest of the front just follows in whatever way is allowed by
the global nonlinear properties of the dynamical equation: {\em (i)} a bifurcation in the structure of the
linear  spreading point is responsible for the transition to coherent
pattern forming fronts in  the EFK equation; {\em (ii)} the quintic
CGL equation illustrates that the limited
range of existence of coherent pattern forming fronts can lead to a
transition from coherent pulled fronts to
incoherent pulled fronts; {\em (iii)} if the state generated by a
front in the cubic or quintic CGL equation is unstable,  the first
front remains a coherent front if the spreading speed of the secondary
instability is less than $v^*$ of the first front, while the first
front  becomes
incoherent if the pulled speed of the secondary front is larger than
$v^*$ of the first front.

The difficulty in making generic statements about pushed fronts is
that apart from the fact that their convergence to the asymptotic
speed $v^\dagger$  is exponential, there are few generic
features. Because of the strong
heteroclinicity requirements (\ref{pushedexpr}) or
(\ref{pushedcoherent}) that the slowest decaying exponential mode 
is absent,  pushed front solutions are technically determined by an analysis which has all the properties
of a  nonlinear eigenvalue problem for an isolated solution. Because
of this structure, the relaxation towards the asymptotic pushed front
speed and shape is generally exponential, and singular perturbation
theory can be applied to pushed fronts (section \ref{sectionmba}).  
Moreover, because of this structure, whether a pushed front solution exists depends on the full
{\em global} and hence {\em nonlinear} properties of the equations. As
a result, virtually every example is special, and there are few
nontrivial examples which one has been able to obtain
analytically. The quintic CGL equation is a remarkable exception in
this regard. 

As we discussed, with few exceptions, most  mathematical investigations of front
propagation into unstable states have focused on uniformly
translating fronts and in particular on systems of equations which are
closely related to the F-KPP equation. It appears to us that many of
the  ingredients of the theoretical framework that we have discussed
here are ready for more rigorous analysis: rigorous developments that
I expect are around the corner are  {\em (i)}
 rigorous bounds for the Green's function of whole classes of 
linearized equations and, in relation with this, rigorous proofs of
the pulled front mechanism of new classes of equations,  {\em (ii)}
proofs of  the universal relaxation of  pulled fronts,  {\em (iii)}
the development of a mathematical framework for the pulled to pushed
transition of coherent pattern forming fronts, and {\em (iv)}
extension of the approach of Sandstede and Scheel \cite{sandstede,sandstede2} to analyze the
interaction between front dynamics and the dynamics of the state
behind the front, or to classify defects.

\section{Experimental and theoretical examples of front propagation
into unstable states}\label{sectionexamples}

In this section  we discuss a number of examples of front propagation
into unstable states. We first review experiments   on fronts or in
which   front propagation plays an important  role, and then move
on to discuss examples of fronts in theoretical models. The examples
concern mostly fronts in one dimension. The 
difficulty of analyzing patterns whose dynamics is governed by the
propagation of pulled fronts is discussed later in section \ref{sectionmba}.

At first sight, one might well wonder why front propagation into unstable states 
is a relevant problem anyway. After all, it seems counterintuitive
that a physical system  would naturally end up in an {\em unstable}  steady state and
wait for a nice front to develop! On closer inspection, however, there 
are many reasons why front propagation into unstable states is not an 
esoteric problem:
\begin{enumerate}
\item  A natural way in which a system can stay in a
  self-sustained unstable state is when an overall flow or motion
  makes the instability convective (see sections
  \ref{sectionconvversusabs} and \ref{sectionnonlinearconv2abs}). This 
  happens in particular in many fluid dynamic systems. Likewise, in
  plasma physics the strong asymmetry between the mobility of
  electrons and ions usually makes plasma instabilities convective. 
Examples where these considerations apply are discussed in sections
\ref{sectionstreamers}, \ref{sectionparallelshear},
\ref{sectionvonkarman}, \ref{secfrontsnoise1}, \ref{secfrontsnoise2},
and \ref{sectionmullinssekerka}. 
\item
Sometimes a natural way to probe a system is to quench it into an
unstable state with a laser pulse, ramping or flipping a voltage, flipping a
plate, dropping a temperature, scraping a skin tissue, etcetera. In all these 
cases, the relevant question is whether we can change the system on a
timescale shorter than the natural  time scale of the evolution  of the state into
which we quench it.  Examples are encountered in sections
\ref{sectiontcrb}, \ref{sectionpropagatingrt},
\ref{secpropRayleigh}, \ref{sectionstreamers}, \ref{sectiondebunching}, \ref{secspinodal},
\ref{sectiondynamicaltest}, \ref{sectionsmectic}, and
\ref{sectionwounds}.
\item We sometimes encounter intrinsically chaotic
  systems where the dynamics itself continuously generates states
  which are transient because they 
   themselves are unstable. We discuss a clear example of this in
   section \ref{sectionkupperslorz}.
\item  In thermodynamic systems with first order
  transitions, interfaces play a predominant role because the bulk
  nucleation rate is exponentially small near the transition, while
  interface velocities  usually vary  linearly  with the distance from 
  the transition. Although fluctuations near a second order
  thermodynamic transition are usually too large to allow clear fronts to
  develop, this is not necessarily true for polymer systems (section
  \ref{secspinodal}). Moreover, near supercritical (``second order
  like'') bifurcations in pattern forming systems, where fluctuation effects  are
  usually small \cite{ch}, a similar argument holds as well: as we
  shall discuss more explicitly in the next subsection, if
  $\varepsilon$ is the dimensionless control parameter measuring the
  distance from the instability threshold,  the front speed
  scales as $\sqrt{\varepsilon}$ while the growth rate of bulk modes
  scales as $\varepsilon$. Hence, front propagation in principle
  always dominates for sufficiently small $\varepsilon$
  \cite{fineberg}. Of course, the range over which fronts do dominate
  the dynamics depends on the prefactors, i.e., on the time and length
  scales of the problem under investigation. Such considerations play
  a role in the examples of sections \ref{sectiontcrb},
  \ref{sectionpropagatingrt}, \ref{secpropRayleigh},
  \ref{sectionstreamers}, \ref{secspinodal}, \ref{sectionsupercond},
  and \ref{sectiondynamicaltest}.
\item Front propagation into an unstable state sometimes
  emerges theoretically from an unexpected angle through mapping a seemingly
  unrelated problem onto a front propagation problem --- see sections
  \ref{meanfieldgrowth}, 
  \ref{sectionvanzon}, \ref{sectionphasetrans}, \ref{sectioncarpentier} and \ref{sectionsearchtrees}.
\end{enumerate}

     \subsection{Fronts in Taylor-Couette and Rayleigh-B\'enard experiments}\label{sectiontcrb}
Soon after Dee and Langer \cite{dl} and Ben-Jacob and co-workers
\cite{bj} drew the
attention of the physics community  
  to the special simplicity of what we now refer to as pulled fronts,
two  experiments on pattern forming systems were done that 
are still very illuminating: One by Ahlers and
Cannell \cite{ahlers}  on vortex fronts in a Taylor-Couette cell, and one
by Fineberg and Steinberg \cite{fineberg} on front propagation in a
Rayleigh-B\'enard cell.

A Taylor-Couette cell consists of two concentric cylinders which can
rotate independently. The gap between the two cylinders is filled with
a normal Newtonian fluid. When the inner and outer cylinder rotate slowly, the flow
between the two cylinders is laminar and in the azimuthal
direction. At higher rotation rates,  one finds an amazing number of fluid dynamics
instabilities when both the inner and the  outer cylinder are rotating
with sufficiently high frequencies  \cite{ch,swinney1}.  
The experiments that are of interest to us  probed fronts
that lead to  the first nontrivial patterned state that emerges
if the rotation rate of the inner cylinder is increased while the
outer cylinder does not rotate  \cite{ahlers}. At some critical
rotation rate, there
is  a bifurcation to a state with so-called Taylor vortices:
a   pattern which is periodic in the direction along the axis of the cylinders
 emerges. This flow is referred to as Taylor vortex flow, since the
velocity in the $r,z$ plane has the appearance of alternating
clockwise and counterclockwise rotating vortices that span the whole
gap between the two cylinders. 

A Rayleigh-B\'enard cell consists of a fluid sandwiched between two
parallel plates at a fixed but different temperature. When the  bottom
plate is hotter than the top plate, the fluid density is largest at
the top. For small enough temperature differences, there is no
convection --- there is just heat conduction  through a quiescent fluid.
However, when the temperature difference and hence the density
difference becomes  larger than some threshold value, this simple
 state   is unstable and convection
sets in spontaneously \cite{ch,fauve,gc,manneville}. The
convection patterns that emerge just above threshold are often referred
to as rolls, but if we look at the velocity pattern in a vertical
cross section, they look like the vortex-type patterns in the $r,z$
cross-section of the Taylor-Couette cell.  

The transition to Taylor vortex patterns and to convection patterns in
the Rayleigh-B\'enard case are both examples  of forward
(supercritical) bifurcations to a stationary finite wavelength
patterns.  Just above onset, the patterns can be described by a
so-called amplitude expansion by writing the physical fields in terms
of the complex amplitude as \cite{ch,gc,hakim,newell2,nishiura}
\begin{equation}
\mbox{ physical fields} \propto A e^{ik_cx} + A^* e^{-ik_c x} . \label{physicalfields}
\end{equation}
For simplicity, we restrict the analysis to one-dimensional patterns
as this is the relevant case
for the front experiments; $k_c$ is the wavenumber of the
pattern at onset, and $x$ is the coordinate along which the patterns
develop. The amplitude expansion  simply   implements 
the observation that just above onset the amplitude of the patterns is
small, and that the amplitude $A$ varies on slow space and time
scales. In lowest order, $A$ then obeys the so-called real   Ginzburg-Landau amplitude
equation\footnote{The equation --- used as the amplitude equation (\ref{ampeq}) in
the discussion of fronts in Rayleigh-B\'enard and Taylor-Couette
experiments in section \ref{sectiontcrb} below --- is  called the
real Ginzburg-Landau 
equation because the prefactors of all the terms are real.  The
amplitude  equation for traveling wave patterns is the CGL
 equation; as discussed in section \ref{sectioncglcubic}, in principle
all its terms  have complex prefactors
\cite{ch,gc,walgraef}. } \cite{ch,fauve,gc,newell2,nishiura,vhhvs,walgraef}
\begin{equation}
\tau_0 \partial_t A = \epsilon A + \xi_0^2 \partial^2_x A - g |A|^2 A.
\label{ampeq}
\end{equation}
Here $\epsilon$ is the dimensionless control parameter, which for
$\epsilon >0$  is a
measure for how far one is above the instability threshold.
The time scale $\tau_0$ and length scale $\xi_0$  depend on the
particular system, and  can be calculated explicitly from the linear 
dispersion relation of the system under study. This has been done  both for the
transition to Taylor vortices and for the transition to
Rayleigh-B\'enard convection patterns. The nonlinear parameter $g$,
which follows from the amplitude expansion, is also
known for both systems, but it, of course, does not play a role for the
pulled fronts of interest here. 

The real amplitude or Ginzburg-Landau equation should in general not be thought of as just a simple
straightforward generalization of the F-KPP equation to a complex
field --- while the F-KPP equation only allows for stable homogeneous
solutions $u=\pm 1$, the amplitude equation admits stable  phase winding
solutions of the form $A=a e^{iqx}$ with $a$ and $q$ constant \cite{ch,fauve,gc,nishiura,walgraef}. As
(\ref{physicalfields}) illustrates, these
describe physical patterns with a wavenumber $k_c+q$ different from
the critical wavenumber. The interaction and competition between the
various modes in general makes the dynamics of the amplitude $A$ much
more interesting and complicated than that of the F-KPP
equation. Nevertheless, since the dynamics of pulled fronts is
dominated by the linear terms in $A$, and since these are the same for 
the two equations, both the asymptotic pulled front solutions and the
convergence to them are the same for the two equations. 

The attractive feature of the experiments is that because $\xi_0$ and $\tau_0$ are known
explicitly, they offer the possibility to study  front propagation
{\em quantitatively}. Indeed, for the amplitude equation in the form
(\ref{ampeq}) we get for the pulled front  speed
\begin{equation}
v_{\rm ampeq}^* = 2 \frac{\xi_0}{\tau_0} \sqrt{\epsilon} . \label{vwithunits}
\end{equation}
This result explicitly confirms the assertion made already in point 
{\em(iv)} at the beginning of section \ref{sectionexamples}
that near a supercritical bifurcation, the front speed increases
as $\sqrt{\epsilon}$, while the growth rate of bulk modes according to
(\ref{ampeq}) increases linearly in $\epsilon$. As stressed by
Fineberg and Steinberg \cite{fineberg}, front propagation into
unstable states can therefore generally be observed at small enough
$\epsilon$, i.e. just above the onset of the transition.

In order to compare the measured front velocities to the 
F-KPP equation in the standard form in which the prefactors of the linear terms are equal
to 1, the  experimental results have been reported in terms of the
scaled velocity
\begin{equation}
\tilde{v} = v \frac{\tau_0}{\xi_0 \sqrt{\epsilon}}.  \label{vscaled}
\end{equation}

\begin{figure}[t]
\begin{center}
{\tt (a)} \hspace*{-3mm} \epsfig{figure=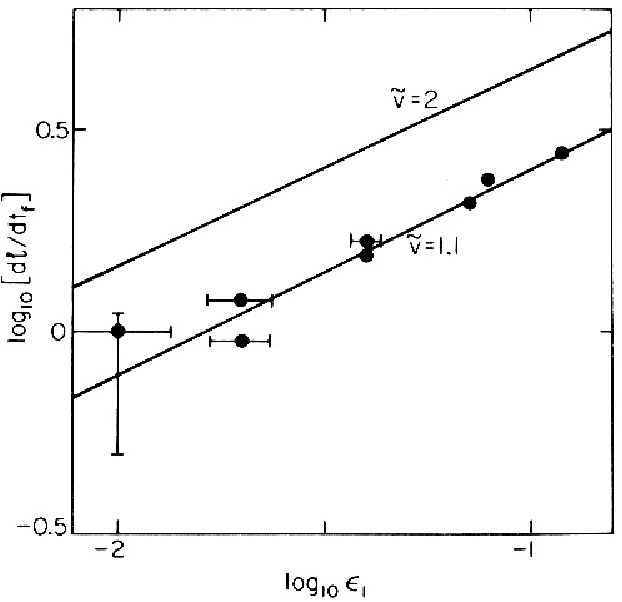,width=0.45\linewidth} \hspace*{0.3cm}
{\tt (b)} \hspace*{-3mm} \epsfig{figure=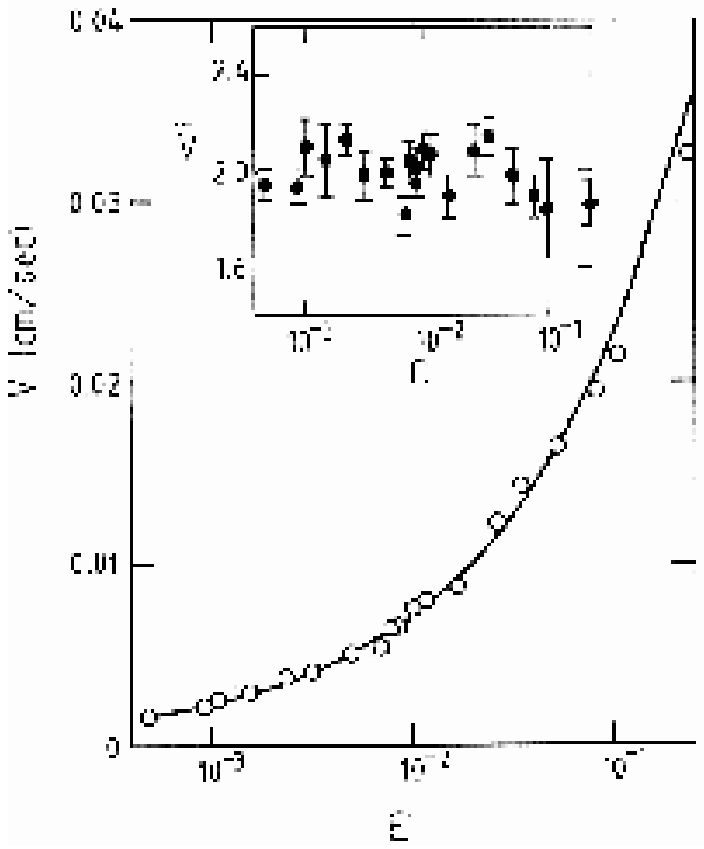,width=0.40\linewidth}
\end{center}
\caption[]{ (a) Velocity data of fronts in the Taylor vortex
experiments of Ahlers and Cannell \cite{ahlers}. Data are plotted on a
log-log scale as a function of $\epsilon$. The solid lines have slope
$\frac{1}{2}$ and hence illustrate the $\sqrt{\epsilon}$ scaling of
the velocity. The lower line shows that the data are consistent with a
scaled velocity $\tilde{v} \approx 1.1$ instead of the asymptotic
values $\tilde{v}^{\rm as}= 2$.  (b)
Experimental data of Fineberg and Steinberg \cite{fineberg} on fronts
in the Rayleigh-B\'enard system on a log-linear scale. The solid line
is the theoretical prediction (\ref{vwithunits}). The inset shows the
scaled velocity $\tilde{v}$ which is consistent with the theoretical
asymptotic value 2, within the error bars.}\label{figvdataexp}
\end{figure}

Of course, front speeds should only converge to $v^*_{\rm ampeq}$ if the initial
condition is sufficiently localized,  as explained in section
\ref{sectionleadingedgedominated}. In both experiments, the system was initially below the
instability threshold, corresponding to $\epsilon <0$ in the amplitude
equation (\ref{ampeq}), and then brought above threshold ($\epsilon
>0$) at time $t=0$. In the Taylor-Couette experiment, even below
threshold, when the bulk of the cell is in the stable laminar 
flow state, there are
so-called Ekman vortices at the ends of the cylinders. These play the role of
localized nonzero initial conditions for the amplitude $A$ when the
system is brought above threshold at time $t=0$. For the 
Rayleigh-B\'enard experiment, Fineberg and Steinberg used a long cell,
so as to get one-dimensional fronts, and they turned on an extra
heater at the cell ends as well,  when they brought the system to
$\epsilon>0$. This made the convection patterns nucleate at the ends
and then propagate into the bulk of the cell.

Fig.~\ref{figvdataexp}{\em (a)} shows the  velocity of the Taylor
vortex fronts as they propagate into the Taylor-Couette cell over a
distance of up to 10 to 15 vortex diameters \cite{ahlers}, plotted on
a log-log scale. As the two solid lines indicate, the $\sqrt{\epsilon}$-scaling of the
vortex velocity is confirmed, but the prefactor is only about 55\% of
the predicted asymptotic values $\tilde{v}^*_{\rm ampeq}=2$:
the data are consistent with  $\tilde{v}\approx
1.1$ instead of 2.

\begin{figure}[t]
\begin{center}
{\tt (a)} \hspace*{-3mm} \epsfig{figure=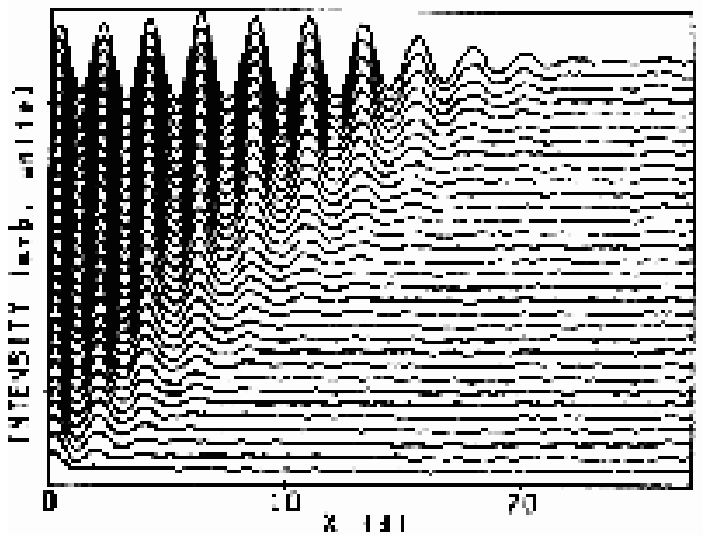,width=0.445\linewidth} \hspace*{0.3cm}
\hspace*{3mm}
\epsfig{figure=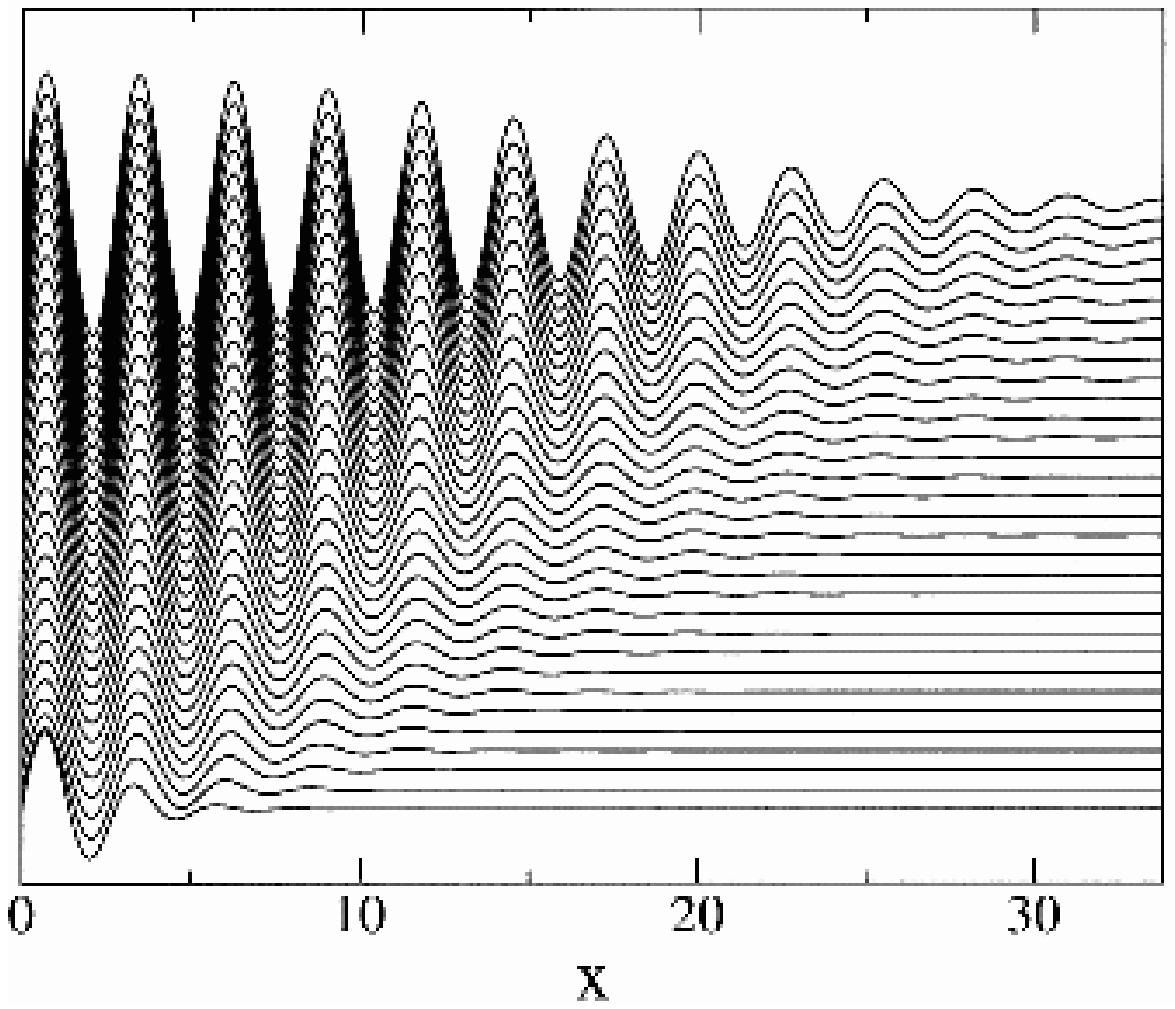,width=0.39\linewidth}\hspace*{-6.2cm}
{\tt (b)}\hspace*{5.8cm}
\end{center}
\caption[]{ (a) Space-time plot of a propagating front in the Rayleigh-B\'enard
experiment  of Fineberg and Steinberg \cite{fineberg} for
$\epsilon=0.012$, made using the shadowgraph technique. The time
between successive traces is 0.42 $t_v$, where $t_v$ is the vertical
diffusion time in the experiments, and distances are measured in units
of the cell height $d$. (b) A space-time plot obtained from a
numerical simulation of the Swift-Hohenberg equation at the same value
$\epsilon=0.012$ and length and time scales adjusted to correspond to
those in the experiments of (a). After Kockelkoren {\em et al.} \cite{kockelkoren}.
}\label{shadowgraph}
\end{figure}

The fact that the measured front velocity was significantly below the
asymptotic  value in this 1983 Taylor-Couette experiment made many
researchers wonder whether  there was some\-thing cru\-cial\-ly
wrong,\footnote{It was realized that if the measured velocity had been too high, a possible
candidate for the discrepancy could have been that the initial
convection profile associated with the Ekman end vortices was not
sufficiently localized, i.e., that at $t=0$ the amplitude did not fall
off to the right faster than $e^{-\lambda^*x}= e^{-\sqrt{\epsilon}
x/\xi_0 }$. However, as we discussed in section \ref{sectiontwofold}, asymptotic front
speeds below $v^*$ are impossible.}
and sti\-mu\-la\-ted Fine\-berg and Stein\-berg a few years later to
perform experiments on fronts in Rayleigh-B\'enard convection. Actual
space-time plots of their  convection front  profiles as they propagate to the
right along their cell, are shown in Fig.~\ref{shadowgraph}{\em (a)}.
These traces illustrate how these fronts (as well as those in the
Taylor-Couette cell)  are examples of true {\em
pattern forming fronts} propagating into an unstable state, like those
observed in simulations of the Swift-Hohenberg equation
(\ref{swifthohenberg}), see the central panel of Fig.~\ref{F1},
and Figs.~\ref{figsh}{\em (b)} and \ref{shadowgraph}{\em (b)}.

The data of Fig.~\ref{figvdataexp}{\em (b)} for the scaled front
velocity $\tilde{v}$, calculated from the actual data by using the
theoretical values for $\xi_0$ and $\tau_0$ in (\ref{vscaled}) were
found to be consistent to within a few percent with the asymptotic
value $\tilde{v}^{\rm as}=2$. At the time, this was considered to be an
important experimental confirmation of the quantitative prediction of
the pulled front speed, and of the idea that the concept of a pulled
front applies more generally to pattern forming systems. But, as we
shall see, there is a new twist to the story.

The reason for the apparent discrepancy between theory and experiment
on the Taylor vortex fronts remained unresolved till Niklas {\em et
al.} \cite{niklas} performed an explicit numerical simulation of such
fronts which showed that the front speed  had not yet relaxed
to its asymptotic value, in contrast to what was found in numerical
studies of the propagating fronts in the Rayleigh-B\'enard cell
\cite{luecke}.
Indeed,  over the range of times probed
experimentally,  the transient speeds they found for the Taylor vortex
fronts were quite consistent with the experimental data. Instead of showing these
results, we will approach this issue by extrapolating back from the
asymptotic formula (\ref{v(t)relaxation}) as this allows us to discuss both experiments
in a unified way.

When (\ref{v(t)relaxation}) is applied to the amplitude equation, one finds for the
long time rate of approach to the asymptotic speed $\tilde{v}^*=2$
\begin{equation}
\tilde{v}(t)=2 - \frac{3}{2 \epsilon t/\tau_0 } + \frac{3
\sqrt{\pi}}{2  (\epsilon t /\tau_0)^{3/2}} + \cdots .\label{asympscaled}
\end{equation}
This formula shows that in any  experiments on fronts just above the
onset of a finite-wavelength supercritical bifurcation, one can a
priori only
hope to observe   front speeds very close to the asymptotic value at times
\begin{equation}
t \gg t_{\rm co} \equiv \tau_0  \pi / \epsilon.
\end{equation}
Note that we have defined the crossover time $t_{\rm co}$ as the time at which the
third correction term is equal in magnitude to the first correction
term  in (\ref{asympscaled}).
Although this asymptotic formula is only valid in the long-time 
regime when the third term is much smaller
than the second one, and although for times of ${\mathcal O}(t_{\rm
co})$ one is in the non-universal crossover regime, the important
message is that the velocity always approaches the asymptotic value {\em from
below} and that transient effects are  very significant, even for
times of order several $t_{\rm co}$. For
example, even at times of  ${\mathcal O}(3  t_{\rm co})$ when the
formula becomes reasonably accurate \cite{evs2}, the velocity is still
some 8 percent below its asymptotic value, and at $t_{\rm co}$ both
corrections terms, though of opposite sign,  are about 24\% of the
asymptotic value in magnitude.

When an actual experiment is suddenly brought into the unstable
parameter range $\epsilon >0$, bulk modes also start to grow at a rate
proportional to $\epsilon$. Since one can only study fronts properly as long as
bulk  fluctuations have not grown enough that bulk ``nucleation''
starts to become visible, in practice experiments like those we
consider here can only be done up to a time which scales as
$\epsilon^{-1}$ \cite{ahlers,fineberg}. In other words, {\em
experiments can in practice only be done up
to a time which is a finite multiple of $t_{\rm co}$!}  

In the Taylor vortex front experiments, most measurements  for $\epsilon
=0.02$ were done in the time interval of 1-2 $t_{\rm co}$. It is
therefore not surprising that transient effects did play an important
role in these experiments \cite{niklas,wvsinbusse} and that the measured
velocity was significantly below the asymptotic value. However, from our present
perspective, it {\em is} important to turn the question around and
ask whether   the fronts observed in
the Rayleigh-B\'enard experiments  were indeed propagating with
approximately the asymptotic speed, and if so, why  \cite{kockelkoren}.

As it turns out, in the experiments of Fineberg and Steinberg, growth
of bulk modes also limited the observation of fronts to times of order
$t_{\rm co}$ \cite{fineberg}. Kockelkoren {\em et al.}
\cite{kockelkoren} therefore reanalyzed the experiment from this
perspective, and concluded that most likely the actual measured front
speed {\em was} some 15\% below the asymptotic one. In their
interpretation, the theoretical value for $\xi_0$ used in converting
the observed front speeds to dimensionless front speeds $\tilde{v}$
may have been about 15\%  smaller than the actual one.\footnote{Indications that this may
be the case come from fact that the observed wavelength at onset was
also about 13\% bigger than the theoretical one. See
\cite{kockelkoren} for further discussion of this.}  If so, the
velocity had been 
overestimated by about this amount in the interpretation of the
experiments, thus hiding a possible transient effect. 

\begin{figure}[t]
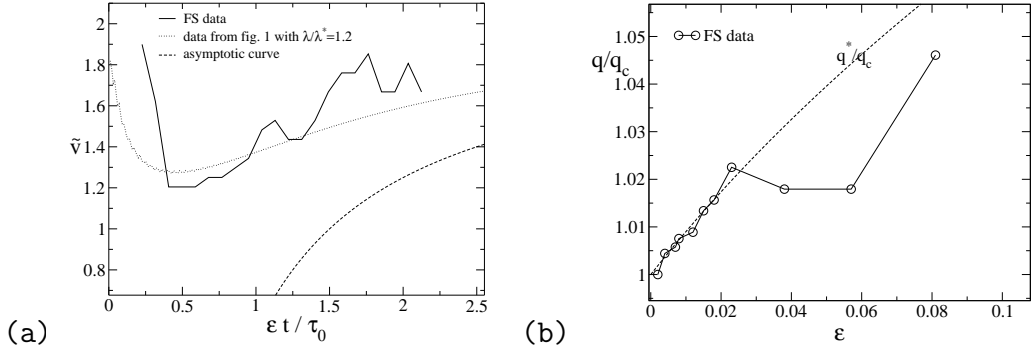

\begin{center}
{\tt (a)} \hspace*{-0.5mm}
\epsfig{figure=fig_vkockel.eps,width=0.40\linewidth} 
\hspace*{3mm}{\tt (b)} \hspace*{0.1mm}
\epsfig{figure=fig_wavelkockel.eps,width=0.42\linewidth}
\end{center}
\caption[]{ (a) Full line: scaled velocity versus time in the Rayleigh-B\'enard
front \cite{fineberg} whose space-time plot  is shown in
Fig.~\ref{shadowgraph}{\em (a)}. The velocity is obtained by
interpolating the maxima of the traces. As explained in the text,
a value of $\xi_0$ which is 15\% bigger than used by Fineberg and
Steinberg to convert the data. A different value would not change the
shape of the curve, only the absolute numbers on the vertical
axis. The dashed line shows the measured velocity in a simulation of
Eq.~(\ref{ampeq}) with an exponential initial condition $A(x,0)\sim
e^{-1.2 \lambda^* x}$ and the dotted line the asymptotic result
(\ref{asympscaled}) with the $t^{-3/2}$ term excluded. Note that the curves are not fitted, as only the
absolute vertical scale is affected by our choice of $\xi_0$.
   (b)  Open circles:  data-points for the wavenumber
selected by the front in the experiments of Fineberg and
Steinberg. Dashed curve:  $q^*/q_c$ for the Swift-Hohenberg equation,
 with parameters corresponding to
the Rayleigh-B\'enard experiment. The fact that the data follow the
line $q^*/q_c$ for small  $\varepsilon$, indicates that just above
threshold the experiments probe the leading edge of the front.
After Kockelkoren {\em et al.} \cite{kockelkoren}.
}\label{figkockel}
\end{figure}

We stress that only new
experiments can settle whether this interpretation is the correct
one. However, there are also indications in the actual traces of the
fronts in Fig.~\ref{shadowgraph}{\em (a)} that the velocity in this run
was slowly  increasing in time. The solid line in
Fig.~\ref{figkockel}{\em (a)}  shows the experimental velocity as a
function of time, extracted by measuring successive front positions
with a ruler and using a value of $\xi_0$ which is 15\% larger than
the theoretical value \cite{kockelkoren}. For comparison, a plot of the velocity as
function of time in a simulation of the real Ginzburg-Landau equation (\ref{ampeq})
is  shown with a dashed line. 

One important point  has not been included in the theoretical
analysis so far. When growth of bulk modes in
practice becomes important for times of order $t_{\rm co}$, then at
the latest times the growth of the bulk fluctuations may become
important. This will have the effect of {\em increasing} the velocity
above the one studied analytically and numerically with localized
initial conditions and  in the absence of
fluctuations. With this caveat, we nevertheless tentatively conclude
that both experiments on fronts in pattern forming systems illustrate
that such fronts are prone to  transient behavior associated with the
slow power law relaxation of pulled fronts from below to $v^*$.

As a final note, we point out that both experiments also addressed the
{\em wavelength selection} by the front. Wavelength selection goes
beyond the lowest order amplitude equation (\ref{ampeq}), since the
real F-KPP-type fronts in this equation simply yield a wavenumber
equal to the critical wavenumber $k_c$ throughout the front, to order
$\sqrt{\epsilon}$. For the changes in order $\epsilon$, one needs to
start from the  linear dispersion relation of the problem, correct to
order $\epsilon$. 
In the original paper on the Rayleigh-B\'enard experiment, the
data were  compared to the results for the selected wavelength in
the Swift-Hohenberg equation (\ref{swifthohenberg}). Strong deviations were found in
this case. The most likely explanation for this is the following
\cite{kockelkoren}: for small $\epsilon$, the full front whose width
scales as $1/\sqrt{\epsilon}$ is too wide to
even fit in the experimental cell; hence one then can not measure the
selected wavenumber {\em behind}  a
fully developed pulled front, but instead only an effective wavenumber in the leading edge  of
the front which is close to  the local wavenumber $k_{\rm r}^*$ given in
(\ref{v*sh}). Fig.~\ref{figkockel}{\em (b)} shows that indeed the
Rayleigh-B\'enard data for small $\epsilon$ are consistent with this.

More recently, the problem of fronts in the Rayleigh-B\'enard and
Taylor-Couette system has been analyzed in great  detail by L\"ucke and
co-workers \cite{buchel,luecke8,luecke1,luecke3,luecke4,luecke5,luecke,luecke9,luecke6}, both
in the case discussed here and in the 
presence of a throughflow (see section \ref{secfrontsnoise1}).  In these
approaches, the front predictions based on the lowest order amplitude
equation are compared in detail with those for pulled fronts obtained
by determining the linear spreading point of the full Navier-Stokes
equations for this system, and with full numerical simulations. We
refer to these papers for a detailed discussion of how the full result
starts to differ from the lowest order amplitude result as
$\varepsilon$ increases. In particular, the effect that the nodes of
the fields drift in the front region before coming to rest behind the
front is a characteristic effect that goes beyond the amplitude
equation (the Swift-Hohenberg equation also shows this, of course).

     \subsection{The propagating Rayleigh-Taylor instability in thin films} \label{sectionpropagatingrt}
The name Rayleigh-Taylor instability refers to gravitational
instabilities  of the interface between two fluids or of stratified
fluids \cite{chandrasekhar}. A simple example 
is when one has two immiscible fluids separated by a planar interface,
in which the heavier fluid is on top of the lighter one. More
generally the instability arises in density-stratified fluids, when the
acceleration is directed from the heavier to the lighter fluid. The
instability is important in a variety of practical situations,
ranging from explosives to the spreading of paint or coating on a
solid surface.

\begin{figure}[t]
\begin{center}
{\tt (a)} \hspace*{-0.5mm}
\epsfig{figure=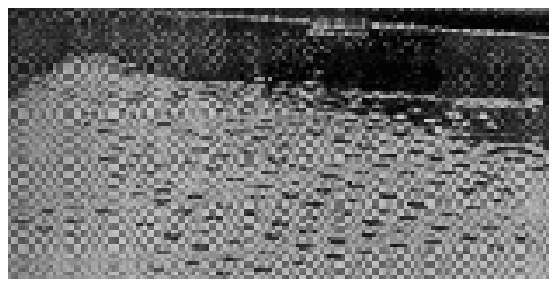,width=0.45\linewidth} 
\hspace*{3mm}{\tt (b)} \hspace*{0.1mm}
\epsfig{figure=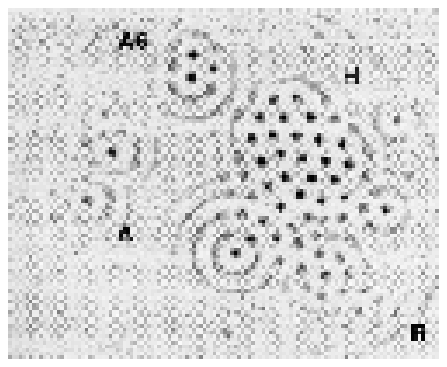,width=0.30\linewidth}
\end{center}
\caption[]{ (a)  Photograph of a well-developed array of pendant drops
at the end of the experiments of Fermigier, Limat, Wesfreid  {\em et
al.} \cite{fermigier1,fermigier2,limat}. The 
drops are seen from above through the glass plate across which the
film was spread before the system was made unstable by flipping the
plate. From \cite{limat}. (b) Two-dimensional patterns observed in the
Rayleigh-Taylor instability of a thin layer in the same experiments
\cite{fermigier2,limat}. Note that a ``roll'' or 
``stripe'' type pattern generally appears near the boundary of the
regions. The two structures labeled $A$ and $A6$  were initiated
by small dust particles on the interface. When they grow, they usually
develop into a six-field symmetric hexagonal pattern like in H. 
The pattern near R was initiated by the thickness gradient close to
the edge of the fluid layer.
}\label{figlimat1}
\end{figure}

\begin{figure}[t]
\begin{center}
\epsfig{figure=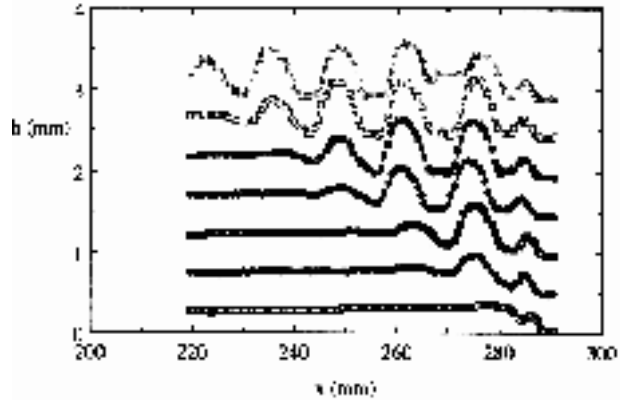,width=0.60\linewidth}
\end{center}
\caption[]{ Time evolution of the thickness profile as a
Rayleigh-Taylor front propagates into an unstable state in the
experiments of Fermigier, Limat, Wesfreid and co-workers \cite{fermigier1,limat}. The profiles
have been shifted up for clarity and are separated by time intervals of
30 seconds. $x$ designates the coordinate perpendicular to the
stripe-type pattern in the front region. 
}\label{figlimat2}
\end{figure}

About a decade ago, Fermigier, Limat, Wesfreid and co-workers
\cite{fermigier1,fermigier2,limat} performed a series of experiments
on the Rayleigh-Taylor instability of thin films. They studied pattern
selection by first spreading  thin films of silicon oil on a flat solid
plate, and then making the system unstable by flipping this plate
upside down. For films of suitable thickness, the instability develops
slowly enough that it can easily be studied. The photo of figure
\ref{figlimat1}{\em (a)} illustrates that the typical hexagonal patterns that
one often observes after the instability is well-developed. The
experiments illustrate several  basic issues in pattern
formation,\footnote{For example, the generic appearance of hexagonal
patterns is associated with the fact that  the thin film equations (\ref{thinfilm2})
below for the deviation $\zeta$ of the film thickness from the uniform
value, are not invariant under a change of sign of
$\zeta$ (see \cite{bodenschatz} and references therein). In the Rayleigh-B\'enard instability hexagonal
patterns generically arise for the same reason very close to
threshold, since very small non-Boussinesq effects break the symmetry that the
Boussinesq equations happen to exhibit. Furthermore, the fact that
the pattern looks stripe-like near the edge, even though the
pattern is hexagonal in the bulk --- see Fig.~\ref{figlimat1}{\em (b)}
---  is a generic phenomenon (also known
in convection, see e.g. \cite{bodenschatz}) which can be
understood simply from  symmetry considerations \cite{pismen}: When one considers  a
front or domain wall solution in the amplitude
equations  for the three modes necessary to describe a
hexagonal pattern, symmetry considerations dictate that the prefactor
of the spatial second derivative term of the modes whose wavevector is
normal  to the front is a factor two larger than those of the
other two modes whose  wavevector makes an angle of 60 degrees with
the front normal. This means that the ``coherence length'' or
correlation length of this first mode is larger than that of the other
two modes, and that the pattern in the front region looks
stripe-like \cite{pismen} (see also
\cite{csahok,doelman}). }\label{stripefootnote}  but we
will focus here on the propagating fronts. Right after the plate has
been flipped, a film of uniform thickness is linearly unstable. In many cases,
the instability first developed near the boundary of the sample and
then propagated into the bulk of the sample. Figure \ref{figlimat2}
 shows  experimental traces of such a front as it propagates into
the unstable state \cite{limat}.

The equations that describe the evolution of a thin liquid film are
well known: the separation of length scales allows one to simply
integrate out the coordinates perpendicular to the film layer
(``lubrication approximation'')
\cite{oras}. If we denote for the experimental case of the free film
flow the film height $h$ as the thickness of the film measured from the
plane, the 
dynamical equation for $h$ becomes  \cite{oras,fermigier1}
\begin{equation}
\partial_t  h + \frac{1}{3\eta} \nabla_{\perp} \cdot [ h^3 \nabla_{\perp}
(\rho g h + \gamma \nabla_{\perp}^2 h ) ] =0 , \label{thinfilm}
\end{equation}
where $\eta$ is the viscosity, $\rho $ the density of the liquid, $g$ the gravity, $\gamma$ the surface
tension, and $\nabla_{\perp} $ the gradient operator in the direction
along the surface. To study the stability of a film of uniform
thickness $h_0$, we note that if we substitute $h=h_0 + \zeta$ into
(\ref{thinfilm}), the equation can be written in the form 
\begin{equation}
\label{thinfilm2}
\partial_t \zeta + \frac{h_0^3}{3 \eta} (\rho g \nabla_{\perp}^2 \zeta
+ \gamma \nabla_{\perp}^4 \zeta )  = N(\zeta, \nabla_{\perp}\zeta,
\nabla_{\perp}^2 \zeta) .
\end{equation}
Here $N$ denotes all terms which are nonlinear in $\zeta$ and its
derivatives. As noted in \cite{fermigier1}, the linear terms in
(\ref{thinfilm2}) are exactly the same as the linear terms appearing
in the various model equations discussed in section \ref{section4thorder}
(the Kuramoto-Sivashinsky equation, the Cahn-Hilliard equation  or the
Swift-Hohenberg equation at $\epsilon=1$). After a proper rescaling
of space and time units, the pulled front velocity
$v^*$ for the Rayleigh-Taylor fronts in the thin film experiments is
therefore just given by Eq.~(\ref{v*sh}).
 
 The experimentally observed front speed was indeed found to be
consistent with the pulled speed $v^*$ within the uncertainty  of the
experimental measurements and the theoretical estimate  (which was of
order 30\%). In addition,   there seemed to be a clear trend for the
front patterns to have a wavelength  some 5\% smaller than
the wavelength corresponding to the most unstable mode. This is also
what would be expected from the results for  pulled fronts with the
above linear dispersion relation.

Although the results of the experiments thus seem to indicate that the
Ray\-leigh-Tay\-lor fronts in the experiment are examples of pulled
fronts,  we do want to express an important  caveat: just
like the bifurcation to hexagonal convection patterns is generally
subcritical (except  when a symmetry imposes the equations to be
invariant under a change of sign of the dynamical fields), the
bifurcation to hexagonal Rayleigh-Taylor patterns should be expected to have a
subcritical character too; indeed this is indicated by the weakly
nonlinear results in \cite{fermigier1}. This suggests  that
when the instability is weak, one will actually get pushed rather than
pulled fronts. However, to my knowledge this possibility has neither been
explored theoretically nor experimentally.

\subsection{Pearling, pinching and the propagating  Rayleigh instability}\label{secpropRayleigh}

Another instability  with which  Rayleigh's name is associated  is the
instability of a cylindrical body of fluid \cite{chandrasekhar}. In daily life, we encounter this
phenomenon  when a stream of water from a tap breaks up into
drops or when the paint  on  a thin wire or hair of a brush breaks up
into droplets. The instability is caused by the surface tension:
for a  ``peristaltic'' perturbation of the radius of the
cylindrical fluid,  the capillary pressure due to surface tension is
enhanced  in the narrower region and reduced in the wider regions;  this
pressure difference tends to enhance the perturbation even more.

\begin{figure}[t]
\begin{center}\hspace*{-2mm}
{\tt (a)}  \hspace*{-2mm} 
\epsfig{figure=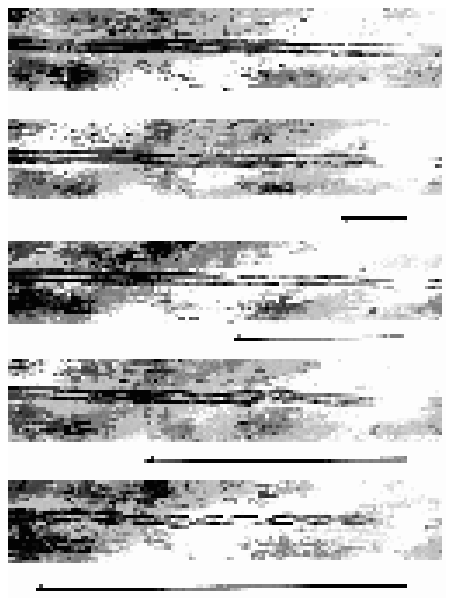,width=0.375\linewidth}  \hspace*{0.3cm}
{\tt (b)} \hspace*{-3mm}
\epsfig{figure=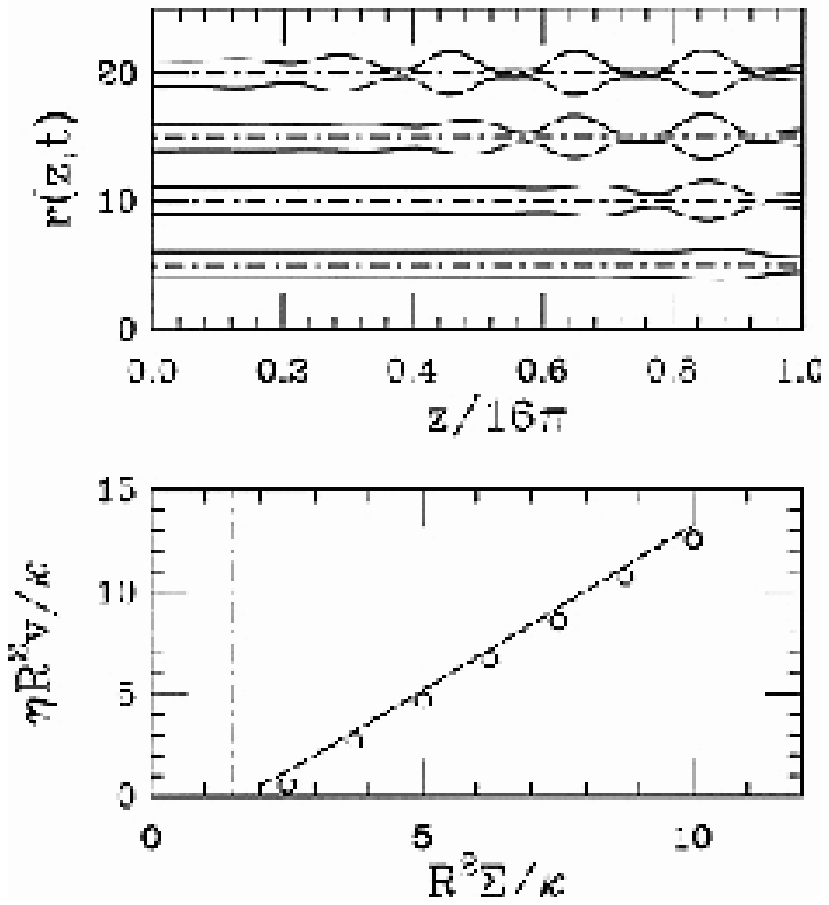,width=0.47\linewidth}
\end{center}
\caption[]{(a) Photographs showing the  pearling
front  along   a tubular membrane formed by lipid bilayers
in the  experiments of Bar-Ziv, Tlusty and Moses
\cite{moses,moses2},  as it propagates out from the bright laser spot
on the right.  In this experiment the dimensionless control parameter
$\varepsilon = (\Sigma -\Sigma_c)/\Sigma_c =6.5$.   From top to bottom the 
times are 0.14, 0.68, 0.86, 1.04 and 1.22 seconds after the laser was
turned on. The arrow indicates the leading edge as it was determined
by the authors. The height of each photographs corresponds to a
distance of about 10 $\mu$m. (b) Results by Powers
and Goldstein  \cite{powers1}
for the simplified model (\ref{sigmakappaeq}) 
for the pearling instability. The top panel shows four snapshots
of the pearling obtained from  numerical simulation of the model. The higher snapshots are
obtained at later times, so the front is traveling to the left. Bottom panel: dimensionless front velocity
as a function of the ratio of surface tension and bending modulus. The solid line denotes the pulled
front velocity $v^*$, the dots the data obtained in the numerical simulations. Note that the data
points lie slightly below the curve for $v^*$. Undoubtedly, this is due to the slow convergence
of the front speed to the asymptotic speed $v^*$.  }\label{figpearling}
\end{figure}

In tubular membranes, the instability normally does not arise, since
membranes generally have a small surface tension but a high bending
rigidity, which measures  the resistance of the membrane
against changes in the curvature. However, Bar-Ziv and Moses \cite{moses} observed
that when they applied laser tweezers to their tubular
membranes, a sinusoidal instability developed, see Fig.~\ref{figpearling}{\em (a)}. This instability
propagated out at a constant velocity in both directions from the point of application of the
tweezers.  Presumably, the instability is due to the fact that the tweezers pull  some lipid molecules
 into the trap, and that as a result the surface tension
increases, rendering the   cylindrical geometry  unstable. Indeed, as
pointed out in \cite{moses}, for a tube with surface tension $\Sigma$
and bending modulus $\kappa$, the free energy ${\mathcal F} $ can be
written as
\begin{equation}
{\mathcal F} = \int \ds S \, [ \Sigma + 2 \kappa H^2 ] . \label{sigmakappaF}
\end{equation} 
Here $H$ is the mean curvature  and  the integral is over the surface
$S$. For a tube  with cylindrical symmetry and the $z$-coordinate along
the axis, we can write the terms in ${\mathcal F}$ in terms of $r(z)$ as
\begin{equation}
\ds S = \ds z \, 2 \pi r \sqrt{1+ r^2_z} ,\hspace*{1.5cm} H = \frac{r_{zz}}{
(1+r_z^2)^{3/2}} - \frac{1}{r (1+r_z^2)^{1/2}} .
\end{equation}
As pointed out by Bar-Ziv and Moses a tube of constant radius $R$ exhibits
the Rayleigh instability when the surface tension is above a critical
value $\Sigma_c \approx 3 \kappa/R^2$. Above this value the free energy is lower for a periodically
modulated tube than for a tube of constant radius.  As $\kappa \to 0$ one recovers
the pure Rayleigh instability.  Thus,  the experimental scenario is that  the laser tweezers
pull in the lipid molecules, and that the increase in surface tension that this entails is large enough
for  the membrane to become unstable.

The pearling instability through propagation of the Rayleigh fronts
that the experiments suggest have been studied theoretically  
by Powers and Goldstein and co-workers \cite{powers1,powers3}, who pointed out that in  the
approximation that the flow profile through the tube remains a
parabolic Poiseuille profile,\footnote{This approximation, which is in
the spirit of the  the thin-film equations described in the previous
section, is justified in the limit in which distortion of the tube
diameter happens on length scales much longer than the tube
diameter. This is not really true for the Rayleigh instability. Nevertheless,
the equation is expected to capture the essentials of the pearling instability.} the  dynamical 
equation for the tubular interface becomes
\begin{equation}
\partial_t r^2 = \frac{1}{4\eta} \partial_z \left(r^4 \partial_z\left[
\frac{\delta {\mathcal F}}{\delta r} \right] \right) , \label{sigmakappaeq}
\end{equation}
where ${\mathcal F}$ is the interfacial free energy given in
Eq.~(\ref{sigmakappaF}) above.  The numerical results from
\cite{powers1} for this equation are shown in
Fig.~\ref{figpearling}{\em (b)}. The top panel shows four snapshots as
the Rayleigh front propagates along the tube, generating strongly
nonlinear bead-like
undulations  behind it.  In the lower panel the velocity of these fronts is
compared to the pulled front velocity $v^*$ for this model (full line)
as a function of the dimensionless parameter $R^2 \Sigma /\kappa$. Note
that this velocity vanishes at a critical value  ---
as already noted above, below this value the cylindrical tube is
stable. Note that the data points fall slightly below the pulled speed
$v^*$. Although not enough data are provided in \cite{powers1} to
check this explicitly, this is undoubtedly due to the slow convergence
of the front speed to the asymptotic value $v^*$ --- as mentioned
before, numerical data at long but finite times will alway approach $v^*$ very slowly
from  below,
and indeed  a discrepancy like the one in this plot is what one often
encounters if a careful extrapolation of the speed to its asymptotic value is not made.

\begin{figure}[t]
\begin{center}
\epsfig{figure=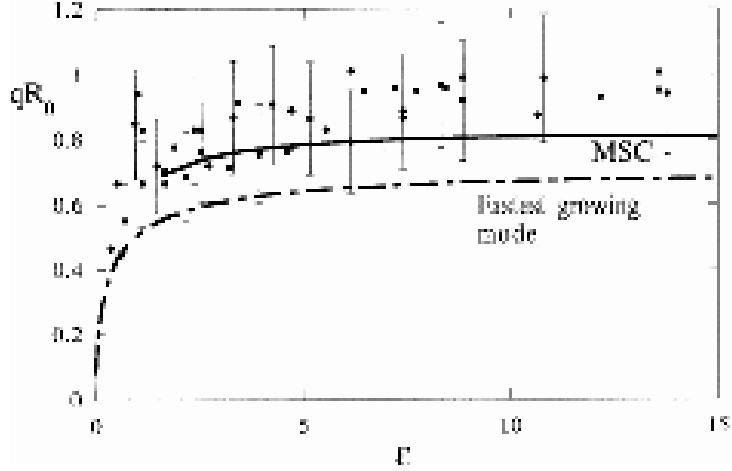,width=0.7\linewidth}  
\end{center}
\caption[]{Experimentally measured values of the
wavenumber of the pearling instability as a function of the
dimensionless control parameter $\varepsilon=
(\Sigma-\Sigma_c)/\Sigma_c$ in the experiments of Bar-Ziv, Tlusty and
Moses \cite{moses2}. The line labeled MSC is the wavenumber selected
by the pulled front in the model discussed in the text. }\label{figmosesdata}
\end{figure}

It is instructive to realize that although the nonlinear behavior of
this pearling model is quite different from those encountered before,
the dispersion relation for small linear perturbations of the radius
is not!  Note that when ${\mathcal F}$ is expanded to quadratic order
in the deviations from a cylinder shape, the highest derivative terms
it contains are of order $r_{zz}^2$. Upon taking the functional
derivative in  (\ref{sigmakappaeq}), we immediately see that the
dispersion relation $\omega(k)$ is a polynomial in $k$ of sixth
degree \cite{powers3,moses2}. Thus, from the perspective of the linear dispersion relation
and the calculation of $v^*$, the equation can be viewed as an interesting
generalization to
sixth order of the fourth order models equations of section \ref{section4thorder}!

In more recent experiments,  Bar-Ziv, Tlusty and Moses \cite{moses2} have tested the
predictions from the above theory in detail, both for the propagation
velocity and for the wavelength of the pattern selected by the
front. Their data for the pattern wavenumber as a function of the
dimensionless control parameter $\varepsilon =
(\Sigma-\Sigma_c)/\Sigma_c$  are reproduced in
Fig.~\ref{figmosesdata}. The full line in this plot (labeled MSC for
``Marginal Stability Criterion'')  shows that  the wavenumber of the pattern  selected by a
pulled front in this model is slightly larger than the wavenumber
corresponding to the fastest growing mode. The experimental data are
consistent with this trend although they lie 
somewhat  above the predicted values. Whether this slight discrepancy
is real is not clear. In any case, however, the experiment illustrates 
an important point: the data shown in Fig.~\ref{figmosesdata} cover a
large range of values of $\varepsilon$: if one knows the dispersion
relation for a given problem, predictions for the properties of a
pulled front can be made for any value of the parameters, not just
close to threshold ($\varepsilon \ll 1$) where an amplitude equation
description with F-KPP-type fronts might be appropriate. 

\begin{figure}[t]
\begin{center}
\hspace*{-4mm}
{\tt (a)} 
\hspace*{4mm} 
\epsfig{figure=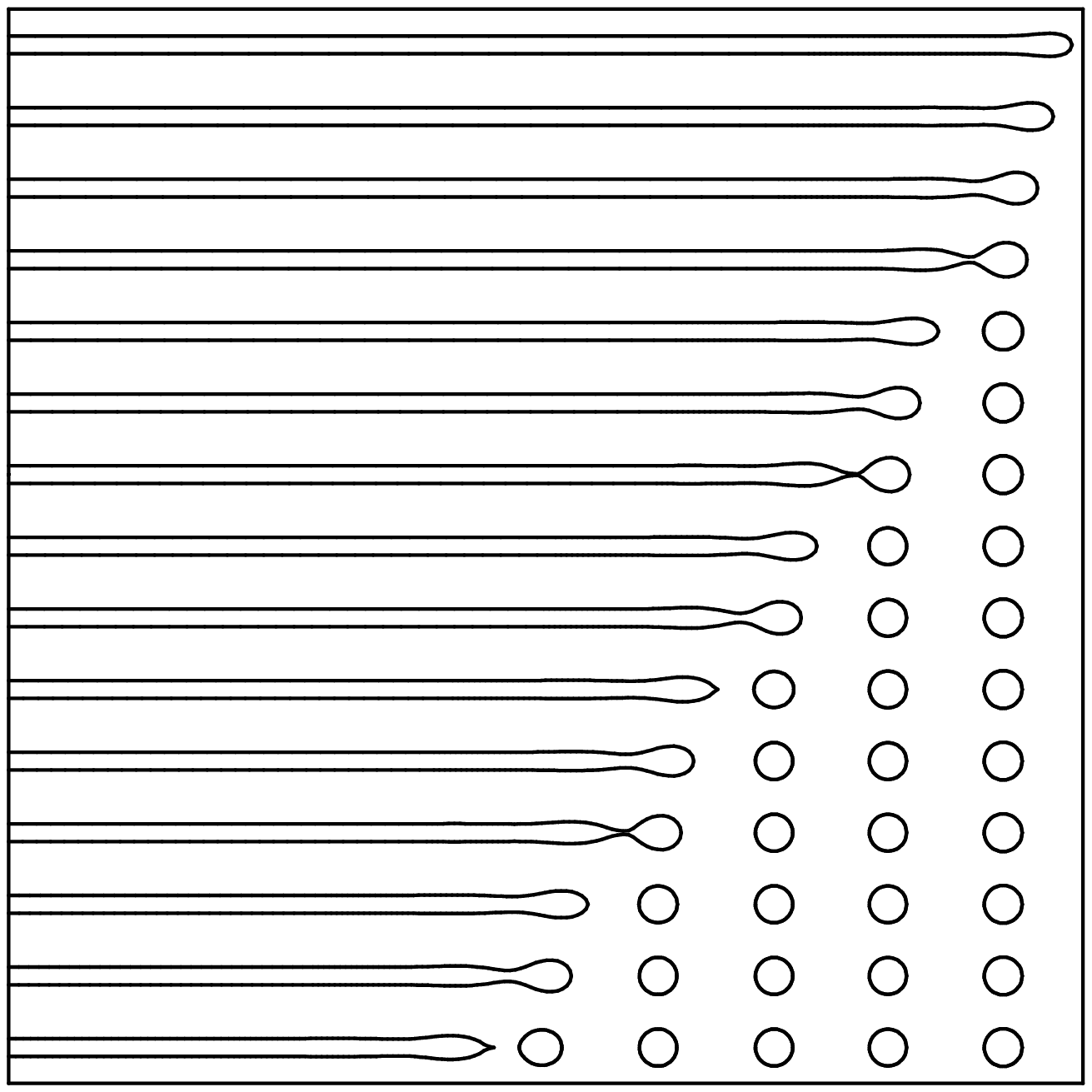,width=0.33\linewidth,bbllx=120pt,bblly=80pt,bburx=502pt,bbury=502pt}
\hspace*{0.3cm}
{\tt (b)} \hspace*{1mm}
\epsfig{figure=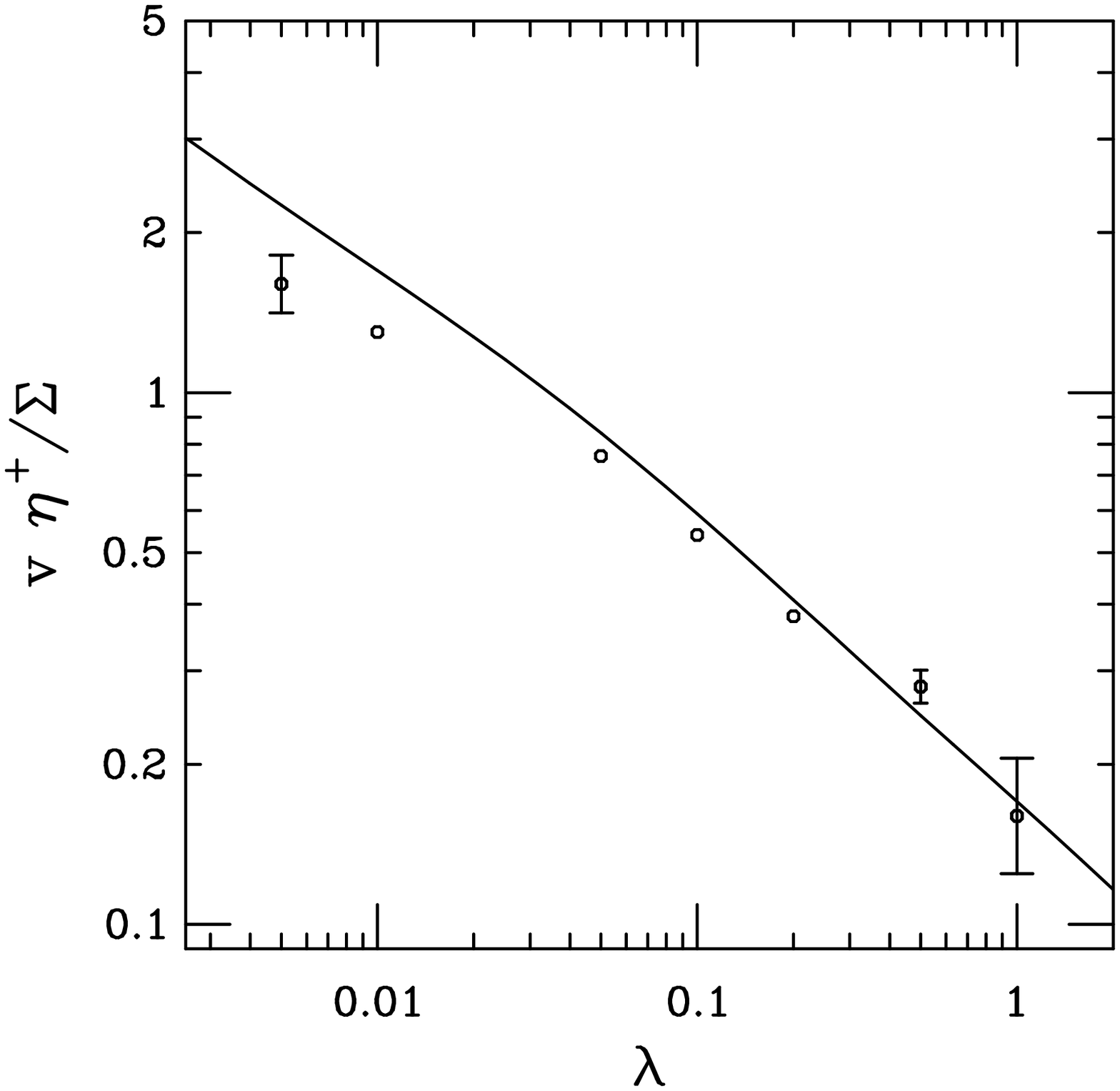,width=0.38\linewidth,bbllx=6pt,bblly=45pt,bburx=532pt,bbury=562pt}
\end{center}
\caption[]{Results from the full hydrodynamical  simulations of Powers {\em et al.} \cite{powers2}
of the propagating Rayleigh instability
in the absence of a bending modulus. (a) Sequence of drop shapes for viscosity ratio $\lambda\equiv
\eta^{\rm inner}/\eta^{\rm outer} = 0.05$ at times $t_n = 6.67n \eta^{\rm
outer} R/ \Sigma$, $n=1,2,3,...,15$ from top to bottom. The evolution
of each connected component was computed independently of the others.
(b) The dimensionless Rayleigh front velocity as a function of the
viscosity ratio $\lambda$. For $\lambda >1 $ the smoothly propagating
front behavior is lost.  }\label{figpowersfull}
\end{figure}

Although the driving force is different --- gravity in one case,
surface tension in the other --- there are some clear similarities
between these propagating Rayleigh fronts and the Rayleigh-Taylor
fronts in the thin film equations discussed in the previous
section. The analogy even  extends  to the following issue: as
we remarked at the end of our discussion of the Rayleigh-Taylor
fronts, that  instability is expected to be (weakly?) subcritical;  
therefore when the linear instability is weak, one expects there to be
a transition to pushed fronts. The same might happen here: as pointed
out by Bar-Ziv and Moses \cite{moses}, there is a small regime where
the transition to modulated  states of the tube is subcritical. In
this regime, one might therefore also expect pushed front solutions,
but to my knowledge this issue has not been studied here either. 

Powers and co-workers \cite{powers2} have also studied the pure
Rayleigh instability (without bending rigidity for the interface)
numerically using the full hydrodynamic equations using a boundary
integral technique. As their calculations reproduced in
Fig.~\ref{figpowersfull}{\em (a)} illustrate, in this case droplets
pinch off when the viscosity $\eta^{\rm inner}$ of the fluid inside the tube is smaller
than $\eta^{\rm outer}$  of the outer
fluid. Fig.~\ref{figpowersfull}{\em (b)}
shows their data for the front velocity as a function of the viscosity
contrast $\lambda= \eta^{\rm inner}/\eta^{\rm outer}$. The full line
is again the pulled front speed $v^*$ for the full problem, while the
symbols mark the numerical data. At the smallest values of $\lambda$
these are some 30\% lower than $v^*$, but there is every reason to
believe that this is again due to finite simulation time and system
size. Provided this is true, these simulations are one of the most
convincing ones that illustrate that pulled fronts have no other
choice than to  propagate for large times with
the linear spreading speed,  even if the dynamics behind them is highly
nonlinear --- what could be more nonlinear and nontrivial than the
pinching off of a droplet?

     \subsection{Spontaneous front formation  and chaotic domain structures in 
ro\-ta\-ting Ray\-leigh-B\'e\-nard con\-vection}\label{sectionkupperslorz}

\begin{figure}[t]
\begin{center}
{\tt (a)} \hspace*{-3mm} 
\epsfig{figure=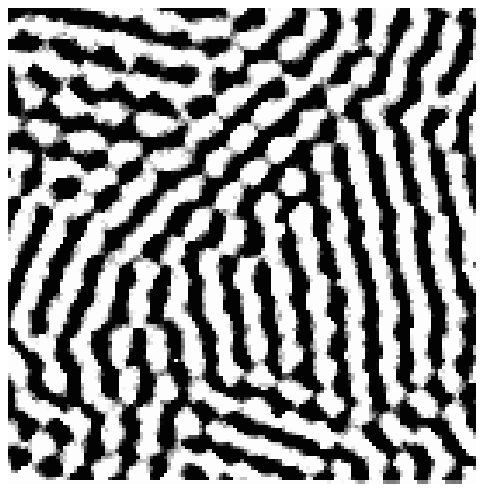,width=0.4\linewidth} \hspace*{0.3cm}
{\tt (b)} \hspace*{-3mm}
\epsfig{figure=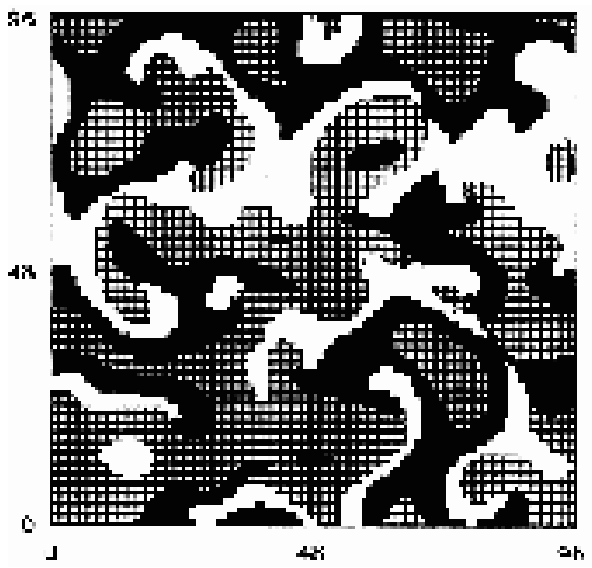,width=0.423\linewidth}
\end{center}
\caption[]{(a) Snapshot of a convection pattern in a rotating
Rayleigh-B\'enard cell in the Kuppers-Lorz unstable regime, 
made with the shadowgraph technique. One
 identifies domains with essentially three orientations of
rolls. From Bodenschatz {\em et al.} \cite{bodenschatz}. (b)  A snapshot of the spatial pattern
in the statistically steady state in simulations of Tu and Cross
\cite{tu} of
Eqs.~(\ref{tucrosseqs}) with $g_+=2$ and $g_-=0.5$. The dark, grey and light regions
represent the domains occupied by $A_1$, $A_2$ and $A_3$,
respectively. The  occupancy is determined by which one of
the three amplitudes is largest.  }\label{figtucross1}
\end{figure}

In section \ref{sectiontcrb} we already encountered a Rayleigh-B\'enard experiment: a 
 fluid heated from below exhibits a transition to convection
patterns beyond some critical value of the temperature difference
between the top and bottom plate. When a Rayleigh-B\'enard cell is
rotated as well, the interplay between the thermal buoyancy and the
rotation-induced Coriolis force gives rise to a whole plethora of new
effects, including traveling wave patterns \cite{ecke,kuo,vanhecke2}. One of the
novel phenomena is the so-called Kuppers-Lorz instability
\cite{kuppers}: above a critical
rotation rate $\Omega_c$, a standard stripe pattern of straight rolls
looses stability to a stripe  patterns which make an angle of about $2\pi/3 $ with
the original ones. However, the new pattern that emerges is in turn unstable
to stripes which make again an angle of about $2\pi/3$ with it, and so
on: no homogeneous stationary stable pattern exists. In practice, the
systems settles into a statistical steady state of domains of stripes of
roughly  three orientations, separated by domains walls or fronts invading
these domains: new domains are created and invaded incessantly. A
snapshot from an experiment is shown in Fig.~\ref{figtucross1}{\em (a)}.

The interpretation of this statistical state was put forward by Tu and
Cross \cite{tu}. They analyzed the amplitude equations for the three
roll amplitudes $A_1$, $A_2$ and $A_3$ in this system,
\begin{eqnarray}
\label{tucrosseq1} \partial_t A_1 &=& \partial^2_{x_1} A_1 + A_1\left(
1-A_1^2-g_-A_2^2-g_+A_3^2\right), \nonumber\\
\label{tucrosseq2} \partial_t A_2 &=& \partial^2_{x_2} A_2 + A_2\left(
1-A_2^2-g_-A_3^2-g_+A_1^2\right), \label{tucrosseqs}\\
\label{tucrosseq3} \partial_t A_3 &=& \partial^2_{x_3} A_2 + A_3\left(
1-A_3^2-g_-A_1^2-g_+A_2^2\right).\nonumber
\end{eqnarray}
Here 1,2 and 3 label the three orientations whose normals $\hat{\bf
e}_1$, $\hat{\bf e}_2$ and $\hat{\bf e}_3$ make an angle of $2\pi/3$
with each other, and $x_i$ are the spatial coordinates projected along
these directions,\footnote{The geometrical projection factors in
$\partial_{x_i}= \hat{\bf e}_i \cdot \partial_{\bf r} $ in the
gradient terms in (\ref{tucrosseq1})-(\ref{tucrosseq3}) are precisely
those which are responsible for the stripe-type appearance of fronts
or domain walls at the edge 
of a hexagonal pattern --- see  footnote \ref{stripefootnote}.}
$x_i= \hat{\bf e}_i \cdot {\bf r}$. The amplitudes in
Eqs.~(\ref{tucrosseqs}) are taken to be real; this
means that we can study the competition between domains but not the
variations of the wavelength of the convection patterns in each domain.

For $g_-=g_+$ this system of
equations is symmetric in all the amplitudes $A_i$, and can be derived
from a Lyapunov function. This implies that the dynamics is then
relaxational, the dynamics tends to drive the system to the minimum of
the Lyapunov function.\footnote{By saying that  a system has a Lyaponov
  function $\mathcal F$,  we mean that it can be written in the form $\partial_t A= 
  -\delta {\mathcal F}/\delta A$, where the term on the right hand
  side is a functional derivative. An example is given in  Eq.~(\ref{tucrosspotential})
  below, and we a also briefly encountered a Lyapunov function in the
discussion of the the Cahn-Hilliard equation in section
\ref{sectionch}.  The derivative form implies
  that $d{\mathcal F}/dt = - \int  (\delta {\mathcal F}/\delta A)^2
  \le 0$ so that ${\mathcal F}$ either decreases  under the dynamics
or stays constant.}\label{footnotelyapunov}
However, for $g_- \neq g_+$ (without loss of
generality one can take $g_-<g_+$) the system has a cyclic permutation
symmetry only and it is non-potential: it can {\em not} be derived from
a Lyaponov functional. For $g_+$ large and $g_-$ small, the $A_2$
amplitude  is strongly suppressed by the $A_1$ amplitude but not vice
versa, so it is intuitively clear that an $A_1$ domain will invade an
$A_2$ domain, etcetera. 

The snapshot of the simulation of the above equations 
\cite{tu} shown in Fig.~\ref{figtucross1}{\em (b)} illustrates that at
any given time, the pattern consists of domains of three
roll-orientations, separated by domain walls or fronts: the dark
domains invade the grey ones, the grey ones the light ones, and the
light ones the dark ones. Note also the similarity with the
experimental picture of Fig.~\ref{figtucross1}{\em (a)}. 

\begin{figure}[t]
\begin{center}
{\tt (a)} \hspace*{-3mm} 
\epsfig{figure=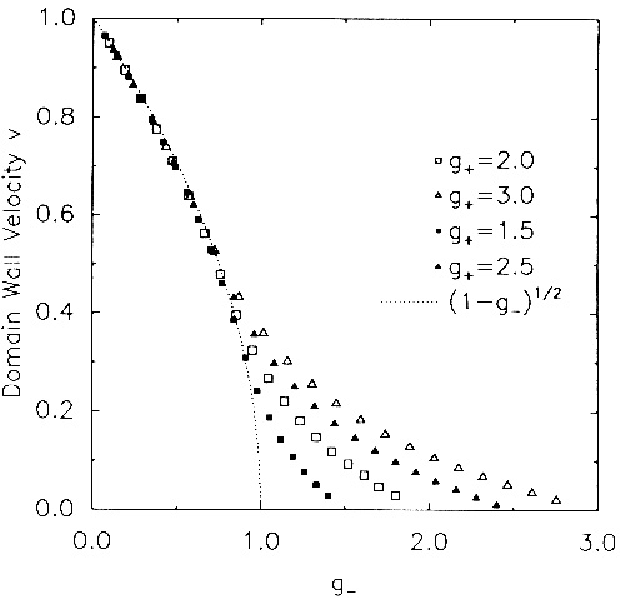,width=0.423\linewidth} \hspace*{0.3cm}
{\tt (b)} \hspace*{-3mm}
\epsfig{figure=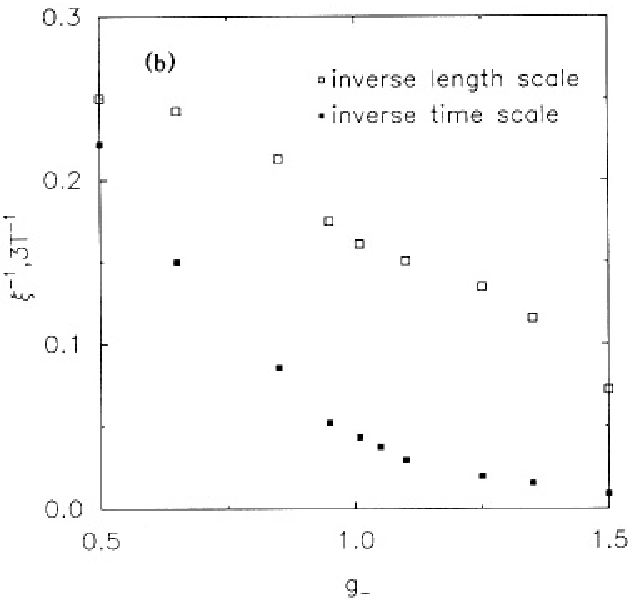,width=0.423\linewidth}
\end{center}
\caption[]{  (a) The front velocity $v$ versus $g_-$ for
various values of $g_+$ as determined numerically by Tu and Cross
\cite{tu} from the
one-dimensional equations (\ref{tucrosspotential}). The dotted line
indicates the pulled front velocity $v^*_{\rm KL}$ given in
(\ref{v*kl}). For sufficiently small $g_-$ the fronts are pulled. From
\cite{tu}. (b) The inverse correlation length $\xi$ and inverse
correlation time $T$ (defined in terms of the amplitude correlation
functions) as a function of $g_-$ in the  numerical simulations  for $g_+=2$. }\label{figtucross2}
\end{figure}

When Tu and Cross \cite{tu} analyzed the statistical properties of
these steady states, a surprising feature was found: while the
Kuppers-Lorz instability sets in when $g_-$ decreases below 1, the
correlation length $\xi$ and correlation time $T$ did {\em not}
diverge as $g_- \uparrow 1$ --- see Fig.~\ref{figtucross2}{\em (b)}. 
The clue to understanding this was identified \cite{tu} to be the
behavior of the fronts. For $g_- <1$  a homogeneous 
state is linearly unstable to the growth of one of the other two
amplitudes; indeed, it is easy to see from Eq.~(\ref{tucrosseq1})
that the linear spreading speed of an $A_1$ perturbation into a domain where
$A_2=1$ equals 
\begin{equation}
v^*_{\rm KL} = 2 \sqrt{1-g_-} . \label{v*kl}
\end{equation}
On the other hand, even though an $A_2$ domain is linearly stable to
small perturbations in all the amplitudes, for $1< g_- < g_+$ it is only
{\em metastable}: an $A_1$ domain will invade an $A_2$ domain
even though it is linearly stable. We can show this by using the
curious feature that two-mode  subdynamics of these equations {\em is}
actually of potential nature, even though the full three-mode dynamics
{\em is not}. Let us consider the one-dimensional 
dynamics of two modes only, 
\begin{eqnarray}
\label{tucrosseq4} \partial_t A_1 &=& \cos^2\theta_1 \, \partial^2_{x} A_1 + A_1\left(
1-A_1^2-g_-A_2^2\right),\\
\label{tucrosseq5} \partial_t A_2 &=& \cos^2 \theta_2\, \partial^2_{x} A_2 + A_2\left(
1-A_2^2-g_+A_1^2\right).
\end{eqnarray}
Here $\theta_1$ and $\theta_2$ are the angles between the $x$-axis and
the vectors ${\hat{\bf e}_1}$ and $\hat{\bf e}_2$. Upon transforming
to the variables $\bar{A}_1= \sqrt{g_+} A_1$ and $\bar{A}_2 =
\sqrt{g_-}A_2$ we can write the equations in the potential form
\begin{equation}
\frac{\partial \bar{A}_1}{\partial t} = - \frac{ \delta {\mathcal
F}}{\delta \bar{A}_1} ,\hspace*{1.5cm}
\frac{\partial \bar{A}_2}{\partial t} = - \frac{ \delta {\mathcal
F}}{\delta \bar{A}_2} ,\label{tucrosspotential}
\end{equation}
where the ``free energy'',  which plays the role of a Lyapunov
functional, is given by
\begin{eqnarray}
{\mathcal F} =  \half \int  \ds x & &~ \left[ \cos^2 \theta_1 \left( \partial_x
\bar{A}_1 \right)^2   -\bar{A}_1^2 + \frac{1}{2 g_+} \bar{A}_1^4 +
\right.  \nonumber \\
 &  & +  \left. \cos^2 \theta_2 \left( \partial_x
\bar{A}_2 \right)^2
-\bar{A}_2^2 + \frac{1}{2 g_-} \bar{A}_2^4 + \bar{A}_1^2\bar{A}_2^2
\right].
\end{eqnarray}

For $g_1 >1$, this free energy functional has local minima at
$\bar{A}_1= \sqrt{g_+}, \bar{A}_2=0 $ and $\bar{A}_1=0,
\bar{A}_2=\sqrt{g_-}$. For $g_-<1$
the second point becomes a saddle, in agreement with the fact that the
state $A_2$ state is unstable to the growth of the
$A_1$-mode. However, even though the free energy has two local minima
for $g_- >1$, for any $g_-<g_+$ the second one corresponding to the nonzero $A_2$-mode
always has a higher free energy than the first one, which
corresponds with the $A_1$-phase. These two minima are separated by  a
free energy barrier (very much like two stable phases near a first
order transition), but  the relaxational nature of this subdynamics
implies that any initial condition which corresponds with an
$A_1$-domain on the left and an $A_2$-domain on the right will develop
into a coherent front which moves to the right. Such ``bistable''
fronts which connect two linearly stable states are well known for such
types of equations;  they are like pushed fronts and their speed approaches an asymptotic value
exponentially fast.

Indeed, as Tu and Cross found --- see Fig.~\ref{figtucross2}{\em (a)} ---
the scenario which emerges is that for any $g_- < g_+$ the dynamical
attractor is a statistically steady chaotic domain state. If $g_-$ is
reduced below $g_+$,  the
invasion of domains is due to pushed fronts  up to some critical
value of $g_-$ below which the fronts are pulled.\footnote{Note 
that the numerical data-points in the pulled regime in
Fig.~\ref{figtucross2}  are slightly below $v^*_{\rm KL}$; as always,
this is a sign of the slow power law convergence to the asymptotic
pulled front speed.} The Kuppers-Lorz
instability is therefore not only a nice illustration of how fronts
propagating into an unstable state can be generated dynamically, but
it is also one of the few realistic examples we know of where the
front propagation mechanism changes from pushed to pulled upon
changing a parameter. In this particular experiment this parameter
can even  be tuned easily by changing the rotation rate.\footnote{For
  fronts between two locally stable states  the
front  velocity will behave linearly in $g_+-g_-$  for $g_-\approx
g_+$. Since the domain dynamics is driven by the front motion, we
therefore expect the correlation length $\xi$ and time scale $T$ to
scale as $\xi/T\sim (g_+-g_-) $ close to the point $g_-=g_+$. The
correlation length $\xi$ in 
Fig.~\ref{figtucross2}{\em (b)}  does seem to vanish indeed faster
than $T$, but there are insufficient data to test this hypothesis.}

We finally note that as as we will discuss in section \ref{sectionmba}, pulled fronts
are not amenable to the usual sharp interface approximation or moving boundary approximation.
Whether this gives rise to any noticeable difference between the transient domains in the
pushed regime just above the Kuppers-Lorz instability and those in the pulled regime at
higher rotation rates,  has apparently not been explored. However, I consider it unlikely that
it does, since as Fig.~\ref{figtucross2}{\em (a)} shows, there is only a large separation of
scales between the front thickness and the domain size in the regime near threshold.

     \subsection{Propagating discharge fronts: streamers} \label{sectionstreamers}

\begin{figure}[t]
\begin{center}
{\tt (a)} \hspace*{-8mm} \epsfig{figure=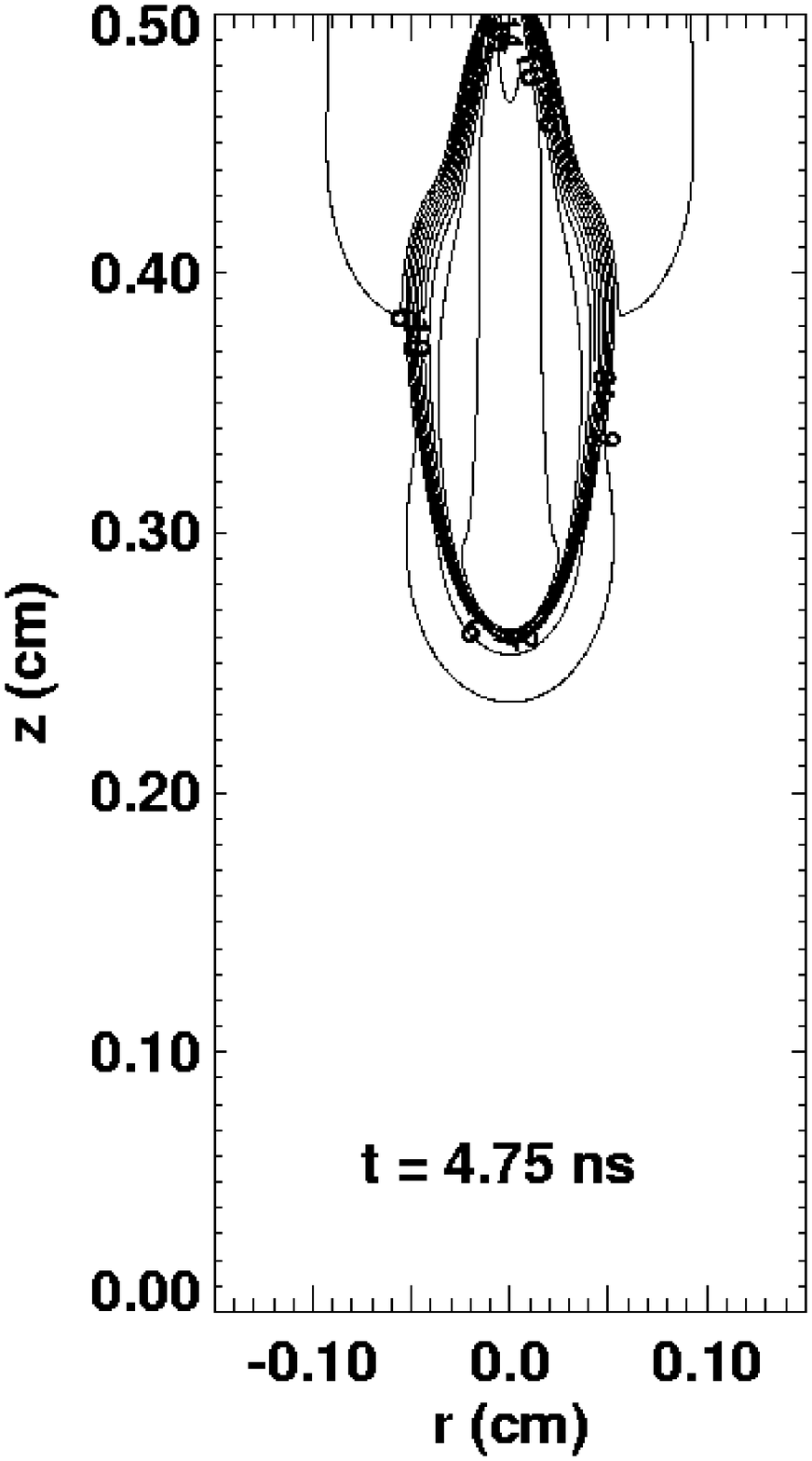,width=0.255\linewidth} \hspace*{0.1cm}
{\tt (b)} \hspace*{-8mm} \epsfig{figure=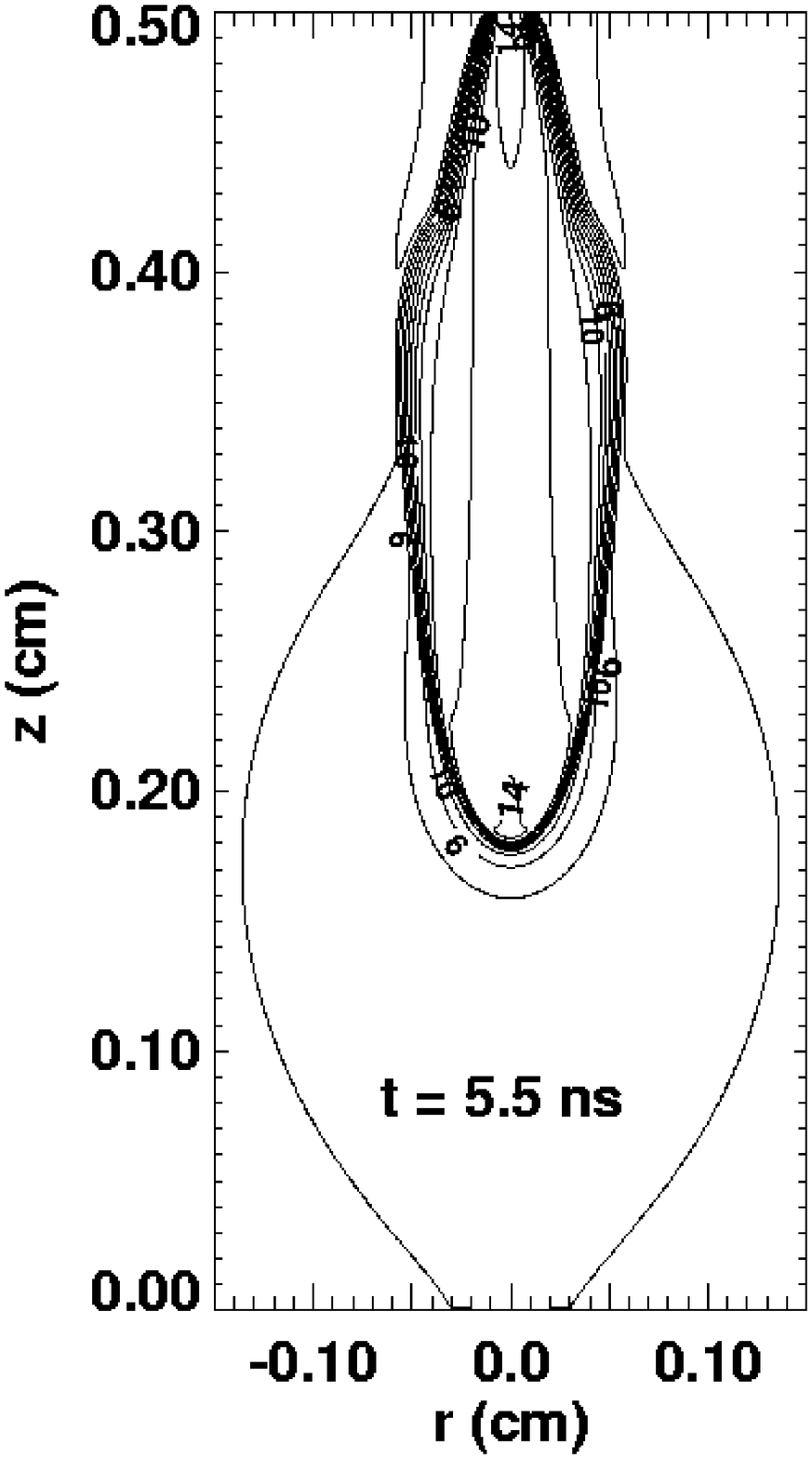,width=0.255\linewidth}\hspace*{0.4cm}
{\tt (c)} \hspace*{-2mm} \epsfig{figure=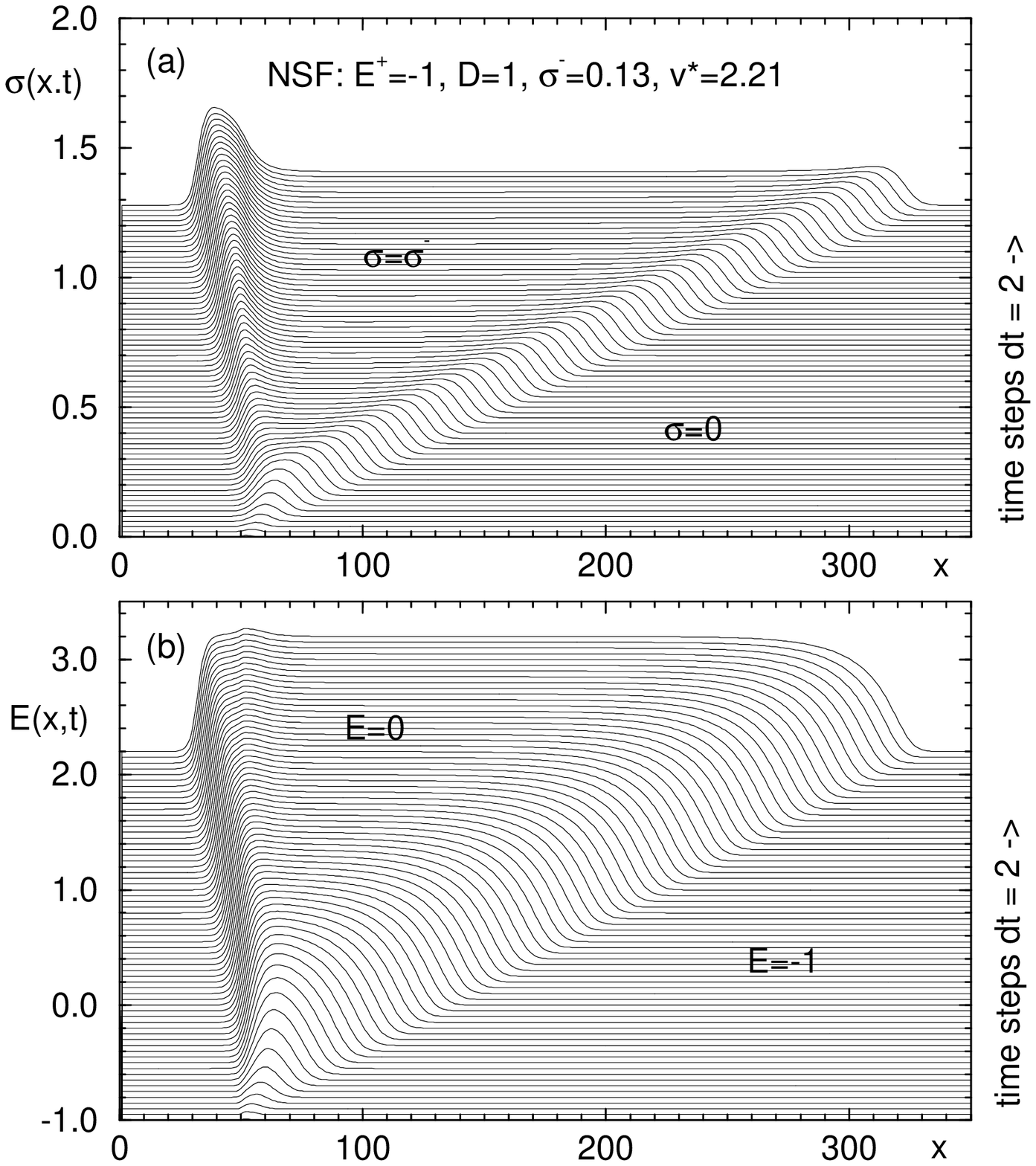,width=0.31\linewidth}
\end{center}
\caption[]{(a) and (b): results of the numerical simulations of
Vitello {\em et al.} \cite{vitello} of the
full three-dimensional simulation of equations
(\ref{streamers1})-(\ref{streamers3}) for parameters corresponding to
nitrogen gas and a potential difference between the planar electrodes
of 25 $kV$. Each line indicates an increase of $n_{\rm e}$ by a factor
of 10. Shown are snapshots of the electron density 4.75 and 5.5
nanoseconds after an initial ionization seed was placed at the upper
electrode. (c) Space-time plot of a simulation of the equations by
Ebert {\em et al.} \cite{streamers2} for
the case of a planar streamer front  in the case
$D=D_{\rm e}\alpha_0/(\mu_{\rm e} E_0)=1$ \cite{streamers2}. Upper panel:
the dimensionless electron density $\sigma= n_{\rm e} e/(\epsilon_0
\alpha_0 E_0)$ as a function of time and space. Lower panel: the
electric field  measured in
units of $E_0$. The initial condition was a homogeneous field $E=-E_0$
and a small, charge-neutral ionized region with  Gaussian electron
and ion density around $x=55$. The front propagating to the right in
the same direction as the electron drift velocity  is  called a 
negative streamer front; it is a pulled front with asymptotic speed
$v^*_{\rm str} $ (=2.21 in dimensionless units), given by
Eq.~(\ref{streamer4}).  The front propagating slowly to 
the left is a pushed positive streamer front.}\label{figstreamers}
\end{figure}

When the electric field is large enough, free electrons in a gas
accumulate sufficient energy  inbetween  collisions 
 that they can knock out an electron from a neutral gas molecule in a
collision. In air, the ionization of nitrogen is dominant, and in this
case the ionization reaction can be summarized as 
$e^- +N_2 \Longrightarrow 2 e^- + N_2^+$. This  type of
avalanche phenomenon can  naturally lead to to the formation of
discharge patterns (sparks!)  whose dynamics is dominated  by the
propagation of fronts  into the unstable
non-ionized  state,  as is
illustrated  in Fig.~\ref{figstreamers}. Figures {\em (a)} and {\em
(b)} show two snapshots of simulations of Vitello {\em et al.}
\cite{vitello} of the formation of a discharge
pattern in nitrogen between two planar electrodes across which a
potential difference of $25 kV$ is applied. Initially, at time $t=0$,
the gas between the electrodes is non-ionized, except for a small
region near the upper electrode. Due to the large field, the electrons
immediately  get accelerated downward into the non-ionized region,
ionizing neutral molecules along the way.  Figs.~\ref{figstreamers}{\em
(a)} and {\em (b) } show the electron density level lines  4.75 and 5.5
nanoseconds later. The regions inside  the finger-like regions of
these so-called streamer patterns are weakly ionized plasmas;  the
regions where the level lines crowd mark the zones where the electron
density drops  quickly to a very small value, and where most of the
ionization takes place. The negative space-charge in this transition zone
effectively shields the outer field in the non-ionized region from that
in the streamer body where both  the space-charge and the field are
small.

The propagation of streamer patterns is well described by the
following set of equations for the electron density $n_{\rm e} $, the ion
density $n_+$ and the electric field ${\bf E}$
\cite{streamers2,vitello},
\begin{eqnarray}
\partial_t n_{\rm e} 
& = & \alpha_0 |n_{\rm e}  \mu_{\rm e}  {\bf  E} | e^{-E_0/{|{\bf
E}|}}  - \nabla \cdot (-n_{\rm e}
\mu_e {\bf E} - D_{\rm e}  \nabla n_{\rm e} ), \label{streamers1}
\\
\partial_t n_{\rm + }  &= & \alpha_0
 |n_{\rm e}  \mu_{\rm e}  {\bf E} | e^{-E_0/{|{\bf  E}|} },
\label{streamers2} \\
{\bf \nabla} \cdot {\bf  E} &=& \frac{e}{\epsilon_0} (n_+-n_{\rm e} ). \label{streamers3}
\end{eqnarray}
 The first term on the right hand side of
the first two equations is the impact ionization term. As  $\mu_{\rm
e}$ is the electron mobility,  $-\mu_{\rm e}{\bf 
E}$ is the electron velocity, so the prefactor is just the impact rate
of electrons on the ions, while  the exponential factor accounts
for the ionization rate as a function of the field ---  small fields
give a small ionization rate, as 
the electron speed at impact is small. The other two terms on the
right hand side of (\ref{streamers1})  account for the drift and
diffusion of the electrons. Similar terms are absent in the equation
for the ion density $n_+$ since on the time scales of interest drift
and diffusion of the much heavier ions can be ignored.  Finally,
Eq.~(\ref{streamers3}) is the Maxwell  equation relating the field
strength to the charge density.

Clearly the impact ionization term in these equations  makes
the non-ionized state with $n_{\rm e}=n_+=0$ linearly {\em
unstable}. Moreover,  since a  nonzero charge density tends to screen the field,
the field is largest just ahead of a front. The exponential factor in
the impact ionization term is an
increasing function of the field strength, hence the growth 
is gradually decreased from a large value ahead of the front to a
small value in the plasma behind the front where the field is
screened. At first sight, it may therefore come as no surprise that
discharge fronts which propagate 
in the same direction as the electron drift velocity --- so-called
negative streamers --- are examples of pulled fronts
\cite{streamers1,streamers2}.  Now, apart from the drift term, 
Eq.~(\ref{streamers1}) has, to  linear order in $n_{\rm e}$,  the same terms
as the F-KPP equation (\ref{fkpp}), a linear growth term and a diffusion
term. We therefore immediately obtain the asymptotic speed of planar streamer
fronts 
\begin{equation}
v^{\rm as}_{\rm str} = v^*_{\rm str}  = - \mu_{\rm e} E^+  + 2 \sqrt{ D_e
|\mu_{\rm e} E^+|  e^{-E_0/E^+} }, \label{streamer4}
\end{equation}
where $E^+$ is the value of the field just ahead of the front.
This prediction has been fully confirmed by extensive numerical
simulations by Ebert {\em et al.} \cite{streamers1,streamers2},  an example of which is shown
in Fig.~\ref{figstreamers}{\em (c)}.

The streamer problem is   very instructive from a more general point
of view. First of all, in the F-KPP equation, it is rigorously known \cite{aw2,depassier1,depassier2} 
that one is always in
the pulled regime if the growth function $f(\phi)/\phi < f^\prime |_{\phi=0}$.
The streamer problem nicely illustrates that such simple results
generally do not hold in more complicated cases. As we mentioned above,
the ionization term in the streamer equations rapidly decreases when
the field strength drops; hence the ionization rate per electron
$\alpha_0 \mu_{\rm e} |{\bf E}| \exp{-E_0/|{\bf E}|} $, which is the analog of
the growth ratio $f(\phi)/\phi$ in the F-KPP equation, decreases
monotonically from the front side to the back side of the front. In
spite of this, for each set of parameter values, the streamer
equations not only admit {\em pulled} negative streamer front solutions, but
also {\em pushed}  positive streamer front solutions which propagate in the
opposite direction. Fig.~\ref{figstreamers}{\em (c)} illustrates
this. The reason for this is that streamer front propagation arises through
the interplay of ionization {\em and} screening of the field.
Screening is a  nonlocal phenomenon mediated through
Eq.~(\ref{streamers3}), and so the monotonic behavior of the field
dependence of the local ionization term is just half of the story.

Secondly, the negative streamer patterns like those of
Figs.~\ref{figstreamers}{\em (a,b)} are  nice  examples of
interface-like growth patterns 
whose dynamics is associated with propagating pulled fronts.
 Indeed, if we write the electric field in terms of the electrical
potential $\Phi$ as ${\bf E} = - \nabla \Phi$, we see that  outside the streamer body, where the charge
density is negligible, the potential  obeys to a good approximation the Laplace equation $ \nabla^2
\Phi =0$. Moreover, Eq.~(\ref{streamer4}) shows that to a good
approximation the normal velocity of a streamer equals $\mu_{\rm e}
\nabla \Phi$. These are precisely the two equations for viscous
fingering \cite{bensimon,langerchance,pelce} or for thermal plumes in
two dimensions \cite{zocchi}, so if  we  think of the streamer pattern as a
moving interface, we expect their dynamics to have a number of
similarities with  the dynamics of viscous fingers or plumes
\cite{streamers2,vsalt}. 

However, the story is not that simple! The problem is that pulled fronts can not
straightforwardly  be mapped onto a moving boundary problem, even when their width
is much smaller than the pattern scale [as is clearly the case for the
streamer patterns of Figs.~\ref{figstreamers}{\em (a,b)}].
As we will discuss more generally in section \ref{sectionmba}, the fact
that the dynamically important region of pulled fronts is {\em ahead}
of the front itself entails not only a power law convergence to the
asymptotic speed but also a breakdown of the standard moving boundary
approximation. The precise implications of this for streamers are
still under active investigation --- very much like dendrites, they do
show a tip-splitting instability \cite{streamersinst},  but the
dispersion relation of small perturbations of a planar discharge front
does appear to be different from what one would expect based on the
analogy with viscous fingering or dendrites.

     \subsection{Propagating step fronts during  debunching of surface
steps}\label{sectiondebunching}

When a crystal is cut with an angle slightly different from one of the principle crystal
facets,  the resulting ``vicinal surface''  contains a lot of 
steps. The predominant mechanism during  
vapor growth is then that adatoms which have landed on the surface diffuse towards these
steps and attach there. The crystal growth is thus accompanied by the propagation of steps 
along the surface. 

It has been known since long that growing mono-atomic steps can ``bunch'': instead
of staying  equidistant,  on average, they bunch  together and form
macro-steps. Many of these and other step instabilities can be
understood in terms of the following simple model of step flow
\cite{gilmer,frank,kandel1,vandereerden},
\begin{equation}
\partial_t X_n = f_+(W_n) + f_-(W_{n-1}) + \gamma \partial^2_y X_n.  \label{stepfloweq}\label{stepbunchingeq}
\end{equation}
Here $X_n(y,t)$ is the position of the $n$th  step measured along
$y$, the coordinate along the step. The terms $f_+(W_n)$ and
$f_-(W_{n-1})$ describe the growth due to attachment of atoms from the 
terrace of width $W_n=X_{n+1}-X_n$ in front of the step and the
terrace of width $W_{n-1}$ just behind the step. The last term is a
curvature correction analogous to the surface tension type term that
occurs in almost all interfacial problems. Note that from a
mathematical point of view, Eq.~(\ref{stepfloweq}) is of mixed
character: it is a partial differential equation with respect to the
$y$-variable, but a difference equation with respect to the step index 
$n$. 

A straightforward stability analysis shows that  an  equidistant
array of steps
(all $W_n = W$) is unstable when the attachment kinetics is such that
steps predominantly incorporate atoms from the terrace behind it,
i.e. if  $f_-^\prime (W)> f_+^\prime (W)$. This is in full agreement with the
intuitive idea that when a step lags  a bit behind, the terrace behind
it  becomes smaller and hence the step will capture fewer atoms from this
terrace.\footnote{Early work on step bunching by Frank
  \cite{frank} described it on a coarse-grained scale in terms of the
  Burgers equation. If the step velocity decreases as a function of
  step density, the well-known formation of shocks in the Burgers
  equation implies step bunching. From this perspective, the phenomenon 
  is similar to the formation of traffic jams in elementary models for 
  traffic flow.} This instability leads to step bunching in the
nonlinear regime.

\begin{figure}[t]
\begin{center}
{\tt (a)} \hspace*{-3mm} 
\epsfig{figure=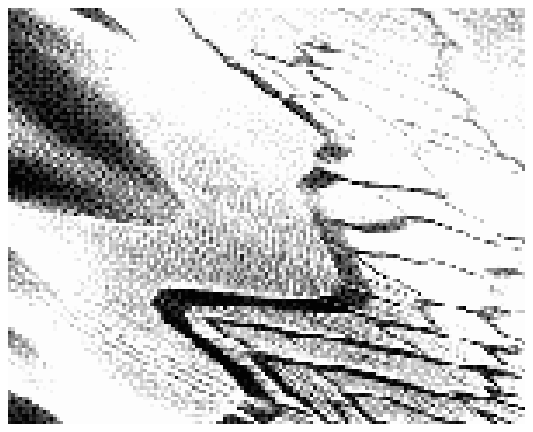,width=0.36\linewidth} \hspace*{0.3cm}
{\tt (b)} \hspace*{-3mm}
\epsfig{figure=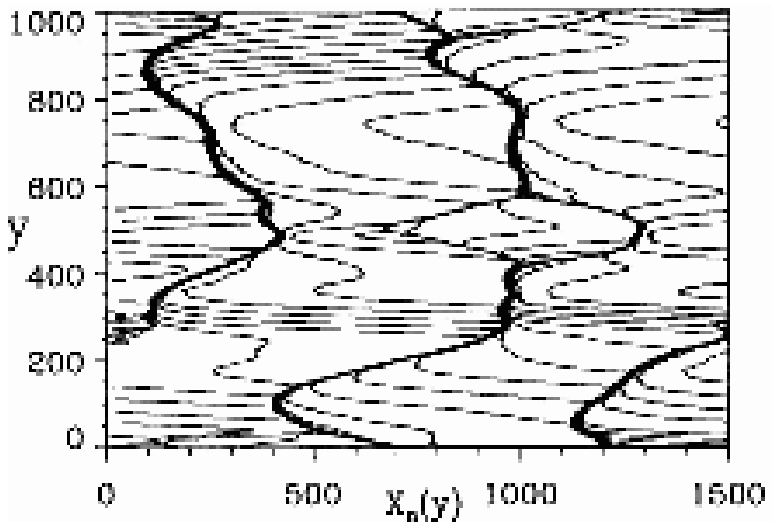,width=0.44\linewidth}
\end{center}
\caption[]{(a) Reflection electron microscopy image of a vicinal Si(111) surface 10 seconds after
the a change of direction of the current direction; a series of steps
``peel off'' from a large macro-step in  a characteristic pattern:
steps ``debunch'' from the macro-step that is visible as the thick
line meandering through the middle of the figure.
 From Latyshev {\em et al.} \cite{latyshev}.
 (b) Snapshot of a Monte Carlo simulation by Kandel and Weeks
\cite{kandel2} of the stochastic version
of the step model (\ref{stepfloweq2}). Note the similarity with the experimental observations
shown in panel figure (a).}
\label{figstepbunching1}
\end{figure}

It has been found in various experiments that  application of  a current
along the surface can induce very complex patterns. Indeed, one of the
new phenomena that occurs is shown in Fig.~\ref{figstepbunching1}{\em (a)}:
right after reversal of the current, steps ``peel off'' from a
macro-step in a characteristic pattern.

This experimentally observed ``debunching'' behavior is not found in
the above model for the step dynamics, but 
Kandel and Weeks \cite{kandel2} have shown that it can be understood
in terms of the following extension of it. The shortcoming of model
(\ref{stepfloweq}) is that the growth functions $f_\pm (W)$ only depend on
the terrace width right ahead of and behind the step. However, when steps 
bunch they get so close that this simple approximation breaks down:
capture of adatoms by steps in a bunch is strongly suppressed because
steps cannot move any closer. Instead, surface diffusion over the
entire bunch becomes more probable. Kandel and Weeks \cite{kandel2} therefore replace 
(\ref{stepfloweq}) by
\begin{equation}
\partial_t X_n = f_+(Z_n^{\rm (f)} ) + f_-(Z^{\rm (b)}_{n}) + \gamma \partial^2_y X_n, \label{stepfloweq2}\label{stepbunchingeq2}
\end{equation}
  for $W_n >W_{\rm min}$, while  $\partial_t X_n=0$ for $W_n<W_{\rm min}$. Here
  $Z_n^{\rm (f)} $ ($Z_n^{\rm (b)}$) is the width of the first terrace 
  in front of (behind of) the $n$th step that is larger than $W_{\rm
    min}$.  Note that this introduces a dynamically generated
  non-locality in the model: the terraces from which a step captures
  atoms depend on the dynamical state of the step configuration
  itself. As a result, as discussed in detail in \cite{kandel2}, these
  equations also lead to a {\em debunching instability} near the edge of a
  step bunch: one or a series of steps can ``peel off'' in a
  characteristic fashion from the bunch, while the other steps in the
  bunch remain virtually immobilized together.  A snapshot of a Monte
  Carlo simulation \cite{kandel2} of  a stochastic version of this model with 
\begin{equation}
f_-(W) = k_- W,  \hspace*{1cm} f_+(W) = k_+ W , \label{stepbunchingeq3}
\end{equation}
is shown in Fig.~\ref{figstepbunching1}{\em (b) }. Clearly, the type of step patterns found in this
model is remarkably similar to that seen experimentally. 

\begin{figure}[t]
\begin{center}
\epsfig{figure=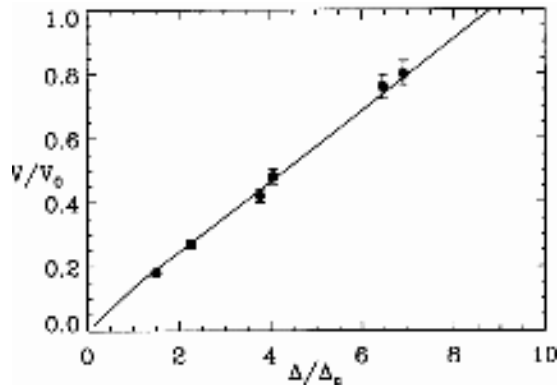,width=0.52\linewidth} 
\end{center}
\caption[]{Comparison of the prediction of $v^*_{\rm debunch}(\Delta)$
  (full line)
  with the velocity data (symbols) from the simulations  shown in
  Fig.~\ref{figstepbunching1}{\em (b)}. $V_0$ is the initial step
  velocity $V_0= (k_++k_-)W$ and $\Delta_c= \sqrt{2\gamma/(k_- -
    k_+)}$. From Kandel and Weeks \cite{kandel2}.}
\label{figstepbunching2}
\end{figure}
From our perspective, the interesting aspect is the arrays of steps
--- the almost parallel lines in the figure --- which cross from one
step bunch to the other. As we mentioned above, the step configuration 
near the edge
of a bunch is unstable to the peeling off of an array of steps, and so 
the propagation of the detachment point can be viewed an an example of 
a front propagation into an unstable state problem. Indeed, by
linearizing the dynamical equations (\ref{stepbunchingeq2}) and
(\ref{stepbunchingeq3}) about an array of straight steps separated by
a distance $\approx W_{\rm min}$ and assuming a behavior for
perturbations $\delta X_n$ of the form
\begin{equation}
\delta X_n(y,t)  \sim e^{-\lambda(y-n\Delta) + \sigma t}
\end{equation}
one obtains a linear dispersion equation of the form
\begin{equation}
\sigma(\lambda) = k_+\left( e^{\lambda \Delta} -1\right) + k_-
\left(1-e^{-\lambda\Delta} \right) + \gamma \lambda^2.
\end{equation}
Note that the various terms in this expression do reflect  the mixed
character of the equation: the exponential terms are characteristic of a
finite difference equation, the last term of a partial differential
equation.
The linear spreading speed $v^*_{\rm debunch} $ can of course be
obtained straightforwardly by determining the minimum of the curve
$\sigma(\lambda)/\lambda$. The comparison of $v^*_{\rm debunch}$  with 
the simulation data for various
 $\Delta$  is shown in 
 Fig.~\ref{figstepbunching2}. The data clearly show that the debunching 
  point propagates  along the bunch with the linear spreading velocity,
 so apparently the debunching process is an   example of a  pulled
 fronts.  Pulled fronts were also found to describe certain aspects of 
 the initial bunching process in the original model (\ref{stepbunchingeq}) \cite{kandel1}.

Recent experiments on this debunching instability
\cite{ellenwilliams,ellenwilliams2} are consistent with 
the relation between growth length $\lambda$ and step
spacing $\Delta$ found by Kandel and Weeks \cite{kandel2}
but do not test the predictions for the front velocity directly. 
More recent theoretical work \cite{kandel3} shows that a strain-induced step bunching
instability can be convective.

\subsection{Spinodal decomposition in polymer mixtures} \label{secspinodal}

When a homogeneous mixture is quenched within the so-called spinodal region,
the homogeneous state is unstable to composition fluctuations. The
lowest free energy state towards which the system evolves at long
times is one in which it is demixed into two homogeneous phases of
different composition. On the way towards this demixed state the
system is spatially inhomogeneous on mesoscopic scales. This
spatio-temporal demixing process is called spinodal decomposition \cite{bray,gunton}.

At intermediate times the dynamics is normally
dominated by motion of interfaces between domains in which the
composition is close to one of the equilibrium compositions. The
initial phase, however, depend on the system under study. For most
systems (like ordinary liquids), the fluctuations are large enough that right after the quench
these fluctuations grow out due to the instability of the bulk
mode. Moreover, this regime is often difficult to probe
experimentally, since it happens on too short time scales. For
sufficiently long polymers, however, the dynamics is slow
enough that this regime becomes experimentally accessible
\cite{wiltzius}. Moreover, 
 the longer the polymers the smaller the fluctuations, so that
the coarse-grained  mean-field like models which have mostly been
studied in the literature become more appropriate for polymer systems. It is also
conceivable  that the short-time  dynamics  is
then sometimes dominated by the
propagation of composition modulation fronts into the unstable
homogeneous state. If so, it is likely that such fronts
typically start at the walls of the sample, since a wall is normally
preferentially wetted by one of the compositions \cite{ball}. 

A simple model equation for spinodal decomposition is the
Cahn-Hilliard equation which we already introduced in section
\ref{sectionch}.  The front propagation problem in
this equation in the presence of noise was studied from the above
perspective by Liu and Goldenfeld \cite{liu2} and by Ball and Essery
\cite{ball} who both found that  the composition modulation fronts
in this equation are pulled fronts  which lead to a incoherent
pattern which continues to coarsen.\footnote{The data points for the numerical front
velocities obtained shown in  \cite{liu2} are as usual slightly below the value
$v^*$: As we remarked already several times, this is true for almost all published
data, and reflects the slow convergence of the speed of a pulled front
to its asymptotic value.}  The simulations of Fig.~\ref{figch} also
showed this.

Experimental evidence  for the above surface-induced front-dominated spinodal
decomposition scenario in polymers was found by Jones {\em et al.}
\cite{bates}, but no direct quantitative comparison was made with the
predictions from the theory of front propagation.

We finally note that the competition between bulk growth and
front propagation in  a model with a non-conserved order parameter was
also studied in \cite{noise1,noise2}.

     \subsection{Dynamics of a superconducting front invading a normal state} \label{sectionsupercond}

The equilibrium and dynamical behavior or vortices in type II superconductors is a vast and active
field with many ramifications, which range from fundamental statistical physical studies 
to applications \cite{blatter}.  Even in the area of vortex dynamics, interesting questions concerning
the behavior of propagating fronts separating domains with different vortex properties (e.g. a
different density of vortices) come up \cite{duran,marchevsky}, but most of the issues that
arise in this area are different from those which are 
the main focus of this paper.  We therefore limit  our discussion
here to some  interesting theoretical findings concerning fronts
propagating into an unstable state of a type I superconductor, which
also raise some new fundamental questions. Some indirect evidence for
such fronts have been found from measurement of the magneto-optical response 
of thin films \cite{freeman} but the time-resolution has been insufficient to study
the dynamics directly.  Recent
advances  in magneto-optical techniques \cite{huebener,wijngaarden} to visualize vortex patterns
may open up the possibility to do so in the near future. 

It is well known \cite{tinkham} that in so-called type I
superconductors the normal state is linearly unstable\footnote{Tinkham
\cite{tinkham} formulates this  slightly differently, but if the
the problem is formulated as a  stability analysis, i.e., as the problem of finding the 
growth rate $e^{\sigma t}$ of the linear eigenmodes, one immediately
sees that below $H_{\rm c2}$ the normal state is unstable.} to the
superconducting state when the magnetic field $H$ is reduced below the
critical value $H_{\rm c2}$.  For fields $H_{\rm c2} < H < H_c$, the
normal state is linearly stable, but has a higher energy than the
superconducting state; the superconducting state can then only form
through nucleation. The barrier for this nucleation of superconducting
domains vanishes as $H$ approaches $H_{\rm c2}$ from above.
The dynamical behavior that can result
if a type I superconductor is quenched into this unstable state was
studied a number of years ago by Liu, Mondello  and Goldenfeld \cite{liu}
and by Frahm, Ullah and Dorsey \cite{dorsey2,frahm}, who pointed out the
analogy of these two regimes to the spinodal and nucleation regimes  of
phase separation (see section \ref{secspinodal}). Moreover, they drew
attention to the fact that
when the superconducting phase propagates into the normal phase, the
effective long-wavelength equations show a strong analogy with the
diffusion equations describing diffusion limited growth. As a result, one
expects these fronts to have  a similar Mullins-Sekerka like
long-wavelength instability as crystal growth problems 
(see \cite{caroli,kassner,kessler3,langerrmp,langergodreche,pelce,pomeau} and section
\ref{sectionmullinssekerka}).

In time-dependent Ginzburg-Landau theory  the dynamics
of a superconductor is governed by the dynamical equation for the complex
superconducting wave function $\psi(r,t)$ \cite{dorsey2,frahm,dorsey2,tinkham}
\begin{equation}
\frac{\partial \psi}{\partial t} = - \frac{\delta {\mathcal F}}{\delta
\psi^*},
\end{equation}
where the dimensionless free energy functional ${\mathcal F}$ is given
by
\begin{equation}
{\mathcal F} = \int \ds{\bf r} \left\{ - |\psi|^2 + \half |\psi |^4 +
\left| \left[ (i\kappa)^{-1} {\vec \nabla} -{\bf A}\right] \psi \right|^2 +
({\vec \nabla}
\times {\bf A})^2 \right\}. \label{fsupercond}
\end{equation}
The Ginzburg-Landau parameter $\kappa= \lambda(T)/\xi(T)$ in this
expression  is the
ratio of the  penetration depth $\lambda$ and the coherence length
$\xi$; type I superconductors are characterized by $\kappa <
1/\sqrt{2}$. Furthermore, ${\bf A}$ in (\ref{fsupercond}) is the
magnetic vector potential which is related to the magnetic field ${\bf
B} $ through ${\bf B}= {\vec \nabla} \times {\bf A}$; using Amp\`ere's
law and Ohm's law, the 
dynamics of ${\bf A}$ is governed by
\begin{equation}
\Sigma \, \partial_t {\bf A} = \left[ \kappa^{-1} {\rm Im}\, \left(\psi^* {\vec
\nabla} \psi \right) -|\psi |^2 {\bf A} \right]    - {\vec \nabla}\times({\vec
\nabla}\times {\bf A}).
\end{equation}
Here $\Sigma $ is the dimensionless conductivity \cite{frahm}
 and the term between square brackets in this expression is the
supercurrent contribution in the Ginzburg-Landau formulation.  
The gradient term in this  term describes the generation
of  supercurrents in fronts or  interfacial zones and near
surfaces, while the second term is the Meissner term which is
responsible for flux expulsion from the superconductor state where
$|\psi |^2 \neq 0$ (note that it acts like a linear damping term for
${\bf A}$ which drives the field to zero).

\begin{figure}[t]
\begin{center}
{\tt (a)} \hspace*{-3mm} 
\epsfig{figure=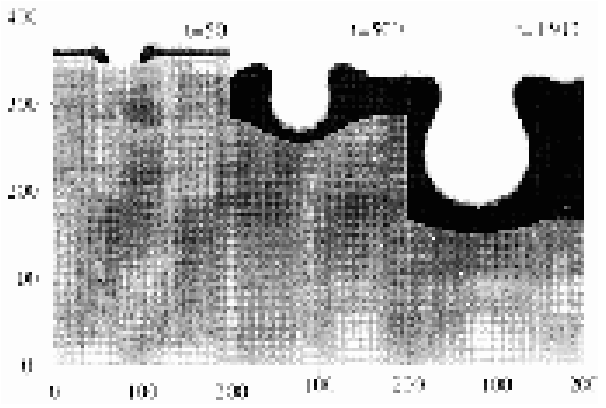,width=0.42\linewidth} \hspace*{0.2cm}
{\tt (b)} \hspace*{-3mm}
\epsfig{figure=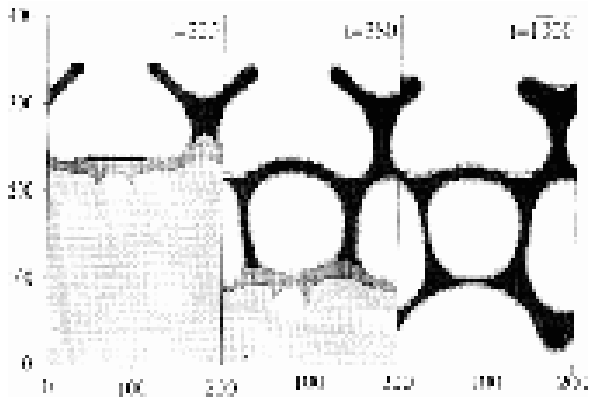,width=0.42\linewidth}
\end{center}
\caption[]{(a) Penetration of the propagation of a superconducting
domain into a normal domain in numerical calculations by Frahm {\em et
al.} \cite{frahm}. The penetration is visualized through the strength of the
magnetic field shown in greyscale, with black corresponding to
$B\approx H_c$ and white corresponding to $B\approx 0$. The field
strength in the normal phase corresponds to  a value of about 0.57
$H_{\rm c}$ which corresponds to the nucleation regime where the
normal phase is linearly stable, while $\kappa=0.3$ and
$\Sigma=0.1$. Times are measured in units of the order parameter
relaxation time \cite{frahm,liu}. (b) As in (a), but now for a field $H=0.28 H_{\rm c}$ which
corresponds to the spinodal regime where the normal phase is locally
unstable.  }
\label{figsuperconductors}
\end{figure}

Figure \ref{figsuperconductors} shows two examples of the simulations of
Frahm {\em et al.} \cite{frahm}  of a superconducting state
propagating into the normal state of a two-dimensional sample with the
magnetic field ${\bf B} = {\vec \nabla} \times {\bf A}$ perpendicular
to the sample.  Since the magnetic field is
expelled from the superconductor a good way to represent the dynamics
is by plotting the strength of the magnetic
field in greyscale. Figure \ref{figsuperconductors}{\em (a)} shows an example of a
superconducting front penetrating the normal state in the nucleation
regime ($H_{\rm c2}<H<H_{\rm c}$) where the normal state is linearly
stable. Due to the flux expulsion, the superconducting region shows up
as white in the figure, while the black zones 
illustrate the field enhancement due to the supercurrents in the
interfacial region. Note also that the size of the protrusion
increases in time, in agreement with the analogy with
diffusion-limited growth problems noted above which suggests that the
interface should be unstable on long length scales.

Figure \ref{figsuperconductors}{\em (b)} shows that the dynamics is very
different when the front propagates into a linearly unstable normal
phase (the spinodal regime). In this regime the front itself appears
to become very complex, and the dynamics is accompanied by the
occurrence of phase slips in the superconductor order parameter
$\psi$ (points in space and time  where the amplitude $|\psi |$ vanishes
so that the  phase of $\psi$ can change by a factor
$2\pi$). 

The scenario that one expects  for superconductor fronts propagating into
a normal state is somewhat similar to the one discussed for the
quintic CGL equation in section \ref{sectioncglquintic}: for fields $H_{\rm c2}<H<H_{\rm
c}$ the behavior is subcritical, and the dynamics of the fronts
separating the two phases is like that of a bistable front or a pushed
front. When upon reducing the field the value $H_{\rm c2}$ is crossed,
the normal state becomes linearly unstable --- this corresponds to
$\varepsilon$ becoming positive in the quintic CGL. The linear
spreading speed $v^*$ then becomes nonzero, but clearly for fields
just below $H_{\rm c2}$ the dynamically relevant fronts will remain
pushed. The question then arises, however, whether upon decreasing the
field even more the fronts may become pulled for some range of
parameters, just like it happens in the quintic CGL equation.
 If so, this would be of great advantage, since it might
provide a handle at calculating some of the properties of the complex
patterns of Fig.~\ref{figsuperconductors}{\em (b)} analytically, just
like the propagating Rayleigh instability of section \ref{secpropRayleigh} allows
one to obtain most essential features of the pattern analytically.

To our knowledge, the question whether the full two-dimensional complex patterns in
the spinodal regime are governed in some cases by propagating pulled
fronts has not been studied yet. Some indication that a transition to
pulled front propagation might be possible comes from the work of Di
Bartolo and Dorsey \cite{dorsey}. They studied the propagation of
one-dimensional fronts in the absence of the possibility of phase slip
generation (as  $\psi$ was taken to be real and equal to $f$ below) and in the case that the
external field in the normal phase vanishes. In dimensionless units, the equations in this case
reduce to 
\begin{eqnarray}
\partial_t f & = &\kappa^{-2 }\partial_x^2 f - a^2 f +f-f^3,\\
\Sigma\, \partial_t a &=& \partial^2_x a - f^2 a.
\end{eqnarray}
The authors studied the case in which the front propagates into the
normal state with $f=0$ and $a=a_{\infty}$. Since the magnetic field
$H$ is this case is simply related to $a$ by $H=\partial_x a$, and
since $a=0$ in the superconductor behind the front,
$\int_{-\infty}^{\infty} H = a_{\infty} $ is simply the amount of
magnetic field trapped in the front region and the field in the normal
phase far ahead of the front vanishes. Di Bartolo and Dorsey
\cite{dorsey} found that fronts in this restricted equation are pulled
for small enough values of $a_{\infty}$ but pushed for larger
$a_{\infty}$.  However, because of the possibility of the
generation of phase slips was suppressed by taking $f$ real, it is conceivable that
fronts at some point  are unstable to the splitting off of a
localized normal region with some magnetic field trapped and a faster
growing growing front with a smaller amount of trapped field. Clearly, various
issues regarding these superconductor fronts still remain to be resolved.

We finally note that apart from the practical significance of studying
whether fronts might be pulled in the two-dimensional case, this
question is also relevant from the following perspective. As we will
discuss in section \ref{sectionmba}, pulled fronts are not amenable
to the usual moving boundary approximation. The full implications of
this are not know, so the superconductor front problem might be a good
one to explore this issue, if they admit pulled fronts. The only other realistic example known
to me where this issue appears to have immediate relevance for the pattern formation
are the streamer discharge patterns analyzed in section \ref{sectionstreamers} (See also the 
remark at the end of section \ref{sectionkupperslorz}  on the Kuppers-Lorz instability). The recent
advance in increasing the time resolution of magneto-optic techniques \cite{wijngaarden} may open
up the possibility that these dynamical issues will become experimentally relevant in the near future.

\subsection{Fronts separating laminar and turbulent regions in parallel shear flows: Couette and
Poiseuille flow } \label{sectionparallelshear}

In this and the next three sections we will discuss 
hydrodynamic instabilities in which both front dynamics and
fluctuation effects or turbulence play an important role.

Two of the  basic textbook examples of hydrodynamic flow states are Couette
flow and Poiseuille flow. We have already encountered Couette flow in
section \ref{sectiontcrb} where we considered flow between two
concentric cylinders. For that setup, Couette flow refers to the basic
laminar flow state. The many instabilities that are found in this
system are due  to the interplay of the inertial effects and the
Coriolis force in a rotating system. In this
section, we will be concerned with planar Couette flow, flow between
two plates sheared in opposite directions. Experimentally, this setup
is realized by moving a transparent plastic band between two glass
plates; the fluid in between is then sheared by the plastic band
\cite{dauchot2,daviaud,dauchot}.   In planar Couette flow of a normal
Newtonian fluid, the basic flow state for flow in the $x$-direction
between plates separated by a distance $2d$ in the $y$ direction  is simply a linear velocity
profile $v_x = v_{\rm plates} y/d$; thus the shear rate $\partial
v_x/ \partial y$ is constant.  The other classic example of a
shear flow  is flow through a pipe or between plates (the
planar case) driven by a pressure gradient: Poiseuille flow. In this
case, the basic velocity profile is parabolic, so the shear rate is
zero on the center line and increases linearly towards the walls.

\begin{figure}[t]
\begin{center}
\epsfig{figure=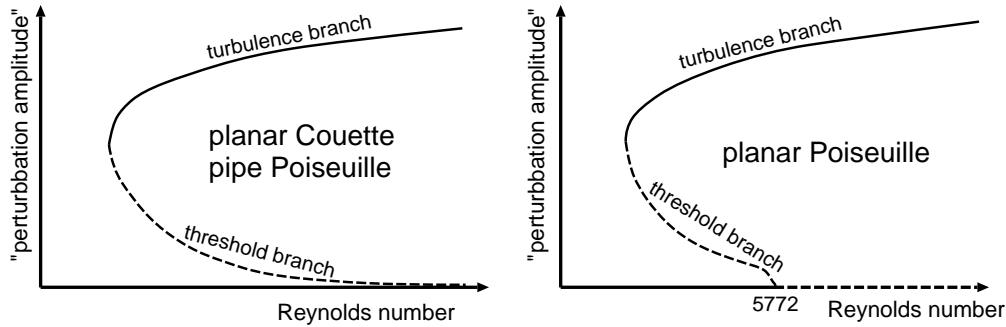,width=0.95\linewidth}
\end{center}
\caption[]{ Schematic illustration of the subcritical bifurcation structure of
the transition to turbulence in Couette and Poiseuille flow
\cite{huerre2}.  For
planar Couette flow and pipe Poiseuille flow (left plot), the laminar base flow is
linearly stable for {\em any}  finite Reynolds number, but the threshold to
turbulence decreases as $Re^{-\gamma}$ with some exponent $\gamma$ as
$Re \to \infty$ \cite{chapman}. Planar Poiseuille flow is linearly
unstable for $Re> 5772$ but  the transition is subcritical and the
linear instability does not play much of a role in practice for $Re$
of order 2000 or less, where turbulence is often already observed in
practice. Note that the drawing is schematic only: although only a single curve 
is drawn in the left plot, there are important differences between planar Couette flow
and  pipe Poiseuille flow. }\label{figturbulent0}
\end{figure}

\begin{figure}[t]
\begin{center}
{\tt (a)} \hspace*{-2mm} \epsfig{figure=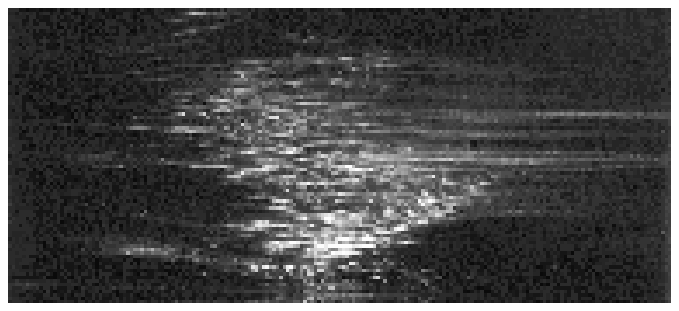,width=0.47\linewidth} 
{\tt (b)}
\hspace*{-1mm}
\epsfig{figure=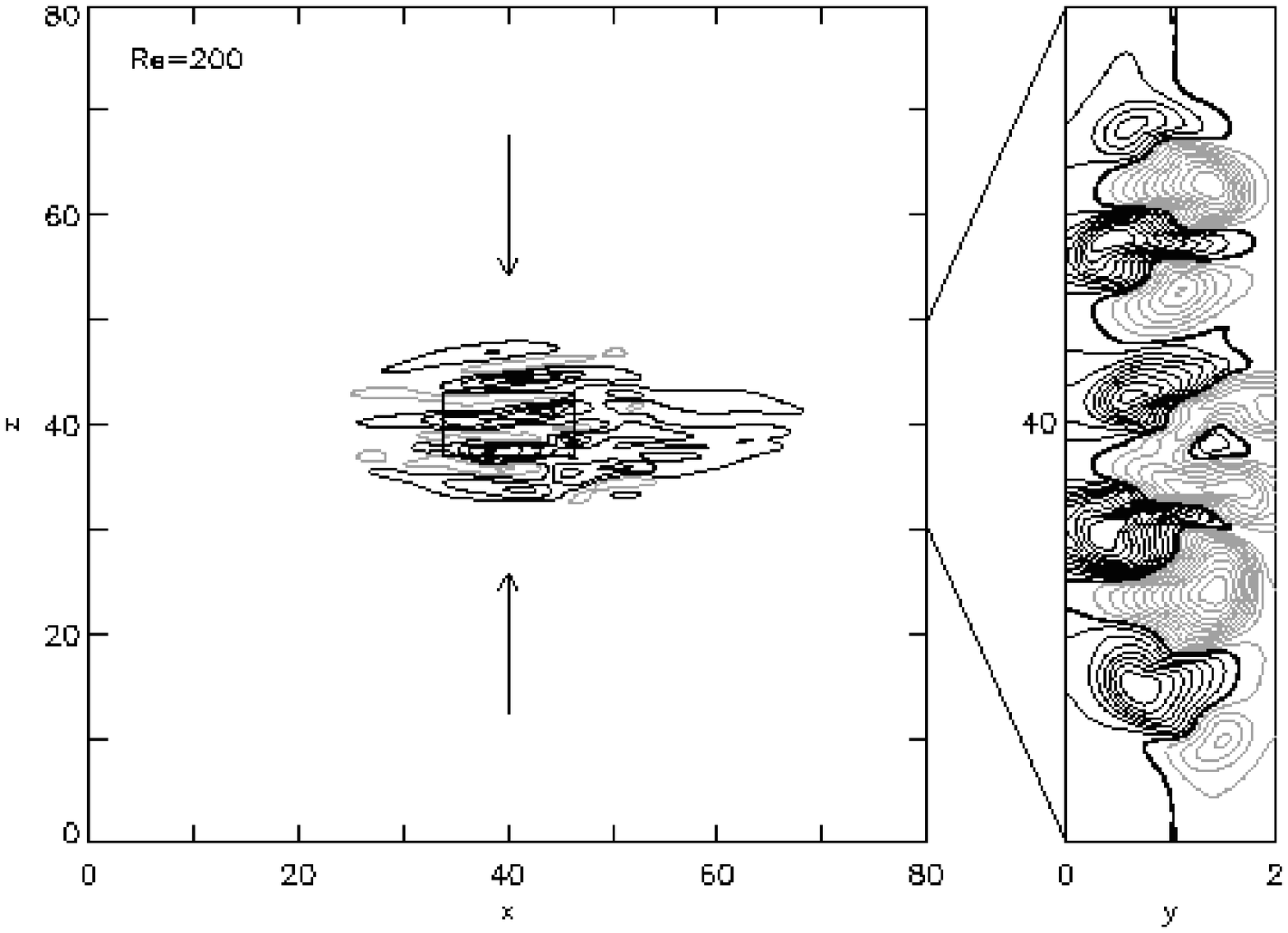,width=0.40\linewidth}
\end{center}
\caption[]{ (a) Example of a turbulent spot in planar Couette
flow, taken  from the review of Manneville and Dauchot \cite{dauchot}. The seeding of the fluid with
small Kalliroscope flakes  makes the turbulent regions show up brightly, while
laminar regions remain dark (the flakes are lined up by the shear
gradient in laminar regions).  The spreading of such turbulent spots depends on the  Reynolds
number of the base flow \cite{dauchot2,daviaud,dauchot}. (b) Snapshot
of a turbulent spot in the simulations of Schumacher and Eckhardt
\cite{schumacher}. The flow is in the $x$-direction. On the right, a cross-section in the direction
normal to the planes is shown. See Fig.~\ref{figturbulent2} for
further details. }\label{figturbulent1}
\end{figure}

The control parameter for Couette flow or Poiseuille flow is the
Reynolds number $Re= U d/\nu$ where $U$ is a typical velocity of the
flow (the velocity at the plates in the case of Couette flow), $d$ the
half spacing of the plates or  the radius of the pipe, and $\nu$ the
kinematic viscosity of the fluid. For large enough Reynolds numbers,
Couette flow and Poiseuille flow become turbulent. However, the
transition to turbulence in such systems is not a supercritical
bifurcation, in which a nontrivial mode grows gradually in amplitude
beyond a threshold at which the basic flow becomes unstable. Instead,
the transition is subcritical \cite{chapman,grossmann,huerre2}, as sketched in
Fig.~\ref{figturbulent0}. This implies that over some range of $Re$
linearly stable laminar flow can coexist with turbulent 
 regions\footnote{It is interesting to note that also in turbulent
superfluid  pipe flow, there are strong experimental
indications for the coexistence of turbulent and laminar domains
\cite{vanbeelen}.}   (very much like near a first order phase
transition). An example of a turbulent spot in a
planar Couette experiment is shown in Fig.~\ref{figturbulent1}{\em
(a)}. The competition between such   turbulent and laminar regions as
a function of Reynolds number has been explored in detail only in the last
decade \cite{dauchot2,daviaud,dauchot}. 

Actually, as Fig.~\ref{figturbulent0} indicates, although the
subcritical behavior is common to all three cases, there is an
important conceptual difference as well:  planar Poiseuille is
{\em linearly unstable} beyond $Re=5772$ while planar Couette and pipe
Poiseuille flow are {\em stable} for any finite Reynolds number ---
the critical amplitude in the latter cases decreases as $Re^{-\gamma}$
for $Re\to \infty$ \cite{chapman,grossmann} and hence we can think of this case
as an instability which has been pushed to infinity.  In the latter
two cases, the absence of a true instability also implies that there
can not be any pulled fronts: the fronts that separate a turbulent
region from a laminar region must always be pushed.

For the case of planar Poiseuille flow, one might wonder
whether  the propagation  of a turbulent region into the laminar flow
state could correspond to a pulled front, as the situation is somewhat
similar to the one  found for the quintic complex Ginzburg-Landau equation
discussed in section \ref{sectioncglquintic}.\footnote{In the quintic complex
Ginzburg-Landau equation, expanding chaotic spots were observed in
numerical simulations in some  parameter ranges \cite{vsh}. Their behavior
is very much like what one would expect intuitively for turbulent
spots. See \ref{figcglquintic1}.} However,  a priori this possibility is already  unlikely 
 for realistic Reynolds numbers: since the generic behavior of
planar Poiseuille flow is so close to that of planar Couette and
pipe Poiseuille flow, where fronts  definitely have to be pushed, we
similarly expect the fronts separating turbulent and laminar domains
to be pushed as well. It was indeed found \cite{deissler3}
that the linear instability of the planar Poiseuille profile
always is always convective, so that pulled fronts could never propagate upstream.
As we shall see below, however,  the simplified picture
based on a straightforward linear
stability analysis of the unperturbed flow does not suffice for
studying the spreading of a spot, because a coupling with induced
cross flow has to be taken into account.

The point at issue here has recently been studied numerically for a closely related
situation by
 Schumacher and Eckhardt \cite{schumacher}. These authors performed
direct simulations of the Navier-Stokes equations for the case of
shear flow between planes. However, unlike in the case of plane
Couette flow, for numerical convenience they imposed  free-slip
boundary conditions on the bounding plates and drove the bulk flow by
 bulk force which varies as a cosine in  the $y$-direction normal to
the planes, and which  forces the flow  in opposite directions 
near the two plates.

Fig.~\ref{figturbulent1}{\em (b)} shows a snapshot of a turbulent spot in the simulations,
which evolved from a localized perturbation of the laminar profile in
the center of the simulation cell. Note that qualitative similarity
with the turbulent spots observed experimentally: Both spots shows a
number of streaks in the stream-wise (horizontal) direction.  These
streaks can also be recognized in  
cross-sections in the $y$-direction perpendicular to the planes shown on
the right. 

\begin{figure}[t]
\begin{center}
\epsfig{figure=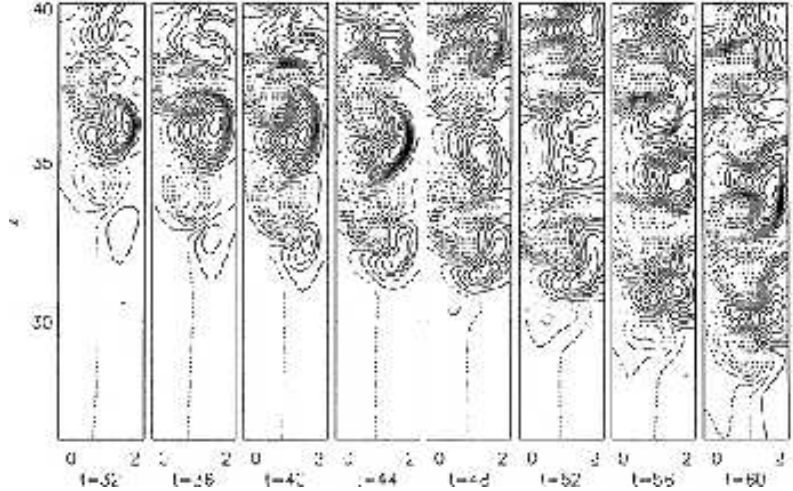,width=0.75\linewidth} 
\end{center}
\caption[]{ Snapshots of the stream-wise velocity $v_x(z,y)$ as it
spreads downward in the simulations of Schumacher and Eckhardt
\cite{schumacher} for $Re$=200. Dotted lines denote contour lines
for positive velocity values, solid lines the contour lines with
negative velocity values. Note that the turbulent region propagates
downward with a roughly constant speed.}\label{figturbulent2}
\end{figure}

The spreading of the turbulent spot in time in the ``span-wise''
(vertical) direction in these simulations is
shown in Fig.~\ref{figturbulent2}. It is immediately obvious that the
turbulent region spreads vertically through some kind of front
structure which advances with a rather well-defined speed. Moreover,
the local turbulent energy ${\bf v}^2$ averaged over the normal
direction was found to decay {\em exponentially} into the laminar region,
very much as one would expect for a well-defined front solution and as
has been found in some incoherent regimes of the quintic CGL equation \cite{vsh}. Since Schumacher
and Eckhardt found that the turbulent spot also gives rise to an
overall ``span-wise'' outflow $U_z $ in the vertical direction, they
analyzed the linear spreading velocity $v^*$ by linearizing the flow
equations about a laminar state which is the sum of the base flow and
this outflow $U_z$ \cite{schumacher}, using  the so-called
Orr-Sommerfeld equation.\footnote{The Orr-Sommerfeld equation is the
linearized equation governing the stability of nearly parallel
flows. It is based on linearizing the Navier-Stokes equations in 
the deviations about a base flow profile $U$.  \cite{drazin}. In the
inviscid limit the equation reduces to the so-called Rayleigh
stability equation used in some of the work discussed in section
\ref{sectionvonkarman} on the formation of B\'enard-Von Karman vortices.}
 The measured front speeds  in the
numerical simulations were about a factor 10 larger than the value of
$v^*$ obtained this way, leaving no doubt that in the parameter regime
studied the span-wise spreading of a turbulent spot is {\em not} governed by
a pulled front: These fronts are pushed.
 An earlier analysis along the same lines of the spreading of turbulent
spots in plane Poiseuille flow lead to the same conclusions
\cite{henningson}.

We would like to stress that the above observations  are {\em just empirical}:
 the simulations  do give strong indications that it does make
sense to associate the expansion of a turbulent spot with a rather
well-defined coherent front, but, as we already pointed out in section
\ref{sectionincoherentpffs}, not much is known about the
properties of such fronts. In fact, also in the quintic CGL
equation, spreading of chaotic spots was found to proceed with a more
or less constant speed larger than $v^*$ (see
Fig.~\ref{figcglquintic1}{\em (a)}), but to our knowledge it is a complete mystery
why and how this happens, and whether there are still common
mechanisms at play. 

From this perspective, it is  very
intriguing that the simulations show that the turbulence front has a
spatial decay rate $\lambda$ which is about twice as
large\footnote{According to Fig.~8 of \cite{schumacher}, in
dimensionless units the exponential
decay rate $\lambda\approx 1.05$ at $Re$=300. From the numerical
values listed after Eq.~(19) in this paper, one finds $\lambda^*=
\sqrt{\varepsilon/D} \approx 0.55$ at $Re$=300.} as
$\lambda^*$:  these pushed turbulent fronts apparently also 
 fall off with an exponential decay rate larger than $\lambda^*$,
just as  uniformly translating pushed fronts or coherent pattern
forming fronts do!

     \subsection{The convective instability in the wake of bluff
bodies:  the B\'enard-Von Karman vortex street} \label{sectionvonkarman}

Another classic hydrodynamic  instability which was analyzed by some of the founding fathers of 
fluid dynamics, B\'enard and Von Karman, is the formation of a ``vortex  street'' behind a 
cylinder in a flow: for Reynolds numbers $Re$ less than about 4 the
fluid flow around a cylinder is laminar. When $Re$ increases past this 
value,  two symmetric eddies form behind the cylinder, but when $Re$
become larger than about 50, vortices are shed from the cylinder in an 
asymmetric pattern. This well known instability is also of immediate
technological  interest, since  the  formation of these
vortices can cause serious damage to suspension
bridges if they are in resonance with the eigenfrequencies of the
bridge. Much of the original work focused on the
formation and dynamics of these vortices, their spacing, etc., in
other words on  the well-developed strongly nonlinear flow regime. In
the last two decades, it has become clear from experimental
\cite{provansal1} and
theoretical work \cite{triantafyllou,yang}, however, that the onset of
the instability is associated with a linear instability in the wake of 
the cylinder.

In 1984, Mathis, Provansal and Boher \cite{provansal1} studied the
velocity fluctuations in the region behind the cylinder (the ``wake''
of the cylinder) using laser Doppler velocimetry. It was found that
beyond a threshold which for long cylinders approaches a value
$Re_{\rm c}=47$, the root mean square velocity fluctuations grow as
$\sqrt{Re-Re_{\rm c}}$. This is already a strong indication that the
transition to vortex formation is associated with a supercritical
(``forward'') bifurcation in the wake, and that the behavior there
could be modeled by a CGL type of amplitude
equation.

Further evidence for this scenario came soon from an analysis of the
stability of the flow in the wake
\cite{huerre0,monkewitz0,monkewitz1,monkewitz2,triantafyllou,yang}. In this
region, the flow is nearly parallel. It varies rapidly in the
$y$-direction normal to the cylinder and the overall flow direction, but very slowly in the
stream-wise direction itself. If one then ignores in a first approximation the
variation of the flow in the stream-wise $x$-direction,\footnote{At this
point, of course, there is a strong connection with the discussion of
so-called ``global modes'' discussed in section \ref{sectionglobalmodes}. In principle one should take the
 equation linearized about  an overall flow pattern $U(y,x)$ whose
 spatial $x$-dependence in the stream-wise direction is kept. Then 
 the linearized equation has spatially dependent coefficients, so one
 can not do a Fourier-type analysis. Usually, a 
 a WKB-type analysis of the resulting global mode is done instead at
 this stage. In the limit where the region of instability is large
 enough, the latter type of analysis reduces to the local one used
 here \cite{chomaz2}. See section \ref{sectionglobalmodes} for further
discussion of this point. } one can then decompose
the flow perturbation in this direction in terms of Fourier modes by writing
\begin{equation}
\mbox{flow field} \sim f(y; k) e^{ikx-i\omega(k)t},
\end{equation}
where the eigenfunction $f(y;k)$ and the dispersion relation
$\omega(k)$ are determined by the Orr-Sommerfeld equation. The latter
  is based on linearizing the Navier-Stokes equations in $v_y$
and $v_x(x,y)-U(y) $ about a  solution for the  velocity profile $v_x= U(z)$ in the wake
\cite{drazin}. Once $\omega(k)$ is determined from the eigenvalue
problem of the Orr-Sommerfeld equation (or the Rayleigh equation to
which it reduces in the inviscid limit), the linear spreading speed
$v^*$ can easily be determined numerically. An analysis along these
lines shows that the flow becomes linearly convectively unstable for value of
$Re$ around 20, and that for $Re$ around  40 the wake becomes linearly
absolutely unstable. This means that for $Re$ above this value, a
pulled front in the wake will propagate upstream, towards the
cylinder.  The frequency of the mode at the transition point,
Re$\, \omega(k^*)$, then gives the frequency of  vortex shedding.

The above results imply the following scenario for the
B\'enard-Von Karman vortex instability. At some critical Reynolds
number $Re_{\rm c}$, which the experiments indicates to be about 47 \cite{provansal1},   the region
behind the cylinder changes over from convectively unstable to
absolutely unstable. The saturation of the velocity fluctuations as
$\sqrt{Re-Re_{\rm c}}$ indicates that one can describe the behavior
close to threshold with a cubic CGL equation for the
variation of the  unstable mode along the stream-wise $x$-direction. According to the
linear dynamics of the flow equations in the wake, the instability is
convective below $Re_{\rm  c}$ and  absolute above $Re_{\rm c}$: in
the latter regime a perturbation also spreads upstream towards the
cylinder.   If a perturbation of 
the wake flow field
grows  large enough that nonlinearities become important, then the
perturbation
develops into a front. This front is pulled (as fronts which propagate 
into the zero-amplitude state of the cubic CGL
equation are always pulled) and hence its velocity $v^*$ behaves just
as discussed already above: for $Re>Re_{\rm c}$ the pulled front moves 
upstream towards the cylinder. In this picture, the vortices arise as
the highly nontrivial and strongly nonlinear structures behind the
pulled front! The situation is thus  analogous to the one for the
propagating Rayleigh instability discussed in section
\ref{secpropRayleigh}, where the propagating Rayleigh front left
behind pinching droplets.

For a  detailed study and discussion of this scenario of vortex shedding, we
refer to the recent work by Pier and coworkers \cite{pier1,pier2}. 
We also  note that the ideas have been taken further by Le\-we\-ke and
Pro\-van\-sal and co\-workers: they show that secondary transitions in the
wake of bluff bodies (like a cylinder) in the regime $180 < Re < 350$ 
are both qualitatively as well
as quantitatively described by a CGL equation for
the behavior in the wake \cite{albarede,provansal3,provansal2}. Some
of the incoherent behavior of the
fluctuations is attributed to a Benjamin-Feir instability \cite{ch,gc}  of
the nonlinear mode. 

     \subsection{Fronts and noise-sustained struc\-tu\-res in con\-vec\-tive sys\-tems I: the Tay\-lor-Cou\-ette
sys\-tem with through flow} \label{secfrontsnoise1}

In almost all hydrodynamic systems with advection due to  an
overall flow, front propagation is an important ingredient of the dynamics. We already
encountered an example  in the previous section on the B\'enard-von Karman instability
in the wake  of a cylinder. There are also many pattern-forming
systems in which the primary bifurcation is a Hopf-bifurcation  to
traveling waves. The issue  whether the instability is convective or 
absolute then immediately arises, and if the instability is strong
enough that nonlinear saturation effects play a role --- as is
generally the case near a supercritical (forward) bifurcation ---
fronts   often play an   important role in   the dynamics and in  the 
emergence of noise-sustained structures.  E.g., the so-called
``blinking'' traveling wave state  in binary mixtures can be
understood quantitatively in terms of fronts which propagate back and
forth in the experimental cell \cite{cross1,fineberg2}, and also incoherent
pulse dynamics in traveling wave systems has been analyzed in such
terms \cite{vanhecke3}. In this and the next section, we focus on two
examples where front dynamics is intimately connected with the
transition from coherent to incoherent pattern dynamics. In this
section we focus on an example of flow-induced advection, and in the
next section we discuss aspects of the  role of fronts in systems
exhibiting a  Hopf bifurcation to traveling wave states.

\begin{figure}[t]
\begin{center}
{\tt (a)} \hspace*{-3mm} \epsfig{figure=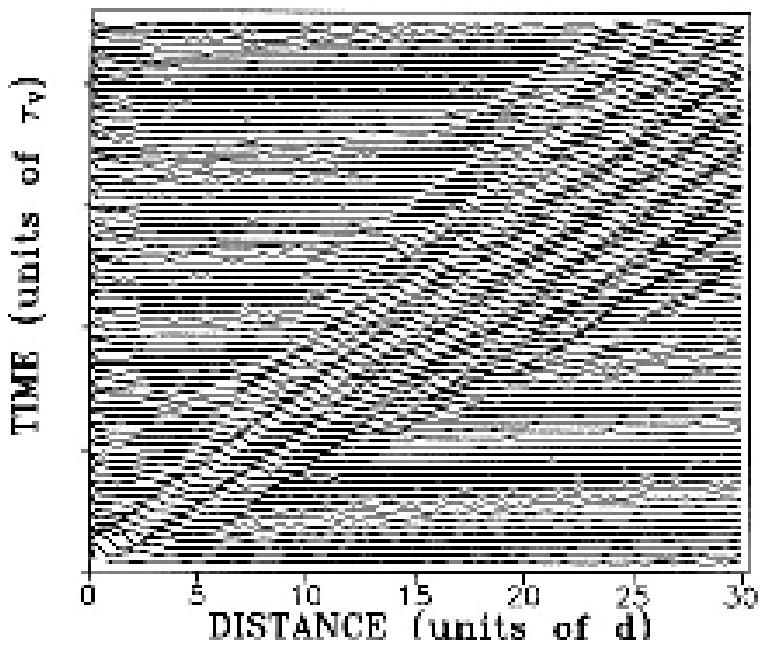,width=0.4\linewidth} 
{\tt (b)}
\hspace*{-3mm}
\epsfig{figure=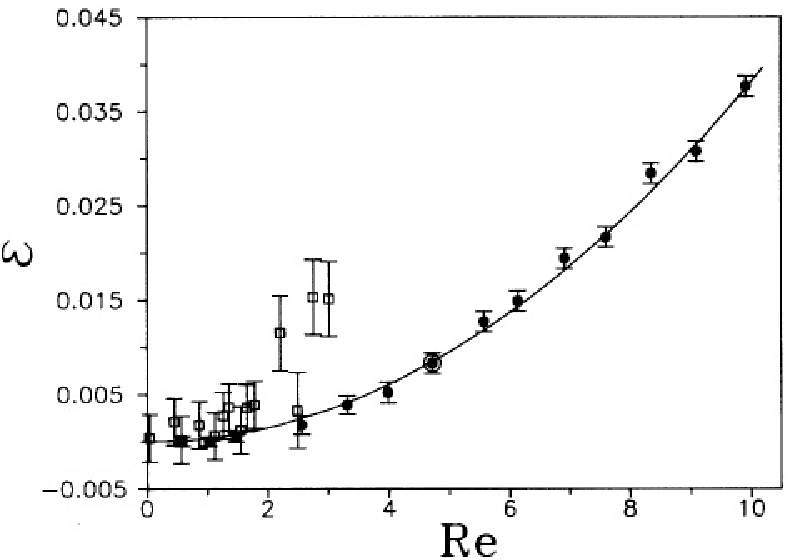,width=0.48\linewidth}
\end{center}
\caption[]{ (a) The growth of a perturbation in the convectively
unstable region of the phase diagram for $Re=6$ and
$\varepsilon=0.0477$  ($\varepsilon -\varepsilon_{\rm c}(Re) \approx
0.032$)    in the experiments by
Tsameret and Steinberg \cite{tsameret} on Taylor-Couette flow with
though flow. The pulse was generated by a
sudden movement of the inlet boundary back and forth. Times are
measured in units of the vertical diffusion time $\tau_{\rm v}$
($=d^2/ \nu$ with $\nu$ the  kinematic viscosity of the fluid)  and
the distance along the axis is measured in units of the gap spacing
$d$. (b) The
threshold to the convective instability $\varepsilon_c (Re)$ as a
function of Reynolds number $Re$ in similar experiments by Tsameret
and Steinberg \cite{tsameret}. The solid line marks the threshold
according to a full theoretical stability calculation, and the solid 
circles the threshold values determined experimentally by tracing the
evolution of a pulse generated near the inlet. The open squares are
data obtained by extrapolation from a fit to the CGL
 equation down from higher $\varepsilon $ values. At small
enough values of $Re$, these agree to within experimental error. 
  }\label{figtcwithflow1}
\end{figure}

We already introduced the Taylor-Couette cell in section
\ref{sectiontcrb}: two concentric cylinders whose ``gap'' between
them is filled with a fluid. When the inner cylinder is rotated, at
some critical rotation rate  $\Omega_c$ a stationary Taylor vortex pattern is
formed; this pattern is periodic in the direction along the axis of
the cylinders. Babcock, Ahlers and Cannell \cite{babcock,babcock2} and Tsameret
and Steinberg \cite{tsameret2,tsameret} studied the behavior near threshold in
the presence of a through flow in the axial
direction.\footnote{Similar effects will happen in a Rayleigh-B\'enard
  cell with through flow \cite{mueller2}, but this system does not
  appear to have been studied experimentally in as much detail.} Such a through
flow has two effects. Firstly, it changes the onset of the instability
to Taylor vortex patterns. Secondly, and more importantly, it changes
the nature of the instability, as the through flow obviously advects
perturbations away from the inlet of the flow at a finite rate. Hence just above the
instability threshold, when the growth is small, the system is only
convectively unstable. This is illustrated in the space-time
diagram of 
Fig.~\ref{figtcwithflow1}{\em (a)}, which show the evolution of
a perturbation which was initiated near the inlet. Clearly, the initial
perturbation grows out and spreads (the pattern widens in time) while
being advected away down the axis. The fact that also the left flank
of the perturbation is moving to the right --- hence is retracting ---
 confirms that the system is
convectively unstable in this case.  

By studying whether the center of such a wave packet grows or decays
in the co-moving frame, both groups have extracted the threshold
$\varepsilon_c= [\Omega_c(Re)-\Omega_c(0)]/ \Omega_c(0)$  as a function
of the Reynolds number $Re$ of 
the through flow. The results from Tsameret and Steinberg
\cite{tsameret} are reproduced in Fig.~\ref{figtcwithflow1}{\em
(b)}. The experimental variation of the threshold is in excellent
agreement with the results of a direct  stability analysis of
the Couette flow with axial through flow. For values of $\varepsilon$
above the line in the diagram, the flow in an infinitely long system is unstable to the formation
of Taylor vortices.

\begin{figure}[t]
\begin{center}
{\tt (a)} \hspace*{-3mm} \epsfig{figure=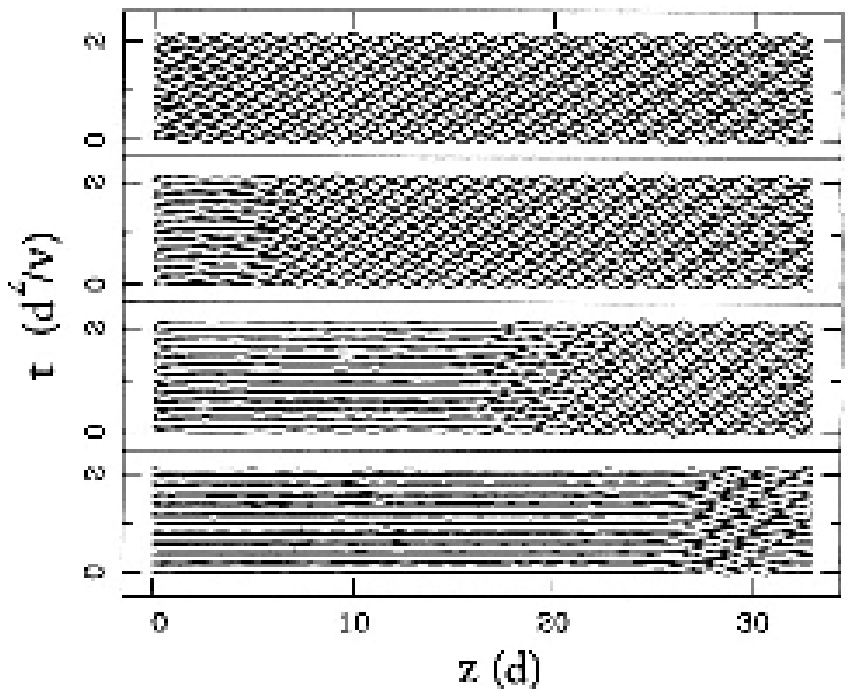,width=0.473\linewidth} 
{\tt (b)}
\hspace*{-3mm}
\epsfig{figure=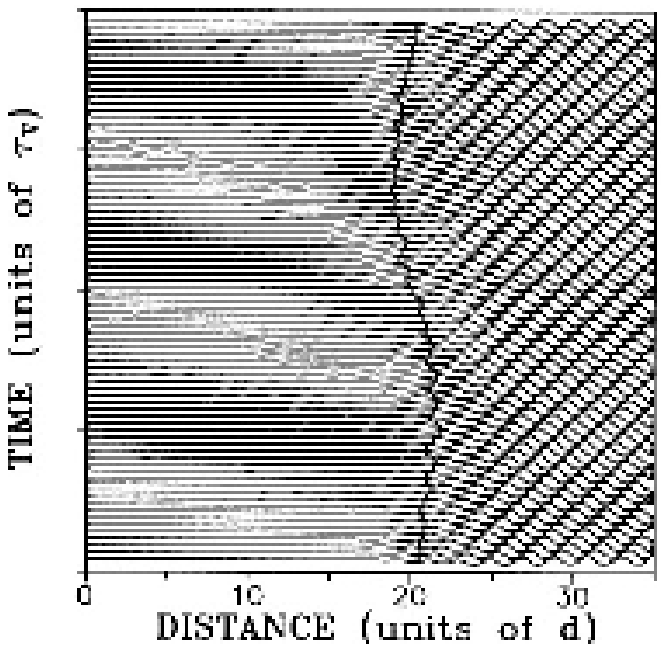,width=0.397\linewidth}
\end{center}
\caption[]{ (a) Space-time plots of the patterns at $Re=3$ for
increasing values of $\varepsilon$  in the experiments by Babcock
{\em et al.} \cite{babcock}, from bottom to top $\varepsilon
=0.0347, 0.0632, 0.0822$ and $0.1020$.
Only the first quarter  of the
apparatus behind the inlet  is
shown. (b) A longer space-time plot  in the
convectively unstable regime from the  experiments of Tsameret and
Steinberg \cite{tsameret} at $\varepsilon=0.04$ and $Re=3$. The
full vertical  line marks the position of the front defined by the position
where the pattern amplitude reaches a fixed value.  }\label{figtcwithflow2}
\end{figure}

If there were no noise in the system, one would not observe patterns
in a finite system in the convectively unstable regime just above the
line in Fig.~\ref{figtcwithflow2}{\em (b)}. In practice, there is of 
course always some noise, and the type of patterns
this gives rise to as $\varepsilon$ is increased above
$\varepsilon_c(Re)$ is illustrated in Fig.~\ref{figtcwithflow2}{\em
(a)} from Babcock {\em et al.} \cite{babcock}. 

The lower panel corresponds to the case just above threshold
($\varepsilon-\varepsilon_c \approx 0.031$). In this case, incoherent
Taylor vortex patterns are found a distance of order 25  down the
cylinders away from the inlet --- small perturbations and fluctuations
near the inlet are amplified while they are advected away, and only
at this point have they grown large enough in size that they are
measurable. Behind this region, they saturate (the plot actually shows
only the first quarter  of the total cell). In this region, the combination of the
fluctuations and perturbations near the inlet and  the
convective effects give rise to an incoherent front-like state. As
Fig.~\ref{figtcwithflow2}{\em (b)} from the other group
\cite{tsameret} illustrates, the effective front position (defined by
tracing the point where the pattern amplitude reaches a certain level)
is indeed slowly wandering back and forth on a  longer time scale. 

As we go up in Fig.~\ref{figtcwithflow2}{\em (a)}, $\varepsilon$ is
increased. The pattern fills more and more of the cell, and at the
same time become more coherent, while the width of the front
separating the two regions decreases. Clearly, in the upper panel for
$\varepsilon=0.102$ the Taylor vortex pattern fills the whole cell,
and one is in the absolutely unstable regime. The pattern selection in
this regime and near the transition was studied numerically in \cite{buchel,luecke8,luecke7,luecke2}.

\begin{figure}[t]
\begin{center}
\epsfig{figure=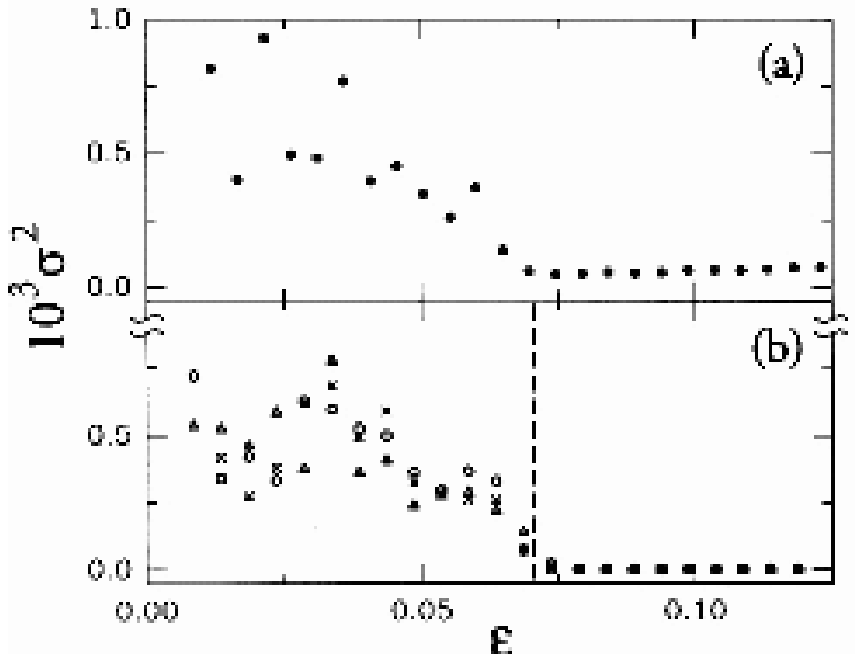,width=0.415\linewidth} 
\hspace*{0.1cm} 
{\tt (c)}
\hspace*{-3mm}
\epsfig{figure=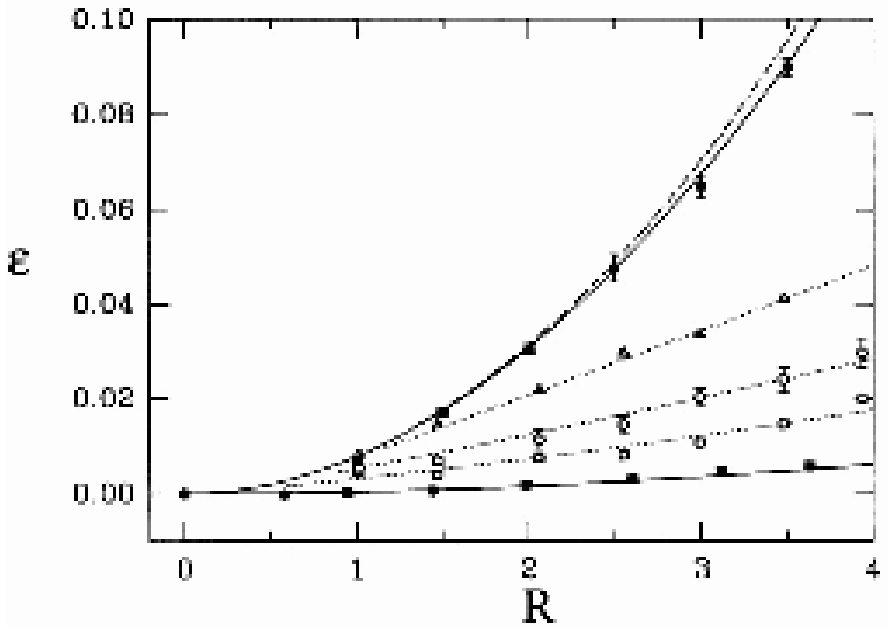,width=0.465\linewidth}
\end{center}
\caption[]{ Various data from the experiments of Babcock {\em et al.}
\cite{babcock}. (a) (Upper left panel) Normalized mean squared width $\sigma^2$ of the peak in the
frequency  
spectrum as a function of $\varepsilon$ at $Re=3$  obtained from the
measurement of the pattern amplitude as a function of time at a
position 100$d$ behind the inlet (note that this is three times
further from the inlet than the region shown in
Fig.~\ref{figtcwithflow2}{\em (a)}). The sharp transition at
$\varepsilon=0.065$ indicates the change from the small-$\varepsilon$ regime where the
patterns are  incoherent due to
 phase noise  to a coherent large-$\varepsilon$ regime where pattern
are coherent. This value is therefore associated with the transition
from the convectively unstable to the absolutely unstable regime.  (b) (Lower left panel) Corresponding results for the 
CGL equation (\ref{cgltc}) with three different noise levels,
$10^{-6} (\times), 10^{-5} (\Delta ),$ and $10^{-4} ($O$ )$. The
dashed line locates the transition $\varepsilon_{\rm ca}^{\rm CGL}
=0.071$ \cite{babcock}. (c) Stability diagram for axisymmetric Taylor
vortex patterns. The lower line and data points mark the line
$\varepsilon_c(Re)$ where the instability sets in. The upper data
points locate the transition $\varepsilon_{\rm ca}$ from the convectively unstable regime to
the absolutely unstable regime as determined from the change-over from
coherent to incoherent patterns via measurements of the spectrum like
those in (a), while the dashed line indicates $\varepsilon_{\rm
ca}^{\rm CGL}$. Between the lower line and the upper line, the system
is convectively unstable, and noise-sustained structures like those in
the one but lowest panel of Fig.~\ref{figtcwithflow2} occur. Open
symbols indicate the boundaries where noise sustained structures
occur in simulations, from top to bottom, at  $z=25d$, $50d$, and $100d$. The dotted
lines indicate estimates of these boundaries from the deterministic
CGL equation (\ref{cgltc}) with a fixed value of
$A(z=0)$ imposed.
 }
\label{figtcwithflow3}
\end{figure}

The transition from the convectively unstable to the ab\-so\-lu\-te\-ly
un\-sta\-ble re\-gi\-me is re\-flected in the coherence of the pattern. A good
way to quantify this is to measure the frequency of the pattern at a
fixed position. The more coherent the pattern, the sharper the peak
associated with the pattern frequency (the position of the peak is
simply fixed by the wavelength and phase velocity of the pattern). The
upper panel of Fig.~\ref{figtcwithflow3}{\em (a)} shows the mean
squared width $\sigma^2$ of the measured peak.

The experimental data on this system have been compared in detail
\cite{babcock,tsameret} with
the results of    numerical studies \cite{niklas2} 
 and with the analytical predictions based on the amplitude
equation.
Just above the threshold of supercritical (forward) bifurcation to a
finite-wavelength pattern, the pattern dynamics can quite generally be
described by an amplitude equation \cite{ch,gc,hakim,walgraef}.
Because of the advection of the patterns by the through flow, the
appropriate amplitude equation in this case is the cubic CGL equation 
\begin{equation}
\tau_0 \partial_t A + s_0 \partial_z A = \xi_0^2 (1+ ic_1) \partial_z^2
A + \varepsilon(1+ic_0) A -g_0 (1-ic_3) |A|^2 A, \label{cgltc}
\end{equation}
where the second term on the left hand side is the group velocity term associated with
the advection of the pattern, and
where $z$ is the fixed coordinate frame along the axis of the
cylinders. For an impression  of the various types of behavior of the incoherent structures in
the CGL equation with an advection term in a
finite system with noise we refer to the work by  Deissler
\cite{deissler1,deissler3,deissler4,deissler2}, Deissler and
Kaneko\footnote{Deissler and Kaneko \cite{deisslerkaneko,kaneko} have
introduced in particular the 
notion of velocity-dependent  Lyapunov exponent. This concept will be
touched upon briefly in section 
\ref{sectionerrorpropagation}. }
\cite{deisslerkaneko} and Proctor {\em et
al.} \cite{proctor}. In the present case, all parameters
associated with the terms linear in $A$ in this equation were
known directly from the stability analysis 
\cite{niklas2}, while 
$c_3$  was determined via   the standard method  of  calculating the nonlinear
terms in an amplitude expansion \cite{niklas2}
(the parameter $g_0$ only sets the amplitude scale and hence is not of
importance in comparing theory and experiment). This therefore allowed a
direct comparison between theory and experiment to be made with only
one adjustable parameter, the effect of the noise at the
inlet.\footnote{Several features  of the scaling behavior are actually
captured by the deterministic equation if one takes the value of $A$
at the inlet at $z=0$ fixed at a given
value. \cite{babcock,tsameret}.} 
Fig.~\ref{figtcwithflow3}{\em (b)} shows the
results of simulations of Babcock {\em et al.} \cite{babcock} 
in which only a random noise term of strength
$\Delta$ was added to Eq.~(\ref{cgltc}). As can be seen, the behavior
of the mean squared width of the spectrum of $A$ in the simulations,
sampled at the {\em same} distance $z=100d$ as in the experiments, quantitatively
reproduces the experimental data of Fig.~\ref{figtcwithflow3}{\em (a)}!
Moreover, it is seen that the behavior is quite independent
 of the value of the noise strength $\Delta$, so the comparison
between theory and experiment is effectively made without 
adjusting any
crucial  parameter! 

For the CGL equation (\ref{cgltc}), the transition
from the nonlinearly convectively unstable to the nonlinearly absolutely unstable regime  can
easily be calculated  explicitly since the fronts are pulled. For zero
group velocity, $s_0=0$, the linear spreading 
velocity $v^*$ is simply
\begin{equation}
v_{\rm CGL}^*= 2 \frac{\xi_0}{\tau_0} \sqrt{ \varepsilon (1+c_1^2) } ,\label{v*cgleq}
\end{equation}
compare Eq.~(\ref{v*cgl}) for the case $\xi_0=1$, $\tau_0=1$.
In the presence of the advection term with $s_0$, a pulled front
connecting the $A=0$ state on the left to the saturated state on the
right moves with velocity $s_0-v_{\rm CGL}^*$. When this velocity is
positive,  the front is convected away and the
system is convectively unstable, while when $s_0-v^*_{\rm CGL}<0$ the
front moves upstream to the inlet and the system is absolutely
unstable. The transition from the convectively unstable to the
absolutely unstable regime therefore occurs when $s_0-v^*_{\rm CGL}=0$,
i.e., for
\begin{equation} 
\varepsilon_{\rm ca}=  \frac{s_0^2 \tau_0^2}{4 \xi_0^2 (1+c_1^2)} .\label{epsilonca}
\end{equation}
For the parameters corresponding to the experiments of
Figs.~\ref{figtcwithflow3}{\em (a)} this gives the value marked with
the dashed line  in panel {\em (b)} --- the remarkable agreement confirms
that the transition from coherent, virtually noiseless patterns at
larger values of $\varepsilon$ to noise-induced fluctuating structures
coincides with the transition from the absolutely unstable to the convectively
unstable regime.

Finally, Fig.~\ref{figtcwithflow3}{\em (c)} from \cite{babcock} shows
the full phase diagram  of the present system as a function of
$Re$. The full symbols along the upper line mark the transition from
the coherent patterns to the incoherent noisy patterns determined
experimentally from the spectrum, as in panel {\em (a)} for $Re=3$,
and the dashed line marks $\varepsilon_{\rm ca}$ according to
(\ref{epsilonca}). The open symbols denote the values of $\varepsilon$
where noise-sustained structures arise for three different downstream
values of $z$.

A detailed theoretical study of the Taylor-Couette system with through
flow, which goes beyond the amplitude equation, can be found in
\cite{luecke8,luecke9,luecke6}.
Also the Rayleigh-B\'enard system with through flow has been 
analyzed along similar lines \cite{luecke1,luecke3}. 

In conclusion, these experiments on Taylor-Couette patterns in the
presence of through flow nicely illustrate  several conceptual issues:
{\em (i)} the distinction between nonlinearly absolutely unstable and convectively
unstable regimes; {\em (ii)} the change  from coherent to
incoherent patterns that this transition implies; {\em (iii)} the
fact that in realistic systems true front structures arise where saturation
behind the front is important;  {\em (iv)} 
 since fronts in the cubic Complex 
Ginzburg-Landau equation are pulled, all essential properties can
still be calculated explicitly from the linear dispersion relation; 
{\em (v)} the importance of noise sustained structures in general for
convective systems.

 The effect of noise on convective systems has recently been studied systematically for the CGL equation 
by Proctor {\em et al.} \cite{proctor}, who map out the full phase
diagram (see also \cite{luecke7,luecke2}). In line with our discussion
of the behavior of fronts in the CGL equation in sections \ref{sectioncglcubic} and 
\ref{sectioncglquintic}, the behavior as a function of the
control parameter $\varepsilon$ depends strongly on whether or not the state selected by the front 
is unstable to the Benjamin-Feir instability. 

We finally note that recently the effect of through flow on chemical
reactions has also been studied experimentally and theoretically \cite{kaern,mcgraw}.

    \subsection{Fronts and noise-sustained structures in convective systems II: coherent and incoherent
sources and the heated wire experiment} \label{secfrontsnoise2}

In the previous section, we discussed the relation between
noise-sustained structures in pattern forming system in which an
advection of the patterns is induced by externally imposing a
flow.  In pattern forming systems which exhibit a Hopf-bifurcation to
spatially and temporally periodic patterns,
 the ensuing traveling wave patterns intrinsically have a  nonzero group
 velocity, and close  to the threshold the instability to a single
 mode is always convective. If the transition is supercritical
 (forward), then close to the threshold in the convectively unstable
 regime the emergence of noise-sustained patterns is again intimately
 connected with the  dynamics of pulled fronts. 

We focus here on the discussion of defects in one-dimensional systems which exhibit
a supercritical transition to traveling-wave patterns, as these are most intimately connected
with the front propagation issue of interest in this paper. One should note, however, that
the motion of defects has experimentally also been studied in great
detail in binary mixtures \cite{kaplan,kolodner}. The fact that transition
to traveling wave states in this system is subcritical rather than
supercritical gives rise to a number of  interesting additional effects, like the ``locking''
of a defect to the underlying period of the pattern \cite{bensimon2,pomeau3}. We refer to the
papers by Kaplan and Steinberg \cite{kaplan} and Kolodner
\cite{kolodner} for examples and for an entry into the literature on these issues.

A new feature, in comparison with the
 discussion of the previous section, is that if the underlying system is spatially reflection
symmetric, two types of traveling-wave states will be possible,
left-moving waves and right-moving waves. If each mode suppresses the
other, the long-time
dynamical state of the system is often dominated by sources and
sinks. A source is a solution which sends out left-moving waves to the 
left, and right-moving waves to the right, while a sink absorbs a
right-moving wave from the left and a left-moving wave from the
right.   Since sources are the active generators of the traveling
waves, their behavior is most important for the dynamics of a 
traveling wave pattern. As we shall see, sources induce a 
sharp crossover from coherent to incoherent dynamics, which is closely 
related to the one discussed in section \ref{secfrontsnoise1} above.

The appropriate amplitude equations for the pattern dynamics just
above onset of a Hopf bifurcation to traveling wave patterns are \cite{ch}
\begin{eqnarray}
   \tau_0 ( \partial_t A_{\rm R} + s_0\partial_x A_{\rm R}\ ) & = &
   \varepsilon A_{\rm R} +\xi_0^2 (1 + i c_1) \partial_x^2 A_{\rm R}
   - g_0 (1 - i c_3) |A_{\rm R}|^2 A_{\rm R}  \nonumber \\ \label{CCGLE1}& &
   \hspace*{2cm} - g_2 (1 -
   i c_2) |A_{\rm L}|^2 A_{\rm R},\\ 
\tau_0  ( \partial_t A_{L} -
   s_0\partial_x A_{\rm L}\ )  & = &  \varepsilon A_{\rm L} +\xi_0^2 (1 + i
   c_1) \partial_x^2 A_{\rm L}  - g_0 (1 - i c_3) |A_{\rm L}|^2
   A_{\rm L} \nonumber \\ & &
   \hspace*{2cm}  - g_2 (1 - i c_2) |A_{\rm R}|^2 A_{\rm L}.\label{CCGLE2}
\end{eqnarray}
Here $A_{\rm R}$ and $A_{\rm L}$ are the amplitude of the right and left moving
waves, and $s_0$ is the linear group velocity.\footnote{We stress here
  that
  the term ``right-moving'' and 
 ``left-moving'' refers to the group velocity, in particular also in the definition
 of a source. In principle, it is possible that the phase velocity of
 the traveling waves is opposite to the group velocity. If this is the 
 case, to the eye it appears that the pattern runs into the source
 instead of being sent out by it.  See \cite{vanhecke4} for further
 discussion of this. }  The terms on the first line 
of each equation are the same for each individual mode as in the single CGL equation
(\ref{cgltc}) for the advected Taylor vortices. The last term on the
second line describes the coupling of the modes. We will be interested 
in the regime $g_2>g_0$ where one mode suppresses the other one
sufficiently strongly that standing wave patterns do not form \cite{ch,vanhecke4}. 

\begin{figure}[t]
\begin{center}
\epsfig{figure=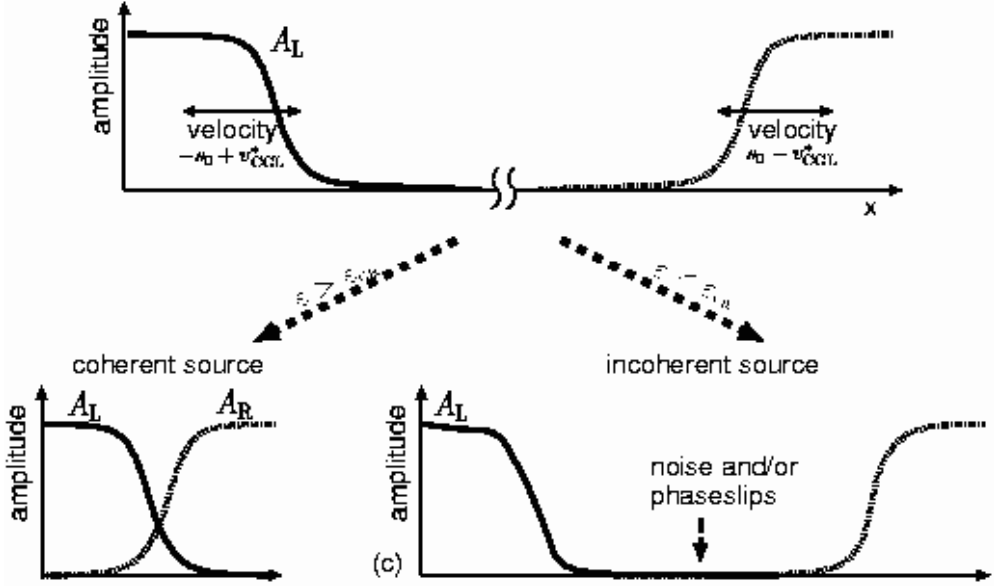,width=0.95\linewidth} 
\end{center}
\caption[]{Schematic illustration of  how incoherent and coherent
sources in traveling wave systems are built from  front solutions. In
the upper panel, we start from two isolated  front solutions which are
so far apart that they are non-interacting. In the absolutely unstable
regime, for $\varepsilon > \varepsilon_{\rm ca}$, the two fronts move
together. The coherent source solution they then form can be thought
of as a bound state of the two fronts (lower left panel). In the
convectively unstable regime, $\varepsilon < \varepsilon_{\rm ca}$,
the two fronts would move apart in the absence of interactions and of
noise. In practice,  a wide incoherent fluctuating source
results. Whether the average source width is determined by the noise
or the interactions, is at present unknown. Possibly, this also
depends on the initial separation of the fronts.
 }
\label{figsources1}
\end{figure}

To understand how the dynamical behavior or sources is intimately
related to that of fronts and to the transition from the convectively
unstable to the absolutely unstable regime,
 consider first the case in which there are  no left-moving 
patterns, $A_{\rm L}=0$. In this case the amplitude equation
(\ref{CCGLE1}) for $A_{\rm R}$ reduces to the single CGL equation
(\ref{cgltc}) considered in the case of the advected Taylor vortex
pattern.  As we  already mentioned there, fronts in the cubic CGL equation are
pulled, and hence  the velocity of the rightmost  front in the top
panel of Fig.~\ref{figsources1} (it is sketched with a dashed line and 
 connects  the unstable state at $x\to -\infty$ to the saturated finite
amplitude state for $x\to \infty$) is simply $s_0 -v^*_{\rm CGL}$,
where $v^*_{\rm CGL}$ is the linear spreading velocity of the single
CGL equation, given in (\ref{v*cgleq}). Furthermore, $\varepsilon_{\rm 
  ca}$ given by (\ref{epsilonca}) marks the transition from the
absolutely unstable regime for larger $\varepsilon$  to convectively
unstable regime for smaller $\varepsilon$. Thus, for $\varepsilon <
\varepsilon_{\rm ca}$ the right front actually recedes in the positive
$x$-direction, for $\varepsilon >\varepsilon_{\rm ca}$ the growth is
strong enough that it moves upstream. See the top panel of Fig.~\ref{figsources1}.

The picture of how sources are built from fronts, as it has emerged from theoretical studies
\cite{coullet,vanhecke4}, is the following. Consider two widely
separated fronts  in $A_{\rm   L}$ and $A_{\rm R}$, as sketched in 
Fig.~\ref{figsources1}{\em (a)}. When $\epsilon> \varepsilon_{\rm ca}$ 
the two fronts move towards each other; once they get close they 
 form a stationary  coherent source solution which sends out waves of
 the same wavelength and frequency to both sides  \cite{vanhecke4}.\footnote{We
 can thus think of the sources in the absolutely unstable regime
 $\varepsilon > \varepsilon_{\rm ca}$ as bound states of two pulled
 fronts. This is somewhat reminiscent of the pulse solutions
   in the single quintic CGL equation, which can be thought of as a
   bound state of two pushed front solutions \cite{vsh}.}
 Very much like what we saw in the previous
 section, sufficiently deep into the  absolutely unstable regime $\varepsilon >
 \varepsilon_{\rm ca}$ the coherent sources give thus rise to coherent 
 traveling wave patterns with only very small fluctuations.

In the convectively unstable regime $\varepsilon < \varepsilon_{\rm
  ca}$ the two  fronts of Fig.~\ref{figsources1}{\em (a)} would move
infinitely far apart if there were no fluctuations and no
interactions. The discussion of the previous section would lead one to 
expect incoherent pattern dynamics in this regime, which originates
from the center of the convectively unstable region between the two
fronts, as indicated in Fig.~\ref{figsources1}{\em (c)}. This {\em is
  indeed} what is found in numerical studies \cite{vanhecke4}, but the 
origin of the fluctuations is not completely understood in this
case. Clearly, if a sufficiently large random noise is added to the
coupled equations (\ref{CCGLE1}) and (\ref{CCGLE2}), one will enter a
similar incoherent
fluctuation-dominated regime like the one we discussed in  section
\ref{secfrontsnoise1}. The numerical studies give reason to believe,
however, that in the small noise limit, another  intrinsic incoherent 
dynamical regime exists where the fluctuations result from the
interaction of the two fronts in their tails, presumably 
via the generation of phase
slips which give rise to intrinsically chaotic behavior which in turn
is advected away towards the front-like regions. 

\begin{figure}[t]
\begin{center}
\epsfig{figure=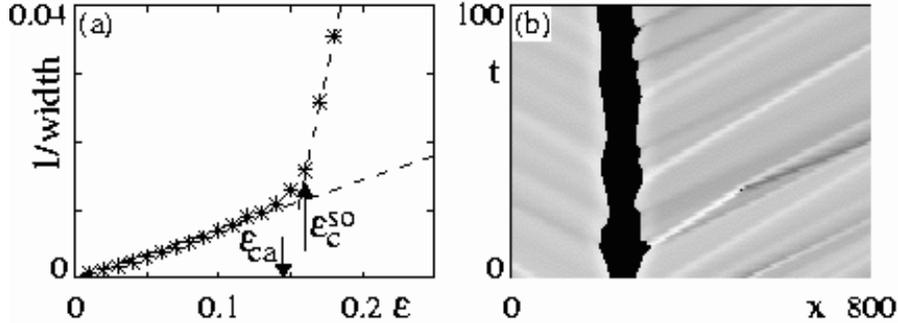,width=0.85\linewidth} 
\end{center}
\caption[]{Results of  numerical simulations of the behavior or
  sources in the coupled CGL equations (\ref{CCGLE1}) and
    (\ref{CCGLE2}) in the absence of external noise. (a) Inverse
    average source width as a function of $\varepsilon$ for the
    coupled equations with $s_0=1.5$, $c_1=-1.7$, $c_2=0$, $c_3=0.5$, 
   $g_2/g_0=2$, and $\xi_0=\tau_0=1$. Note the crossover  at
   $\varepsilon_{\rm c}^{\rm so}$ just above
   $\varepsilon_{\rm ca}=0.14 $, and the fact that the width diverges
   inversely proportional to $\varepsilon$ for small $\varepsilon$.  (b) Space-time plot of the local
   wavenumber of the fluctuating source for $\varepsilon=0.11
   <\varepsilon_{\rm ca}=0.14$, illustrating the fluctuations of the
 width and the incoherent dynamics it entails in the traveling wave
 domains away from it. In the black region the sum of the modulus of
 the two
 amplitudes have fallen below 10\% of the saturated value; the light
 and dark streaks correspond to hole-like wavenumbers packets sent out 
 by the source. From Pastur {\em et al.} \cite{pasturshort}.
 }
\label{figsources2}
\end{figure}

One way in which the crossover between the two results shows up is in
the width of the sources, defined as the distance between the two
points where the two amplitudes reach a fixed fraction of the
saturated value.
In Fig.~\ref{figsources2}{\em (a)} we show the results for the inverse 
of the  (average)  source width in numerical
simulations of Eqs.~(\ref{CCGLE1}) and (\ref{CCGLE2}). When $\varepsilon $ 
is decreased towards $\varepsilon_{\rm ca}$, the width of the coherent 
source solutions increases rapidly: the width of the coherent source
solutions appears to diverge at $\varepsilon_{\rm ca}$. Just before
$\varepsilon_{\rm ca}$, however, the width becomes so large that the
fluctuation effects from the region where both amplitudes $A_{\rm R}$
and $A_{\rm L}$ are small, take over. This happens at the point marked
by $\varepsilon_{\rm c}^{\rm so}$ in  Fig.~\ref{figsources2}{\em
  (a)}. For smaller values of $\varepsilon$ sources are incoherent and 
their width scales as $\varepsilon^{-1}$. Fig.~\ref{figsources2}{\em
  (b)} shows a space-time plot of an incoherent source in this parameter 
range: the  core (black region) fluctuates in time, and these
fluctuations are reflected in the grey regions where the amplitudes
are close to their saturation values.\footnote{Not only is the phase of the
amplitudes fluctuating in these regions, the source also sends out
coherent structures at irregular intervals. These are visible as the
light streaks, and  appear to correspond to so-called homoclon
solutions \cite{vanhecke5,vanhecke6}.}   As noted before, whether the incoherent
source fluctuations arise from numerical noise in the simulations or
from intrinsic dynamics in their center, is at present not completely
clear. It is possible that there is no unique answer: when we generate 
source solutions starting from initial conditions with two widely
separated fronts,  it is conceivable that it   depends
on the initial separation of these fronts, which mechanism dominates.

\begin{figure}[t]
\begin{center}
\epsfig{figure=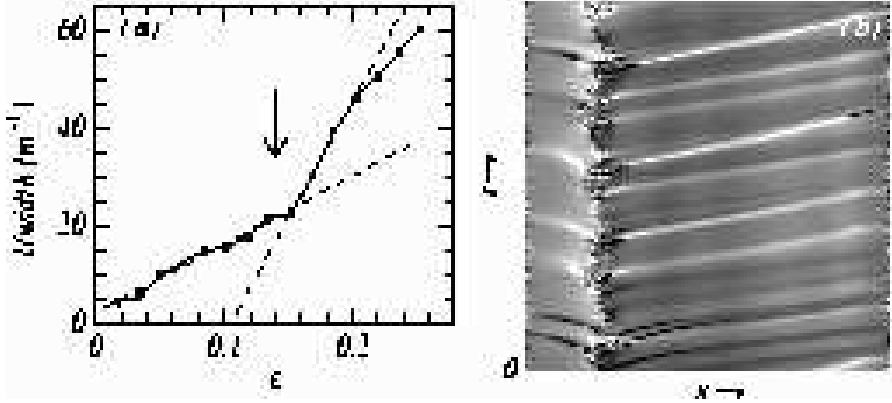,width=0.85\linewidth} 
\end{center}
\caption[]{(a) Dependence of the inverse of the average width of a source on the
reduced control parameter $\varepsilon$ which measures the distance
from the threshold at which traveling waves appear in the heated wire
experiment of Pastur {\em et al.} \cite{pasturshort,pasturlong}. The value of
$\varepsilon_{\rm ca}$ which was obtained independently from
measurements of the group velocity $s_0$, the correlation length
$\xi_0$ and the time scale $\tau_0$, is indicated by an arrow. At this
point, a crossover in the inverse source width as a function of
$\varepsilon$ is observed in the same way as in the numerical
simulations, see Fig.~\ref{figsources2}{\em (a)}. (b) Space-time
diagram of the local wave number of a source in the experiments for
$\varepsilon=0.11$, i.e.,  in the incoherent source regime. The extent
of the $x$-axis is 42 cm, the total time is 10485 s.
 }
\label{figsources3}
\end{figure}

Recent experiments on traveling waves near a heated wire\footnote{In a related
  experiment on hydrothermal waves in a cell heated from the side, the 
  transition from the convectively to absolutely unstable regime was
  recently also verified quantitatively \cite{chiffaudel2}.}  have
confirmed these theoretical predictions \cite{pasturshort,pasturlong}.
 In such experiments, a wire is
suspended a few millimeters below the surface of a fluid, and heated
with the aid of an electrical current \cite{vince3,vince1,vince2}. Above some critical heating,
the temperature pattern near the wire exhibits a bifurcation to waves
which travel along the wire; by detecting the deformation of the
surface, they can be detected with high precision. Measurements of
this type not only confirm that the bifurcation is supercritical, and
hence that the above coupled CGL equations are the appropriate
amplitude equations, but also yield all parameters entering formula (\ref{epsilonca})
for the crossover value \cite{pasturshort,pasturlong}. The value obtained from such measurements is
$\varepsilon_{\rm ca} = 0.14 \pm 0.02$. In agreement with the results
of the theoretical and numerical studies,  a crossover in the inverse
width of the sources is found precisely at this value, see
Fig.~\ref{figsources3}{\em (a)} \cite{pasturshort,pasturlong}. Furthermore,
as Fig.~\ref{figsources3}{\em (b)} illustrates, 
space-time traces of the dynamics in the convectively unstable regime
below $\varepsilon_{\rm ca}$ show that the sources are wildly
fluctuating in this regime. Very much like what was found for the
width of the frequency peak in the traveling Taylor patterns in
Fig.~\ref{figtcwithflow3}{\em (a)},
the peak  in the spectrum rapidly broadens as $\varepsilon$ is
reduced below $\varepsilon_{\rm ca}$ \cite{pasturlong}. 

We finally stress again that just above a supercritical transition, the
situation is generally rather simple because the fronts are
pulled: For this reason, the transition from convective to absolute
instability is given by the linear criterion (the sign of
$s_0-v^*$). When a transition is subcritical (inverted), the fronts
near the transition are usually pushed (see section \ref{sectionwhenpushed}), and then the
distinction between the two regimes is determined by the velocity of
the pushed front, for which no general results are know. As an example 
of this, we may mention that  a secondary bifurcation  observed in  a
cell  heated from the sides  is subcritical, and
that a  pushed-front mediated transition from the convectively to
absolutely unstable state was recently observed in such an experiment 
\cite{chiffaudel}.

The discussion in this session clearly illustrates that fronts are  
important building blocks  of the dynamics of traveling wave
systems. Although this behavior has been mapped out sufficiently well
that quantitative predictions can be made in realistic cases, on a
more fundamental level many issues are poorly understood. We already
mentioned that the origin of the fluctuating sources is not understood 
very well --- is external noise necessary or is the deterministic
interaction between the tails of the fronts sufficient to give the
incoherent behavior? Can one understand some of the behavior from
studying the (non)existence of coherent source solutions following the
methods of \cite{sandstede}? Do sources send out homoclon solutions
\cite{vanhecke6,vanhecke7}? Why do sources seem to conform
experimentally to those arising in amplitude equations, while sinks
do not?

\subsection{Chemical and bacterial growth fronts} \label{chemicalbacterial}
The issue of front propagation into unstable states often plays a role
indirectly in theoretical analysis of waves  and fronts in coupled reaction-diffusion
equations --- e.g., fronts are an important building block of
spirals. Nevertheless,  clean examples of single fronts in realistic experimental
situations do  not appear to be abundant. In this section we will
first briefly discuss  few of the results of a series of experiments
designed specifically to study fronts, and then briefly touch on the 
broader implications of the difference between pushed and pulled fronts  for
pattern dynamics in coupled reaction-diffusion systems. 

In the last decade, experimentalists have been able to develop
chemical reactors in which the Turing instability and other chemical reaction
patterns  could be probed
\cite{dekepper1,dekepperreview,epstein2,epstein1,swinney3,meron,swinney2}. 
The Turing instability is the
stationary bifurcation to periodic patterns in coupled reaction
diffusion systems, which may occur when the activator (the component
with autocatalytic characteristics) has a diffusivity which is
significantly smaller than that of the inhibitor \cite{epstein2,murray,walgraef}. The suppression of the
activator diffusion coefficient is experimentally achieved by reversibly
binding to an immobile species attached to an aerogel; this same gel also allows the 
continuous feeding of the reactants without inducing convection.

A few years ago, Horv\'ath, Lagzi  and T\'oth \cite{horvath2,toth1} and De Kepper and coworkers
\cite{dekepper2,dekepper3} have introduced
variants  of such experiments that allow them to systematically
produce and study fronts and the two-dimensional patterns they give
rise to. The basic reaction is a chlorite oxidation
of the $S_4O_6^{2-}$ ion. In this reaction, the
hydrogen $H^+$ ion plays the role of the autocatalyst, and its
diffusion is suppressed in a controlled way by incorporating carboxylate groups which
reversibly bind it  in the
polymer gels (even an electric field may change the effective
diffusion ratio \cite{toth2}). Initially planar reaction fronts were created by cutting
the polymer gel with the reactants into two, treating one of
the parts so as to induce a reaction, and then putting the two parts
carefully back together again in a sealed cell  to prevent
evaporation.

The dominant reactions in this system can to a good approximation be
described by the following dimensionless  coupled reaction-diffusion
equations \cite{dekepper2,horvath2}
\begin{eqnarray}
\partial_t \alpha &=&  \nabla^2 \alpha - \alpha \beta^2(\kappa + 7
\alpha) ,\\
\partial_t \beta &=& \delta  \sigma^{-1}\nabla^2 \beta + 6 \sigma^{-1} \alpha
\beta^2 (\kappa+7\alpha),
\end{eqnarray}
where $\alpha$ is the dimensionless $S_4O_6^{2-}$ concentration and
$\beta$ the dimensionless $H^+$ concentration, $\delta$ is the ratio
of diffusion coefficients of  these two components in the absence
of the binding of the $H^+$ to the carboxyl groups, and $\kappa $ is a
parameter which depends on the relative chlorite excess. The
coefficient $\sigma$ accounts for the reversible binding of the
hydrogen: as $\sigma$ increases, both the effective hydrogen diffusion
and the effective reaction rate is  suppressed. 

Note that the state $\beta=0$ (no hydrogen) in this system of
equations is unstable: as soon as $\beta$ becomes nonzero the
autocatalytic reaction term in these equations makes $\beta$
increase. However, since this reaction term is quadratic in $\beta$,
to linear order in $\beta$ there is no instability. Hence there is no nonzero
linear spreading speed $v^*$, and fronts in this equation are 
always of the pushed type.

\begin{figure}[t]
\begin{center}
{\tt (a)} \hspace*{-3mm} \epsfig{figure=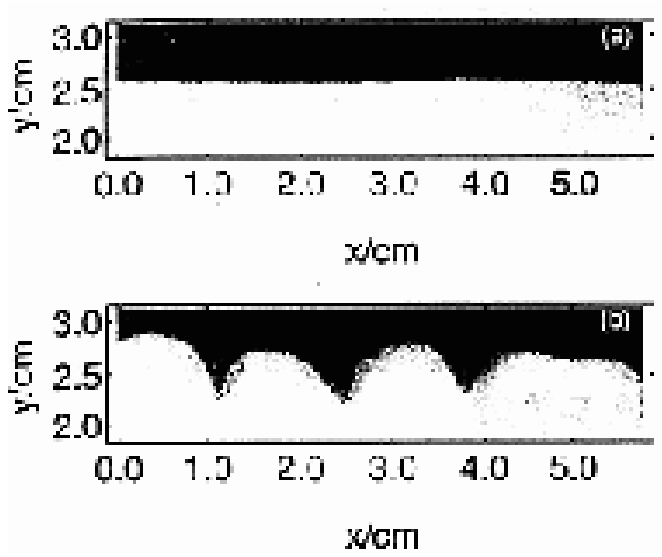,width=0.39\linewidth} 
{\tt (b)}
\hspace*{-3mm}
\epsfig{figure=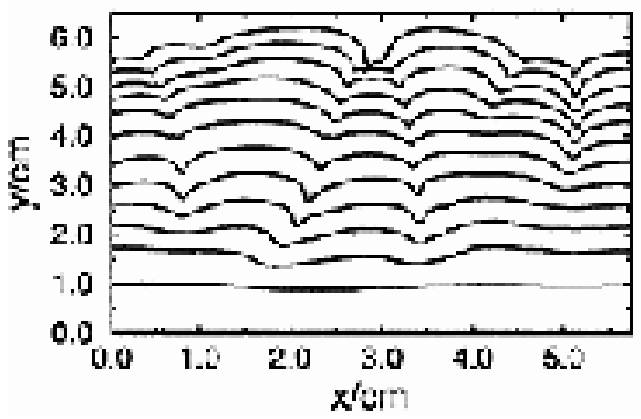,width=0.485\linewidth}
\end{center}
\caption[]{ (a) Front images in the experiments by T\'oth and Horv\'ath
\cite{toth1}. The upper image corresponds to the regime where the
front is stable, while the lower one to the regime where the effective hydrogen diffusion
coefficient is so small that the planar interface is unstable. The
fronts are advancing upward. (b) Traces of the front positions in
these experiments at time
intervals of 120 minutes in the regime corresponding to the lower image of
(a). The dynamics of the interfaces in this regime is very similar to that found in the
Kuramoto-Sivashinsky equation, the generic lowest order equation just above threshold of
a long-wavelength interface instability.  }\label{fighorvath}
\end{figure}

Theoretical studies \cite{horvath1,kuramoto1,kuramoto2} of fronts in
such coupled reaction-diffusion models have shown that when the
effective diffusion coefficient $\delta$ of the autocatalytic $\beta$ component
is sufficiently small, the fronts exhibit a long-wavelength
instability of the type found for the Kuramoto-Sivashinsky equation
discussed in section \ref{sectionks}. In agreement with this, T\'oth and Horv\'ath
\cite{toth1} found that when the hydrogen ion diffusion was
sufficiently suppressed, the fronts exhibited the lateral type of 
structures  familiar from the Kuramoto-Sivashinsky equation
\cite{kuramoto1,kuramoto2,malevanets,pomeau2,shraiman2,sivashinsky},
in agreement with the theoretical predictions. An 
example of  some of their observations
is shown in Fig.~\ref{fighorvath}.

To what extent are these results of interest from a more general
perspective?  The answer lies
in the deep connection with the issue discussed later in section
\ref{sectionmba}: because the fronts in this reaction-diffusion system
are pushed, the spectrum of the linear stability operator of a planar
front is gapped, and the dynamically important region for the
stability modes is the front region itself. Because of this, in the
thin interface limit the front behavior can be described by a moving
boundary  or effective interface  approximation, in which the front is
viewed as a line (in two dimensions) or a sheet (in three dimensions) of
vanishing width. As we shall argue below, this immediately implies
that for sufficiently small ratio of diffusion coefficients, the
reaction front exhibits a long-wavelength instability, and this in turn
means that just above the
instability threshold, the interface dynamics maps onto that of the
Kuramoto-Sivashinsky equation (\ref{kseq}).\footnote{The  prefactor of the linear
terms in the Kuramoto-Sivashinsky equation are fixed by the
coefficients in the dispersion relation of the weakly unstable
interface mode. The prefactor of the nonlinear term is fixed by
the fact that for an isotropic system, the projection of the velocity
onto the overall propagation direction
of a piece of  interface of height $h$ whose normal makes an angle $\theta$ with the
propagation direction, is simply $v_{\rm planar}/ \cos \theta = v_{\rm
planar}(1 + \half (\nabla h)^2 +\cdots)$. For a more explicit derivation
of the Kuramoto-Sivashinsky equation and a study of the chaotic behavior 
in the context of the present type of reaction-diffusion equations,
see, e.g., \cite{malevanets}.}

\begin{figure}[t]
\begin{center}
\epsfig{figure=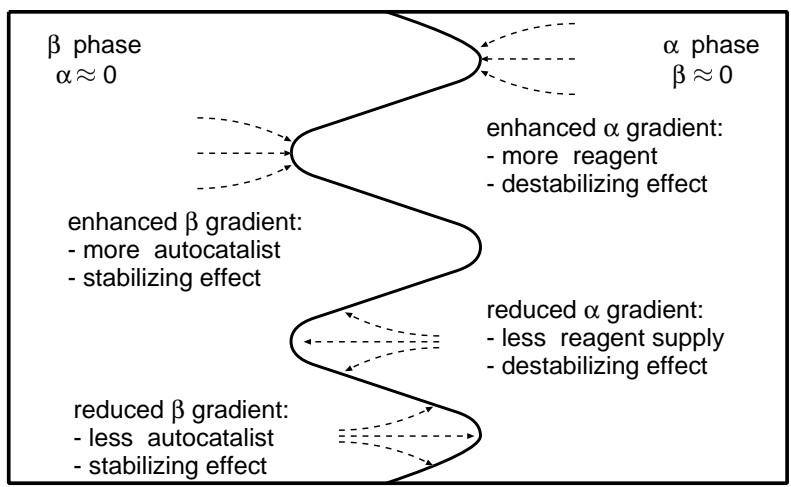,width=0.75\linewidth}
\end{center}
\caption[]{Schematic illustration of the origin of long-wavelength
lateral instabilities of a fronts  in coupled reaction diffusion equations which describe
the propagation of the domain of an ``autocatalytic'' substance $\beta$
(e.g. bacteria in a bacterial growth model) into a region with
abundant reactant $\alpha$ (nutrient in the bacterial case). The front
is indicated by the solid line and is moving to the right. If it is a 
pushed front, its dynamics  can be mapped onto an effective interface model or moving
boundary problem for perturbations on length scales much larger than
the interface width. For  long-wavelength perturbations of this
effective 
interface or boundary (drawn with a solid line),  the supply of the
reactant is enhanced at forward protrusions and decreased for the
protrusions staying behind. This effect  clearly tends to enhance
the instability. On the back side, the situation is the opposite: the
diffusion of $\beta$ towards the interface is reduced behind the
forward protrusions, and this tends to stabilize the interface. Which
of the two effects dominates, depends the ratio of diffusion
coefficients.  If the diffusion coefficient of $\beta$ is sufficiently
large relative to that of the $\alpha $ reactant, then the stabilizing
effect wins, while if the diffusion coefficient of $\beta$ is small
enough, the destabilizing effect dominates and the interface is
unstable. For pulled fronts, the argument does not hold because they
can not be described in an effective interface approximation.}\label{figinstability}
\end{figure}

The point now  is the following.  As we will discuss in section
\ref{sectionmba} {\em pulled fronts can not simply be viewed as a
  moving boundary or interface
in the limit when their nonlinear transition zone is thin}, because
their dynamically important region is the semi-infinite region ahead
of the front. On the other hand,  when a front is pushed, the
situation is very different and actually much simpler:  in the limit
where their width is much smaller than the  
typical length scale of the pattern (e.g., the typical front radius of
curvature), {\em pushed fronts are amenable to a moving boundary or
  effective interface
  approximation} in which the front is treated as a line or sheet of
zero thickness at which the outer fields obey certain boundary
conditions. As we illustrate in Fig.~\ref{figinstability}, a
propagating reaction diffusion front which effectively leads to a
replacement of one species ahead of it  by another one on the back, {\em will
generically   exhibit  a long-wavelength instability of the
Kuramoto-Sivashinsky type in the limit in which the diffusion
coefficient of the species behind the front (the autocatalyst in the
language used above) is much smaller than that of the one in front
(the reactant)}.\footnote{For moving fronts, the long-wavelength
instability that arises for sufficiently small diffusion coefficients
of the phase behind the front  is intimately related to the
Mullins-Sekerka instability discussed in section
\ref{sectionmullinssekerka}. This may not be obvious at
first sight, since   the dispersion relation
(\ref{mullinssekerkadispersion}) is not analytic in $k$, while the
actual dispersion relation  near threshold of the long-wavelength
instability of an interface has of course an  expansion in   $k^2$.
 The reason for the difference  is that the usual
Mullins-Sekerka dispersion relation is derived in  the limit that
 the interface growth velocity is small enough that the diffusion
 equation of the outer phase ahead of the front can be approximated by 
 the Laplace equation.  Deep into the unstable region of the phase
 diagram, the dispersion relation of bacterial growth models like the
 one specified by equations (\ref{kmmodel1}) and (\ref{kmmodel2})
 becomes more like the Mullins-Sekerka form
 (\ref{mullinssekerkadispersion}) --- see e.g. \cite{judith}.}

We need to stress that the above analysis does not give a prediction
concerning the lateral instability of  fronts which are pulled.
Indeed, the discharge fronts
discussed in section \ref{sectionstreamers} are examples of pulled
fronts which {\em do}  exhibit a long wavelength instability, while  for the 
 fronts in coupled chemical reaction-diffusion 
equations of the form
\begin{eqnarray}
\partial_t \alpha &=&  \nabla^2 \alpha - \kappa \alpha \beta - \alpha \beta^2 ,\\
\partial_t \beta &=& \delta  \nabla^2 \beta +  \kappa \alpha \beta + \alpha \beta^2 ,
\end{eqnarray}
Horv\'ath {\em et al.} \cite{horvath1} have found strong numerical indications 
that the fronts were always weakly stable in the pulled regime
$\kappa> \half$. As
soon as they were pushed for $\kappa < \half$, they were found to exhibit a long-wavelength 
instability, in accord with the above arguments. From a different
angle, Kessler and Levine \cite{kessler2} have also found results
which are consistent with our scenario: they pointed out  that if a set of
continuum reaction-diffusion equations with pulled fronts is simulated 
with a discrete lattice model, the resulting model does obey a
long-wavelength lateral front instability  of the type discussed
above if the ratio of diffusion coefficients is sufficiently
small. This is consistent with the observation, to be discussed in 
section \ref{sectionstochasticfronts}, that the effective cutoff
provided by the particles makes the front (weakly) pushed, and hence
that it should have a long-wavelength instability in the limits in
which the diffusion coefficient of the phase on the back is much
smaller than the  one in the phase ahead of the front.

The  general theme of this section is also of immediate relevance for
biological growth models. E.g., bacterial colonies  can exhibit 
growth patterns which are reminiscent of Diffusion Limited Aggregation 
clusters \cite{meakinbook} and other growth patterns
\cite{benjacob1,benjacob2,czirok,benjacob3,matsushita}. Various
reaction diffusion models have been suggested to  
explain some of this behavior
\cite{benjacob4,czirok,benjacob3,mimura}. In particular, it has been argued
 \cite{benjacob3,kitsunezaki,benjacob5}  that in cases in which the bacteria secrete a 
fluid which acts like a lubricant for their motion, a model with a
nonlinear diffusion coefficient of the type
\begin{eqnarray}
\partial_t \alpha &=&  \nabla^2 \alpha -  \alpha \beta^m,\label{kmmodel1}\\
\partial_t \beta &=& \delta  {\vec{\nabla}}\cdot (\beta^k \vec{\nabla}
\beta) +   \alpha \beta^m , \label{kmmodel2}
\end{eqnarray}
with $m=1$ would be appropriate.  In this case, $\beta$ is the
dimensionless bacterial density, and $\alpha $ plays the role of the
nutrient field. As we already mentioned above, fronts in this model
are pulled in the case $k=0, m=1$, and for any $m>1$ the fronts are
pushed and do have a long-wavelength lateral instability. Likewise,
with nonlinear diffusion, $k>0$ but bilinear kinetics ($m=1$),
 fronts are pushed and  have a long-wavelength lateral
 instability for small enough $\delta$ \cite{judith}, again  in accord with our
 general arguments.\footnote{Technically, the situation for the
   bacterial growth model (\ref{kmmodel1}), (\ref{kmmodel2}) with
   $m=1$ is
   somewhat more complicated than we present it here: in the limit
   $\delta \downarrow 0$ the growth fronts have all the properties of
   an interfacial growth model with a Mullins-Sekerka like
   instability, but the moving boundary approximation is probably not
   rigorously correct, as it is not quite justified  on all physically 
   relevant length scales \cite{judith}.}  Thus, these considerations are an important
 ingredient for chemical and biological model building:  the above 
 reaction-diffusion model (\ref{kmmodel1}), (\ref{kmmodel2}) 
with $m=1, k=0$ is a bit of a singular case,\footnote{Actually, to my
  knowledge,  neither the singular behavior in the limit
  $k\downarrow 0$ nor the one in the limit
  $m\downarrow 1$ has been worked out; these limits appear to be
  interesting technical challenges \cite{judith}.} as for any
  $m>1$ or $k>0$ the fronts are pushed and exhibit a long
 wavelength instability for small $\delta$.

It is important to stress that although I believe the above  scenario to be
generally true on the basis of the strength of the arguments of
section~\ref{sectionmba}  and those illustrated in
Fig.~\ref{figinstability}, this powerful line of argument is, to my knowledge,
relatively unexplored --- I have not seen it discussed explicitly in the
physics or mathematics\footnote{Apparently, in some cases which we
  would refer to as an example of a coupled reaction-diffusion problem 
  with pulled fronts, it has been noted that standard analysis trying
  to prove convergence to a moving boundary description breaks down
  (D. Hilhorst, private communication), but the connection with the
  general scenario advanced here appears not to have been made.}
literature,  and it should be considered as an interesting
line of future research to investigate or (dis)prove this
argument. The considerations are clearly an important ingredient for
chemical or biological model building: 

We finally note that the fronts we have discussed here connect domains where
the  system itself does not have a finite-wavelength instability. Fronts which generate
a state which is unstable to a Turing or Hopf instability, or which propagate into a state
with a pattern due to this instability, are discussed in \cite{sandstede}.

     \subsection{Front or interface dynamics as a test of  the order
of a phase transition}  \label{sectiondynamicaltest}

When  the
thermodynamic phase transition  between two phases is of first order,
both individual phases are stable to small perturbations of the order
parameter in the neighborhood of the transition. This immediately
implies that there is then a nucleation barrier for the formation of a
droplet or nucleus of one phase in the other. Since this nucleation
barrier is large near the transition (as it proportional to a power of
the ratio of the finite surface tension of the interface between the
two phases and the difference in chemical potentials, which becomes
arbitrarily small near the transition), the nucleation rate for such
droplets of the other phase is small. On the other hand, once an
interface exists, it can usually grow with a speed which is linear in
the driving force, i.e., the difference in chemical potentials or free
energy $\Delta F$.\footnote{An exception are faceted crystals which in the absence of
dislocations grow via the nucleation of new layers on top of the
existing one. This nucleation rate is again small for small driving
forces. If there are screw dislocations at the surface, these can act
as sources for step motion, and the growth rate is quadratic in the
chemical potential difference \cite{gilmer}.}
\begin{equation} 
v \sim   \Delta F \sim T-T_{\rm c} \label{dynamicaltesteq0},
\end{equation}
where $T_{\rm c}$ is the transition temperature. Here
 we used the fact that  near  first order transition the difference in
free energy is linear in the temperature difference. The fact that the
nucleation rate is exponentially small near the transition, while
interfaces respond linearly to the driving force, is the reason for
the ubiquity of interfacial growth phenomena in physics.

Near a second order transition, the situation is very different. For
our purposes the important point is that the response is very
asymmetric. If the system is rapidly heated up from below $T_{\rm c}$ where
the system is ordered to above $T_{\rm c}$, the driving force for the order
to disappear is finite everywhere in the bulk of the system, so the order parameter 
relaxes homogeneously in the bulk of the system --- one normally will
not see a propagation of domains of the disordered phase into ordered
domains: Instead, the order just dies away homogeneously. On the other
hand, when the system is quenched from above to 
below $T_{\rm c}$, the most common situation is that patches of the stable
ordered phase grow out; usually the order parameter has a different
sign (say magnetization up or down in the cases of systems with an
Ising symmetry) or direction (in cases where the order parameter is a
vector) in different patches, and so the initial phase is then followed by
one where the domains   coarsen in size or where defects anneal out. 
However, if the fluctuations
are small it may happen that the ordering in this case occurs mostly
through the growth of domains of the stable phase into the disordered
phase. The velocity of these interfaces will scale as $\xi/\tau$ where
$\xi$ is the correlation length which diverges as $|T-T_{\rm c}|^{-\nu}$ at a
second order transition, and $\tau $ is the characteristic correlation
time which diverges as $|T-T_{\rm c}|^{-\nu z}$ \cite{hohenberg}. So we
generally get a power-law scaling $v\sim |T-T_{\rm c}|^{\nu (z-1)}$. In mean
field theory, with $\nu=\half$ and $z=2$ for diffusive systems, this
gives $v\sim |T-T_{\rm c}|^{1/2}$.

\begin{figure}[t]
\begin{center}
\epsfig{figure=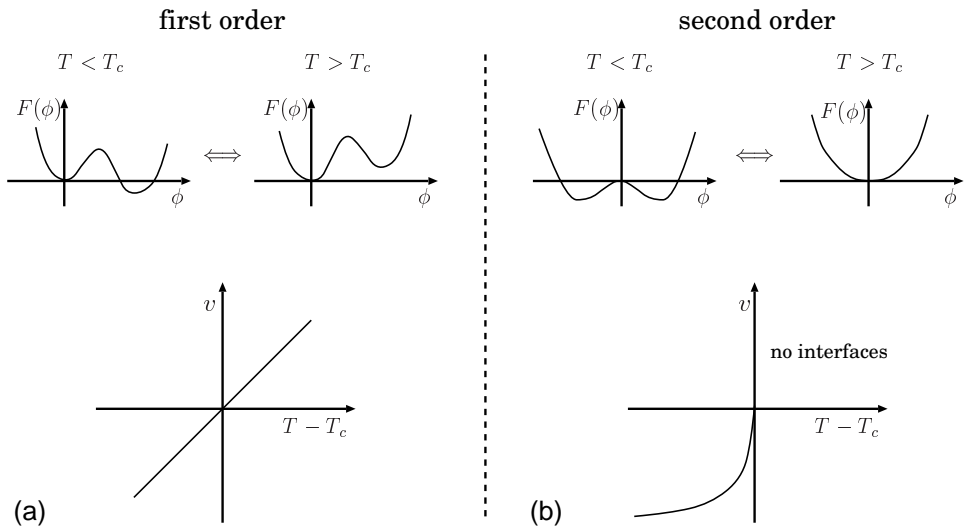,width=0.95\linewidth} 
\end{center}
\caption[]{Schematic sketch of the relation between the order of a
phase transition and the behavior of the interface velocity near the
critical point. The qualitative behavior is indicated within a Landau
picture, but the behavior holds more generally for non-mean-field
systems. (a) The case of a second order phase transition. The upper
part indicates the behavior of the free energy density as a function
of the order parameter below and above $T_{\rm c}$, while the lower part
illustrates  that the interface velocity is linear in the deviation of
the temperature at the interface from the critical value. (b)
Similarly for the case of a second order phase transition. Note that
the steady state propagation of an interface into a
domain of the ordered state is not possible, as there is a finite  driving force on
the bulk order parameter to relax to zero. On the ordered side, the
front velocity depends nonlinearly on the deviation from the critical
point. In Landau theory this dependence is as $v \sim \sqrt{|T_{\rm c}-T|}$. }\label{figdynamicaltest}
\end{figure}

These considerations can be illustrated simply in a Landau mean field
picture for an order parameter $\phi$ described by a  the dynamical
equation
\begin{equation}
\frac{\partial \phi}{\partial t} = - \frac{\delta {\mathcal
F}}{\delta \phi},\hspace*{1cm} \mbox{with} \hspace*{1cm} {\mathcal F}
= \int \ds {\bf r}\, \left[
\half ({\vec \nabla} \phi)^2 + F(\phi) \right]. \label{phasetranseq1}
\end{equation}
The dynamics implied by this equation is such that the free energy
${\mathcal F}$ is non-increasing in time (${\mathcal F} $ acts like a
Lyapunov function, see footnote \ref{footnotelyapunov}), so it tends
to drive $\phi$ to the value at which the free energy density ${F}(\phi)$ is minimal.

In the case of a first-order transition, the function $F(\phi)$ has
the form sketched in Fig.~\ref{figdynamicaltest}{\em (a)}: there are
two local minima (or three if the order parameter is symmetric under a 
change of sign). At high temperatures, the one corresponding to the
disordered phase ($\phi=0$) is the absolute minimum, but as the
temperature is lowered, the other one corresponding to the ordered
phase comes down, and at $T=T_{\rm c}$ the value of $F$ at the two minima is 
the same. Below $T_{\rm c}$, the latter one is the absolute minimum. 
If one consider a planar interface solution $\phi=\phi(\zeta)$ in a
frame $\zeta=x-vt$ moving with velocity $v$ in this case,
Eq.~(\ref{phasetranseq1}) reduces to
\begin{equation}
-v \frac{\partial \phi}{\partial \zeta} = - \frac{\delta {\mathcal
F}}{\delta \phi},
\end{equation}
which upon multiplying with $\partial_\zeta \phi$ gives
\begin{equation}
v= \frac{ \int_{-\infty}^{\infty} \ds \zeta \, \partial_\zeta  \delta
  {\mathcal F}/ \delta \phi }{  \int_{-\infty}^{\infty} \ds \zeta \, (\partial_\zeta
  \phi)^2 } =  \frac{ F(\phi(\infty)) - F(\phi(-\infty)) } {
\int_{-\infty}^{\infty} \ds \zeta \, (\partial_\zeta
  \phi)^2 }.
\end{equation}
The term in the numerator is a unique number for
functions $F(\phi)$ like those sketched in
Fig.~\ref{figdynamicaltest}{\em (a)}. For stationary interfaces, when
the minima of the free energy in the two phases is the same, this
term is related to the excess energy associated with the presence of
the interface \cite{bray,gunton,vsalt}: we can then interpret it as
the surface tension. Since this term is positive, the above expression 
confirms that the interface velocity is linear in the difference in
the free energy densities of the two phases it separates, and hence
linear in the temperature, as
sketched in the figure and anticipated already in Eq.~(\ref{dynamicaltesteq0}). Moreover,  it
propagates in such a direction that the domain with the lowest free
energy density expands. 

The function $F(\phi)$  corresponding to a second order transition is sketched in
Fig.~\ref{figdynamicaltest}{\em (b)}, and the prototypical dynamical equation for
this case is of course the F-KPP equation (\ref{fkpp})  with $f(u) = \varepsilon u -
u^3$ and $\varepsilon \sim T-T_{\rm c}$. For $\varepsilon>0$, i.e., below 
$T_{\rm c}$, we have indeed the possibility of fronts propagating into the
unstable state $u=0$ with speed $v^* = 2 \sqrt{\varepsilon}$.  As
was already stated above, for $\varepsilon <0$, (above $T_{\rm c}$),  any
nonzero initial condition for $u$ is driven rapidly to zero, and
hence no front propagation is possible. Thus, as illustrated in the
lower part of Fig.~\ref{figdynamicaltest}{\em (b)}, near a second order
transition, propagation of fronts is very asymmetric relative to
$T_{\rm c}$, on one side fronts or interfaces are possible, but their
growth velocity is a nontrivial power law of the distance from
criticality, $T-T_{\rm c}$, while on the other side interface motion does
not occur.

Normally, one does not study the motion of an interface in order to
determine the order of a phase transition, of course. However, these
considerations have been of use in at least one case, the study of the 
so-called Halperin-Lubensky-Ma \cite{halperin} effect near a nematic-smectic A
transition. In the nematic phase, the orientation of the long liquid crystal molecules
acquire  a directional order: on average they all point in the same
direction \cite{degennes}.  This directional order is the slow mode of a nematic phase 
that is responsible for much of the special behavior and applications
of these materials. In the smectic phase, the molecules also obtain a
layered ordering, and the smectic A phase is the one where the
molecules point on average along the normal to these layers. In a
classical Landau theory,  the nematic to smectic-A transition can be both of first
and of second order: if we think of $\phi$ in the discussion above as
the smectic layering order parameter, then  in a Landau theory, the
free energy density  $F(\phi)$ can have either of the forms sketched 
in Fig.~\ref{figdynamicaltest}. However, Halperin {\em et al.}
\cite{halperin} showed that when
the coupling to the director fluctuations is taken into account, and
when subsequently the director fluctuations are integrated
out,\footnote{The analysis of Halperin {\em et al.} \cite{halperin} also applies to
  type I superconductors: the couping to the gauge fluctuations can
  drive the normal to superconductor transition weakly first order in
  some regimes.} the 
Landau expression acquires a cubic term $|\phi|^3$ which renders the
transition weakly first order in the regime where without this term it 
would be of second order. The Halperin-Lubensky-Ma effect is thus an example of a 
fluctuation-driven first order transition.

\begin{figure}[t]
\begin{center}
{\tt (a)} \hspace*{-3mm} \epsfig{figure=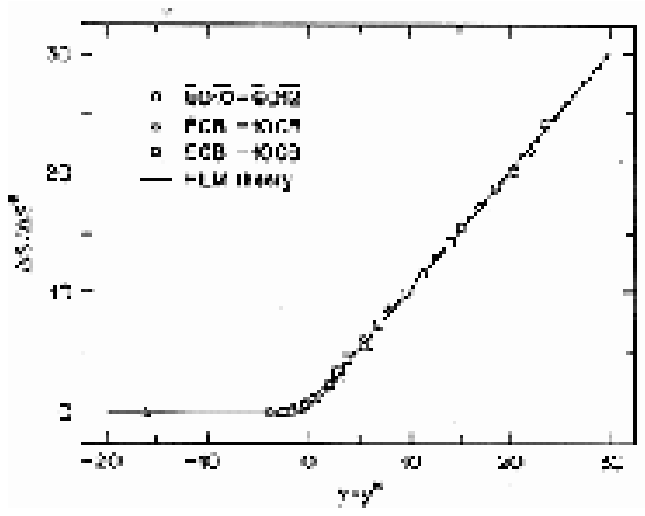,width=0.5\linewidth} 
{\tt (b)}
\hspace*{-3mm}
\epsfig{figure=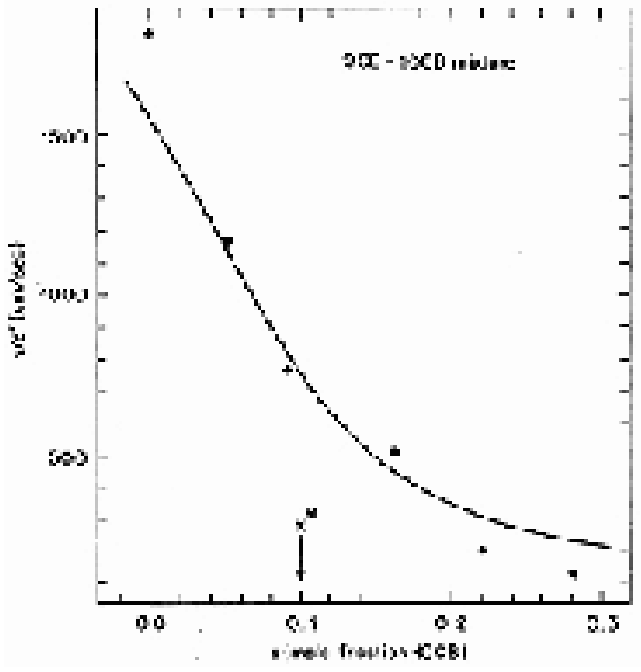,width=0.39\linewidth}
\end{center}
\caption[]{ (a) Normalized universal scaling form for the change of
entropy $\Delta S$ at the transition (which is proportional to the
latent heat) in a three series of liquid crystal
mixtures. The full line is a fit to a scaling form obtained from the
Landau free energy expression (\ref{dynamicaltestlandau}). The
variable along the horizontal axis is a scaled composition
variable. In the absence of the cubic term proportional to $b$ in
Eq.~(\ref{dynamicaltestlandau}), the curve would vanish linearly at
$y-y^*=0$, and would be zero to the left of it: $y^*$ marks the
tricritical point in the scaled parameters and in the absence of the
cubic term. From Anisimov {\em et al.} \cite{cladis3}.
(b)  The measured interface mobility $v /\varepsilon$ plotted as a
function of the concentration in the 9CB-10CB mixtures according to
the analysis by Anisimov {\em et al.} \cite{cladis3}. The full line
is obtained from the Landau expression obtained by fitting the
parameters to latent heat data like in (a) and to measurements of the
correlation length, apart from an overall factor that sets the scale.  }\label{figdynamicaltest2}
\end{figure}

The first dynamical indications for the existence of this fluctuation effect came from
a series of experiments near the nematic to smectic-A transition of
Cladis {\em et al.} \cite{cladis2}. When the temperature was changed 
through $T_{\rm c}$  for a series of liquid crystals mixtures which before had been
concluded to span a tricritical point where the transition changed
from first order to second order,
propagating interfaces were observed, both upon raising the
temperature and upon  lowering the temperature through $T_{\rm c}$. As
explained above, this might be taken as a sign that the transition was
in fact always at least weakly first order for all
compositions.\footnote{The tricky part experimentally is to rule out
the possibility that there were no temperature gradients in the
experiments which induced an effective interface --- after all, if the
temperature is above $T_{\rm c}$ on one side of the sample and below $T_{\rm c}$
on the other side, one will also create an interface if the transition
at constant temperature is second order. Care was taken in the
experiments \cite{cladis3,cladis2} to rule out such gradients, but the
strongest evidence for the fluctuation-induced weakly first order
transition actually comes from the consistency of the dynamical
measurements with the latent heat measurements and the $X$-ray
measurements of the correlation length \cite{cladis3}.} A more
careful analysis by Anisimov {\em et al.} \cite{cladis3} later
confirmed this and showed the consistency of these dynamical
measurements with measurements of the latent heat and the correlation
length near the transition \cite{anisimov2}. These authors took the  Landau free energy
expression 
\begin{equation}
{\mathcal F} (\phi) = \int d{\bf r}\, \left[ \half \xi_0^2 ({\vec
\nabla} \phi)^2 + a \phi^2 + b |\phi|^3 + c \phi^4 + d \phi^6\right] ,
\label{dynamicaltestlandau}
\end{equation}
with the cubic term motivated by the analysis of Halperin,  Lubensky
and Ma, which as mentioned above always renders the transition first
order. They then fitted the parameters $a$ and $c$ in this expression to both the
latent heat data and the $X$-ray measurements of the correlation
length of this series (for $b=0$, the point where $c=0$ marks the
tricritical point). The other parameters were fixed once for the
whole series of mixtures. The fit of the measurements of the cross-over
function for the latent heat (which is always nonzero for $b \neq 0$)
is shown in Fig.~\ref{figdynamicaltest2}{\em (a)}. Once these
coefficients are determined, one can the calculate the slope
$v/\varepsilon\sim v/|T-T_{\rm c}$ of
the interface response near the transition, apart from an overall
factor that sets the scale. As Fig.~\ref{figdynamicaltest2}{\em (b)}
shows, the curve obtained this way from static measurements fits the
experimental data for the interface mobility $v/ \varepsilon$
remarkably well for the 9CB-10CB mixtures. Taken as a whole, the
dynamical measurements together with the theoretical analysis
thus give quite strong evidence for the prediction by Halperin
{\em et al.} \cite{halperin} that the coupling to the director fluctuations drive the
nematic to smectic-A transition weakly first order. 

     \subsection{Switching fronts  in smectic $C^*$ liquid crystals}\label{sectionsmectic}
\begin{figure}[t]
\begin{center}
{\tt (a)} \hspace*{-2mm} \epsfig{figure=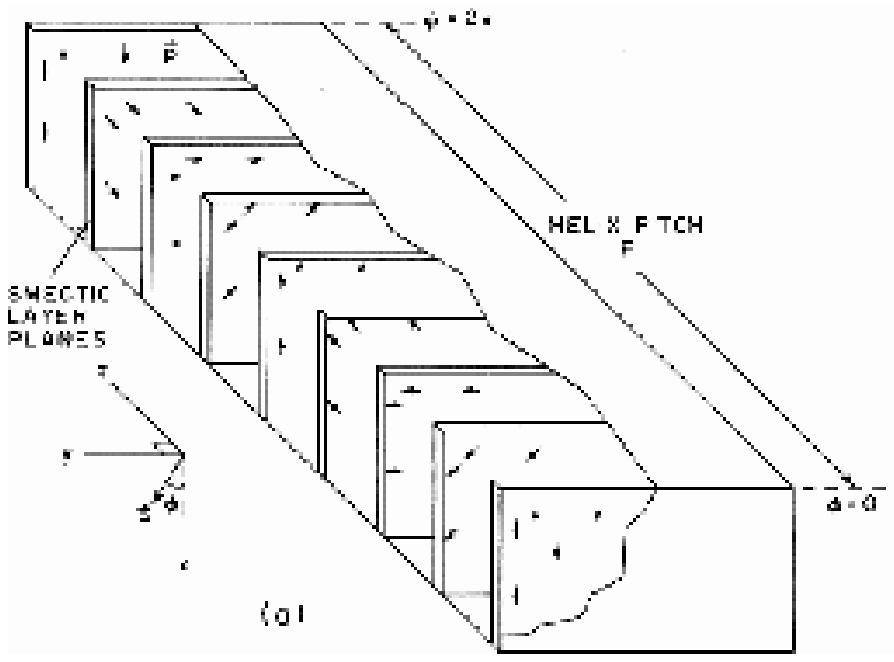,width=0.45\linewidth} 
{\tt (b)}
\hspace*{-1mm}
\epsfig{figure=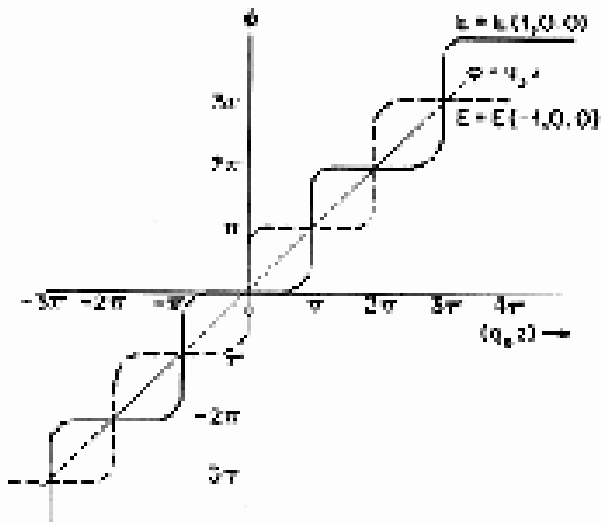,width=0.40\linewidth}
\end{center}
\caption[]{ (a) Sketch of a smectic-C$^*$ phase of a liquid
crystal. In this phase the molecules are layered, but their average
orientation  is tilted relative to the normal to the planes. The
local polarization $P_s$, which is normal to the $c$-director,   is indicated with arrows in
this figure. From layer to layer, this angle $\phi$ makes a small rotation,
so that over a large distance the angle is wound over $2\pi$.  From
Maclennan {\em et al.} \cite{clark}.  (b)
Schematic illustration of the stable angle solution $\phi(z)$ of a
smectic-C$^*$ in a strong electric field $E$ along pointing in the
positive $x$-direction (full line) and in the negative
$x$-direction (dashed line). Note that in the first case, there are
wide plateaus at $0,\pm 2\pi, \pm 4\pi, \cdots$, while in the latter
case these are at $\pm \pi, \pm 3\pi,\cdots$. Upon switching the field
direction, each domain wall where $\phi$ changes rapidly before the
switching splits into two fronts propagating into the two adjacent
plateaus, which have been made unstable due to the reversal of the
field.  From \cite{vanhecke}. }\label{figliquidcrystal1}
\end{figure}

In the previous section, we already encountered the smectic-A phase of
a liquid crystal. In this phase the molecules form layers. Along
 the layers the molecules are fluid-like (no
ordering), but the orientation of the molecules is  aligned on average along the normal of the layer.
In a smectic-C phase, the molecules again have a layered
ordering, but on average they are tilted in each layer with a fixed angle
relative to the normal. The projection of this tilt onto the layers
forms an azimuthal angle, which is an important slow hydrodynamic
variable for the smectic-C phases. In the smectic-C$^*$ phase, finally, this
angle rotates over a small angle from layer to layer, so that it makes
a full twist of $2\pi$ over a mesoscopic distance, the ``pitch'' of
the liquid crystal. This situation is sketched in
Fig.~\ref{figliquidcrystal1}{\em (a)}.  In this figure, the local
polarization (which is normal to
 the average orientation of the molecules) is
indicated with little arrows, and the angle this vector makes with the
$x$-axis is denoted by $\phi$; the $z$-axis is the coordinate  along the normal of
planes.

What makes the smectic-C$^*$ phase especially interesting from our
perspective is that the angle $\phi$ can be oriented both with an
electric field and with a magnetic field. Without a field, the twist
of the director is uniform, i.e. $\phi= 2\pi z/p$, where $p$  is the
pitch length, which depending on the material can range from a few
tenth of a micron to several microns. If we consider for simplicity
the case that the angle $\phi$ is uniform within each layers, then in
the presence of  electric
and magnetic fields perpendicular to the layers,  the free energy per
unit area can
be written as \cite{cladis1,degennes,clark} 
\begin{equation}
{\mathcal F} = \int \ds z\left[ \frac{ K}{2} \left( \frac{\partial
\phi}{\partial z} -
q_0\right)^2  -P E \cos \phi +  \left( \frac{\Delta \varepsilon E^2 }{8\pi} + 
\frac{\Delta \chi H^2}{2} \right) \cos^2 \phi \right] \label{liquidcrystalF}
\end{equation}
Here $P$ is the ferroelectric polarization, which points normal to the
direction in which the molecules are tilted. The second term describes
the
the dielectric and diamagnetic coupling: $\Delta \varepsilon$ is the
dielectric anisotropy and $\Delta \chi$ similarly the diamagnetic
anisotropy \cite{degennes}. Note that the electric and
magnetic field enter the same way. In the theoretical analysis below, we 
therefore simply put $H=0$ and consider  the two cases in which
$\Delta \varepsilon$ is positive or negative. Experimentally, however,
the fact that there are two different contributions to this term is
important. First of all,   it allows one to shift the  importance 
of this term relative to the polarization term. Secondly, if $\Delta
\varepsilon$ and $\Delta \chi$ have opposite signs  it  opens up
the possibility to change the {\em sign}  of this term. As we shall see
below, this has important implications for the front dynamics.

Suppose we start from a case without any fields, so that the 
smectic is in the uniform helical state $\phi= 2\pi z/p$,  and  switch on the
electric field to a large positive value. Then  the free energy density
is lowest for $\phi\approx 0, \pm 2\pi, \pm 4\pi$, etc. Because of the
twist in the initial state, the smectic will  form a series of
domains of length close to the pitch length $p$, separated by thin domain
walls where  $\phi$ rapidly changes by $2\pi$. This situation is
sketched in Fig.~\ref{figliquidcrystal1}{\em (b)} with the full
line. If the field is now rapidly reversed, $E\to -E$, then the free
energy density is lowest in the regions where $\phi= \pm \pi, \pm
3\pi $, etc. In the limit in which the fields are large so that the
domain walls are thin, the original walls will then split into two
fronts  which propagate into the domains which have been made high
energy domains by the field reversal. 

\begin{figure}[t]
\begin{center}
{\tt (a)} \hspace*{-2mm} \epsfig{figure=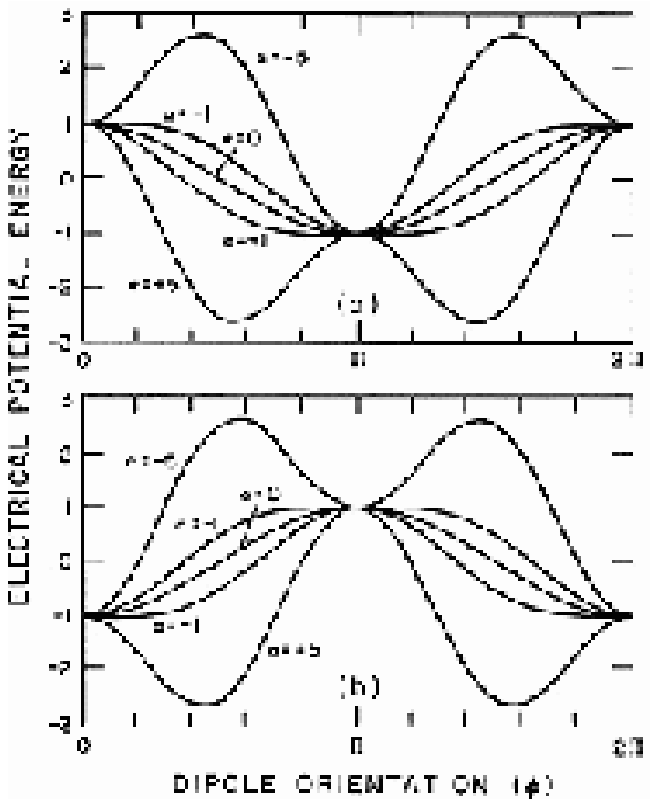,width=0.34\linewidth} 
{\tt (b)}
\hspace*{-1mm}
\epsfig{figure=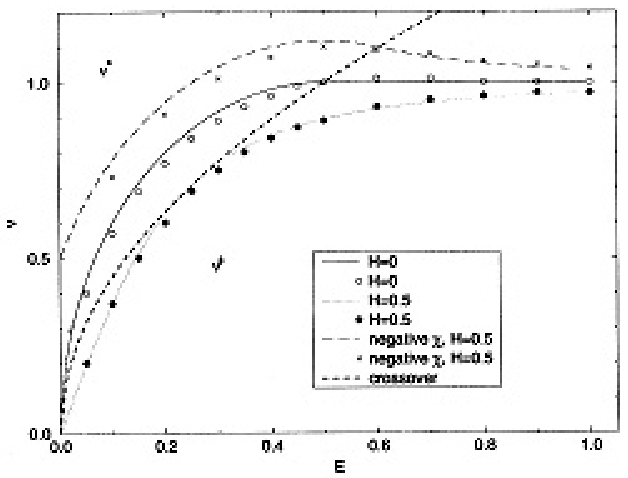,width=0.53\linewidth}
\end{center}
\caption[]{ (a) Electrical free energy density  (\ref{liquidcrystalF}) of 
the smectic-C$^*$ as for various values of 
$\alpha$. The upper panel corresponds to the case $E<0$ and the lower panel
to the same field positive field strength. From Maclennan {\em et al.}
\cite{clark}.    (b) Front velocities  for the 
generalization of Eq.~(\ref{liquidcrystaleq3}) to the case $H=0$ for three values of the field. The full
curve corresponds to the case $H=0$ discussed in the text. The crossover from pulled to pushed
fronts occurs at $E=1/2$ in this case. Symbols indicate the velocity obtained numerically from
simulations in runs of finite length. Note that the data points tend to lie slightly below the 
asymptotic velocities in the small field range, due to the slow power law convergence to the
pulled front speed $v^*$. From \cite{vanhecke}. }\label{figliquidcrystal2}
\end{figure}

In the absence of the dielectric and diamagnetic terms  the situation
is very simple: the switching of the fields makes the state in the
domains unstable, since the prefactor of the $\sin \phi$ term in the
free energy changes sign. Hence  in the high field limit we have a
clear case of pulled 
front propagation into unstable states. The general situation is more intricate as it depends
on the sign of the dielectric and diamagnetic term; let us define for
$H=0$ 
\begin{equation}
\alpha = \frac{\Delta \varepsilon| E| }{4\pi P } ~\hspace*{1cm}  (H=0).
\end{equation}
If  $\alpha$  is negative, then the dielectric field
contribution has extrema  for the same angles where the polarization term is extremal, while
when $\alpha$ is negative, it is minimal for $\phi= \pm \pi/2,
3\pi/2$, etc. The full behavior of the field contributions to the free
energy density as a function of $\phi$ are sketched in 
Fig.~\ref{figliquidcrystal2}{\em (a)}  for various values $\alpha$,
both for field $E <0 $ (upper panel) and for field $E>0$ (lower
panel).  Consider first  the case $\alpha <0$, and take initially $E>
0$. As the lower panel shows, the state with $\phi=0$ then has the
lowest free energy, so in the domains we have $\phi\approx 0$
(modulus $2\pi$). Now, upon switching the electric field, the free
energy becomes the one of the upper panel. As the curves indicate, the
state in the domains is then unstable for $-1<\alpha <0$ and
metastable for $\alpha < -1$. Thus, by increasing the field one can
continuously go from a case of front propagation into unstable states
to a case of a front propagating into a metastable state, in this
regime.

Clearly, the regime $\alpha >0$ is  different, as
Fig.~\ref{figliquidcrystal2}{\em (a)} illustrates. While upon
increasing the field $E$ for negative
values of $\alpha$ the free energy density sharpens at minima, upon
increasing the field for $\alpha >0$ the minimum initially flattens
and then turns into a local maximum. Moreover, upon switching the electric
field for $\alpha >1$ the state which was stable before switching experiences
a finite driving force to the new minimum. This implies that in this regime one will not observe
front propagation.

For smectic-C$^*$ liquid crystals, the moment of inertia of the molecules is normally
small  enough that  the inertial terms can be neglected in comparison with the viscous torque. 
In this regime, the
dynamical equation for the angle $\phi$ then becomes  $\eta \partial_t 
\phi=  - \delta {\mathcal F}/\delta \phi$, where $\eta$ is a twist
viscosity. In appropriate dimensionless variables this gives for $\alpha<0$ and $E>0$
\cite{cladis1,degennes,clark}
\begin{equation}
\partial_t \phi = \partial_z^2 \phi  + f(\phi) \hspace*{1cm} \mbox{with}~ f(\phi) =  E \sin \phi - E^2 \sin \phi
\cos \phi, \label{liquidcrystaleq3}
\end{equation}
which is nothing but the prototype equation for front propagation, the
F-KPP equation (\ref{fkpp}); the only difference with the usual case is that the function $f(\phi)$ is
now a periodic function of $\phi$, which allows for stable arrays of kinks\footnote{In the usual case,
the equation also admits solutions which correspond to a periodic array of kinks  where
$\phi$ switches between the stable states of $f(\phi)$, but these multiple kink arrays are
then unstable to pairing.}  like those sketched
in Fig.~\ref{figliquidcrystal1}{\em (b)}. 

For a  full study of the dynamics exhibited by this equation upon reversal
of the field direction we refer to the work by Maclennan, Clark and Handschy
\cite{clark}; we focus our discussion here on the front dynamics which is relevant in
the parameter range where the front width is much smaller than the
pitch $p$.

 The surprising feature in this case is that the selected front velocity can be obtained
analytically both in the pulled and in the pushed regime, because it turns out that it is possible
to  solve for the pushed
front solutions by making the
Ansatz $h=E \,\sin(\phi)$ in the ``reduction of order method'' discussed briefly in
the example at the end of  section \ref{selectionutfs}. Indeed, for the case $H=0$
this yields pushed solutions for $E>1/2 $  with velocity $v^\dagger = 1$ 
of (\ref{liquidcrystaleq3}),
whose analytic form is given by \cite{cladis1,stewart,vanhecke} 
\begin{equation}
\phi^\dagger (\xi) = 2 \mbox{arctan} [\exp (-\sqrt{| \alpha | } \xi)]  .
\end{equation}
Since $v^*= 2 \sqrt{E-E^2}$, one indeed immediately sees that $v^\dagger > v^*$ for $E>1/2$. 
As mentioned earlier, the results for $H\neq 0$ can all be obtained by appropriate transformations
that translates this case back to the case $H=0$ summarized above. Nevertheless, from a practical
point the possibility to  play with  both $E$ and $H$  may be  quite important, since when $H\neq 0$ the 
front solutions are pushed both for small and for large fields $E$. Fig.~\ref{figliquidcrystal2}{\em (b)} 
shows the selected front speed as a function of $E$ for three values of the magnetic field $H$. 

As is well known, liquid crystals are important for displays; if one would want to use the present
switching effect in applications, the switching time is of course an important parameter, and this
is inversely proportional to the front speed. For a discussion of the transient behavior as well as of
the comparison with experiments, we refer to the review by Maclennan {\em et al.} \cite{clark}.

\subsection{Transient patterns in structural phase transitions in solids}\label{secsalje}
One possible way in which a structural phase transition in a solid can occur is when certain atoms
or ions which have  two different competing sub-lattices  available,
order below some critical temperature.
Often, such ordering phenomena in solids are strongly coupled to strain deformations and lower the
symmetry of the crystal structure. About a decade ago, it was 
 conjectured that certain
transient metastable tweed patterns that are sometimes found near structural phase transitions
 might be due to the propagation of a pattern forming
front into an unstable state \cite{salje}. 

For the transient patterns  of interest
--- e.g. tweed patterns associated with vacancy ordering in YBCO
superconductors or Al,Si ordering in Na   
feldspar --- the equilibrium ordered state is to have have a
homogeneous phase in which the ions  are (partially) ordered on one of
two available sub-lattices. In practice, however, domains with
alternating order are sometimes observed upon shock heating. Salje
\cite{salje} has
suggested that these pattern arise from a pattern forming front
propagating into the structurally unstable homogeneous disordered
phase. In particular, if $\phi$ denotes the dimensionless kinetic
order parameter with $\phi=\pm 1$ indicating the two possible
sub-lattice ordered states,  and $\phi =0$ the disordered state in a
coarse-grained description, Salje \cite{salje,salje2} arrived, on the basis of
a treatment of the kinetics which approximately includes nonlocal
strain effects, at the following dynamical equation for the order
parameter:
\begin{equation}
\partial_t \phi = \partial^2_x \phi -\gamma \partial^4_x \phi + \delta_1
(\partial_x \phi)^2 + \delta_2 \phi^2 \partial^2_x \phi. \label{saljeeq}
\end{equation}
Note that this equation can be viewed as an extension of the EFK
equation discussed in section \ref{sectionefk}. Like that one, the stable lowest
energy states are clearly the homogeneously ordered states $\phi=\pm
1$. Moreover, since the terms linear in $\phi$ are exactly those of the
EFK equation, it follows from the results of section \ref{sectionefk} that indeed
for $\gamma > 1/12$ a pulled front propagating into the unstable
disordered $\phi=0$ state will give rise to transient
patterns. Simulations in \cite{salje} also confirmed this.\footnote{Note that the term proportional 
to $\delta$ in (\ref{saljeeq}) tends to enhance the growth rate, and that it is quadratic
in the amplitude. We therefore would expect a transition to pushed fronts for sufficiently
large $\delta$; to our knowledge, this possibility has not been explored, however.}

To our knowledge, there is at present no direct evidence in support of
the intriguing  conjecture that the transient modulated order patterns are
 caused by pattern forming  pulled fronts --- the difficulty with solid
ordering phenomena is that   it is often hard to rule out other possible mechanisms. 

\subsection{Spreading of the Mullins-Sekerka instability along a growing interface and the
origin of side-branching} \label{sectionmullinssekerka}

When the growth of an
interface is limited mainly by how fast the material necessary for
growth can be transported towards it
from the phase into which it grows, or by how fast heat produced at
the interface can be transported away through diffusion into the phase
into which it grows, then a growing planar  interface of this type is unstable: 
A small protrusion of the interface towards
the phase into which it grows leads to an enhancement of the gradients
in front of the interface. This enhanced gradient leads to an
enhancement of the diffusion and hence to an enhanced growth: the
protrusion will grow larger and larger, and render the interface
unstable. This so-called Mullins-Sekerka instability
\cite{caroli,kassner,langerrmp,mullins1,mullins2}, which we essentially
already encountered in section \ref{chemicalbacterial},
underlies an enormous variety of diffusion-limited growth processes,
ranging from crystal growth from the melt or  electric
discharge patterns like the streamers of section
\ref{sectionstreamers}, to fractal growth phenomena like Diffusion Limited Aggregation
\cite{barabasi,meakinbook}.

\begin{figure}[t]
\begin{center}
\epsfig{figure=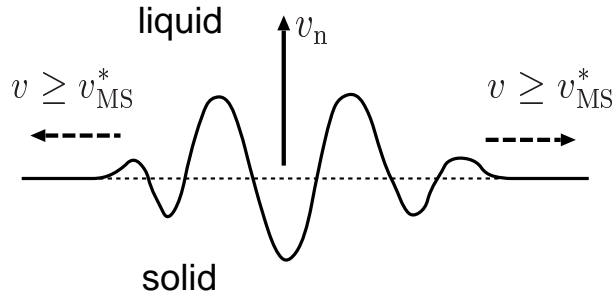,width=0.58\linewidth} \hspace*{0.3cm}
\end{center}
\caption[]{Schematic illustration of the spreading of a localized
perturbation along  a growing crystal melt interface which is unstable
as a result
of the Mullins-Sekerka instability. The dashed line indicates the
unperturbed planar interface, which is  growing in the
vertical direction with a normal velocity $v_n$. The dashed arrows
indicate the propagation  of the perturbation along the
interface. In accord with the discussion of section \ref{sectiontwofold}, the asymptotic
Mullins-Sekerka front speed is larger than or equal to the linear
spreading velocity $v^*_{\rm MS} $ given in
Eq.~(\ref{mullinssekerkav*}). }\label{figdendrite1}
\end{figure}

We focus our discussion on the Mullins-Sekerka instability of a planar crystal interface
growing into an undercooled melt. In this case, the dispersion relation of small
perturbation in the height $h$ of the interface  of the form $h\sim
e^{-i\omega t+ ikx}$ is \cite{caroli,langerrmp,mullins1}
\begin{equation}
\omega = i v_{\rm n} |k| (1- d_0 \ell_{\rm D} k^2). \label{mullinssekerkadispersion}
\end{equation}
Here $v_{\rm n} $ is the normal velocity of the planar interface,   as sketched in
Fig.~\ref{figdendrite1}, $d_0$ is the capillary length which is proportional to
the surface tension of the interface, and $\ell_{\rm D}$ is the
diffusion length on the liquid side of the interface.

The linear spreading velocity associated with this dispersion relation 
is easy to determine; one simply finds
\begin{equation}
v^*_{\rm MS} = \sqrt{3} v_{\rm n}, \label{mullinssekerkav*}
\end{equation}
with
\begin{equation}
\lambda^* \equiv  k_{\rm i}^* =  \frac{1}{\sqrt{6 d_0\ell_{\rm D}}}  ,
\hspace*{1cm} k_{\rm r} ^*= \frac{1}{\sqrt{2 d_0 \ell_{\rm D}} } ,
 \label{ratioofks}
 \end{equation}
where, as before, $k^*_{\rm r}= \mbox{Re}\,k^*$ and $k^*_{\rm i}
=\mbox{Im}\, k^*$.
To our knowledge, the spreading of the Mullins-Sekerka instability
along the interface has not been studied directly, neither
theoretically nor experimentally.\footnote{The experimental  difficulty 
  is that it is very hard to
prepare an unstable interface in this case; if one starts with a
stable interface and then tries to bring the interface to the unstable 
regime, the buildup of the diffusion boundary layer usually gives rise 
to  long transients. Often, the instability already arises during this
transient regime. }  We therefore do not know whether the
propagation 
of the Mullins-Sekerka along the interface corresponds to a pushed or pulled front, 
although since we know that dendritic growth is such a strong
instability without saturation,  we might  intuitively expect it to be 
pushed in the case  of a flat interface illustrated in
Fig.~\ref{figdendrite1}. Thus, in line with the arguments of section
\ref{sectiontwofold}, we simply conclude that the front will propagate sideways with
asymptotic velocity $v\ge v^*_{\rm MS}$. Note that if the front would
happen to be 
pulled, the relation (\ref{ratioofks}) shows that the amplitude of the
oscillations grows exponentially by a factor $e^{r^*}$ over one wavelength $2\pi/k^*_r$, with
\begin{equation}
 r^* \equiv \frac{2\pi k_{\rm i}^*}{k_{\rm r}^* } =  \frac{2\pi}{\sqrt{3}} \approx
3.62.
\end{equation}
 
\begin{figure}[t]
\begin{center}
{\tt (a)} \hspace*{-3mm} \epsfig{figure=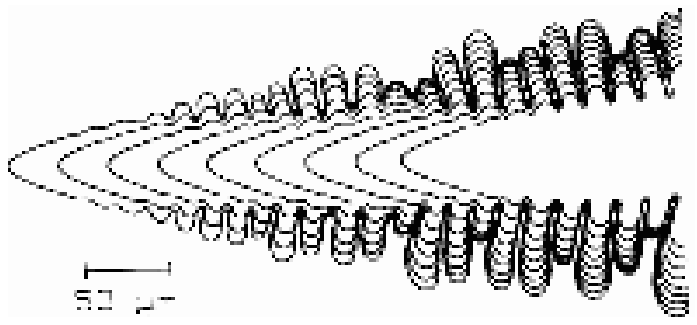,width=0.45\linewidth} 
{\tt (b)} \hspace*{-3mm} \epsfig{figure=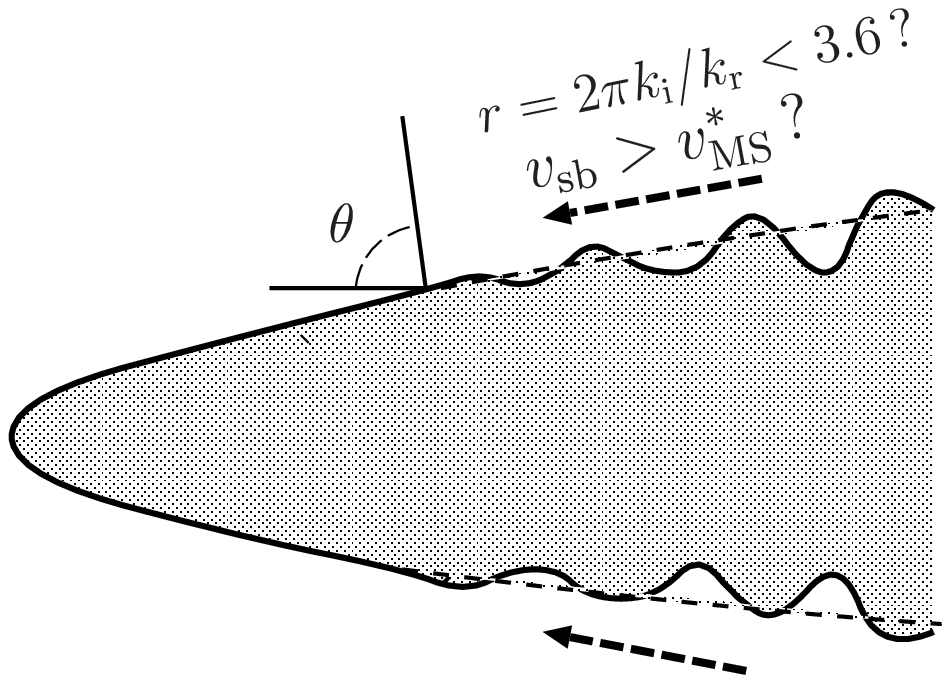,width=0.45\linewidth}
\end{center}
\caption[]{ (a) Snapshot of a $NH_4Br$ dendrite in the experiments by
Dougherty {\em et al.} \cite{dougherty1,dougherty2}. (b) Sketch of a
dendrite tip identifying the various  quantities discussed in the
text. }\label{figdendrite2}
\end{figure}

As was also discussed in section \ref{sectionleadingedgedominated}, even if a front is pulled for
sufficiently localized initial conditions, it can move faster than
$v^*$ if the initial conditions are falling off slower than
$e^{-\lambda^* x}$ in space (this case was referred to as ``leading
edge dominated dynamics'' in section \ref{sectionleadingedgedominated} and in \cite{evs2}). For
pattern forming fronts like these, it was argued in \cite{vs2} that
for a given spatial decay rate $\lambda = k_{\rm i} < \lambda^*$,
the wavenumber would be the one maximizing the growth rate
$\mbox{Im}\, \omega$, i.e., that $k_{\rm r} $ would be determined
implicitly by the condition $\partial \omega_{\rm i} (k) / \partial
k_{\rm r}$=0. In the present case this gives 
\begin{equation}
k_{\rm r}^2 = k_{\rm i}^2 + \frac{ 1 } {3 d_0 \ell_{\rm D}},
\end{equation}
and for the velocity $v$ of the leading edge of such profiles this gives
\begin{eqnarray}
v  & \equiv &  \frac{\omega_{\rm i}}{k_{\rm i}} =  v_{\rm n} \frac{k_r}{k_i}
\frac{2 d_0 \ell_D}{r}  =   v_{\rm n} \, \frac{2 \pi }{3 r
  \left(1-r^2/(4\pi^2)\right)} \label{vofsidebranches} \\  & > &  v^*_{\rm MS}= v_{\rm n} \sqrt{3} \hspace{1.2cm} (\mbox{for}~r\equiv 2\pi k_{\rm i}/k_{\rm r} <
r^*= 2\pi/ \sqrt{3}) \nonumber
\end{eqnarray}

Let us now explore a simple  implication  of this idea for
dendrites.  Dendrites are the tree-like growth structures of the type
sketched in Fig.~\ref{figdendrite2}{\em (a)} which the Mullins-Sekerka
instability gives rise to. Understanding the shape and velocity
selection of a dendritic tip was one of the outstanding problems in
the field of interfacial pattern selection in the 1980-ies. As
discussed in various reviews
\cite{benjacob1,benjacob2,caroli,kessler3,langergodreche,pelce,pomeau}, it is now
accepted by most workers  that the size and 
velocity of the near-parabolic tip  are determined by the nonlinear
eigenvalue problem for a uniformly growing near-parabolic tip, and that
although surface tension effects are typically small, they act as a
singular perturbation. The picture that has emerged is thus that the
occurrence of ``sidebranches'' on these dendrite tips is not important
for the dynamical mechanism that ``selects'' the tip shape and
velocity.

The accepted view in the field is that  the sidebranches occur due to
selective amplification of noise from the tip: small perturbations and
fluctuations occur at the tip of the dendrite; these grow out, while
being convected away in a frame moving with the tip itself. These
ideas have been put forward on the basis of WKB-type analysis of the
spreading, growth and advection of side-branch ``wave packets''  which start
near the tip \cite{barber,pieters,pieters2}. Though there is some  evidence for this
behavior \cite{cummins}, experiments on this issue have turned out to
be hard in general; moreover, it is not clear how realistic a WKB-type
calculation is for the experimentally relevant regime where tip radius and
sidebranch spacing are comparable.  However, our
analysis above gives us a very simple different way to verify this picture from the data, and in 
addition  suggests a useful way to analyze experimental data on sidebranches. The
discussion is a slight reformulation of  \cite{scc}.

The above picture that sidebranches arise from the amplification of noise near the tip
is based on the idea that the flanks of the smooth needle solution underlying the dendritic tip
region are unstable to the Mullins-Sekerka instability, but that the instability is 
convective there. Although the sidebranches do not form  a coherent front due to the fluctuations,
 their envelope must propagate on average with a velocity whose projection $v_{{\rm sb}||}$
along the growth direction
is equal to the tip velocity $v_{\rm tip}$.  For a region on the side illustrated in 
Fig.~\ref{figdendrite2}{\em (b)} where 
the underlying tip profile  makes an angle $\theta$ with the growth direction, this velocity  can
be expressed in terms of the  velocity $v_{\rm sb}$ of the sidebranch front along the interface and the
normal velocity:
\begin{equation}
v_{\rm tip} = v_{{\rm sb}||} =  v_{\rm sb} \sin \theta + v_{\rm n}
\cos \theta\nonumber =
  v_{\rm sb} \sin \theta + v_{\rm tip}  \cos^2 \theta ,
\end{equation} 
where we used the fact that $v_{\rm n} = v_{\rm tip} \cos \theta$.  
Equating the two sides gives simply 
\begin{equation}
v_{\rm sb} = v_{\rm tip} \sin \theta .
\end{equation}
 This equation immediately
allows us to compare measured
quantities  and infer the underlying dynamics from it. Indeed,
if we take the interface on the sides as locally planar by ignoring
the curvature of the underlying needle solution in this  
region and assume that the average wavenumber of the incoherent
sidebranches can be associated with $k_{\rm r}$ above,    $v_{\rm sb}$
is nothing but the velocity given in (\ref{vofsidebranches}). If
 $v_{\rm sb}$ is indeed significantlybigger than $v^*_{\rm MS}$, the
sidebranch instability is indeed convective, and the selective
amplification of noise scenario is corroborated by  this analysis
too. Since $v^* = \sqrt{3} v_{\rm n}  = \sqrt{3}  v_{\rm tip} 
\cos\theta$, in our lowest order approximation  we may conclude:
\begin{equation} 
  \begin{array}{c} \mbox{sidebranch instability convective}\\
    \mbox{sidebranches = amplified tip noise} \end{array}
  \Longrightarrow \left\{ \begin{array}{c}  \theta > 
      60^o\\   r< 3.62 \\
     v_{\rm sb}   > v^*  ~[v_{\rm sb} \mbox{obeys}~ (\ref{vofsidebranches})]\\
\end{array}  \right.
\end{equation}

\begin{figure}[t]
\begin{center}
\epsfig{figure=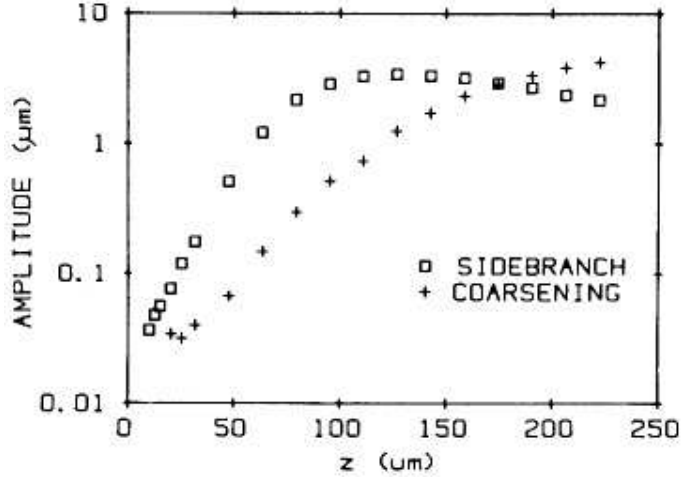,width=0.65\linewidth} 
\end{center}
\caption[]{Measured root mean square amplitude (squares) of the sidebranches of the  dendrite shown in 
Fig.~\ref{figdendrite2}{\em (a)} as a function
of the distance $z$ from the tip, in the experiments by 
Dougherty, Kaplan and Gollub \cite{dougherty1}. The crosses give a
measure of the amount of coarsening of the sidebranches; they are based on the 
amount of spectral power at low frequencies \cite{dougherty1}.}\label{figdendrite3}
\end{figure}

To illustrate this, let us apply the above line of argument to the
dendrite shown in Fig.~\ref{figdendrite2}{\em (a)}. First of all, the
angle on the flanks of the tip region where the sidebranch amplitude
is small, is bigger than $60^o$. Furthermore, the amplitude of
the sidebranch amplitude in this experiment was measured explicitly
by Dougherty, Kaplan and Gollub \cite{dougherty1,dougherty2}. As
Fig.~\ref{figdendrite3}
shows, this amplitude initially does rise exponentially; the 
spatial exponential growth rate $k_{\rm i}$ is easily estimated to be about $ 1/(16
\mu\mbox{m})$ from these data. This is an indication
that indeed in this case, nonlinearities lead to saturation of the sidebranch instability, and hence
that if there were no fluctuations, the sidebranch front would
correspond to a pulled front.
Since the wavelength is about 13 $\mu$m in this 
experiment, we find $r\approx 0.8 < r^*$. In addition, the sidebranch velocity
$v_{\rm sb}$ is also found to obey (\ref{vofsidebranches}) to a good
approximation, which indicates that the consistency of this line of argument.  Hence, both the angle and
the spatial growth rate confirm that the sidebranches in this
experiment are consistent with the scenario \cite{barber,pieters} that they emerge from the
amplification and convection of small fluctuations near the tip. 

Many dendrites appear very much like the ones shown in
Fig.~\ref{figdendrite2}{\em (a)}  at first sight. It  therefore does appear that  in
most cases the sidebranches are consistent with the above amplified
tip-noise scenario.  Unfortunately,  the dimensionless growth
rate $r$ is hardly ever mentioned in  experiments, but judging by the 
eye this quantity does appear to vary significantly from experiment to
experiment. 

An interesting question remains whether in some cases, the sidebranch
instability might become absolute, in that the projection of the  linear
spreading velocity $v^*_{\rm MS}$ becomes comparable to the tip
velocity (such a scenario was also envisioned in \cite{langermk,martin}).
One would expect that tip and sidebranches then become
strongly coupled, and possibly that the tip region would show
appreciable oscillations. That might happen in some parameter range
was already
suggested by
simulations of the so-called Boundary Layer Model by Pieters
\cite{pieters}. Recently, Sakaguchi and Tokunaga \cite{sakaguchi} observed such
behavior in phase field model calculations but the data were not
correlated with the parameter $r$ introduced above. A good way to classify sidebranch regimes
experimentally and to search for this possibility
 is to 
measure the dimensionless growth factor $r$ --- if this value is found 
to increase towards  $r^*\approx 3.6$ then the possibility of such a
regime becomes more likely. Of course, for a realistic comparison with experiments,
effects of interface kinetics and crystalline anisotropy may have to be included, but at
the level of the approximations discussed here this poses no technical
problems.

The above considerations concerning the absolute or convective nature
of the sidebranch instability in my view help us understand how
two competing theories might actually emerge as different limits
within a more general framework. The usual ``solvability
theory'' which focusses on the existence of featureless needle solutions, amounts,
within a WKB approximation, to the requirement that a smooth tip
solution without sidebranch-type modes on the flanks is stable
\cite{davidandboris}. In his ``Interfacial Wave Theory'', on the other
hand, Xu \cite{xu} develops an approach in which he allows,
essentially within the same WKB approximation, a tip mode to match to
divergent sidebranch modes on the sides.\footnote{The turning point in
Xu's WKB analysis becomes, in some limit, the linear spreading point
of our analysis, and in principle 
the ``interfacial wave theory''   predicts the value of the parameter $r$
introduced above. Also note that the differences and similarities between
Xu's approach and the usual solvability theory are brought out most
clearly if one thinks of solvability theory in the spirit of the
formulation of \cite{davidandboris}.} To my knowledge, the stability
of these solutions, constructed this way, has not been fully
investigated. Based on the above analysis, I would expect such
dendrite-type solutions to be actually unstable in substantial parts
of the parameter range --- if so, they would presumably, in the
parameter range where this happens,  give way to needle-type solutions
which are convectively stable in a deterministic approach. In other
words, in the range where this happens the usual solvability scenario
would hold, both for the selection mechanism and for the amplification
of noise. In other parts of parameter space (roughly identified by
generalizations of the ideas presented in this section), needle
solutions would be absolutely unstable to sidebranch modes. I would
expect Xu's picture to emerge naturally in this regime. 

Unfortunately, the dendrite problem is a hard one, and at present we
have to consider the above scenario as speculative. In addition, the
reader should keep in mind that the above view on the dendrite problem
is a minority viewpoint --- most workers consider the various theories
as mutually exclusive, rather than as two extreme limits of a more
general framework.

    \subsection{Combustion fronts and   fronts in periodic or turbulent media} \label{sectioncombustion}
    
    The classic work of Landau \cite{landau}, who analyzed the stability of flame
    fronts in terms of a flame sheet of infinitesimal thickness,
    separating unburned and burned gases, provides a very early
    example of the clever use of a moving boundary approximation for a
    problem which more fundamentally is formulated in terms of
    continuity equations for the temperature and composition of a
    combustible mixture.  The justification for this approximation was
    later shown to be that the activation energy of the relevant
    reactions is normally very high, so that the reaction rate is a very
    steep function of the temperature
    \cite{buckmaster,clavin,joulin,williams,combustion}. Inherently,
    realistic flame 
    fronts are therefore in practice almost always examples of {\em
      pushed} fronts: the reaction rate is very strongly (essentially
    exponentially) suppressed ahead of the flame sheet. In other
    words, the linear spreading speed is essentially zero.
    
    Although the minimal model for flames amounts to two coupled
    partial differential equations for the temperature and
    composition, simple model equations have always played an
    important role in combustion too.  We mention two interesting
    examples coming from the combustion literature.  In fact, an early
    examples of the F-KPP equation (\ref{fkpp}) with a nonlinearity $f(u)$
    which vanishes identically in a finite interval (e.g. $f(u)=0 $
    for $u < u_c$, $f(u) = (u -u_c)(1-u)$ for $u
    \ge u_c$ is due to Gel'fand \cite{gelfand}. This form was motivated by the
    above observation that the reaction rate drops so fast with
    decreasing temperature that it is virtually zero below some
    critical value.

Two other interesting classes of problems have  also emerged from
combustion theory: propagation of fronts in periodic media or turbulent media.  We first discuss
the case of fronts in periodic media. The
simple model problem which has often been studied in this  context
is the following extension of the F-KPP equation \cite{berestycki1,berestycki2}
\begin{equation}
\partial_t u = \partial^2_x u+ q(x) \partial_x u + f(u,
x) , \label{periodic}
\end{equation}
and generalization to higher dimensions.  In this equation, the
advection rate $q(x)$ and nonlinear growth rate $f$ are periodic
functions of $x$ with period $L$: $q(x+L)=q(x)$, $f(u,
x+L)=f(u,x)$.   

Two classes of growth functions $f$ have been studied, those analogous 
to the one by Gel'fand mentioned above, and those which give rise to
{\em pulled} fronts in the F-KPP equation
[e.g. $f(u)=a(x)u (1-u)$]. The latter type of problem of a
pulled front in a periodic medium is
especially interesting from our perspective: It should be possible to
obtain  the asymptotic propagation speed $v^*$ of such pulled fronts
by extending  our general approach using Floquet-Bloch theory for the
linearized equation. To be specific, consider the linearized version
of (\ref{periodic})  with $f(u,x)$ given as above,
\begin{equation}
\partial_t u = \partial^2_x u + q(x) \partial_x u + a(x) u,
\label{periodic2}
\end{equation}
where $q(x)$ and $a(x)$ are $L$-periodic,
\begin{equation}
 q(x+L)=q(x), \hspace*{0.5cm} a(x+L)=a(x) . \label{periodic3}
\end{equation}
According to Bloch's theorem in the language of a physicist or 
Floquet theory in the more general setting, the generalization of the Fourier
transform become the ``Bloch waves''
\begin{equation}
\tilde{u}(k,t) = e^{-i\omega(k) t +ikx } U(x), 
\end{equation}
where $U(x)$ is a periodic function of $x$, $U(x+L)=U(x)$. Just as we
did in section \ref{sectionv*} for 
the Fourier transform, once the dispersion relation $\omega(k)$ is 
obtained, one can analytically continue $k$ into the complex
plane. This immediately leads to the conclusion that the linear 
spreading velocity $v^*$ is again given by the same saddle-point
equations (\ref{saddlepoint}) --- the periodicity of the medium is simply encoded in
the dispersion equation $\omega(k)$. For solid state physicists this
is no surprise: $\omega(k)$ is the analog of the band energy
$\varepsilon(k)$ of Bloch electrons, and as they know, the behavior of 
free electrons in a solid is completely determined by the band
structure.

To the best of my knowledge, the above line of analysis has unfortunately not been
tried yet; we do hope such an approach will be explored in the near
future, as it would probably be the most direct route to obtaining the 
asymptotic speed of pulled fronts and as interesting new phenomena
might arise from the band structure. In particular, for the simple case
$a(x)=1+ a_0\, cos(2\pi x/L)$ the linear equation reduces to the Mathieu
equation, for which it should be possible to obtain a number of
(semi-)analytic results.

Another class of front problems which has emerged from combustion theory is the propagation
of a front in  a spatially and/or temporally random medium  as a model for
turbulent combustion \cite{kerstein3,kupervasser,peters,xin2}. From an
applied point of view, an important question 
is to understand the enhancement of the combustion rate due to turbulent advection. One way
in which this issue has been approached recently within the context of model problems is to to
take the function $f$ in Eq.~(\ref{periodic}) $x$-independent but the 
advection field $q$ in this equation a space- and time-dependent random
variable. This makes the field $u$ into a
stochastic variable as well, and recently several workers  
\cite{abel1,abel2,eyink,freidlin2,freidlin3,kerstein3,tretyakov} have derived 
results for various probability distributions of the advected $u$
variable. One reason that progress can be made on this complicated
problem is that for a pulled front, the linear $u$-equation
captures the essential elements of front propagation (more
mathematically: bounds can be derived for the nonlinear equation,
using the linearized advected dynamics).  It was recently also found
that if the $q$-variable is  a stochastic Levy process, the front
propagation can change drastically\cite{vulpiani}  --- even an exponentially
increasing speed is possible.\footnote{We speculate that this may be
related to the fact that an F-KPP equation with power law initial
conditions can give rise to an infinite speed \cite{needham3}, and that the
probability distribution function of a  Levy
process has  power law tails.} 

There is one important issue concerning fronts in turbulent or random media that to our
knowledge has not been discussed explicitly in the literature. Real combustion fronts are 
pushed fronts, because the combustion rate decreases very rapidly with decreasing temperature.
In the limit of large activation energies, flame fronts are  very thin and they can be analyzed
with a moving boundary approximation \cite{peters,williams}
 --- this amounts to the approximation in which the
flame sheet is treated as an interface of zero thickness. As we discuss in section \ref{sectionmba}, 
for pushed fronts this approximation is indeed justified, but for a pulled fronts it is {\em not}: For
these, the dynamically important region is the semi-infinite domain ahead of the nonlinear front
region.  It is therefore conceivable that simple models like (\ref{periodic}), 
whose dynamically important fronts
are pulled, are not necessarily good models for turbulent combustion. Probability distribution
functions in simple combustion models  might well be very different depending on whether fronts
are pushed or pulled.

  \subsection{Biological invasion problems and time delay equations}\label{sectionbiologicalinvasion}
As mentioned in the introduction, the first studies of front
propagation into an  unstable state were done in the context of
population dynamics and biological invasion problems \cite{fisher,kpp}. Not
surprisingly, this has therefore remained an active field of research
within mathematical biology. The main focus of a  large fraction of
the literature is still
 on proving existence, stability and uniqueness of traveling wave
solutions in population dynamics models which can be considered as
extensions of the F-KPP equation. We refer to chapter 14 of the
books by Britton \cite{britton} and Murray \cite{murray} or to the recent review of Metz {\em et
al.} \cite{metz} for an introduction and overview  of this
subfield. The book by Shigesada and Kawasaki \cite{shigesada} 
not only gives a good review of the theory, but also discusses in
detail a number of applications of the theory to practical invasion
problems. Pulled front propagation into unstable states plays a dominant role in a
recent model of infection in the Hantavirus epidemics \cite{abramson}.

Also in the bio-mathematical literature  the slow power law convergence of pulled fronts
to their asymptotic speed has been discussed \cite{metz},
but I am not aware of examples in that field where this
phenomenon has played a significant role. In real life, the fact that
the entities in invasion problems are normally discrete may play a
more important role --- as we shall discuss in section \ref{finiteparticles}, the cutoff in 
the growth function that the discrete nature of the spreading
population gives rise to, alters a pulled front in a continuum
equation to a (weakly) pushed front in a model for discrete
variables. 

A recent development in this field, reviewed recently by Fort and Mend\'ez \cite{fort2}
 has been to consider the effects of time-delay on  a
front\cite{fedotov1,fort,gallay1,gallay2}. The motivation for such
terms 
in the case of the spreading of viruses into an uninfected population
is that the time 
which elapses between the moment a cell gets infected and the moment a cell dies and
the virus begins to spread is an inherent part of the dynamics which
can not be neglected.  Within the context
of an F-KPP type equation, these effects would seem to lead naturally to equations with
a memory kernel, like\footnote{In cases in which the growth term is
most naturally modeled by delay kernel, it would of course make more
sense to also take a delay-type term for the nonlinear term describing
saturation \cite{metz0,metz}. Of course, as long as the fronts remained pulled, this
does not affect our conclusions.}
\begin{equation}
\partial_t u =  \partial^2_x u  + \int^t \ds t^\prime \, K(t-t^\prime)\,
u (t^\prime) -  u^n
\end{equation}
or equations with a finite delay time $\tau$ which are obtained when
$K(t-t^\prime)=\delta(t-t^\prime- \tau)$. However,  in practice most work has
concentrated on second order equations of the type \cite{fort2,gallay1,gallay2}
\begin{equation}
\tau \partial_t^2 u  + \partial_t u = \partial_x^2 u + u -u^n \label{secondordereq}
\end{equation}
which after appropriate rescalings are obtained by assuming
that the time delay is sufficiently short  that the memory
kernel $K$ can be expanded for short times as
\begin{eqnarray}
\int^t  \ds t^\prime K(t-t^\prime)\, u (t^\prime) & = &  \tau_0 u(t) + \tau_1
\partial_t u (t) + \tau_2 \partial^2_t u(t) \cdots, \label{truetimedelayeq}\\
 \tau_m & = & \frac{1}{m!} \int_0^\infty \ds \tau\, \tau^m
\, K(\tau ) . \label{truetimedelayeq2}
\end{eqnarray}
Of course,  the transition from a
parabolic (first order in time)  partial differential equation to a
hyperbolic partial differential equation like (\ref{secondordereq}) poses new mathematical
challenges  concerning existence, uniqueness and convergence
\cite{fort2,gallay1,gallay2}.  However, from our 
pragmatic more ``applied'' point of view, we want  to stress the
following. In practice 
most of the cases  that are considered in the literature on delay
effects on front propagation into an unstable state, are pulled. Furthermore,
 the front velocity is the most important property one needs to
know in practice. In view of this, it is important to realize
 there is absolutely no reason to approximate a true
time delay equation like (\ref{secondordereq}) by a hyperbolic
equation by expanding the delay kernel as in
(\ref{truetimedelayeq}).  After all,  as explained in sections \ref{sectionmoregeneral}, the
spreading velocity $v^*$ and the associated parameters $\lambda^*$ and
$D$ which govern the shape and convergence of a pulled front,
can straightforwardly be determined from the more general class of
models as well.  An example was given at the end of section \ref{sectionmoregeneral}.

\subsection{Wound healing as a front propagation problem} \label{sectionwounds}

\begin{figure}[t]
\begin{center}
\epsfig{figure=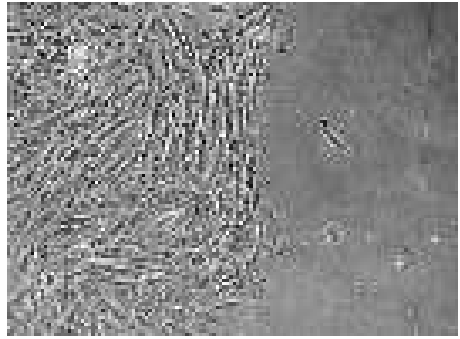,angle=-90,width=0.43\linewidth} 
\end{center}
\caption[]{ Photograph of a typical human cell front 10 hours after wounding. The cells are
mesothelial cells from the peritoneum, the cells that form the superficial layer of the membrane that
covers the abdominal cavity and its organs. From Maini {\em et al.}
\cite{maini,maini2},
courtesy of S. McElwain and
B. MacGillavray.
}\label{figwound1}
\end{figure}

\begin{figure}[t]
\begin{center}
{\tt (a)}
\epsfig{figure=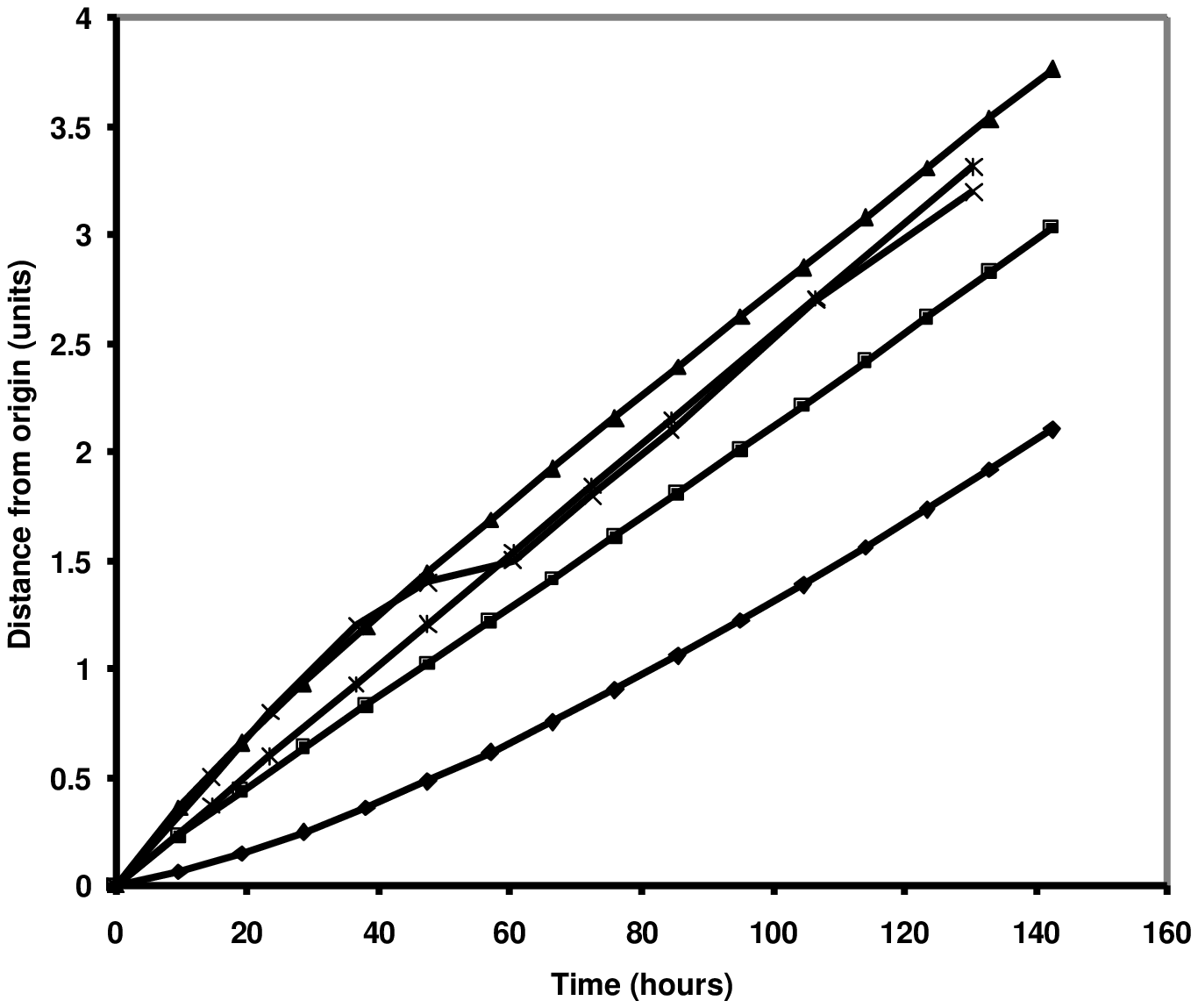,width=0.40\linewidth,bbllx=97pt,bblly=427pt,bburx=493pt,bbury=755pt}
\hspace*{0.3cm}
{\tt (b)} \hspace*{-3mm} \epsfig{figure=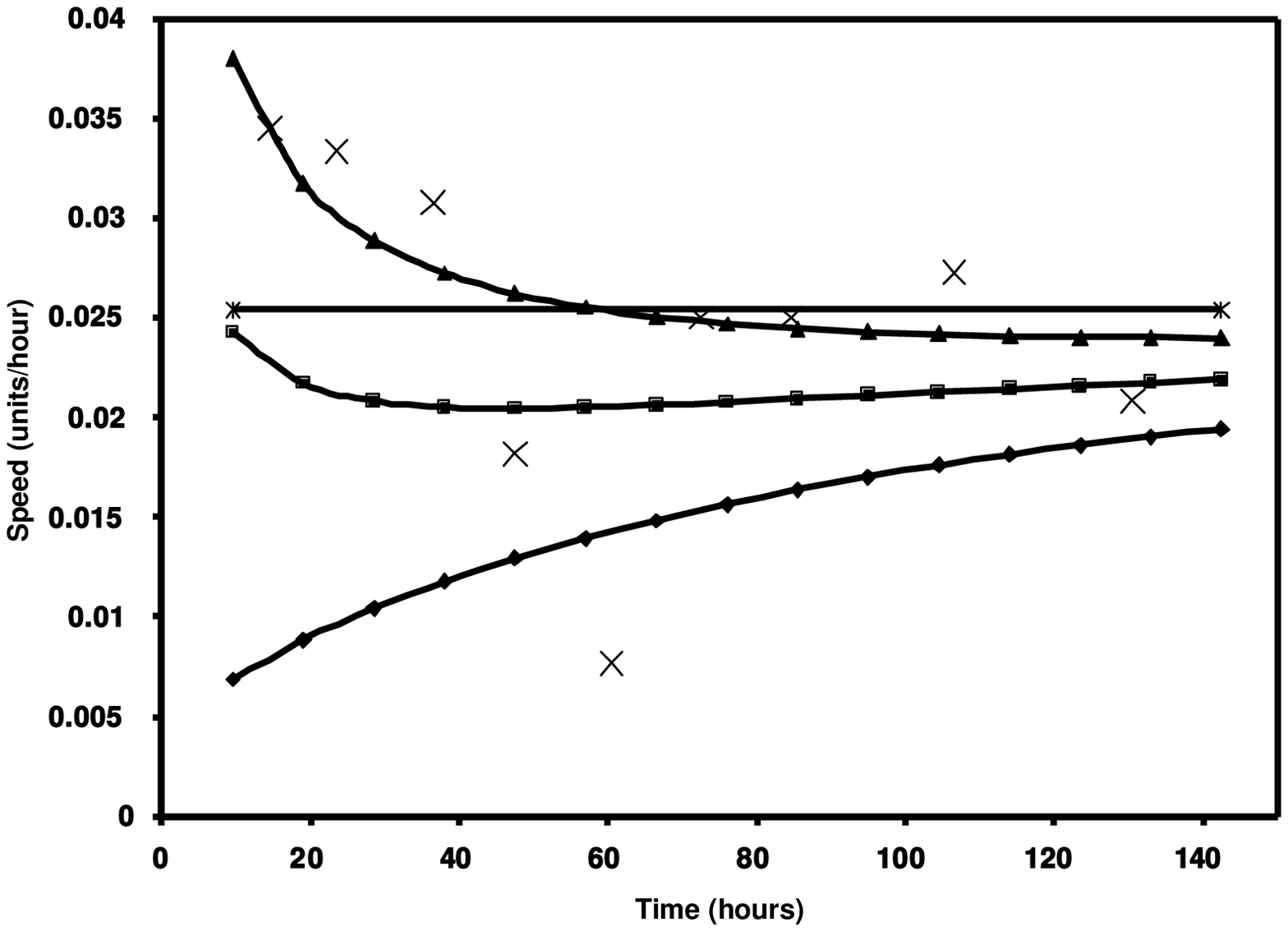,width=0.46\linewidth}
\end{center}
\caption[]{ {\em (a)} Crosses: typical plots from experimental data for the position of the invading cell
front over different substrates. The distance units are 0.25 mm;  the origin is taken as the
position at 9.5 hours and time is measured from that moment on. The other symbols denote
results from simulations of the F-KPP equation with a quadratic nonlinearity, in which the
front position is tracked at 1\% (triangles), 10\% (squares), and 50\% (diamonds) of the
asymptotic value. {\em (b)} Crosses: velocity 
extracted from the front position data of figure {\em (a)}. 
  From Maini {\em et al.} \cite{maini,maini2}, courtesy of S. McElwain and
B. MacGillavray.}\label{figwound2}
\end{figure}

In the previous sections we already mentioned that front propagation into an unstable
state features in various models that have been studied in the context of biological growth
and invasion problems, as well as in the context of pulse  propagation in nerves.
Since there have, to our knowledge,
  been few {\em experiments}  in the life sciences which are directly
  aimed at testing some of the specific front
propagation predictions, we want to draw attention to a very
nice recent   experiment \cite{maini} on the healing of epidermal  wounds,
i.e., wounds on the outer layer of the skin. 
In the experiments, a 4mm
scrape wound was made; after removal of the displaced cells, the remaining 
cells were bathed in a fresh culture medium and the position of the
invading healing front was measured. Fig.~\ref{figwound1} shows 
a photograph of the wound while it is healing: clearly a rather
well-defined front is seen to propagate into the right into the space made
by 
the scrape. The data for the position of the
cell front as a function of time,  extracted from such visual
observations, are shown in  in Fig.~\ref{figwound2} with large solid circles. As
one can see from these data,  the cell front initially moves ahead
fast, and 
then slows down (around a time of order 40-60 hours),  before it
finally settles to a more or less constant speed. 

As discussed in \cite{maini}
several mechanisms of wound healing have been discussed in the
literature. Mathematically, most approaches are formulated as a  system  of nonlinear partial
differential equations for cells, whose dynamics is usually taken diffusive, and whose growth
responds to various chemical signals via chemokinesis and chemotaxis, or to mechanical signals. 
In the most simpleminded approach, one then arrives again at a F-KPP type equation, which
then incorporates the diffusion, but lumps all the coupling to chemical and mechanical signals into 
a simple growth term. For this reason, the results of the experiment on the healing of wounds 
were   compared with numerical results for the  front dynamics in the F-KPP equation, obtained
by tracking the front position at various levels of the dynamical
variable $u$ in Eq.~(\ref{fkpp})
(symbols in Fig.~\ref{figwound2}); indeed, as the figure
illustrates, the velocity data for fronts in the F-KPP equation have
some similarities with  those found experimentally: initially the
velocity is 
relatively high, then there is a transient regime where it is
relatively low, and then it gradually
approaches  the asymptotic speed from below (more detailed examples  of
such  transient  behavior and of the dependence on the level curve
which is tracked, can be found in \cite{evs2}). However, the behavior in
the simulations appears to be more gradual then the rather sharp
crossover seen experimentally after 40-50 hours. 

It is important to keep in mind that the time-dependent velocity of
pulled fronts behaves this way quite generally -- in fact,   we already
encountered the same behavior  in the discussion of fronts in the
Taylor-Couette and Rayleigh-B\'enard systems, see
Fig.~\ref{figkockel}{\em (a)} of section \ref{sectiontcrb}.  As we mentioned already there,
and as will be discussed in more detail again in section \ref{sectionuniversalrel}, for any
pulled front emerging from sufficiently localized initial conditions, the asymptotic speed is
approached from below. The 
precise behavior at small and intermediate times does not only depend on the model, 
but also on the initial 
conditions. In particular, when the initial conditions have large gradients (e.g., when they show
step-function like behavior), the initial front speed at almost all
levels of $u$ is initially large, and then undershoots $v^*$ --- see
Fig.~6 of \cite{evs2}.

If the sharp dip in the velocity of the wound healing is a
reproducible effect, then my own guess is that this is a sign of the
importance of other
effects not included in the model, rather than a transient behavior of
an F-KPP-type model with which the data have been compared so far.

\subsection{Fronts in mean field approximations of growth models} \label{meanfieldgrowth}

In the last two decades it has become clear that many stochastic
interfacial growth processes exhibit scale invariant behavior on long
time and length scales: many growth processes are characterized by
nontrivial dynamical critical exponents for the growth of the 
the interface roughness (the root mean square
interface fluctuations)  with time and system size. There are various types of
approaches that have been used to uncover the various universality
classes that govern the long-wavelength long-time scaling, ranging
from analytical mappings and solution of  lattice models,  to field-theoretic
 renormalization group  calculations and extensive numerical
simulations. The common denominator of almost all of these approaches
is that one starts from an appropriate coarse-grained interfacial
model. Provided such an interfacial description is appropriate, the
topics that concern us in this paper are not directly relevant for
this issue, for a discussion of which we therefore refer to the reader to the various
reviews of this field \cite{barabasi,halpin-healy,krug2,meakinbook}.  However, if
the universal  scaling properties are not the main focus for a given
growth problem, but if instead one wants to analyze  the overall growth
shape   or a growth pattern  or the possibility of a morphological transition,
then a mean-field approximation  is often an indispensable  tool. In this
approximation, the analysis of fronts often does play an important
role. We illustrate this general observation with an old example of a
deposition model 
which at the same time is a nice example of a pulled front in a very
nontrivial difference equation \cite{krug}. The discussion will also
prepare us for an issue  to which we will return later
and which is at present not completely resolved: quite often, in a
mean-field description one arises at pulled fronts. As we shall
discuss in more detail in sections \ref{sectionmba} and
\ref{sectionstochasticfronts},  deterministic pulled
fronts do not converge to the standard type of interfacial
description and as a result a standard type mean-field analysis may
miss some of the essential ingredients of  the underlying stochastic
model, or even exhibit  pathological behavior that the underlying model
does not have.

In ballistic deposition, particles rain down ballistically but at
random positions onto a
cluster and stick to it  with a given probability as soon as  they come to a site neighboring
the cluster. Consider  the special case of a
two-dimensional square lattice with the depositing particles coming
straight from above at discrete times (for simplicity we summarize  a special case of
the analysis of Krug and Meakin \cite{krug}, who consider more general dimensions
and deposition under and angle). Then a particle coming down in the
$i^{\rm th}$ column can stick with probability $p$ at a the first site
in that column 
which is a nearest neighbor of the highest occupied site in that
column or in the two neighboring columns.  In a
mean-field approximation, one ignores all correlations and formulates
this growth process in terms of the probability  $\rho_t (x,z)$ that a
site at $x,z$ is part of the deposit at time $t$. In this
approximation and for one-dimensional profiles, the
appropriate dynamical  equation is \cite{krug}
\begin{eqnarray}
\rho_{t+1} (z)- \rho_t(z) & = & p [1-\rho_t(z)] \left\{ 1-[1-
\rho_t(z-1)][1-\rho_t(z)]^2 \right\} \nonumber \\
&  & \hspace*{1.6cm} \times \prod_{z' = z+1}^{\infty}
[1-\rho(z')]^3 . \label{krugeq}
\end{eqnarray}
The term on the left describes the change in the probability at a site
at height $z$ that is part of the cluster; it changes when a
particle is deposited at that site, and the terms on the right hand
side model this effect in  a mean
field approximation. The first term on the right hand side 
is the probability that that site at height $z$ is empty, the terms
between parentheses  is the probability that at least one of the
neighboring sites (at height $z-1$ in that column or at height $z$ in
the neighboring columns) is occupied, and the product term on the
second line is the probability that none of the sites in the column or
its two neighbors is occupied.

Eq.~(\ref{krugeq}) is a  difference equation in both the
discrete space and time variables; its form is unusual, in that the
change at height $z$ depends on all the probabilities at higher
sites.\footnote{A continuum version appropriate in the limit $p\to 0$
can be found in \cite{bensimonetal}.}  A rather complete study of the
general form of this equation was given by Krug and Meakin
\cite{krug}; their numerical solutions showed that at long times, the
dynamics  leads to front type solutions. Moreover the asymptotic
velocity of these fronts turned out to be $v^*$: empirically, the
fronts are found to be pulled.\footnote{Intuitively, this can be
understood from the observation that the nonlinearities in the
dynamical equation express the suppression of the growth due to
screening.}  Indeed, although as we discussed in
section \ref{sectionmoregeneral} it is now clear that the  linear spreading velocity $v^*$
is given by the same equations for any equation which upon Fourier-Laplace
transform leads to a linear equation of the form (\ref{generalform}), this appears to
be one of the earliest examples where the pulled velocity was
calculated for a nontrivial difference equation, and where the power
law convergence to $v^*$ was also tested.  Indeed, by
linearizing the equation in $\rho$ and substituting $\rho\sim
e^{\sigma t -\lambda z}$, one finds the dispersion relation
\cite{krug}
\begin{equation}
\sigma(\lambda) = \ln \left\{ 1 + p \left[ e^\lambda +2 \right]
\right\}.
\end{equation}
from which $\lambda^*$ and $v^*$ can easily be obtained by calculating
the minimum of $\sigma(\lambda)/ \lambda$. As mentioned above, Krug
and Meakin also verified that the front velocity converged to $v^*$ as
$1/t$, with a prefactor which was consistent with (\ref{v(t)relaxation}) to within
15\% (presumably, this small discrepancy is due to the higher order
$1/t^{3/2}$ correction, which was not known at the time). This slow
convergence in time in this case actually entails a very slow $1/z$ correction of
the frozen-in density profile!

A common feature of mean field approximations of growth models that
(\ref{krugeq}) also exhibits is that the growth is nonzero for
arbitrarily small particle density $\rho$. This makes the state
$\rho=0$ really a linearly unstable state and gives rise to the
existence of a finite  linear spreading speed. In reality however,
lattice models have an intrinsic cutoff for growth: there has to be at
least one particle for a cluster to be able to grow. As we shall see
in section \ref{sectionstochasticfronts}, the fact that the particle
occupation number is ``quantized'' has important consequences for growth
fronts: they are effectively always  pushed rather than pulled.

     \subsection{Error  propagation in extended chaotic systems} \label{sectionerrorpropagation}

In almost all examples we have considered so far, the unstable state
into which a a front propagates is a well-defined state which is
homogeneous in the appropriate variables.\footnote{The unstable
  phase-winding state into which the fronts discussed in section \ref{sectioncglcubic}
  propagate are homogeneous in the amplitude $a$ and wavenumber
  $k$.}  An exception to this which deserves to be mentioned is the
work by Kaneko \cite{kaneko} and Torcini, Grassberger and Politi \cite{torcini1} on fronts
propagating into  extended chaotic systems. The unstable state in this 
chase is itself a chaotic state which is characterized by positive
Lyapunov exponents. Specifically, the authors consider the Coupled Map 
Lattice
\begin{equation}
x_i^{n+1} = f\left( [1-\varepsilon] x_i^{n} + \half \varepsilon [
  x_{i-1}^n+x_{i+1}^n ] \right) ,
\end{equation}
where $i$ and $n$ are the discrete space and time variables. Note that 
the terms in the argument of $f$ proportional to $\varepsilon$ have
the form of a discretized version of the diffusion term. The function
$f$ maps the interval onto itself, as is usual in studies of maps.

\begin{figure}[t]
\begin{center}
\epsfig{figure=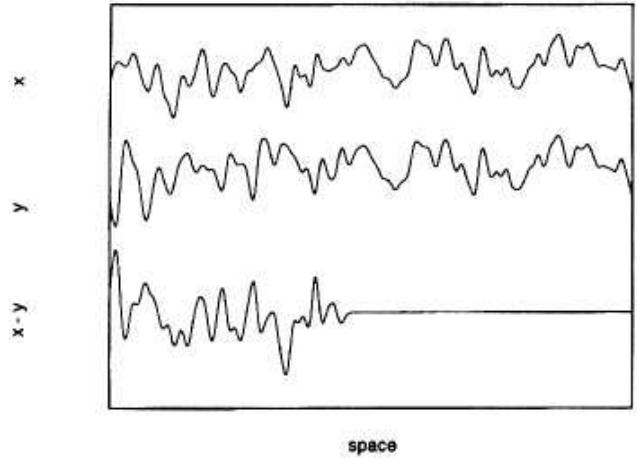,width=0.6\linewidth}  
\end{center}
\caption[]{Two initial realizations $x$ and $y$ of states of the
  coupled map lattice considered by Torcini {\em et al.}
  \cite{torcini1}, together with the difference $x-y$. Note that while 
  each state is chaotic, the difference variable vanished on the
  right. The propagation of the ``error'' $x-y$ into this state is then
  an example of a front propagating into an unstable chaotic state. Both
  pulled and pushed fronts can occur, depending on the form of the
  function $f$.  }\label{figtorcini}
\end{figure}

In general, a Lyapunov exponent of a chaotic system measures how two infinitesimally close initial 
states grow apart. In this sense a chaotic state is an unstable state, since any small perturbation
away from it grows out in time. The notion of a front introduced for such a system is illustrated in
Fig.~\ref{figtorcini} from \cite{torcini1}. One considers two realizations of extended chaotic states 
$x_i^0$ and $y^0_i$ which differ on one side of the system 
(say $i \le0$)  but not on the other. The front 
in this case is thus an ``error propagation front'' in  the difference variable $x_i^n-y_i^n$ --- while
each individual state is chaotic everywhere, the ``error'' between them spreads more and more to
the right.

In the early work of Kaneko \cite{kaneko}, the 
Lyapunov exponent was studied in a frame moving with fixed speed. The
natural speed of the front was then found to be the speed at which this
velocity-dependent Lyapunov exponent was equal to zero. In our words,
this means that for the nonlinearities studied by Kaneko, the ``error
fronts'' were pulled and that the speed $v^*$ was given in terms of
the Lyapunov exponent.  Here we follow the formulation of Torcini {\em
et al.} \cite{torcini1}, who also studied the transition from pulled
to pushed incoherent error fronts (see also \cite{giacomelli2}).
One can indeed
extend the notion of a pulled front to this
case by analyzing the Lyapunov  exponents $\Lambda (\lambda)$ for
infinitesimal perturbations $u_i^n $ of $x_i^n$ of the form
\begin{equation}
u_i^n \sim e^{\Lambda  n - \lambda i} ,
\end{equation}
where the evolution of the $u_i^n$ is governed by the linearized
tangent map
\begin{equation}
u_i^{n+1} = f^\prime\left( [1-\varepsilon] x_i^{n} + \half \varepsilon [
  x_{i-1}^n+x_{i+1}^n ] \right) \left( [1-\varepsilon] u_i^{n} + \half \varepsilon [
  u_{i-1}^n+u_{i+1}^n ]\right) 
\end{equation}
Clearly, $\Lambda(\lambda)$ is the analogue for an extended chaotic system
of the dispersion relation $-i \omega(k)$, and the analogue of the
expression (\ref{saddlepoint}) for the linear spreading speed $v^*$ is
\begin{equation}
v^*_{\rm CML} = \frac{\Lambda (\lambda^*)}{\lambda^*} = \left. \frac{d
\Lambda(\lambda )}{d \lambda } \right|_{\lambda^*}.
\end{equation}
Although the precise justification of this expression has not been studied, it is intuitively clear
that the basis it is that the Lyapunov exponent defines a finite time
scale  $\Lambda^{-1}(\lambda^*)$. Thus, 
after averaging over times sufficiently longer than
$\Lambda^{-1}(\lambda^*)$, one obtains a linear problem characterized
by an  effective dispersion
relation $\Lambda(\lambda)$, in terms of which the asymptotic $(t\to
\infty)$ spreading problem is well-defined. 

In the numerical studies, it was found that for various functions $f$
(corresponding to the logistic map, the cubic map and the tent map),
the error propagation fronts were pulled (in the average sense) while
in other cases (e.g., $f(x) = r x $ mod 1 with $r>1$), the fronts were
pushed in that their average asymptotic speed was larger than
$v^*$. Clearly, these error propagation fronts are examples of
incoherent fronts, and as we noted in section
\ref{sectionincoherentpffs} we have at present no sharp mathematical
characterization of a pushed incoherent front solution --- 
intuitively we expect that enhancement of the  growth  by the
nonlinearities will tend to give rise to a pushed front and that,
 like a coherent pushed front solution, it
falls off spatially faster than the pulled front. But how to identify
a pushed front other than by the empirical observation that it moves
faster than $v^*$ I do not know. Indeed,  Torcini and co-workers
\cite{torcini1,cencini} have  been  able to provide  
a reasonable guidance  into what type of maps give pushed error
fronts, building on the insight from the nonlinear diffusion equation
that pushed fronts are generally associated with growth functions $f$
which increase faster than linear for increasing $u$.  Possibly,
further study of the pushed to pulled transition in these fronts might
help to develop a sharp definition of pushed incoherent fronts. For recent
extensions of such chaotic front studies, including the effective
diffusion of such fronts, the reader is referred to \cite{torcini2}. 

We  note in passing that  Politi  and coworkers
\cite{giacomelli1,pikovsky} have argued that upon coarse-graining the effective
equation for the dynamics of the difference variable $d_i= x_i-y_i$
of the lower graph of Fig.~\ref{figtorcini} becomes a diffusion-type equation with
multiplicative noise, $\partial_t d = \nabla^2 d+ \eta d$ where $\eta$ is a
stochastic noise term. With the Cole-Hopf transformation $d=e^h$ this
equation is equivalent to the KPZ equation \cite{barabasi,halpin-healy,kpz,krug2}. It is
amusing to realize that the arguments of section
\ref{sectionfluctuatinguniclasses}  indicate that
therefore the scaling properties of 
such  fluctuating error-propagation fronts are equivalent to those of the KPZ
equation, but in one dimension higher than one would naively expect:
As explained in that section, 
for pulled fronts the fluctuations in the direction of propagation
continue to contribute to the scaling behavior, instead of being
integrated out. 

    \subsection{A clock model for the largest Lyapunov exponent of the
particle trajectories in a dilute gas} \label{sectionvanzon}
One present line of research in kinetic theory touching on the
foundations  of statistical physics is the study of  the  Lyapunov exponents of the trajectories of
the constituent particles or atoms  in a dilute  gas (the largest Lyapunov exponent
gives the rate with which two nearby initial conditions grow apart
under the dynamics, see section \ref{sectionerrorpropagation}
above). A few years ago  van Zon, van Beijeren and Dellago
\cite{vanzon} were able to calculate the largest Lyapunov
exponent exactly in a low density hard sphere gas  by expressing it in terms of
the speed of a pulled front in a differential-difference equation.  The model is actually very closely
related to a computationally very efficient lattice model introduced by Brunet
and Derrida \cite{derrida,derrida3}
 in the context of the connection with   phase transitions in disorder models discussed 
in section \ref{sectionphasetrans}.

For calculating  the Lyapunov exponents, one has to analyze how the
original  trajectories of a typical hard sphere grow apart from a
shadow one obtained from an infinitesimally different initial
condition. For the largest Lyapunov exponent, the following picture
becomes appropriate in the dilute gas limit. When a sphere collides
with another one, the difference after the collision between the original trajectory of
each sphere and its shadowing trajectory is, on average, equal to the
largest of the two before the collision,  multiplied by a constant
factor that accounts for the enlargement due to the collision. This
implies that the largest Lyapunov exponent is, in appropriate units,
given by the front speed of a  ``clock model'', in which every particle carries a 
clock with a discrete time $k $ which is advanced at
every collision \cite{vanzon}.
This happens according to the following rule:
when two particles collide, they {\it both} reset their respective clock 
values, say $k$ and $\ell$, to either $k+1$ or $\ell+1$, whichever is 
the largest. Thus, if we denote the number of particles with clock value 
$k$ by $N_k$, we obtain the following dynamical equation:
\begin{equation}
{d N_k \over dt} = - \sum_{\ell = -\infty;~\ell \not=k}^\infty R_{k,\ell}
-2R_{k,k}+ 2 \sum_{\ell = -\infty}^{k-1} R_{k-1,\ell}~,
\end{equation}
where $R_{k,\ell}$ denotes the rate by which collisions occur between 
particles with clock values $k$ and $\ell$. In the dilute gas limit correlation effects become negligible;
then  $R_{k,\ell}$ is simply  proportional to $N_k N_\ell / N^2$ when $k \not = \ell$ and to 
$N_k^2 /(2 N^2)$ when two particles with equal clock
value $k$ collide. Then, upon writing $f_k = N_k/N$
and scaling the time appropriately, we obtain
\begin{eqnarray} {d f_k \over dt} 
&=& - f_k + f_{k-1}^2 + 2 f_{k-1} C_{k-2}~, \nonumber \\
&=& - f_k + C^2_{k-1} - C^2_{k-2}~,
\end{eqnarray}
where we have set
\begin{equation}
 C_k = \sum_{\ell = -\infty}^{k} f_\ell.
\end{equation}
Adding the equations for $f_\ell$ for all values of $\ell \le k$ then yields 
\begin{equation} 
\label{ramses2}
dC_j(t) /dt =-C_j(t) +C^2_{j-1}(t).
\end{equation}

Equation (\ref{ramses2}) has two spatially and temporally constant
 solutions: $C_j=0$ and $C_j=1$. The first
one is stable and describes the state where all clocks are set to
a value larger than $j$; the second one is unstable and corresponds to
the case where all clocks are set to values less than $j$. As time
proceeds, all clock values are continuously increased, and hence in
the context of this model it is natural to consider the invasion of
the unstable state $C_j=1$ by the  state $C_j=0$ --- as indicated above, the
front speed then determines the largest Lyapunov exponent of the dilute hard sphere
gas.

Numerical solutions have shown that the fronts propagating into the unstable
state $C_j=1$ are pulled fronts \cite{vanzon}. This in itself is maybe
not so much of a surprise, as it is easy to  
convince oneself that the nonlinearity on the right hand side tends to reduce the growth relative
to that of the terms linearized about the unstable state (very much like in the F-KPP equation 
with a nonlinearity of the form $f(u) =u -u^2$). However,  the equation does illustrate nicely
a number of general points concerning pulled fronts:  
\begin{enumerate}
\item
 The
dynamical equation (\ref{ramses2}) is a difference-differential
equation. Nevertheless, as
we already pointed out so often, the
same equations for $v^*$ and for the rate of approach to $v^*$
(Eq.~(\ref{v(t)relaxation}) of section 
\ref{sectionunirelsimple}) hold. The convergence of the front speed to 
its asymptotic value according to this result was checked explicitly
with high precision numerical simulations 
in \cite{evs2,evsp}.
\item
  Although the dynamics is very
different from that of the F-KPP equation --- note the  very
asymmetric dynamics:  the dynamics
of $C_j$ is unaffected by that of $C_k$ with $k>j$ --- the equation is
simple enough that several of the methods that play an important role
in  proving the existence and convergence of fronts for the F-KPP
equation (like comparison theorems) can be extended to this equation
\cite{evsp}.
\item
 As we will discuss in section
\ref{sectionstochasticfronts},  when a mean field equation is
simulated with a stochastic equation with a finite number of particles
$N$, the convergence as a function of $N$ to the asymptotic  pulled  $v^*$ of the 
 $N\to \infty$ mean-field equation is very slow, logarithmically slow. This  slow
convergence was also found when  the present clock model was simulated
with a finite number of particles \cite{vanzon}. 
\item
  In the
context of the clock model the $C_j$'s are non-decreasing functions of
$j$: $C_j \le C_{j+1}$ for any $j$.  However, if we just take
Eq.~(\ref{ramses2}) as a general dynamical equation in which the $C_j$'s
are allowed to decrease with increasing $j$, then we find that the
state $C_j=1$ is also convectively unstable to a front where the $C_j$
decrease for increasing $j$! This turns out to be a retracting pulled
front, a front which moves to the right in accord with the fact that
any perturbation or coherent structure solution of (\ref{ramses2}) can only move towards
increasing $j$. Such a retracting front corresponds to a negative
value of $v^*$ in our equations,\footnote{Of course,there is only one dispersion relation
 for Fourier modes of the dynamical equation linearized about the
unstable state. The fronts discussed above and analyzed in
\cite{vanzon,evsp} fall off as $\exp (-\lambda^* j)$ with $\lambda^*= 0.768
$ and $v^*=4.311$. The retracting fronts correspond to the solutions of (\ref{v(t)relaxation}) which
have negative $\lambda^*= -1.609$ and negative $v^*=-0.373$.} and has an amusing  property:
whereas the speed of a front which genuinely propagates {\em into} the unstable
state approaches the positive asymptotic speed from {\em below} due to the
negative $1/t$ correction in (\ref{v(t)relaxation}), the {\em absolute value } of the speed of a
retracting front is larger than $|v^*|$ for large times, since $v^*$
itself is negative!
\item
 It is easy to modify the equation according
to the general rules of thumb of section \ref{sectionwhenpushed} so that
the fronts become pushed; an explicit example is discussed in \cite{evsp}.
\end{enumerate}

\subsection{Propagation of a front  into an unstable ferromagnetic state}\label{ferromagnet}

Although it may not be of great practical relevance, we briefly
mention an amusing example whose front dynamics is governed by  two pulled fronts 
separated by a phase slip region, the dynamical
equations for an anisotropic ferromagnet
\cite{elmer1,elmer2}. The width of domain walls and the wavelength
of spin-wave states in ferromagnets
is often large enough that a continuum approximation is justified. For
a ferromagnet with  with an easy-plane  anisotropy the free energy in
dimensionless units reads \cite{llcontmedia}
\begin{equation}
{\mathcal F} = \half \int {\rm d} x \, [(\partial_x \theta)^2 + \sin^2\theta
(\partial_x \phi)^2 + \cos^2 \theta ].
\end{equation}
Here $\theta$ and $\phi$ denote the direction of the magnetization $M$ in
polar coordinates, and we have assumed that the magnetization only
varies in the spatial $x$-direction. The last term in the expression
shows that the energy of states with polar angle $\theta = \pi/2$ is
lowest, so indeed the free energy density describes a situation with
an easy-plane direction.

The dynamics of a ferromagnet is governed by the so-called
Landau-Lifshits equations 
\begin{equation}
\frac{ d {\bf M} }{dt} = \gamma {\bf M} \times \frac{d {\mathcal
F}}{d {\bf M}}  + \lambda
{\bf M} \times \left( {\bf M} \times \frac{d {\mathcal
F}}{d {\bf M}} \right).
\end{equation}
In the high damping limit, the first term which describes the torque
can be neglected; in appropriate time units the equations then become
\begin{eqnarray}
\partial_t \theta &=& \partial_x^2 \theta + [1- (\partial_x \phi)^2]
\sin \theta \cos \theta ,\\
\partial_t \phi & = & \partial_x^2 \phi + 2 (\partial_x \phi) (\partial_x
\theta ) \mbox{cotg} \theta.
\end{eqnarray}
Note that if we could consider $\partial_x \phi$ to be fixed, the 
first equation would be nothing but the F-KPP equation for
$\theta$, with the state $\theta = \pi/2$ stable if
$(\partial_x\phi)^2 <1$ and unstable if $(\partial_x\phi)^2 > 1$. Thus
a phase-winding solution of the form $\theta =\pi/2, \phi = kx$ is
unstable for $k>1$. Elmer {\em et al.} \cite{elmer1,elmer2} considered front propagating
into this unstable state; because of the similarity of the first
equation with the F-KPP equation, it is no surprise that these fronts
are pulled, but the new feature is the coupling to the
$\phi$-variable.\footnote{It is  amusing to  note that the nonlinearity of the $\theta$-equation for fixed
phase gradient $\partial_x \phi$ is
a special case of the dynamical equations for the director angle in the smectic $C^*$ problem discussed
in section \ref{sectionsmectic}.}
 This front leaves behind  a state with $\theta \approx 0$ but with the phase gradient
$\partial_x \phi$ essentially unaltered. Because of the strong coupling between phase and polar angle
due to the cotangent  term in the $\phi$-equation, this phase-winding state is unstable. As
a result, the first front is followed by a region where phase slips occurs. The state which
emerges from here has no appreciable phase gradient: The nonlinear term in the $\theta$-equation then
flips sign and this makes the $\theta\approx 0$ state 
 unstable. As a result,   this region is in turn  invaded by a second pulled
F-KPP-like front.  The propagation speed of the phase slip region
can even be calculated from a conservation-type argument for the phase winding, as all the properties
of the back side of the first front and of the leading edge of the
trailing front are known \cite{elmer1,elmer2}. 

When the dynamical equations are written in terms of a complex
amplitude $A = \sin \theta e^{i \phi}$, $A$ is found to obey a
Ginzburg-Landau type equation with an unusual nonlinearity \cite{elmer2}.  From this
perspective, the problem has some similarity with the problem of front
propagation into an Eckhaus-unstable phase winding solution in the
real Ginzburg-Landau equation that was mentioned briefly in section  \ref{sectioncglcubic}.

     \subsection{Relation with  phase transitions in disorder models} \label{sectionphasetrans}
In hardly any of the situations that we have discussed so far do the front
solutions  which become relevant when the initial
conditions are not sufficiently localized, play a role. There is one
remarkable exception to this: it was discovered by Derrida and Spohn \cite{derrida2}
that there are some statistical physical disorder models which can be
shown to have phase transitions by mapping their generating function
onto the F-KPP equation. The two regimes of pulled fronts emerging
from localized initial conditions and of leading edge dominated
dynamics associated with not sufficiently localized initial conditions 
then translate into two different phases of the disorder model!
An amusing aspect is also that in the first regime, the $1/t$ power
law relaxation and the associated crossover behavior of the front
solutions translates back into detailed knowledge of the scaling
behavior of the statistical model. More recently, the same type of
mapping was applied to the renormalization group analysis of
disordered $XY$ models \cite{carpentier} --- see section \ref{sectioncarpentier}.

\begin{figure}[t]
\begin{center}
\epsfig{figure=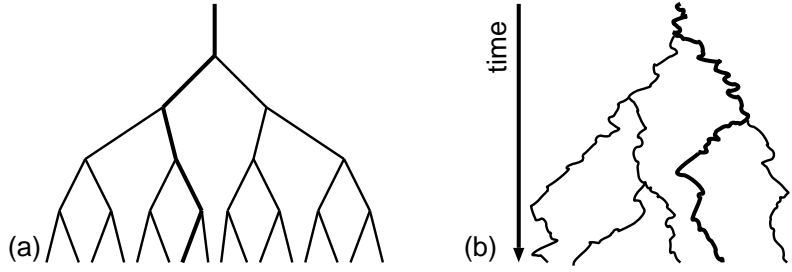,width=0.75\linewidth}  
\end{center}
\caption[]{(a) A ``polymer'' (thick lines) or self-avoiding walk on a
Cayley tree. On each bond an independent random potential is defined.
(b) Illustration of the statistical problem of a polymer on a Cayley
tree with a random potential in the continuum limit.  After
Derrida and Spohn \cite{derrida2}. }\label{figrandompolymers} 
\end{figure}

Let us  sketch the essence of the  argument for the  case of
polymers on a Cayley tree with a random potential \cite{derrida2}.  A
Cayley tree is is a graph which has branches but no loops, as
sketched in Fig.~\ref{figrandompolymers}{\em (a)}. The hierarchical
structure of Cayley trees makes the statistical
physical models defined on them amenable to detailed analysis ---
phase transitions in such models are usually of mean field type, due
to the absence of loops, but apart from this they often catch the
essence of a transition in higher dimensions. We now consider ``polymers'' on such Cayley
trees  --- we can think of them in terms of self-avoiding random walks which step  one 
level down in every ``time'' step --- in the presence of  a random
potential $V$ at the bonds. These values of the potential are
uncorrelated random variables distributed
according to some distribution ${\mathcal P} (V)$.  The statistical
problem for a given realization of potentials
is then defined as follows in terms of  the so-called partition function
\begin{equation}
{\mathcal Z} = \sum_{\rm walks} \exp \left[ -\beta (V_{1i_1} + V_{2i_2} +
\cdots ) \right]
\end{equation}
where $\beta$ is the inverse temperature. The term in the exponents is
the sum of all the potentials in level 1, level 2, etc. It is 
 convenient to consider the vertical direction as a time
coordinate $t$ and to take the time-continuous limit.  The potentials
$V$ on a branch of length $\ds t$ are  then taken as  independent Gaussian variables
with distribution 
\begin{equation}
{\mathcal P} (V) = \frac{1}{(4\pi  \, \ds t)^{1/2} } \exp \left( -
    \frac{V^2}{4\, \ds t} \right) .
\end{equation}
Because of the branching structure of the Cayley tree, the partition
function obeys a simple recursion relation,
\begin{equation}
{\mathcal Z}(t+\ds t) = \left\{ \begin{array}{l} e^{-\beta V} {\mathcal
      Z}(t) \hspace*{2.85cm} \mbox{with probability} ~1-\ds t, \\
e^{-\beta V} \left[ {\mathcal
      Z}^{1}(t) + {\mathcal Z}^{(2)}(t) \right] \hspace*{0.5cm}
  \mbox{with probability} ~\ds t ,
\end{array} \right. \label{zeqs}
\end{equation}
which expresses that in the first small timespan $\ds t$, the walk could
remain a single walk with probability $1-  \ds t$, or have split into
two  with probability $ \ds t$ (the generalization to Cayley trees where 
every branch splits into $n>2$ is obvious). ${\mathcal Z}^{(1)} $ and
${\mathcal Z}^{(2)} $ are the two partition functions on these two
branches.

Because the
potentials are random variables, one actually has to study the
distribution of partition functions. Instead of doing this, it is more 
convenient \cite{derrida2} to study the so-called generating function $G_t(x)$ which
encodes the average and all moments of ${\mathcal Z}$,
\begin{equation}
G_t(x) \equiv \left\langle \exp \left[ - e^{-\beta x} {\mathcal Z}(t) \right]
  \right\rangle ,
\end{equation}
where the brackets indicate an average over the disorder. 
From this definition, one finds immediately that for small $dt$
\begin{eqnarray}
G_{t+\ds t}(x) &=& (1-\ds t) \int \ds V \frac{1}{(4\pi \, \ds t)^{1/2} } \exp \left( -
    \frac{V^2}{4\, \ds t} \right)  G_{t}(x+V) \nonumber \\
& &~~~~  + \ds t \, G_t^2(x) .
\end{eqnarray}
That the branching process simply leads to a term proportional to $G^2$ relies on the fact
that the partition functions ${\mathcal Z}^{(1)} $ and ${\mathcal Z}^{(2)} $ in (\ref{zeqs}) are 
independent, because on a Cayley tree two  branches do not intersect below their common branch
point. 
In the limit $\ds t \to 0$ the above equation for $G$  becomes simply
\begin{equation}
\partial_t G= \partial_x^2 G +  G (G-1 ),\label{gfromfkpp}
\end{equation}
which is nothing but the F-KPP equation (\ref{fkpp})
 with quadratic nonlinearity, if we put $G=1-u$.

The remarkable finding is that the dynamical equation for $G$ does not
depend on the  temperature $\beta^{-1}$ explicitly: the temperature
only enters indirectly through the initial conditions which Derrida
and Spohn considered,
\begin{equation}
G_0(x) = \exp \left( -e^{-\beta x} \right)
\hspace*{0.7cm}\Longleftrightarrow \hspace*{0.7cm}
u  \simeq e^{-\beta x} ~~(x\to\infty).  
\end{equation}
As we know from our discussion of fronts in the F-KPP equation, when $\beta >1$, this corresponds
to localized initial conditions, and the long-time behavior of $G_t(x)$ is then simply given
by the pulled front solution  whose speed approaches $v^*$ according to the
asymptotic formula (\ref{v(t)relaxation}). 
This means that the ``long-time'' asymptotic average properties of the
polymers are completely independent of temperature, and that they approach an asymptotic
scaling behavior independently of temperature as well. Furthermore, when $\beta<1$, the
fronts correspond to fronts whose dynamics is ``leading edge dominated'' and whose velocity
is determined by the initial condition, i.e., the temperature.  This is the high-temperature
regime. The low-temperature ($\beta>1$) regime can be associated with a frozen phase:
roughly speaking the fact that the front speed is independent of initial conditions translates into
the free energy sticking at a constant temperature-independent value --- 
we refer for  a detailed discussion of this and of the translation of the various front results to those for the
random polymer problem to \cite{derrida2}. Below  in section \ref{sectioncarpentier} the connection
to renormalization treatments of disorder models is discussed.

We finally note that a lattice model version \cite{derrida,derrida3}
of the statistical-physical disorder problem discussed
here is closely related to the clock model discussed in section \ref{sectionvanzon}.

\subsection{Other examples}
We finally draw attention to a number of   issues related to front propagation 
which we do not review in detail.

\subsubsection{Renormalization of disorder models via traveling waves}\label{sectioncarpentier}

In the previous section \ref{sectionphasetrans} 
 we  briefly reviewed how the generating function of a disorder model on a Cayley tree obeys 
the F-KPP equation. This connection between traveling waves and
disorder models  is actually more general and powerful, as was discovered by
Carpentier and Le Doussal \cite{carpentier,carpentier2}. Very much
like (\ref{gfromfkpp}) is the evolution equation for the generating function in the 
``time-wise'' direction down the Cayley tree, they found that the
renormalization group equation for the distribution function of the
local fugacity variables of defects in a disordered XY model takes  the form of an F-KPP equation, with
the length scale $\ell $ playing  the role time. Like we saw in
section \ref{sectionphasetrans}, the
universal behavior of this distribution function comes from the
independence of $v^*$ on the details of
the model. Moreover,  the
results for the universal rate of approach to the asymptotic pulled
front speed
$v^*$ in the case of localized intial conditions (corresponding to
temperatures below the glass transition) translate into nontrivial
predictions for the scaling at the transition. We refer to these
papers for an in-depth discussion of this.

\subsubsection{Singularities and ``fronts'' in cascade models for turbulence}\label{sectioncascade}

As is well known,  fully developed turbulence is characterized by an 
energy cascade: energy flows from small wave numbers to larger wave
numbers, till it is dissipated in the viscous range. In a shell model
for this energy flow \cite{eggers},  this energy flow down the cascade  shows
similarities with front propagation. However,   the essential dynamics is more
properly analyzed in terms of a similarity analysis  like that for the
porous medium equation \cite{aronson,goldenfeld,bert2} than in terms of fronts. If interpreted in
terms of front dynamics, these fronts have a nonlinear diffusion
equation reminiscent of the one in  the bacterial growth model
(\ref{kmmodel2}) and are pushed.   

In fact, as stressed in particular by    
by Barenblatt \cite{barenblatt1,barenblatt2}, there is a formal connection between 
similarity analysis and  front propagation, since if one write $x$ and
$t$ of a propagating front solution $u(x-vt) $ in terms of logarithmic variables,
$x=\ln X$, $t=\ln T$, a uniformly translating front solution
$u(x-vt)$ looks like a similarity solution $\tilde{u} (X/T^v)$. 
In this formulation  the propagation  
velocity of the front plays the role of a similarity     
exponent.  This reformulation illustrates that in principle a sharp 
distinction between front propagation and similarity analysis
\cite{barenblatt1,barenblatt2} is difficult to make. In practice, however, it is 
often quite clear what the most natural type of analysis is for a 
given problem: especially if the system admits a linearly unstable
state with a well-defined dispersion relation for the unstable branch
of Fourier modes,  the linear spreading analysis and 
the associated front formulation are the methods of choice.      

\subsubsection{Other biological problems}  \label{sectionotherbio}  
Although the clearest biological examples of front propagation into an unstable
state appear to be the population spreading phenomena mentioned  in
section \ref{sectionbiologicalinvasion} 
and the bacterial growth  patterns discussed in section \ref{chemicalbacterial}, it is
important to realize that there are many  problems in biology
and biophysics which are closely related. E.g.,  much work on
 pulse propagation in bistable systems or excitable media traces
back to the study of the  propagation of pulses in nerves
\cite{kinzel,scott}.  Since these are not examples of front propagation into
unstable states proper, and since excellent books in which these
problems are treated have appeared recently \cite{keener,murray}, we
will not explore the differences and similarities here.

\subsubsection{Solar and stellar activity cycles}  \label{stellar}
There is an extensive literature --- see e.g. \cite{bassom}  for a
brief introduction and  entry into the literature --- associating the
sunspot cycle to the occurrence of so-called dynamo waves.  The
coupled equations for the azimuthal and  radial  magnetic fields
take the form of two reaction-diffusion type equations, and an
important ingredient of the relevant dynamics is associated with the
problem of front propagation into unstable states
\cite{bassom,tobias,worledge}. Of course, for waves on a spherical
rotating object like the sun,  there are many complicating aspects:
not only is there  the question whether the instability is locally
convective or absolute, but  also the fact that the geometry is
intrinsically finite and  that the background of the waves is
spatially varying plays a role. These latter type of issues are
related to the question of the emergence of so-called ``global modes'' 
which we will discuss briefly  in section
\ref{sectionglobalmodes}. For a detailed discussion of the dynamo
waves themselves we refer to the literature cited above.

\subsubsection{Digital search trees}\label{sectionsearchtrees}
After acceptance of this article, we learnt about a very new exciting
line of research: traveling fronts play an important role in computer
science in digital search trees and data compression. Moreover, such 
types of problems can be related to directed DLA  on a
Cayley tree. See \cite{majumdar1,majumdar2} and references therein.

\section{The mechanism underlying the universal convergence towards
$v^*$}\label{sectionuniversalrel}   

In section \ref{sectionunirelsimple} we saw that pulled fronts
converge to their asymptotic  speed and shape\footnote{The shape
  convergence only holds for uniformly translating fronts and coherent 
  pattern forming pulled fronts, of course. Nevertheless, the incoherent
dynamics  of an incoherent pulled front arises from the intrinsic
chaotic behavior of the nonlinear state behind the front (as, e.g., in
panels {\em (a)} of Figs.~\ref{figks}-\ref{figcglquintic2}) or from
the fact that the spreading point dynamics does 
not match on to a fully coherent front profile in the nonlinear region
(as, e.g., in
Fig.~\ref{figcglquintic1}{\em (b)}). In the leading edge the dynamics of incoherent pulled
fronts is still smooth and coherent,  as close inspection of these
 figures  and Fig.~\ref{cglrelaxationfig} below shows quite clearly. This
dynamics is governed by the extension of Eq.~(\ref{largexiexp}) below to pattern
forming fronts.} with  a universal power law
behavior --- e.g., for the velocity we have according to  
(\ref{v(t)2relaxation}) 
\begin{equation}
v(t) = v^* - \frac{3}{2\lambda^*t} +\, \frac{3 \sqrt{\pi}}{2
(\lambda^*)^2 t^{3/2} }\, {\rm Re}\frac{1}{\sqrt{\mathcal D}} + {\mathcal O}\left( \frac{1}{t^2}\right).  \label{v(t)2brelaxation}
\end{equation}
For uniformly translating fronts, for which ${\mathcal D}$ is real,
this expression reduces to  the simpler version
(\ref{v(t)relaxation}).

In the above expression $v^*, \lambda^*$
and ${\mathcal D}$
are given explicitly by Eq.~(\ref{saddlepoint}) in terms of the dispersion relation $\omega(k)$
of the linearized dynamical equation.  
This and  other remarkable features  of this general expression were discussed
extensively in section  \ref{relaxtodincohpfpulled} but we did not
justify or derive the result itself. It is the aim of this section to
review the underlying mechanism and derivation  \cite{evs2} in some
detail. The importance of the derivation does not lie in the technicalities
themselves, but in the fact that the pulled front picture that we have
advanced can be made into a fully explicit and predictive formalism.

\subsection{Two important features of the linear
problem}\label{twofeatures}

Several of the most important insights on which the full derivation of
(\ref{v(t)2brelaxation})   is based \cite{evs2} come from our
understanding of the fully
linear problem discussed in sections \ref{sectionv*} through
\ref{sectionmoregeneral}. In section \ref{sectionv*} we saw that
according to the
saddle point analysis of the fully linear spreading problem  a
 ``steep''  initial condition  gives rise to the 
following long-time expression for a generic  dynamical field
$\phi$,
\begin{equation}
\phi(\xi,t) \sim e^{-\lambda^*\xi +ik_r^*\xi-i(\omega_r^*-k_r^*v^*) t} \,
\frac{e^{-\xi^2/(4{\mathcal D} t)}}{\sqrt{t} }, \label{phiasymp1} 
\end{equation}
where $\xi=x-v^*t$, and where $\lambda^*, k_r^*, v^*$ and ${\mathcal
D}$ are given by the linear spreading point equations
(\ref{saddlepoint}). The first exponential factor is the asymptotic
exponential fall-of with steepness $\lambda^*$ in the frame moving
with velocity $v^*$. The second exponential factor, together with the
$t^{-1/2}$ term, is the first correction term to the asymptotic
behavior --- it arises simply from the Gaussian saddle point
integral. Of course, in this Gaussian term we also recognize the
fundamental similarity solution of the diffusion equation. Thus, if we
define a new field $\psi$ through the transformation
\begin{equation}
\phi(\xi,t ) \equiv e^{-\lambda^*\xi+ik_r^*\xi-i(\omega_r^*-k_r^*v^*)t} \psi(\xi,t), \label{leadingedgetrafo}
\end{equation}
then we expect that the long-time dynamics of $\psi$ is governed by
the equation
\begin{equation}
\frac{\partial \psi}{\partial t} = {\mathcal D} \frac{\partial^2
\psi}{\partial \xi^2}  + \mbox{corrections},\label{leadingedge1}
\end{equation}
since its fundamental similarity solution is the Gaussian form $\psi =
t^{-1/2} e^{-\xi^2/4{\mathcal D}t}$ of (\ref{phiasymp1}). Note that
the transformation (\ref{leadingedgetrafo}), which was called the
``{\em leading edge transformation}'' in \cite{evs2}, acts like a
mathematical magnifying glass: it allows us to focus on the dynamics
of the correction to the dominant exponential behavior. Moreover, do
realize that since the saddle point analysis is quite general (see
section \ref{sectionmoregeneral}) the long-time dynamics of the
leading edge variable $\psi$ is effectively governed in dominant order
by this diffusion equation {\em even if the underlying dynamical
equation is of higher order, a set of equations, a difference
equation, etcetera.} The reason that $\psi$ obeys a differential
equation even when the original dynamical equation is a difference
equation is that for long times $\psi$ becomes arbitrarily smooth in
space and and time.

The above line of reasoning is intuitive and based on working
backward from the general expression (\ref{phiasymp1}). The
diffusion-type form of the equation --- first order in time and second
order in space --- is actually an immediate consequence of the fact
that two roots $k$ coincide at the linear spreading point $k^*$. After
all,
Eqs.~(\ref{saddle1}) and (\ref{saddlepoint}) imply that in the
neighborhood  of the spreading point we have in the
co-moving frame $\xi$
\begin{equation}
\omega -\omega^* = -i {\mathcal D} (k-k^*)^2  + \cdots, \label{omegaexpansion}
\end{equation}
and upon inverse Fourier-Laplace transformation this gives (\ref{leadingedge1}) for
long times. 
In fact, the full generalization of (\ref{leadingedge1}) for $\psi$ is
easy to derive \cite{evs2}: by expanding about the linear spreading
point and taking an inverse Fourier-Laplace transform, we immediately
obtain \cite{evs2}
\begin{equation}
\frac{\partial \psi}{\partial t} =  {\mathcal D} \frac{\partial^2
\psi}{\partial \xi^2}  + {\mathcal D_3} \frac{\partial^3\psi}{\partial
\xi^3} + w \frac{\partial^2 \psi}{\partial t \partial \xi} + \tau_2
\frac{\partial^2 \psi}{\partial t^2} + \cdots ,
\label{leadingedge2}
\end{equation}
where the expansion coefficients ${\mathcal D_3}, w, \tau_2$ etcetera
can all be expressed in terms of the characteristic equation of the
branch corresponding to the relevant spreading point --- see Eq.~(5.64) of
\cite{evs2}. E.g., we simply have ${\mathcal D_3}= (1/3!)
{\rm d}^3\omega/{\rm d} k^3|_{k^*}$.

Since we already know from the saddle point analysis that the relevant
long-time dynamics of the leading edge variable $\psi$ is a
diffusion-type dynamics on slow spatial and temporal scales, the
crucial conclusion from these considerations is that\footnote{There {\em is} actually a surprise here: A
priori, one would expect that the second and third  terms of
(\ref{leadingedge2}) would affect the subdominant $t^{-3/2}$
correction term in the expression (\ref{v(t)2brelaxation}) for
$v(t)$. Actually, they do affect the shape scaling function but {\em
not} the velocity correction --- see Eq.~(5.70) of \cite{evs2}. I do
not have a real intuitive understanding of this remarkable finding
which comes out of the explicit calculation. }

\begin{tabular}{p{0.1cm}|p{13.2cm}}
& {\em the dynamical equation of the leading edge variable $\psi$ that
governs the convergence to the linear spreading behavior in the
leading edge is   a diffusion equation, with
subdominant terms which are determined explicitly in terms of the
dispersion relation $\omega(k)$ of the linear dynamical equation of
the original problem. The appropriate similarity variable for the
long-time expansion of $\psi$ is the similarity variable $\xi/\sqrt{t}$ of
the diffusion equation.}
\end{tabular}

Our second important observation concerning the fully linear problem
is the following. One may well wonder ``how can the leading edge
dynamics ever determine the dominant convergence dynamics of the {\em
full} nonlinear profile?''  To understand this, let us return to
(\ref{phiasymp1}) and write it as
\begin{equation}
|\phi(\xi,t) | \sim e^{-\lambda^*\xi } \,
\left| \frac{e^{-\xi^2/(4{\mathcal D} t)}}{\sqrt{t} }\right| = 
e^{-\lambda^*\xi  - \half \ln t  -\xi^2/(4 D t) }
, \label{phiasymp2} 
\end{equation}
where, as in (\ref{lambda*}),  $D^{-1}= {\rm Re}\,{\mathcal D}^{-1}$. 
If we follow the position $\xi_C$ of the level line
$\phi(\xi_C,t)= C$, then  according to the above expression for the
linear dynamics we have in dominant order 
\begin{equation}
\xi_C =  - \frac{1}{2\lambda^* }\, \ln t + \cdots ~~~~~\Longleftrightarrow~~~~~
\dot{\xi}_C =  - \frac{1}{2\lambda^* t} +\cdots.\label{xClinear}
\end{equation}
We already drew attention to this logarithmic shift in our discussion
of the front relaxation in section \ref{relaxtoutpulled}, but we now
identify  more clearly where this behavior
originates from: 
The $1/t$ relaxation  of the velocity of the level line 
corresponds to a logarithmic shift in the position, and  this
results from the  Gaussian $t^{-1/2}$ prefactor  in combination
with the overall $e^{-\lambda^*\xi}$ spatial decay of the leading edge
variable. In other words, {\em the prefactor of the $1/t$ velocity relaxation is
essentially the exponent of the appropriate similarity solution of the
diffusion equation (\ref{leadingedge1}).}

So why then does the logarithmic shift carry over to the fully nonlinear
profile? The important point to realize is simply that\footnote{The
argument is obvious for the divergent logarithmic term. Actually,
since the front shape relaxation is driven by the $1/t$ term in the
velocity --- see Eq.~(\ref{expansionlarge})  below --- any contribution from the leading edge which gives a
time-dependent shift  which decays slower than the $1/t$ intrinsic shape
relaxation is
universal. This is the reason that even the $1/t^{3/2} $ term in the
velocity relaxation is universal: it corresponds to a shift
proportional to $1/t^{1/2}$. Even
incoherent fronts whose fluctuating width converges faster than $1/t$
to some average value, still exhibit the subdominant $1/t^{3/2}$
velocity relaxation term.}

\begin{tabular}{p{0.1cm}|p{13.2cm}}
& {\em when we track the position of a front whose  width  of the
nonlinear  region is finite, the unbounded logarithmic
shift imposed by the diffusive leading edge dynamics always dominates
the large time behavior over  the relaxation of the front shape itself. }
\end{tabular}

This observation was already illustrated in Fig.~\ref{figfkpprelaxation}. 

\subsection{The matching analysis for uniformly translating fronts
and coherent pattern forming fronts}\label{sectionmatching1}

So we know that the leading edge variable $\psi$,
which measures  the deviation from  the asymptotic exponential
profile $e^{-\lambda^*\xi+ik_r^*\xi-i(\omega^*-k^*v^*)t}$ in the moving frame, obeys a
diffusion-type equation with corrections. To construct the solution to 
the fully nonlinear font relaxation problem, we have to {\em match} the 
behavior in the leading edge (the region where the dynamical equation
can be linearized about the unstable state) to the nonlinear front
region, the region where the nonlinearities in the equation are
important. 

In order to understand the matching behavior, it is clearest to first
consider the case of uniformly translating fronts or coherent pattern
forming fronts: For these the arguments can be made most precise and
for these the matching analysis has been worked out in detail. The case of
incoherent pattern forming fronts will be discussed in the next
section; the analysis given there will give a more intuitive dynamical argument why
the leading edge variable $\psi$ approaches the behavior that we
identify with matching arguments here. 

At the linear spreading point $ k^*$ two roots coincide ---
this is illustrated both by Eq.~(\ref{omegaexpansion}) and the fact
that $v^*$ corresponds to the minimum of the curve $v_{\rm
  env}(\lambda)$ in Fig.~\ref{figvversuslambda}. It is a general
result \cite{arnold1,arnold2} that in the presence of a double root the $\xi\to \infty$ asymptotic
behavior of the uniformly translating pulled front solutions
$\Phi_{v^*}(\xi)$ or the leading coherent pulled front solution
$\Phi^1_{v^*}(\xi)$, is
\begin{equation}\Phi_{v^*}(\xi)  \sim  (a_1\xi+a_2)
e^{-\lambda^*\xi}, ~~~~~
\Phi_{v^*}^1 (\xi)  \sim (A_1 \xi + A_2) e^{-\lambda^*\xi + i(k^*_{\rm r}
\xi - \omega^*_{\rm r}t)},
\end{equation} as we already noted before in   Eqs.~(\ref{Phi*eq}) and (\ref{Phi1*eq}).
As $t\to\infty$, the leading edge dynamics should approach this
behavior, so in dominant order  the matching condition for the leading edge variable
$\psi$ is
\begin{equation}
\psi(\xi ,t\to\infty)  \sim \xi. \label{psimatching}
\end{equation}

At this point, we can already understand the leading term of the
convergence behavior of $v(t)$ in a very simply manner. For simplicity,
we will from now on specialize to  the case of uniformly translating
fronts but the generalization to coherent pattern forming fronts is
straightforward \cite{esvs,willem1,storm1}. The similarity 
solution of the diffusion equation which for large times matches the
above $\psi\sim \xi$ behavior is the ``dipole solution''
\begin{equation}
\psi \sim \frac{\xi}{t^{3/2}} e^{-\xi^2/(4Dt)}
  ~~~~~\Longleftrightarrow~~~~~ \phi  \sim e^{-\lambda^* \xi - 3/2 \ln t + \ln 
    \xi - \xi^2/(4Dt)}.
\end{equation}
Analogously to the discussion following Eq.~(\ref{phiasymp2}), this
result implies that to order $1/t$ the velocity relaxation is
\begin{equation}
v(t)=v^* - \frac{3}{2\lambda^*t} + \cdots,
\end{equation}
which is indeed the leading order term of the full expression
(\ref{v(t)2brelaxation}). 

To go beyond the leading order term, we have to perform a full
systematic matching calculation. To do so, it is crucial to describe
the nonlinear front in the right frame: As we saw in the previous
section, a $1/t$ relaxation of the velocity corresponds to an
ever-increasing logarithmic shift in the position of a level line,
when viewed in the frame $\xi$ moving with the {\em asymptotic} speed
$v^*$.  Thus, as Fig.~\ref{figfkpprelaxation} illustrates so nicely,
{\em if we would attempt to do perturbation theory about the
asymptotic front solution $\Phi_{v^*}(\xi)$, the difference between the
actual transient profile and this asymptotic profile would increase
without bound.} Nevertheless,  the {\em shape} of the transient profile is
always close to $\Phi_{v^*}$ placed at an appropriate position. This suggests
to perturb about the asymptotic front solution in a frame $\xi_X$
which incorporates this shift,
\begin{equation}
\xi_X = \xi- X(t) = x-v^*t - X(t), \label{xiXvariable}
\end{equation}
where  the shift $X(t)$ has the expansion
\begin{equation}
\dot{X}(t)= \frac{c_1}{t} + \frac{c_{3/2}}{t^{3/2} } + \frac{c_2}{t^2} +
\cdots.  ~~~\Longleftrightarrow ~~X(t)=c_1 \ln t - \frac{2 c_{3/2}}{t^{1/2}} +\cdots.
\end{equation}
The expansion in powers of $1/\sqrt{t}$ results from the fact that the
similarity variable governing the long-time dynamics of $\psi$ is
$\xi_X/t^{1/2}$, so powers of $\xi_X$ generate powers of $1/t^{1/2}$.

From here on, the matching expansion is conceptually straightforward
but technically nontrivial. In the leading edge we make a large-$t$ expansion
in terms of the  similarity variable $\xi_X/t^{1/2}$,
\begin{equation}
\psi = \left[ t^\half\,  g_{-1/2}(\xi_X/t^{\half}) + g_0(\xi_X/t^\half) + \frac{
g_{1/2}(\xi_X/t^\half)}{t^\half} + \cdots\right ] e^{-\xi^2_X/(4
{\mathcal D}t)} , \label{largexiexp}
\end{equation}
where  the matching condition (\ref{psimatching})
requires that $g_{-1/2}(\xi_X/t^{1/2}) \sim \xi_X/t^{1/2}$.
In the nonlinear front region we expand the profile in the form
dictated by the above observations,\footnote{Remember that we have written here the
form for the case that the asymptotic front  solution is uniformly
translating. For coherent fronts the expansion  is similar \cite{esvs,willem1,storm1},   we only
have to add the upper indices and a phase relaxation term.}
\begin{equation}
\phi(x,t) = \Phi_{v^*} (\xi_X) + \frac{\eta_1(\xi_X)}{t} +
\frac{\eta_{3/2}(\xi_X)}{t^{3/2}}+ \cdots. \label{expansionlarge}
\end{equation}
These expansions generate hierarchies of equations which can be solved
order by order; in first nontrivial order we straightforwardly recover the $1/t$ term
along the sames lines as above, and in the next order the nontrivial
$t^{-3/2}$ term in the general expression (\ref{v(t)2brelaxation}) for
$v(t)$ is found \cite{evs2}.

The explicit calculation also shows that the functions $\eta_1$ and
$\eta_{3/2}$ are proportional to the ``shape mode'' $\delta \Phi_v /
\delta v$ which gives the change in the shape of the profile under a
change of the velocity $v$: one finds
\begin{equation}
\Phi_{v^*} (\xi) + \frac{\eta_1(\xi_X)}{t} +
\frac{\eta_{3/2}(\xi_X)}{t^{3/2}} = \Phi_{v*} (\xi_X) + \left. \frac{\delta
\Phi_v(\xi_X)}{\delta v} \right|_{v^*} \dot{X}(t) + {\mathcal O}(t^{-2}).
\end{equation}
Since $\dot{X}(t)$ is nothing but the deviation $v(t)-v^*$, this
 implies that to order $t^{-2}$ the profile shape can be written in
the form (\ref{phirelaxshape}), i.e.,
\begin{equation}
\phi(x,t) = \Phi_{v(t)} (\xi_X) + {\mathcal O}(t^{-2}).
\end{equation}
In other words, to order $t^{-2}$ the shape of the profile follows the
shape of a uniformly translating profile with velocity equal to the
instantaneous value $v(t)$.

At first sight, it may appear surprising that such a result could
hold, since on the one hand $v(t)$ always approaches $v^*$ from below, and since on
the other hand uniformly translating profiles $\Phi_v$ with $v<v^*$
approach the asymptotic state $\phi=0$ in an oscillatory manner. There
is no contradiction here, however. The above expression is only the
asymptotic expression for $\xi_X$ fixed, $t\to \infty$. The limits do not
commute: For any {\em fixed}
time, the profile crosses over to the expression (\ref{largexiexp}) for
large enough $\xi_X$. As one might expect from the diffusive nature of
the $\psi$-equation, this crossover region moves to the right
diffusively, as $\sqrt{t}$ \cite{evs2}. 

\begin{figure}[t]
\begin{center}
\epsfig{figure=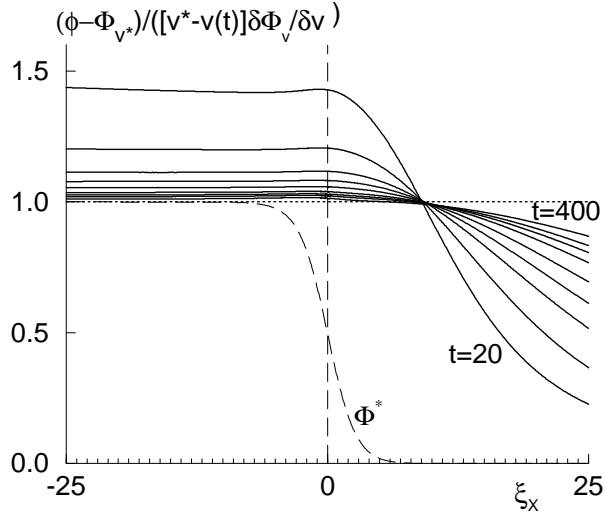,width=0.58\linewidth}
\end{center}
\caption[]{ Numerical confirmation of the analytical prediction for
  the shape relaxation of a transient front in the F-KPP equation
  (\ref{fkpp}) with cubic nonlinearity for various times ranging from
  $t=20$ to $t=400$.  Plotted is the normalized
  deviation from the asymptotic profile $\Phi_{v^*}$ given by
  (\ref{shaperel}). As indicated by the dashed vertical line, the
  co-moving coordinate $\xi_X$ is taken such  
  that $\phi(\xi_X,t)=1/2$. The fact that the ratio (\ref{shaperel})
  converges to 1 confirms that to order $t^{-2}$ the shape of the
  profile follows $\Phi_{v(t)}(\xi_X)$ adiabatically. Note the
  diffusive-type crossover on the far right to the asymptotic behavior 
  (\ref{largexiexp}). Whether it is accidental or significant  that all the lines
  roughly  cross at one point near $\xi_X\approx 10$ is at present not
  understood. From Ebert {\em et al.} \cite{evs2}.  }\label{fkppshape}
\end{figure} 

These analytical predictions for the relaxation of the profile shape
have been confirmed  in great detail numerically for the F-KPP equation (\ref{fkpp})
with a cubic nonlinearity \cite{evs2}. Fig.~\ref{fkppshape} shows a
plot of 
\begin{equation}
\frac{\phi(\xi_X,t) - \Phi_{v^*}(\xi_X) }{\delta \Phi_v/ \delta v |_{v^*}
(v^*-v(t))} , \label{shaperel}
\end{equation}
where the shape mode in the denominator was obtained by numerically solving the 
ordinary differential equation that it obeys. According to
the analytical predictions, the ratio (\ref{shaperel}) should approach unity for long
time; the numerical results clearly confirm this. Note that since the
denominator approaches zero for long times, plots like
Fig.~\ref{fkppshape} are very accurate confirmations of the analytical
expressions --- in some cases the predictions have been verified  to within
six significant figures \cite{evs2,evsp}!

\begin{figure}[t]
\begin{center}
  \epsfig{figure=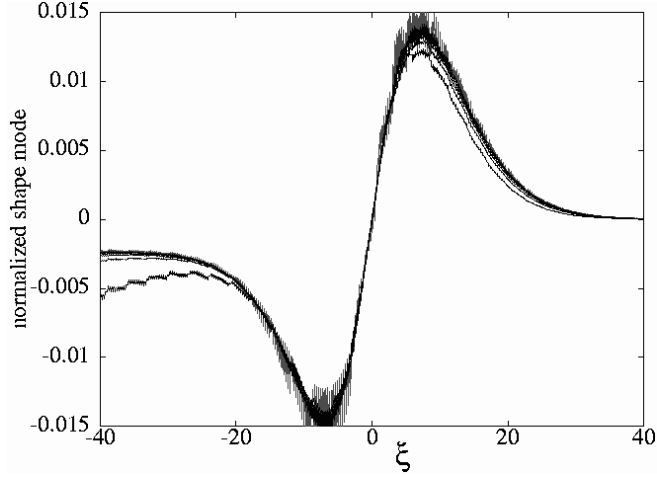,width=0.45\linewidth,angle=-90}
\end{center}
\caption[]{Demonstration of the universal relaxation behavior of the
  shape of a front in the Swift-Hohenberg equation
  (\ref{swifthohenberg}), according to Spruijt {\em et al.}
\cite{esvs,willem1}. Plotted is the ratio (\ref{shshape}) for
  times 20 to 160 in steps of 20 for fronts in numerical simulations
of fronts in for 
  $\varepsilon=1/2$.  The
  fact that the various curves fall almost on top of each other
  confirms that the shape of the fronts follows the relaxation of the
  velocity adiabatically. The curve which deviates somewhat on the
  left side is taken at the earliest time, $t=20$. Up to an overall multiplicative constant, the 
  resulting curve represents the shape mode of the pulled
  Swift-Hohenberg front. Note that since $\varepsilon$ is not small,
  various harmonics contribute to  the expansion of the asymptotic
  profile, and the front solution is {\em not}  close to the front solution
  of the cubic amplitude equation that one can derive in the limit
  $0<\varepsilon \ll 1$. }\label{shshapemode}
\end{figure}

As we already indicated, the matching calculation can be extended to
the case of coherent pattern forming fronts
\cite{esvs,willem1,storm1}. We actually already illustrated in
Fig.~\ref{figcglquintic2}  how the results
of numerical simulations of the Swift-Hohenberg equation
(\ref{swifthohenberg})  confirm the
predicted relaxation behavior of the velocity and the wavenumber just
behind the front.  That the shape relaxation of the front also follows
the velocity
adiabatically  has also been verified in these
studies \cite{esvs,willem1}. For coherent pattern forming fronts an
explicit calculation of the shape modes $\delta \Phi^n_v/\delta v$ is
quite nontrivial, even numerically, since it in principle involves an
infinite set of coupled ordinary differential equations. Therefore in
these studies the quantity 
\begin{equation}
 \frac{ \langle \phi(\xi_X,t) - \sum_{n=0,\pm 1,\cdots} e^{-in\Omega^*t}
  \Phi_{v^*}^n(\xi_X) \rangle_T}{t^{-1}}\label{shshape}
\end{equation}
was studied. Here the brackets $\langle \cdot \rangle_T$ denote an average over one period
$T\approx 2\pi/\Omega^*$ in the co-moving frame. According to the analytical predictions, the various
plots of this ratio should all fall on top of each-other, and the
resulting curve is nothing but the shape mode. As
Fig.~\ref{shshapemode} shows, this expectation is borne out by the
simulations. Given that for such a coherent pattern forming front
obtaining the front profile involves extensive interpolation, it is
remarkable how nicely the scaling works \cite{esvs,willem1}! To our
knowledge this is the only existing explicit demonstration of the
universal shape relaxation of a coherent pattern forming pulled front. 

\subsection{A dynamical argument that also  holds for incoherent
  fronts} \label{sectionmatching2} 

The matching arguments reviewed above were essentially based on the
observation that the asymptotic uniformly translating and coherent
front solutions have a $\xi e^{-\lambda^*\xi}$ behavior, and that the
long-time dynamics in the leading edge should match this
behavior. This essentially gave the requirement that  $\psi \sim \xi$ 
in dominant order for long times. From there on the calculation is in
essence a straightforward matching calculation.

The above argument does not give much insight into how this linear
gradient in $\psi$ builds up {\em dynamically}. Moreover,  the argument
hinges on the existence and behavior of a coherent pulled front
profile. However, the {\em same} relaxation holds 
for {\em incoherent} fronts. Let us now clarify why this is so  and how the linear
gradient builds  up dynamically \cite{storm1}.

\begin{figure}[t]
\begin{center}
\epsfig{figure=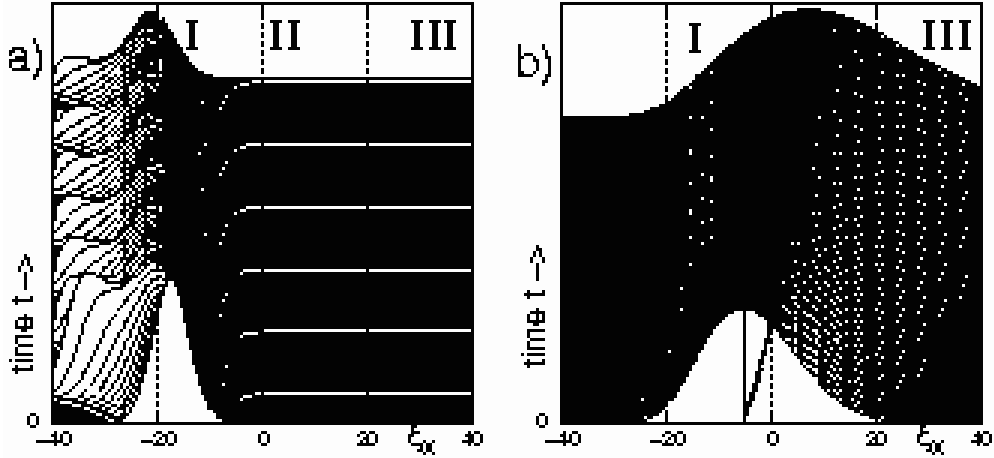,width=0.95\linewidth}
\end{center}
\caption[]{ Results from simulations by Storm {\em et al.}
\cite{storm1} of the quintic CGL equation for
  the same parameter values and run as in Fig.~\ref{figcglquintic2}{\em
    (a)}, shown from time 35 to 144. Panel (a) shows the nonlinear
  function $|N|$ defined in (\ref{Nquintic})  as a function of $\xi_X $. Note that this 
  function falls off exponentially to both sides. (b) The evolution of 
  the leading edge variable $|\psi|$ for the same run as in panel
  (a). The region marked I in the figure is the region where a linear
  gradient  builds up, II marks the crossover region which moves out
  as $\sqrt{t}$  and in region III $\psi$ is falling of in dominant
  order as a diffusive Gaussian with width proportional to
$\sqrt{t}$. The solid line indicates the position of the maximum. The regions 
  I, II and III are also marked in panel (a). Note that the nonlinear
  function $N$ essentially vanishes in these regions and that $\psi$
  builds up to the right of the region where $N$ is
  non-vanishing. Intuitively, we can think of the nonlinearities as
  acting like an ``absorbing wall'' for the  field $\psi$.   } \label{cglrelaxationfig}
\end{figure}

The argument can be formulated quite generally, but let us for
simplicity just  consider the quintic CGL equation
(\ref{quinticcgl}).  If we make a transformation to the leading edge
variable $\psi = e^{\lambda^*\xi -ik_r^*\xi+i(\omega_r^*-k_r^*v^*)t} A$ and write the equation
for $\psi$ in the frame $\xi_X$ defined in (\ref{xiXvariable}), we obtain
\begin{equation}
\partial_t \psi = {\mathcal D} \partial_{\xi_X} ^2\psi + N(\psi) +
{\mathcal O} (t^{-1})  \label{Nequation},
\end{equation}
where the terms of ${\mathcal O}(t^{-1})$ are proportional to
$\dot{X}(t)$ and come from the transformation to the frame $\xi_X$,
and where $N(\psi)$ denotes the nonlinear terms
\begin{equation}
N(\psi) =  (1+ic_3)e^{-2\lambda^*\xi_X} |\psi |^2\psi 
-(1-ic_5) e^{-4\lambda^*\xi_X} |\psi|^4 \psi . \label{Nquintic}
\end{equation}
We normally associate a front with a solution for which the physical 
field, the amplitude $A$ in the present case, {\em saturates} behind the
front.  Because of the exponential term introduced in the leading edge
transformation, this means that the leading edge variable should
vanish exponentially $\sim \exp{\lambda \xi_X}$ as
$\xi\to -\infty$. Thus, to the right, for large
positive $\xi_X$, $|N|$ vanishes exponentially as $e^{-2 \lambda
\xi_X}$ due to the explicit
exponential term in front of the cubic term in  (\ref{Nquintic}), while to the left $|N|$ vanishes
exponentially  as $e^{\lambda \xi_X}$ because of the exponential
vanishing of the leading edge variable $\psi$. In other words, $|N|$ is actually
nonzero only in some limited  spatial range. This is illustrated in the
space-time plot of Fig.~\ref{cglrelaxationfig}{\em (a)},  which shows
$|N|$ for the same simulation of the quintic CGL as in
Fig.~\ref{figcglquintic2}{\em (a)}. The exponential vanishing to the
right is clear from the figure; the time-dependent behavior of $|N|$
on the left reflects the fact that the front is incoherent.

The leading edge of the front is essentially the region where $|N|$ is
negligible. Hence it is the semi-infinite interval to the right of
$N$,  and this nonlinear term acts like a {\em boundary condition} for
the dynamics of $\psi$ in the leading edge! Moreover, since 
in leading order $\psi$ obeys a diffusion equation, let us think of
$\psi$ as a density field obeying a diffusion equation. In this
analogy, the nonlinear term $N$ {\em acts like a localized sink term or an
absorbing wall} which suppresses the density field (it must suppress it
as behind the front $\psi$ vanishes). Now, as is well known, when we
have an absorbing wall at long times the density field will build up a
linear gradient ---  from this analogy we see that the role of the
nonlinearities in the equation is to make the dynamical field $\psi $
build up a linear gradient,  $\psi \sim \xi_X$, at long times! 
In other words, the long-time behavior (\ref{psimatching}) is not
only a matching condition, but it also naturally emerges dynamically from the fact
that the dynamics of  the leading edge variable is
essentially governed by a diffusion equation with an absorbing wall. 
Fig.~\ref{cglrelaxationfig}{\em (b)}  shows a space-time plot of
$|\psi| $ from the simulations of the quintic CGL equation, which fully
confirms  the  gradual buildup of a linear gradient to
the right of the ``absorbing wall'' $|N|$, with a crossover to a
diffusive Gaussian behavior to the right of it.

In conclusion, we finally understand the universal nature of the power
law corrections to the asymptotic pulled front velocity: Only the fact
that there is an absorbing wall matters, its internal structure ---
i.e., the form of the nonlinearities in the equation, or the question
whether the front is uniformly translating, coherent or incoherent ---
is unimportant \cite{evs2,storm1}!

\section{Breakdown of Moving Boundary Approximations of pulled fronts}\label{sectionmba}

So far, we have focused our discussion on the dynamics of fronts in
one dimension. Quite often, however, we are interested in the formation of
patterns in two or three dimensions which naturally can be thought of
as consisting of domains separated by domain walls or fronts. Usually
--- think of viscous fingering or dendritic growth
\cite{brener,caroli,kassner,kessler3,langerrmp,langerma,langergodreche,pelce,pomeau},
 bacterial growth patterns \cite{benjacob1,benjacob2,benjacob4}, late 
stage coarsening \cite{bray,gunton}, flames \cite{buckmaster,clavin,joulin,combustion},
chemical waves \cite{fife,meron,meron2}, thermal plumes \cite{benamar,zocchi} etcetera --- the state of the system inside each
domain is asymptotically close to an intrinsically stable homogeneous
state.  But we have also encountered examples of such type of pattern
formation in higher dimensions which are driven by the motion of a
front into a linearly unstable state: 
streamer patterns of section \ref{sectionstreamers} are formed when
an ionization front propagates into a non-ionized gas,   the unstable normal state
of a superconductor can give way to the superconducting state through
the motion of a front  (see section \ref{sectionsupercond}), the
chaotic dynamics of  domains in the rotating Rayleigh-B\'enard experiments discussed in
section \ref{sectionkupperslorz} is in one regime driven by the propagation of fronts into linearly unstable
states, and if fluctuation effects are sufficiently small, front
propagation into an unstable state  can be relevant to the early stage
of spinodal decomposition as well (see section \ref{secspinodal}).

When the width of the fronts or domain walls separating the domains is
much smaller than  the radius of curvature  which is set by the
typical scale of the pattern, then a natural way to analyze the
pattern dynamics is in terms of a {\em moving boundary approximation}
or {\em effective interface approximation}.
This  approximation amounts to treating these fronts or
transition zones as a mathematically sharp interface or boundary. In
other words, their width is taken to be zero on the outer scale of the
pattern, and their internal
degrees of freedom are eliminated  using singular perturbation
theory. We shall henceforth use the word 
{\em boundary} or {\em interface} to denote this zero width limit and use the word
front when we look at a scale where its internal structure can be
resolved. 

Sometimes, it is numerically advantageous go in the opposite
direction, i.e., to translate a model with
sharp interfaces into what has become known as a {\em phase field
model} in which the order parameter field varies continuously through
the interfacial zone. Examples where this idea was exploited for a variety
of physical problems can be found in
\cite{batesfife,caginalp,gonzalez,karma,kassner2,kobayashi,sakaguchi};
for careful discussions of the derivation of a moving boundary problem
for a variety of different physical systems, we refer to
\cite{benamar,buckmaster,dorsey2,fife,karma,judith}.

Is it true, as one might be tempted to think  when
looking at pictures like the streamer patterns of Fig.~\ref{figstreamers}, that
 a moving boundary approximation or effective interface description
can be derived whenever the fronts in the system under study are much
thinner than the scale of the pattern? Or, to put it more
clearly, {\em can one generally  derive a moving boundary approximation for
front propagation into unstable states in the limit that the width of
the front is much less than its radius of curvature? } The answer to
this question is a definite {\em ``no''} for {\em pulled fronts} and {\em ``yes''} for
{\em pushed fronts} \cite{evs3}. 

How come? The reason for this statement is that for a moving boundary
approximation to hold, we need not only have separation of {\em spatial
scales} but also separation of {\em  time scales}: the main physical idea
underlying the derivation of a moving boundary approximation is that
the front itself can on large length and time scales be viewed as a
 well-defined  coherent structure which responds on a  ``fast'' time
scale --- the internal front relaxation time --- to the ``slow'' change in
the local values of the outer dynamical fields that characterize the
domains. {\em Both elements are missing for a pulled front.} Indeed, as we have seen
in section \ref{secoverview},  the dynamics of a pulled
one-dimensional front is essentially determined in the whole ``semi-infinite'' region ahead of the front,
not  in the nonlinear transition region itself. As a consequence, the velocity
and shape of a pulled front relax very slowly, with a $1/t$ power law, to their
asymptotic velocity and shape. For a pushed front, on the other hand,
both conditions {\em are} fulfilled: in one dimension, a pushed front
relaxes exponentially fast to its asymptotic velocity and shape, and
its dynamics responds essentially only to changes in the nonlinear
front region only.

We will  clarify the above statements below. We first illustrate
how in two and three dimensions the relaxation
correction and the curvature  correction are of the same order for a
spherically symmetric expanding pulled front, and then trace the main
steps of the derivation of a moving boundary approximation to discuss
the breakdown of  singular perturbation theory for pulled fronts.

\subsection{A spherically expanding front}\label{secspherical}

It is well known --- see e.g.  the  classic work of Allen and Cahn
\cite{allencahn} or the reviews \cite{bray,gunton} --- that when the
radius of curvature of a front which connects two stable states of  an
order parameter equation  is much larger than the front width, the
dominant correction to the front velocity is a curvature
correction.  Consider now an a
front expanding radially in two or three dimensions. For large times,
the curvature correction to the front velocity will go as $1/R(t)
\approx 1/ (v_{\rm as} t)$,
where  $v_{\rm as}$ is the asymptotic front velocity.  For a
front connecting two stable states, or more generally for a pushed
front, this is the {\em only} asymptotic correction. This is simply
because such fronts reduce  in this limit to an effective interface --- the
formal derivation is summarized in the next section --- and because for a
weakly curved interface separating two states of a single order
parameter the curvature is the dominant correction. 

For a pulled
front, on the other hand, the universal power law relaxation term is
according to (\ref{v(t)relaxation}) also of order $1/t$, i.e., {\em of the
same order as the curvature correction in this case. }
This illustrates that a weakly curved pulled front can not simply be
viewed as a weakly curved interface with velocity given uniquely in terms of
its local curvature: A moving boundary approximation does not hold. 

\begin{figure}[t]
\begin{center}
\epsfig{figure=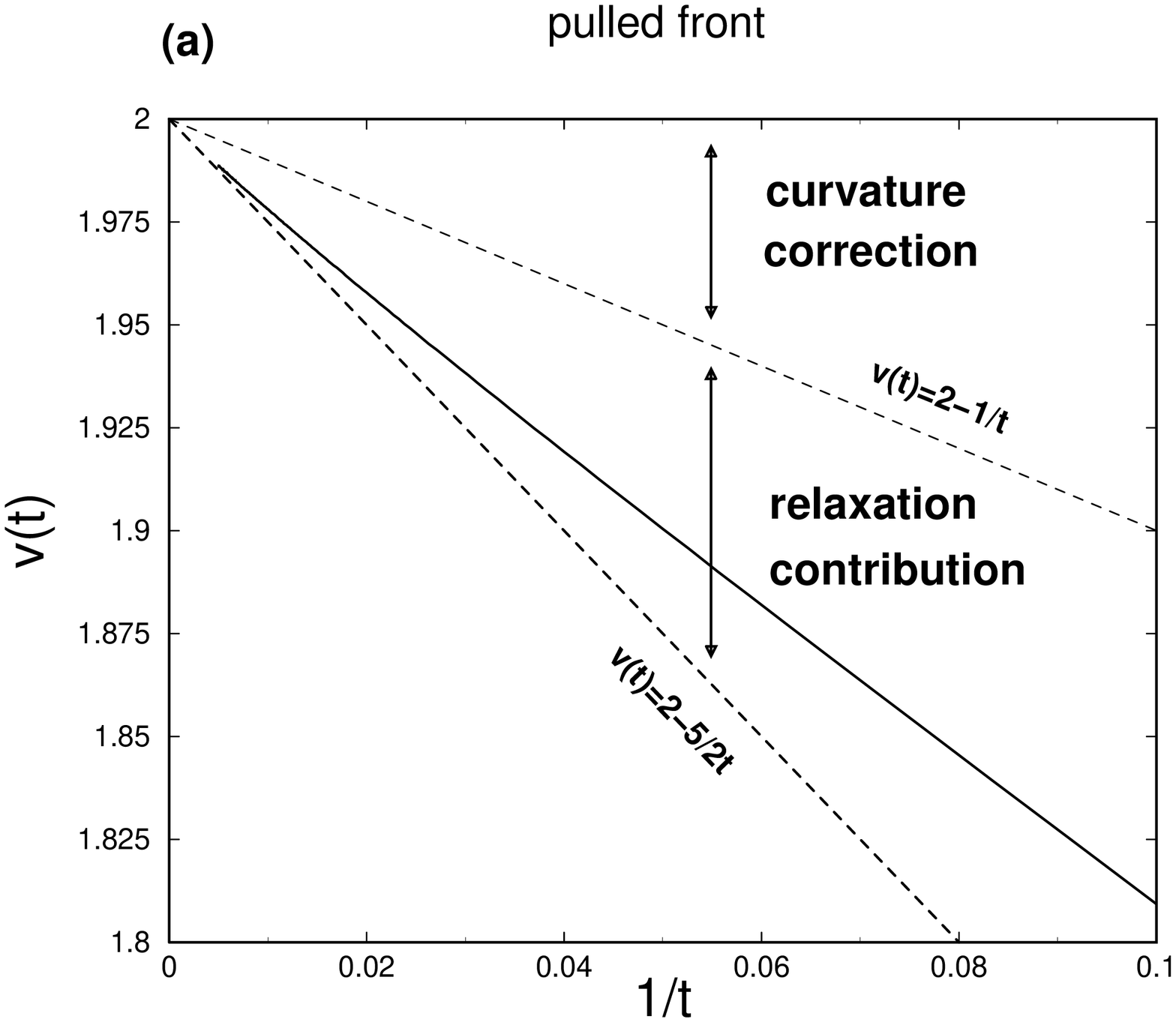,angle=-90,width=0.485\linewidth} 
\hspace*{0.1cm}
\epsfig{figure=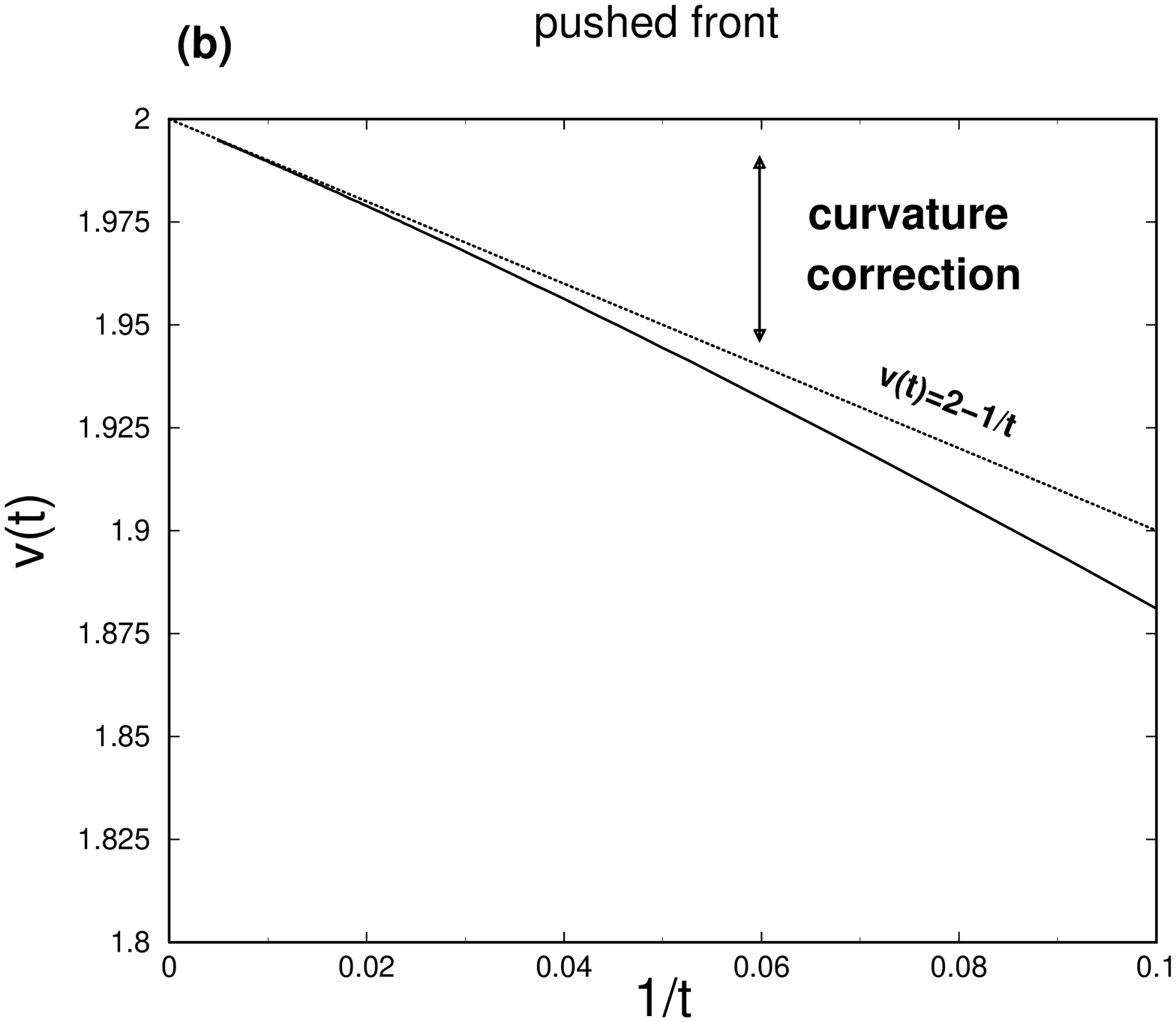,angle=-90,width=0.485\linewidth}
\end{center}
\caption[]{ Illustration of the fact that while a moving boundary
approximation applies to a weakly curved pushed front, an interfacial
description does not apply to weakly curved pulled fronts. 
In both graphs the full lines show the velocity as a function of time of a spherically expanding
front  in the F-KPP equation, starting from initial conditions
$u(r,0)=1 $ for $r<1$ and $u(r,0)=\exp [- (r-1)^2] $ for
$r>1$. (a) In the pulled case, obtained with $f=u-u^3$, the long time
relaxation asymptotically approaches the result $v(t)=2-5/2t $,
illustrating that curvature and relaxation effects are of the same
order. The upper dashed line indicates the relaxation that would result if
only the curvature term would contribute, in other words if a moving
boundary approximation would hold. The slow convergence to the line
$v(t)=2 - 5/2t$ is due to the subdominant $t^{-3/2}$ term in the
velocity relaxation expression (\ref{v(t)relaxation}).  (b) Results for the pushed
case, obtained with $f(u)= \varepsilon  u+du^3-u^5$ with $\varepsilon
=1/4$ and $d= (2 + 2/\sqrt{3})$.  According to the exact result discussed in the example at the
end of section \ref{selectionutfs}, the velocity  of the pushed front for these parameters
  is $v^\dagger=2$ ($v^*=1$ in this case). Note the rapid
convergence to the line $v(t)=2- 2/R(t) = 2-1/t$ which is the result
given by the curvature correction to the velocity, in accord with the
fact that a moving boundary approximation applies to a weakly curved
pushed front. } \label{figspherical}
\end{figure}

We can illustrate these considerations  by considering the F-KPP
equation in three dimensions in
the case of spherical symmetry,
\begin{equation}
\partial_t u (r,t) = \partial_r^2 u(r,t) -(2/r) \partial_r u + f(u).
\end{equation}
For a front at a distance $R(t) \gg1$ from the origin, the second term on
the right hand side which accounts for the curvature of the spherical
front, serves as a small perturbation. When the front is pushed, the
singular perturbation methods discussed below show that indeed it
gives a correction $-2/R(t)$ to the front. Thus, starting from a  localized 
initial condition $u(r,t=0)$ (nonzero only in a ball around
the origin), the velocity of the pushed front for large times is
$v(t)\approx v^\dagger - 2/R(t) \approx v^\dagger -2/(v^\dagger t)
+\cdots$. For a pulled front with $f^\prime (0)=1$, on the other hand, one expects that for
large times one can add the curvature correction and the asymptotic
relaxation term given in (\ref{v(t)relaxation})  with $v^*=2 $
and $\lambda^*=1$, to get\footnote{This result was first  pointed out to me by
B. Derrida (private communication).} 
$v(t)\approx v^* -2/R(t)  - 3/(2t)  \approx v^* - 5/2t
+\cdots$. The
numerical results shown in Fig.~\ref{figspherical} are consistent with this
asymptotic expression  and hence confirm the fact
  that a moving boundary approximation can never hold for 
 a pulled front --- one can not simply characterize the properties of the weakly curved pulled
front in terms of its asymptotic behavior  and curvature correction only. 

\subsection{Breakdown of singular perturbation theory for a weakly
curved pulled front} \label{sectionmbabreakdown}

Let us now trace the main steps of the derivation of solvability type
expressions for a weakly curved or weakly perturbed front, to
illustrate the breakdown of singular perturbation theory for pulled
fronts. 
Our main goal is to discuss how the crucial differences in behavior of
pulled and pushed fronts is reflected in the derivation and in the  behavior of
solvability  expressions that underly the moving boundary
approximation. Therefore,  it suffices here to  consider the 
simplest case, a weakly curved front in the F-KPP equation with a
slowly  varying external parameter, rather than  a more general   ``phase field
model'' of the  type   referred to  in the beginning of section
\ref{sectionmba}. More detailed discussions of the derivation of a
moving boundary approximation can be found in virtually any of the
references cited there. The general and more precise discussion of the
breakdown of singular perturbation 
theory for pulled fronts is given by Ebert {\em et al.} \cite{evs3}.
 
Let us consider the F-KPP equation in two dimensions in the case that
the nonlinear function $f$ depends on a small and  spatially slowly varying
external field $W$,
\begin{equation}
\partial u / \partial t  =  \nabla^2 u  + f( u,W). \label{phase2}
\end{equation}
 A moving boundary approximation consists of first
matching an inner expansion of the problem on the scale where the fast
variable (usually the order parameter variable in phase field models)
varies rapidly to the outer  problem  on large scales, the scale of
variation of the other variables like the temperature etcetera.   In
the present case, when $W$ is a slowly varying external field, the
outer problem is simple for the stable states of $f(u,W)$: $u$ will to
lowest order follow the stable stationary states of $f(u,W)$.    

Let us imagine that the front is also curved on a  large length scale,
so that the curvature $\kappa$ of the front is  a small parameter. 
Since $\kappa$ and $W$ are expected to be small and slowly varying, the velocity and
shape of the profile will differ only weakly from the velocity and
shape of  the straight
(planar) front, and we expect them to vary only slowly in time and space. One
therefore introduces a locally curvilinear coordinate system, with
$\zeta$ measuring distances normal to the front, and $s$ distances along
the  level lines of the front.  In the adiabatic approximation we then
perturb about the reference front solution in the absence of curvature
and $W$, by writing  in the front region
\begin{equation}
  u(x,y,t)  =  U_0(\zeta)+ U_1(\zeta,s,T)+\ldots,                    \label{uansatz}
\end{equation}
where $T$ denotes the slow outer time scale and
where  $U_0$ is the planar front profile for $W=0$ which obeys the
ordinary differential equation 
\begin{equation}
\label{U0}
\frac{\ds ^2 U_0}{\ds \zeta^2}
+v_0\;\frac{\ds  U_0}{\ds \zeta}+f(U_0,W)=0~.
\end{equation}
with $v_0$ the velocity of this profile. Since the strongest variation
of the profile in the front region is in the $\zeta$ direction, in the
weakly curved $\zeta,s$ coordinate system the 
Laplacian becomes to lowest order in the curvature
\begin{equation}
\nabla^2 = \frac{\partial^2}{\partial\zeta^2}
+\kappa \frac{\partial}{\partial\zeta}+\ldots ~.
\end{equation}
Upon substituting (\ref{uansatz}) with this result into the F-KPP
equation (\ref{phase2}), expanding in $U_1$ and $W$, and making the Ansatz that the
velocity correction $v_1$ of the profile follows the change due to the
curvature and potential {\em  adiabatically}, we get
\begin{equation}
L \,U_1 = -  (v_1 - \kappa ) \frac{\ds  U_0 }{\ds  \zeta} - 
\left. \frac{\partial  f(U_0, W)}{\partial   W}\right|_{W=0} W . \label{U1eq}
\end{equation}
Here $L$ is the linear operator 
\begin{equation}
\label{Loperator}
L \equiv \left( {{\ds ^2 }\over{\ds \zeta^2}} + v_0
{{\ds }\over{\ds \zeta}} + \left. \frac{\partial
  f(U,0)}{\partial U}\right|_{U_0} \right)~,
\end{equation}
which results from linearizing about the profile about $U_0$ in the
frame moving with $v_0$. 

If we perform a stability analysis of a
planar front profile in one dimension, the same operator $L$ arises:
for the profile $U_0$ to be stable, its eigenvalues have to be
non-positive. However, since the F-KPP equation with $W=0$ is
translation invariant, $L$ always has  one eigenmode with eigenvalue
zero:  $\ds  U_0/\ds \zeta$ gives the change in the uniformly
translating profile $U_0$ if it is shifted over some small distance
$a$, $U_0(\zeta+a) = U_0(\zeta) + a \ds  U_0(\zeta)/\ds
\zeta + {\mathcal O}(a^2)$, and hence it is a zero eigenmode of $L$,
\begin{equation}
L \, \frac{\ds  U_0(\zeta)}{\ds  \zeta} = 0.
\end{equation}
Since the unknown function $U_1$ in (\ref{U1eq}) is acted on by $L$,
solvability theory (the ``Fredholm alternative'' in more mathematical
terms) requires that the sum of the other terms in (\ref{U1eq}) are orthogonal to
this zero mode. In more technical terms, they have to be orthogonal to
the ``left zero mode''  $\chi_0$ of $L$, the zero mode of the adjoint 
$L^+$ 
\begin{equation}
\label{L+operator}
L^+ \equiv \left( {{\ds ^2 }\over{\ds  \zeta^2}} - v_0
{{\ds }\over{\ds \zeta}} + \left. \frac{\partial
  f(U,0)}{\partial U}\right|_{U_0} \right)
\end{equation}
of $L$, which obeys $L^+\chi_0=0$.
Normally, it is nontrivial to find the zero mode of the adjoint of a
non-self-adjoint operator. In this particular case, $L$ can transformed into the
self-adjoint Schr\"odinger operator with a simple transformation, 
a trick that has often been exploited in the stability analysis
\cite{stability1,evs2,stability7,stability5,stability2,stability4,stability3}.
Here it suffices to note the result of this analysis, which can be
verified directly by substitution, namely that 
\begin{equation}
\chi_0(\zeta) = e^{v_0\zeta} \; \frac{\ds U_0}{\ds \zeta}.
\end{equation}

The solvability condition obtained from taking the inner product of
(\ref{U1eq}) with $\chi_0$ simply reads
\begin{equation}
  v_1 = - \kappa - \frac{\int_{-\infty}^{\infty} {\rm d} \zeta ~
    e^{v_0\zeta} \; {{\ds  U_0 }\over{\ds \zeta}}~ \left. {{\partial
       f(U_0,W)}\over{\partial W}}\right|_{W=0}  } {
    \int_{-\infty}^{\infty} {\rm d} \zeta ~ e^{v_0\zeta} \;\left( {{\ds 
        U_0}\over{\ds \zeta}}\right)^2} \;W. \label{solvab1}
\end{equation}
This equation explicitly gives the change in the velocity of the front
due to the curvature and the variation of the parameter $W$. The first
term is the one we already alluded to in the previous section on
spherical fronts  and gives the
 ``motion by mean curvature'' effect 
familiar from many models for coarsening \cite{allencahn,bray,gunton}. The second one
which gives the response to the changes in the parameter $W$ is of the
form that one typically encounters in a solvability type approach.

The above discussion has been based on  the assumption that
the  curvature and front velocity change on a slow time
scale. Underlying this idea is the assumption that the shape and
velocity of the front relax exponentially  on some time scale of
order unity to the shape and velocity of the planar front we perturb
about. One can show this more explicitly \cite{evs3} by a more careful analysis
in which the adiabatic assumption is not made immediately, and in
which $U_1$ is expanded explicitly in terms of the eigenmodes of the
stability operator $L$. The amplitude of these modes are than of order
$\tau_i/T$, where $\tau_i$ are the relaxation times of these modes  and
$T$ the time scale of the change of $v_1$ and $\kappa$. Thus,  if the
spectrum of the stability operator $L$ is ``gapped'', i.e., if the
relaxation times $\tau_i$ associated with the  stability operator are
bounded, the separation-of-time-scales-condition underlying the derivation of the above
expression is fulfilled. 

There are two ways in which the difference between pushed and pulled
fronts shows up in the context of the above analysis.
First of all, for pushed
fronts  the results summarized in section \ref{stabilityandothermechanisms} do indeed show that the
stability spectrum of a pushed front is gapped, and hence, pushed
fronts obey the conditions underlying the derivation of a moving
boundary approximation following the lines sketched above. For pulled
fronts, however, the conditions are {\em not} fulfilled. That a
separation of scales is not possible is immediately
 clear from the fact that the velocity and shape of a planar
pulled front profile relax as $1/t$ to their asymptotic values; within
the above analysis, this emerges from the fact that the spectrum of
the stability operator $L$ of a pulled front is not ``gapped'', i.e.,
the relaxation times $\tau_i$ of the eigenmodes are arbitrarily
large. 

Secondly, while for pushed fronts the solvability integrals in
expressions like (\ref{solvab1}) are finite
and well-defined, for pulled fronts they are infinite. For the F-KPP
equation discussed above, one can see this as follows. As we
discussed in section \ref{selectionutfs}, a pushed front falls off for large $\zeta$ as
$\exp[-\lambda_2 \zeta]$, where $\lambda_2$ is the second smallest
spatial decay rate. According to the results (\ref{vforfkpp}) for the F-KPP equation given in
the example at the end of section \ref{sectionexponential}, for the
F-KPP equation one has according to (\ref{pushedexpr})
\begin{equation}
\mbox{pushed front:} \hspace*{0.5cm} U_0(\zeta)  \stackrel{\zeta \gg 1}{\sim}  a_2 e^{-\lambda_2 \zeta} \hspace{1cm}\mbox{with}
~\lambda_2  = \frac{ v_0 + \sqrt{v^2_0 -4} }{2} ,
\end{equation}  
with $v_0= v^\dagger  >2$.
Hence  the integrand of the term in the denominator of
(\ref{solvab1}) converges as
\begin{equation}
e^{v_0\zeta} \left( \frac{\ds  U_0(\zeta)}{\ds
\zeta}\right)^2  \sim e^{(v_0 -2\lambda_2 )\zeta} \stackrel{\zeta \gg 1}{=}  e^{-
\sqrt{v_0^2-4} \, \zeta} \hspace*{1cm} (v^\dagger = v_0 > 2),
\end{equation} 
 so that the solvability integral is finite. Likewise, if $\partial
f(U_0,W) /\partial W \sim U_0 $ as $\zeta \to \infty$, the integral in
the numerator of the second term in (\ref{solvab1}) converges. Thus,
for a pushed front, the conditions underlying the derivation of a
moving boundary approximation are fulfilled, and the resulting
expressions for the dependence of the velocity of the front on
slowly varying parameters like $W$ are finite, as it should.

It is easy to see that for pulled fronts, on the other hand, the
integrand in the solvability integrals diverges for $\zeta\to \infty$:
indeed, according to Eq.~(\ref{Phi*eq}) for the asymptotic decay of
the pulled profile $U_0$, we have
\begin{equation}\mbox{pulled front:} \hspace*{0.5cm}
e^{v_0\zeta} \left( \frac{\ds U_0(\zeta)}{\ds 
\zeta}\right)^2  \, \stackrel{\zeta \gg 1}{\sim} \, \zeta^2  e^{(v_0
-2\lambda^* )\zeta}  = \zeta^2  ,
\end{equation} 
with $v^*=v_0=2$. {\em The divergence of the solvability integrals reflects
the fact that for pulled fronts  the dynamically important region is the
semi-infinite region  ahead of the front}, while for pushed fronts
or fronts between two stable states the dynamically important region
for the front dynamics  is the nonlinear front region
itself. At the same time, the divergence signals the
inapplicability  of  singular
perturbation theory for pulled fronts.\footnote{It has been proposed \cite{paquette3,paquette1}
to regularize the integrals by introducing a cutoff $\zeta_{\rm c}$
which is taken to infinity at the end of the calculation. Since then
only the most divergent terms survive, this procedure  reproduces
trivially   the change
in $v^*$ upon changing the linear term in $f$,  but it  does not
cure the inapplicability of a moving boundary approximation for pulled
fronts \cite{evs3}.}

In this section, we have simply illustrated the breakdown of the singular
perturbation theory approach to deriving a moving boundary
 or effective interface  approximation for pulled fronts by
considering the simple cases of the F-KPP equation. But from our
general result that pulled fronts always relax with the same slow
power law to
their asymptotic speed, it is clear that this conclusion holds more
generally. Indeed, even the divergence of the solvability integrals
for pulled fronts can be derived quite generally \cite{evs3}. 

\subsection{So what about patterns generated by pulled fronts?}

So far, we have focused our discussion on the fact that a moving
boundary approximation does not hold for a pulled front. What then are the
implications? To my knowledge, there is at present no clear answer to
this question. Clearly, the fact that moving pulled fronts in higher
dimensions can strictly speaking never be viewed as moving interfaces,
reflects that the important dynamics of a pulled front takes place
ahead of the front itself! This makes these fronts especially
sensitive to changes in the initial conditions or even to slight
changes in the dynamics\footnote{Pulled fronts are therefore also very
sensitive numerically to the scheme and grid size in the region ahead
of the front. Adaptive grid size methods that tend to put a dense grid
in the nonlinear front region may not converge to the right
result. This was pointed out to me by U. Ebert (private communication).} (see section
\ref{sectionstochasticfronts}). Although there are some indications
for this for the streamer patterns discussed in section
\ref{sectionstreamers}, this issue has to my knowledge not been
explored systematically. More fundamentally, the question is whether
in most cases making a moving boundary approximation in patterns
dominated by propagating pulled fronts is ``almost right'' in the
sense that it captures all the gross qualitative features, or whether
there are new effects which are fundamentally due to the pulled nature
of the fronts. To put it concretely: are there qualitative or
quantitative differences
in  the chaotic domain dynamics of the rotating Rayleigh-B\'enard
patterns of section \ref{sectionkupperslorz} between the regime where the fronts are pulled
and when they are pulled? Could it be that the behavior of the domain
correlation function, or the behavior of the 
correlation length and time shown in Fig.~\ref{figtucross2} for the
amplitude model, is fundamentally different  in the pulled and pushed
regimes? Likewise, does the pulled nature of a discharge front show up
in any fundamental way in the shape and dynamics of a propagating
streamer front, or does the difference between the superconductor
patterns of Figs.~\ref{figsuperconductors}{\em (a)} and
\ref{figsuperconductors}{\em (b)} reflect in a fundamental way  the fact
that in the first figure the front is pushed, while in the second one
it might be pulled?
 To my knowledge, these basic questions have not been studied at all.

One  interesting example of a qualitative change in the front behavior,
which is related to this issue, is the work of Kessler and Levine \cite{kessler2}, who
show that the finite particle cutoff effects discussed in section
\ref{finiteparticles} may render a stable
pulled front unstable. Likewise, the behavior of the average
occupation density of an ensemble of DLA fingers hinges on the fact
that the fronts are pushed rather then pulled, as one finds in a mean
field approximation --- see the introduction of section \ref{sectionstochasticfronts}.

\section{Fronts and emergence of  ``global modes''} \label{sectionglobalmodes}
Throughout most of this paper, we have focused our analysis on
systems which were infinite and homogeneous, so that their dynamical
equations were spatially invariant. Especially in fluid dynamics systems in
which there is an overall convective flow,  a crucial feature is often
the fact that the system is
finite and that there is a well-defined inlet  --- think of the 
Taylor-Couette cells with throughflow discussed  in section \ref{secfrontsnoise1}
--- or the fact that there is a well-defined obstacle behind which  the effective parameters are slowly 
varying in space --- think of the wake of the fluid flow behind a body
in the stream-wise direction, where as we discussed in section \ref{sectionvonkarman}
an instability can give rise to a vortex street. Investigations of
these issues  in the field of fluid dynamics have  given rise to
the concept  of ``global mode'' \cite{chomaz2,huerre0,hunt,ledizes}.

\begin{figure}[t]
\begin{center}
\epsfig{figure=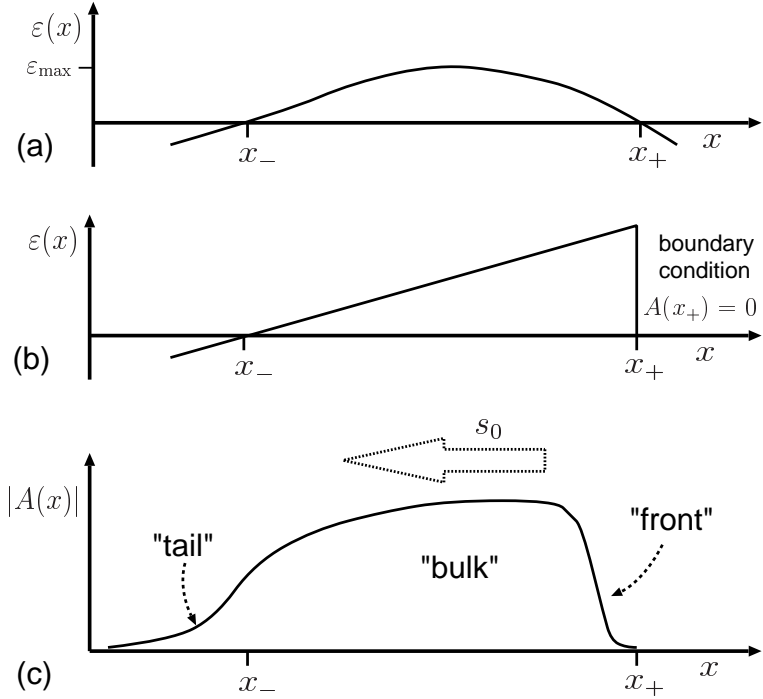,width=0.75\linewidth} 
\end{center}
\caption[]{(a), (b) Examples  of a control parameter $\varepsilon(x)$
which varies in space, and for which the system is locally unstable
only in some finite interval $[x_-,x_+]$.  Case (b) is supplemented
with a boundary condition $A(x_+)=0$. (c) Qualitative sketch of a ``nonlinear
global mode'' in the regime where $\varepsilon(x)$ varies as sketched
in (b) and in which the  interval $[x_-,x_+]$ is large enough that different scaling regimes exist. The steep
region on the right is close to a front solution in a spatially
uniform system. } \label{figglobalmode}
\end{figure}

The idea of a ``global mode'' is essentially the following. Suppose we
consider a system whose properties are slowly varying as a function of
the spatial coordinate $x$ (in the case of the flow behind the
cylinder of section \ref{sectionvonkarman} this would be coordinate in
the stream-wise direction). As a result of this, the dimensionless
control parameter $\varepsilon (x) $ is then a slowly varying function of
$x$. The dimensionless control parameter $\varepsilon$ is defined as the parameter
which marks  that when the system is {\em spatially homogeneous}, the basic
state of the system
exhibits an instability for $\varepsilon >0$, while it is
stable for $\varepsilon <0$. For pattern forming instabilities in
homogeneous systems, the control parameter $\varepsilon$  appears
as the prefactor of the linear term in an amplitude description, see
Eqs.~(\ref{cubiccgl}) and (\ref{quinticcgl}).  When $\varepsilon(x)$
is a slowly spatially varying function which, as sketched in
Fig.~\ref{figglobalmode}{\em (a),(b)},  is
only positive  in some  finite range of coordinates $x$, then we can
think of the system as being locally unstable. Nevertheless,  the
system only turns globally unstable once  the  interval
$[x_-,x_+]$ where $\varepsilon(x)>0$ is large enough: it is clearly
not sufficient that the system is locally unstable ($\varepsilon(x) >0$
for some $x$) ---  instability only sets in when a {\em global mode} goes
unstable.  While for the translationally invariant system the unstable modes are
Fourier modes,  the modes which have been termed  ``global modes''
and which govern the instability of
 inhomogeneous systems are essentially localized to the spatial range
$[x_-,x_+]$ where $\varepsilon(x)>0$. The global modes that govern the
linear instability problem in spatially varying systems have been
studied in particular for the cubic CGL equation (\ref{cubiccgl}) for
general $\varepsilon(x)$ in the limit that the range over which
$\varepsilon(x)$ varies is much larger than the wavelength of the
pattern, so that a WKB analysis can be employed
\cite{chomaz2,huerre0,hunt,ledizes}. In the case sketched in
Fig.~\ref{figglobalmode}{\em (b)}, which might be a reasonable model
for the instability in the wake of a cylinder of section \ref{sectionvonkarman}, the linear
eigenmodes can be calculated explicitly. The nonlinear behavior of such
global modes has in recent years also been studied
\cite{couairon2,couairon3,pier4,pier3,pier1,proctor}. We will  only highlight
here  those aspects of global modes that connect immediately  with the
general theme of this paper.

\subsection{A front in the presence of an overall convective term and
a zero boundary condition at a fixed position}\label{sectionfixedboundary}

As a model problem to investigate the influence of a fixed boundary in
a system with a convective instability, Couairon and Chomaz
\cite{couairon2} have analyzed the F-KPP equation (or Ginzburg-Landau
equation for a real variable)
\begin{equation}
\partial_t u (x,t)  - s_0 \, \partial_x u = \partial_x^2 u(x,t) + f(u),
~~~~~ f(u) = \varepsilon u  + u^3 -u^5,\label{fkppwithU}
\end{equation}
in the semi-infinite axis $(-\infty, L]$, with the boundary condition
\begin{equation}
u(L,t) = 0.
\end{equation}
Note that the second term on the left hand side models an overall
convection with velocity $s_0$ {\em to the left}.\footnote{Like in our
earlier sections, we use the
notation $s_0$ instead of $U_0$ of \cite{couairon2} to denote the
overall group velocity; moreover,  unlike in \cite{couairon2} the overall convection is to
the left rather than to the right.}  As we already
discussed in the example at the end of section
\ref{sectionconvversusabs},  for an infinite system this term can simply
be transformed away by going to a frame moving with velocity $s_0$ to
the left; here, however, the boundary condition  at the fixed position
$x=L$ does not
allow us to do so. The specific choice  for the function $f(u)$ in
(\ref{fkppwithU}) has the advantage that explicit calculations can be
done, since in the pushed regime $\varepsilon < 3/4$ the pushed front
solutions are known explicitly (see the discussion of the ``reduction of
order method''  at the end of section \ref{selectionutfs}), but is
not needed for the  general scaling  arguments below.

Let $v_{\rm sel} $ be the selected front speed in the infinite system for
$s_0=0$. If the advection velocity $s_0$ is larger than the selected
front speed, $s_0 > v_{\rm sel} $,  a front naturally retracts to the
left and  the system is nonlinearly
convectively unstable, as discussed in  section
\ref{sectionnonlinearconv2abs}. 
We then expect  that the asymptotic state is simply $u=0$; the
explicit analysis \cite{couairon2} confirms this. Of course, as we
already saw in section \ref{secfrontsnoise1},
in this regime $s_0>v_{\rm sel}$, the system remains very sensitive to
perturbations. For example, a  finite forcing  at
$x=L$  gives rise to a finite $u$ front-type profile \cite{couairon2}, while  a
small amount of noise at $x=L$ gives rise to incoherent front-type patterns.   The phase
diagram of convective systems with noise has been worked out in
particular by Proctor {\em et al.} \cite{proctor}.

If $s_0 < v_{\rm sel}$ the infinite system is nonlinearly absolutely
unstable: virtually any initial perturbation of the state $u=0$ for
$x<L$ will evolve into a front that propagates to the right, till the
boundary condition at $x=L$ makes itself felt. A full analysis in this
regime confirms our intuitive expectation that the
long-time  asymptotic state is a stationary front-type profile
locked near the right boundary. 

When the advection speed is just slightly less than the selected front
speed in the infinite system, for $0 < v_{\rm sel}- s_0 \ll 1$, we expect that the final
stationary front profile  remains  very close to the profile with
speed $s_0$ just below $v_{\rm sel}$  for  essentially all $u>0$, until a
crossover behavior occurs for very small $u$ of  $ {\mathcal O}(\delta)  \ll 1$. 

Let us first make this more precise for the pushed regime, i.e., when
$v_{\rm sel}= v^\dagger$. The line of analysis we will present below
generalizes many of the results of \cite{couairon2,couairon3} for the specific case considered
there. As we noted in section \ref{selectionutfs},
the pushed front profile  is precisely the one for which the
coefficient $a_1$ in the asymptotic expression (\ref{decayformula})
vanishes: $a_1(v^\dagger)=0$. Since $a_1$ will generally go through
zero linearly in $v$, the uniformly translating front profile
$U_{s_0}(\zeta)$ with
velocity $s_0$ in the infinite system will to  a good approximation in
the leading edge be
given by 
\begin{eqnarray} U_{s_0}(\zeta\gg 1)  &\approx &
a_1^\prime \, (- \Delta v) \, e^{-\lambda_1 \zeta} + a_2 (v^\dagger)
e^{-\lambda_2\zeta}  \nonumber , \\ &=& 
a_2(v^\dagger) e^{-\lambda_2\zeta} \left( 1 + \frac{ |a^\prime_1| \,
\Delta v 
}{a_2(v^\dagger) } \; e^{-(\lambda_1-\lambda_2)\zeta}\right),\label{withuapprox}
\end{eqnarray}
where $a_1^\prime= da_1/dv$ and where
\begin{equation}
\Delta v \equiv v_{\rm sel}- s_0 = v^\dagger -s_0
\end{equation}
is assumed to be positive but small.
As we explained above, the stationary front solution in the presence of the
boundary condition $u=0$ at $x=L$ and the convective term
$-s_0\, \partial_x u$ in the dynamical equation (\ref{fkppwithU}) will  be
the one given by (\ref{withuapprox}) on the outer scale till it
crosses over to different behavior  when $u$ becomes of order
$\delta$. In this crossover range, clearly the overall term $a_2
e^{-\lambda_2\zeta}$ must be of order $\delta$, while the perturbative
correction term will be comparable to the dominant one. The latter
implies that the exponential  term 
between parentheses in the second line must be of order
unity in the crossover region. Upon eliminating $\zeta$ from the two expressions obtained from
these requirements, we obtain the scaling relation
\begin{equation}
\delta  \sim ( \Delta v)^\beta ~~~~\mbox{with} ~~\beta = \frac{\lambda_2}{\lambda_2-\lambda_1}.
\label{betascaling}
\end{equation}
According to this expression the scaling exponent $\beta$ depends on
the properties of the pushed front through the spatial
decay rates  $\lambda_1(v^\dagger)$ and $\lambda_2(v^\dagger)$. Since
outside the crossover region we can to a good approximation use the
term proportional to $e^{-\lambda_2\zeta}$ for the front profile, 
 the distance between the point,  where the front reaches a
finite value, and the point, where it is pinned by the boundary
condition $u=0$, will for $\Delta v \to 0$ scale as
\begin{equation}
\frac{1}{\lambda_2}  \ln \delta  \simeq \frac{\beta}{\lambda_2}  \ln
\Delta v.
\end{equation}

The above scaling behavior (\ref{betascaling}) has also been obtained by Kessler et al. \cite{kessler},
and as they point out this formula gives the scaling of the effect of a finite particle
cutoff on the front speed in the pushed regime (see section \ref{sectionstochasticfronts}).

The scaling behavior of a pulled front is very different. In fact, as
it turns out the perturbation theory about a uniformly translating pulled front in the case
that the front is ``cut off'' at a small value $\delta$ of the field
by a boundary condition that makes the dynamical variable vanish, was
also developed by Brunet and Derrida \cite{derrida} in the context of the stochastic
fronts. We will discuss the rather peculiar logarithmic scaling
dependence in more detail in section \ref{finiteparticles};  the final result for the
relation between $\Delta v = v^*-s_0$ and $\delta$ obtained from Eq~(\ref{bdresult}) below is
\begin{equation}
\Delta v \simeq \frac{ \pi^2 D \lambda^* }{\ln^2 \delta} \hspace*{0.8cm}
\Longleftrightarrow \hspace*{0.8cm}
\delta \simeq \exp \left( - \pi \sqrt{\frac{D \lambda^*}{\Delta v}}
\right), \label{vdeltarelation}
\end{equation}
where as before in this paper $D$ is the effective diffusion
coefficient  which is given in terms  of the linear dispersion
relation by  Eq.~(\ref{lambda*})  and which according  to
(\ref{Dandveq}) is related to the curvature of $v_{\rm env}(\lambda)$
at the minimum $v^*$. Since the front profile rises   approximately as
$\delta e^{-\lambda^*(\zeta-L)}$ away from the boundary in the outer region, the distance between
the right boundary and a point where $u$ is finite will to dominant order
diverge  as $|\ln \delta|/\lambda^*$, i.e., as
\begin{equation}
\pi \sqrt{ \frac{D }{\lambda^*\, \Delta v}} .
\end{equation}
The detailed matching analysis in \cite{couairon2} and in
\cite{couairon3}  is fully consistent
with the above  scaling relations and also yields the prefactors for the
specific nonlinearity $f(u)$ used in (\ref{fkppwithU}). On the other hand, our
arguments show  that these scaling results hold more generally for uniformly translating 
pulled and  pushed fronts which are pinned by a zero boundary condition.  In fact, the arguments
 apply to coherent pattern forming fronts too; indeed the
result of \cite{couairon3} for pulled fronts in the quintic CGL
equation (\ref{quinticcgl}) are consistent with the above dominant scaling
behavior as well.

\subsection{Fronts in nonlinear global modes with slowly varying
$\varepsilon(x)$}\label{globalmodeseps} 

Just above the threshold to instability of a global mode, a
weakly nonlinear description in terms of the single most unstable mode
suffices. However, as soon as  the interval $[x_-,x_+]$ and
$\varepsilon(x)$ are  sufficiently large, it is more natural to think
of the the nonlinear structures as being 
decomposed  into various regions, as sketched in
Fig.~\ref{figglobalmode}{\em (c)}: a front region where the amplitude
rises quickly, a bulk region where a well-developed pattern exists
whose properties follow the local variation of $\varepsilon(x)$ and a
tail region where the pattern amplitude decays back to zero. Most of
the analysis of such type of structures has been done in terms of a
cubic or quintic
CGL equation with a slowly spatially varying $\varepsilon$. To
emphasize the separation of scales, let us denote the slow scale on
which $\varepsilon$ is varying by $X$. Since the
CGL equation  is of second order in its spatial partial derivatives,
there are always two roots $q$ corresponding with a given amplitude
$a$ of a phase winding solution
of the form $A=a\,e^{iqx-i\Omega t}$  of
the homogeneous (translationally invariant) CGL equation ---
see sections \ref{sectioncglcubic} and \ref{sectioncglquintic} and in
particular Eq.~(\ref{phasewindingvalues}). Hence, when
$\varepsilon$ varies on a slow scales $X$ much larger than $q^{-1}$,
it becomes possible to treat the two roots $q_\pm$ as slow variables
$q_\pm (X)$  whose variation can be treated  by  a nonlinear WKB type
analysis --- as is well known \cite{ch,fauve,hakim,vhhvs}, the phase is the slow
variable of the translation invariant CGL equation, while the
amplitude is slaved adiabatically to the phase dynamics.

Intuitively, we expect that when we consider structures of the type
sketched in Fig.~\ref{figglobalmode}{\em (c)}, the region on the right
should be close to a front-type solution of the type we have discussed
in this paper for translationally invariant systems, and that depending on the dynamical equation
under investigation 
these front-like regions could be either pulled-like or
pushed-like. How  does this emerge in the above type of  WKB  analysis? Very
simply: if the translation invariant equation admits pulled front
solutions, the existence of a front type region is signaled by the
two roots $q_\pm (X)$ both approaching the linear spreading point values
$k^*$ in some region \cite{pier4,pier3,pier1}, while if the translation invariant equation
admits pushed fronts \cite{couairon4} there is a region  where the profile is dominated
by one of the roots which is locally close to the mode $k_2$ of the
underlying pushed front  in accord with the defining property
(\ref{pushedcoherent}) of a pushed coherent front solution.

For further discussion of these scenarios and of the  detailed
scaling analysis of the width of the various regions we 
refer to the  literature. In closing it is also useful to point out
that while the idea of a global mode has emerged mostly in the fluid
dynamics literature in recent years, 
the problem of pattern selection through a slowly varying control
parameter --- often called a ``ramp'' --- was already considered
 in the 1980ies both from a more general perspective and for
various specific problems
\cite{eagles,crosskramer,malomedramp,misbahramp,riecke,roth}. Not
surprisingly, there are strong similarities between these approaches
and those mentioned above. 

\section{Elements of Stochastic Fronts}\label{sectionstochasticfronts}
We have so far limited the  theoretical analysis
of front propagation into unstable states to the case in which the front
dynamics is governed by deterministic dynamical equations. In this
last section on special topics we finally briefly address  some elements
of the propagation of stochastic fronts into an unstable state. 

We do want to stress that we will only touch on a few selected topics
that tie in with this review's main emphasis on deterministic fronts.
First of all, in reading this section, one should keep in mind
that  the study of stochastic fronts is a vast  field in itself, to
which we can not do full justice. Our choice will be to focus on those
issues  that are connected with the question of how the deterministic
limit can be approached, and on how the front velocity and front
diffusion coefficient behave in that limit. Secondly, 
as is well known, the critical scaling properties of fluctuating
growing interfaces are an active research subject in itself
\cite{barabasi,halpin-healy,krug2}. We will not directly touch on
these interesting issues, except for a brief discussion at the end of
this section of the question whether fluctuating pulled fronts are in
a different universality class from the normal type of fluctuating
interfaces. 

In section \ref{sectionmbabreakdown} we have seen that while singular
perturbation theory works fine for weakly curved pushed fronts or for
one-dimensional fronts in a slowly spatially varying external field,
it breaks down for pulled fronts. We saw that this implied that while
the motion of a weakly curved pushed front can be mapped onto an
effective sharp interface model, such an interfacial description does
not apply to fluctuating pulled fronts. These findings carry over to
fluctuating fronts propagating into an unstable state: when these are
pushed, their asymptotic  scaling properties are simply those of an
appropriate interface model. For pulled fronts, on the other hand, the
situation is more complicated: as we shall discuss, fluctuating
variants of fronts which in the mean field limit are pulled are very
sensitive to noise or to the cutoff effects introduced inherently in
simulations with discrete particles executing stochastic jumps. We
will focus our discussion on these effects which are special to
fluctuating fronts propagating into a linearly unstable state. A
 more detailed review of several of these issues is given by Panja \cite{debreview}.

It is amusing to note that 
on hindsight, it is fair to say that  the first clear evidence of the  strong
finite-particle cutoff effects  on fluctuating fronts already surfaced
in 1991 \cite{brener2} in the context of a mean-field analysis of Diffusion Limited
Aggregation (DLA) fingers. It had been found empirically that when one
grows many DLA fingers in a channel, the level lines of the average
occupation density of these fingers is very close to the shape of
viscous fingers. When researchers made attempts to study this issue
with the aid of the  mean-field equations for this density, they ran
into the problem that the naive theory, which was based on the
existence of pulled fronts in these equations \cite{nauenberg}, leads
to divergences in two dimensions. Brener {\em
et al.} \cite{brener2} then showed that if the nonlinear growth term
is modified, the fronts behaved properly and the theory reproduced the
observed behavior. The broader implications of this empirical trick
seem not to have been realized at the time. In hindsight, 
this was the first sign that in a finite-particle model like DLA, the
fact that there is no growth if there is no particle at all makes the
front intrinsically pushed and that this is crucial for the model to
be well-behaved. This will be a recurrent theme in our discussion.

In the next section, we will first discuss the slow convergence to the
asymptotic spreading speed $v^*$ as a function of $N$, the average
number of particles in the saturated state behind the front in a stochastic discrete particle
model. Then we briefly review some of the field-theoretic formulations
of stochastic fronts, and we close with a short discussion of the
implications or our findings for the 
asymptotic scaling properties of fluctuating fronts in more than one
dimension.

  \subsection{Pulled fronts as limiting fronts in diffusing particle models: strong cutoff effects}\label{finiteparticles}
The most natural deterministic dynamical equation that arises in the
mean field limit from a stochastic model is the F-KPP equation. After
all, this equation embodies the two essential 
ingredients of a lattice model with discrete particles $A$ which make
diffusive hops to neighboring sites, and which have some reaction of
the type $A \rightleftharpoons 2A$. The latter type of reaction would
after a proper scaling lead to a nonlinearity of the type
$f(u)=u-u^2$ in the F-KPP equation (\ref{fkpp}). In line with  the
literature on this subject we will in this section take the F-KPP equation as our
reference deterministic front equation, but one should keep in mind
that all our conclusions hold more generally. 

One can arrive at the full continuum F-KPP equation from a stochastic
lattice model with discrete particles by taking the limit in which the
length $\ell_D$ which a particle diffuses before giving birth to
another particle, becomes much larger than the mean inter-particle
spacing. A very nice pedagogical discussion of this limit can be found
in \cite{moro2}.  Note that while the convergence to
a F-KPP type equation in such limits is rather obvious, the precise
way in which the front diffusion coefficient $D_{\rm front}$ vanishes
 is less trivial --- stochastic particle
models generally give rise to fronts which themselves exhibit a
stochastic wandering around the average front displacement,  but
$D_{\rm front}$ generally depends in an intricate  way on the total
front structure (see also below). We will not review the
details of such type of scaling limits here as they essentially rely
on advanced methods of probability theory, and refer to the
literature \cite{ba2,bramson2,demasi1,doe1,doe2,kerstein1,kerstein2,moro2} for
details.

The approach to the mean field limit that will interest us here in
particular is the case in which we allow an arbitrary number of
particles per lattice site, but in which $N$, the average number of
particles per lattice site (or correlation length) in the region
behind the front,  becomes large. Again, as $N$ increases, fluctuation
effects are suppressed and we expect  a mean field front equation
to emerge from the underlying stochastic model as $N\to \infty$. Note
that in this case we keep $D_{\rm part}$ fixed (and often of order
unity), so we can {\em not} expect the limit to be governed by a  continuum
F-KPP equation: The mean field limit will be an equation for a
{\em continuum} density variable $u$ on a {\em discrete}  lattice. Luckily, this is no
problem for our discussion as the concepts of pulled and pushed fronts
are equally well defined for difference equations, as we discussed in
section \ref{sectionmoregeneral}. 

Even though earlier simulations \cite{noise4} had already given hints of a very slow convergence
of the speed with $N$ to its asymptotic value, Brunet and Derrida
\cite{derrida} were the first to clearly identify this issue and to
recognize that the $N\to \infty$ convergence to the  mean field limit  is
very different depending on whether the limiting front is pulled or
pushed. For pushed fronts, the nonlinear front region where the
particle density $ u$ is finite determines essentially the front
speed. This is also reflected in the fact that the integrands of the solvability
integrals that arise in the singular perturbation theory of section \ref{sectionmbabreakdown}
converge exponentially.    In this finite-density region the relative importance of fluctuations 
decreases rapidly with $N$, typically as $N^{-1/2}$. Hence the
convergence to the mean field limit  is power law fast if the
asymptotic front is pushed. For pulled fronts, on the other hand, the
situation is different: The fact that they are pulled by the growth
and spreading of arbitrarily small perturbations about the unstable
state implies that they are very sensitive to small changes in the
dynamics in the region of very small particle densities. {\em At the very
tip of any such front is always a region where there are  very
few particles per lattice site or correlation length, and where the
mean field approximation breaks down: Pulled fronts are very
sensitive to the unavoidable existence of this tip region where the
particle density is of order $1/N$.}

Following Brunet and Derrida \cite{derrida}, we can understand the
effect of this on the front speed as follows. For F-KPP type
equations, the asymptotic front profiles are  uniformly
translating, i.e., of the type $u(\zeta)=u(x-vt)$.\footnote{Remember
that as in section \ref{sectiontypesoffronts}, we use the co-moving coordinate $\zeta=x-vt$ for
arbitrary velocities, while $\xi=x-v^*t$.} This is true even for
pulled fronts on a lattice. As we have discussed in section
\ref{sectionexponential}, following Eq.~(\ref{uniform1}, the
uniformly translating fronts with velocity $v>v^*$ are monotonic and
fall off simply as $e^{-\lambda \zeta}$ (in other words, $k_{\rm r}=0$), but for $v<
v^* $ the solutions in the leading edge are of the form 
\begin{equation}
u_{v_{\rm N}} \approx {\rm Re} \, c ~e^{-\lambda\zeta+ ik_{\rm r}\zeta} = c_1 e^{-\lambda \zeta}
\sin \left( k_{\rm r}\zeta + c_2\right), \label{vbelowv*}
\end{equation}
where $c_1 $ and $c_2$ are real coefficients. 

Let us now consider the case in which as $N\to \infty$ the front is
pulled, and normalize the density field $u$ such that $u\approx 1$ in
the region behind the front. Let us furthermore assume that the stochastic model we
consider is such that in the very tip region of the fluctuating front
--- we refer to  the region where there are only  very few
particles per site so that the density field $u$ is of ${\mathcal
O}(1/N)$ at the {\em tip} of the front --- the stochastic rules are
such that the growth is significantly suppressed.  This is what one
typically expects intuitively, as there needs to be at least one
particle in order to have any growth at all. In the case of
suppression of the growth, one expects that 
the asymptotic average front velocity $v_{\rm N}$ approaches  $v^*$
from below as $N\to \infty$, and
hence that for finite but large $N$ the profile in  most of the
leading edge is given to good approximation by Eq.~(\ref{vbelowv*})
above. 

For uniformly translating profiles, we already saw in section
\ref{sectionexponential}   that by 
expanding about the linear spreading point we can for $v_{\rm N}<v^*$ express
 $k_{\rm r}$ as a function of the velocity as
\begin{equation}
k_{\rm r} \approx \sqrt{\frac{\lambda^* (v^*-v_{\rm N})}{D}}, \label{kreq}
\end{equation}
where we have used Eqs.~(\ref{velocityexpansion}) and (\ref{Dandveq}) 
to write the result in terms of  the effective diffusion
coefficient $D$ associated with the
linear spreading point. The variation of the steepness $\lambda$ is of 
higher order in $v^*-v_{\rm N}$ and to obtain the leading behavior we
can therefore take $\lambda\approx \lambda^*$; likewise we can in the
analysis below replace the co-moving coordinate $\zeta$ by
$\xi=x-v^*t$.

\begin{figure}[t]
\begin{center}
\epsfig{figure=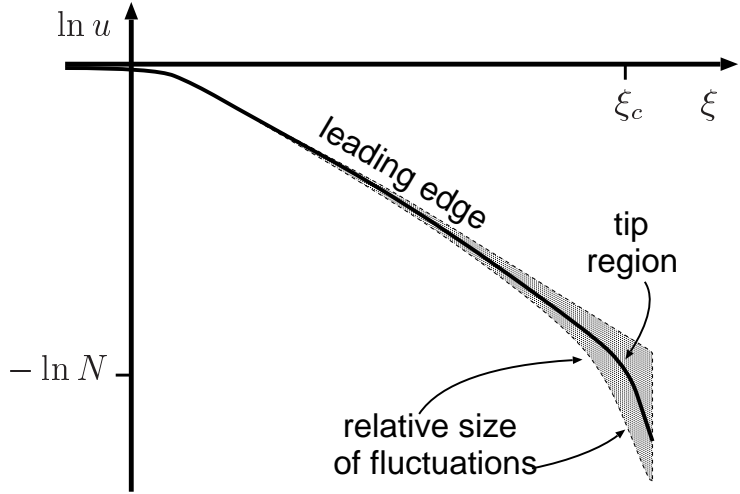,width=0.75\linewidth} 
\end{center}
\caption[]{Schematic sketch of the front profile for a stochastic particle model for very large $N$.
Along the vertical axis, we plot the logarithm of the density $u$ as a function of the co-moving
spatial coordinate $\xi$
plotted along the horizontal axis.  The leading edge of the front, where $u\gg 1$, corresponds to
the behavior on the right for positive  $\xi$. The mean field profile
given by (\ref{secondueq})  is plotted as a full line, while the dotted
area gives a qualitative idea of the importance 
of fluctuation effects. For very large $N$, the fluctuation effects
are small and can be treated perturbatively throughout most of the
leading edge, except near the tip where $\xi_c\simeq (\lambda^*)^{-1}
\ln N$. As the increase in slope of the full line in this region
indicates, the finite particle effects suppress the growth. }  \label{figfluctuatingfront}
\end{figure}

Let us position this coordinate system $\xi$ suitably so that the
nonlinear front region is roughly near $\xi\approx 0$. As $N$ becomes
large, the overall exponential behavior $e^{-\lambda^*\xi}$ of the density field $u$ then
implies that the crossover region where $u = {\mathcal O}(1/N)$ is
located at
\begin{equation}
\xi_c \simeq \frac{1}{\lambda^*} \ln N. \label{xic}
\end{equation}
Moreover, as Fig.~\ref{figfluctuatingfront} illustrates, upon increasing
$N$ the leading edge of the asymptotic profile increases
logarithmically in size, as it spans the region from $\xi$ small to
$\xi_c$ defined above. On the other hand, the region where
fluctuations are important is roughly {\em confined to a range of
  finite size near the tip}. E.g., a distance of order $5/\lambda^*$
behind the tip the typical number of particles per lattice site or
correlation length is already of order 100, so their effect is already 
relatively small. Stated more clearly, for any finite $N$ the leading edge
exhibits a crossover at $\xi_c\simeq (\lambda^*)^{-1} \ln N$ to  a
fluctuation-dominated regime but as $N\to \infty$ the fraction of the
leading edge where the fluctuating front profile is accurately
approximated by the mean field expression (\ref{vbelowv*}) approaches 100\%!

The asymptotic large-$N$ correction to the velocity now follows
directly from simple matching arguments \cite{derrida}. For $\xi $ positive but
small, at the left side of the leading edge, we know that the profile
should converge to the $\xi e^{-\lambda^*\xi}$ behavior as $v_{\rm N}
\uparrow v^*$ and hence $k_{\rm r} \to 0$. This implies  $c_2=0$
and $c_1 \simeq 1/k_{\rm r}$, so that
\begin{equation}
u_{v_{\rm N}} \simeq \frac{1}{k_r} \sin (k_{\rm r}\xi)
\, e^{-\lambda^*\xi}~~~~~~~~~(k_{\rm r} \to 0) . \label{secondueq}
\end{equation}
Let us now look at the crossover region at $\xi_c$;
Eq.~(\ref{secondueq}) immediately implies 
\begin{equation}
\left. \frac{\ds u_{v_{\rm N}}/\ds \xi }{u_{v_{\rm N}}} \right|_{\xi_c}
\simeq  k_r \, \mbox{cotg} (k_r\xi_c) - \lambda^*. \label{uratio}
\end{equation}
As we stated above, near and beyond the crossover scale $\xi_c$ the
deviations from the mean field expressions are significant, say of order
unity. In order that the above expression in the leading edge is
consistent with this, {\em the matching condition we have to impose} is that 
the right hand side of the above expression (\ref{uratio}) deviates
more than infinitesimally from the value $1/\xi_c-\lambda^*$ it has in 
the limit $k_r\to 0$, $\xi_c $ fixed. This is only possible if the
argument of the cotangent approaches $\pi$ in this
limit,\footnote{This is the smallest value of $k_r$ satisfying this
  constraint. Higher values are associated with profiles which have
  oscillations in the leading edge itself, and therefore rejected.}
hence if  in dominant order
\begin{equation}
k_r \simeq \frac{\pi \lambda^*}{\ln N} ,
\end{equation}
so that according to (\ref{kreq}) we finally have 
\begin{equation}
v_{\rm N} \simeq v^* - \frac{\pi^2 \, \lambda^* D }{\ln^2 N}
\label{bdresult} + \cdots.
\end{equation}
This is the central result for the dominant large-$N$ correction that
was first obtained by Brunet and Derrida \cite{derrida}.\footnote{It
  is amusing to note that  from a technical point of view the same result emerged
the same year in the work of  Chomaz and Couairon \cite{couairon2} 
on global modes from a very different perspective --- see
Eq.~(\ref{vdeltarelation}). } As our discussion shows, this expression
holds generally  for fluctuating pulled fronts which converge to a
pulled front as $N\to \infty$ and for which  the
fluctuation-dominated region is in the large-$N$ limit confined to a
finite range near the tip.\footnote{We shall see in section \ref{sectionfluctuatinguniclasses}
 below that when the strength
  of fluctuations scales with the density field itself (rather than
  with the square root of it), then the fluctuations throughout the
  whole leading edge contribute. The fluctuation behavior of such
  fronts is in a different universality class.} Just like the power
law convergence to $v^*$ of a deterministic front is universal and
independent of the details of the model (see sections
\ref{sectionunirelsimple} and \ref{sectionuniversalrel}), so is this 
large-$N$ convergence: like the expressions for the temporal
convergence it only depends on $\lambda^*$ and $D$, quantities which 
are completely determined by the linearized 
equations of the deterministic (mean field) limit! Note also that the
universality is intimately related to the logarithmic $N$-dependence:
the {\em only input condition} for the derivation is the fact that the 
region where the strength of the fluctuations is non-vanishing  {\em
  remains finite} in the
limit $N\to\infty$. Since $\ln N/a\simeq \ln N$ in dominant
order, this implies that the asymptotic behavior is independent of the 
precise behavior in the fluctuation region  --- it does not matter
whether the cutoff is effectively at $100/N$ or at $1/N$ --- the only
thing that matters is that it scales as $N^{-1}$!

The above logarithmic correction to the front velocity contrasts with the power
law dependence $\delta v \sim N^{-1/ \beta}$ derived for the the pushed regime in Eq.~(\ref{betascaling}) above.  This scaling behavior 
was first pointed out and verified by Kessler {\em et al.} \cite{kessler}.

\begin{figure}[t]
\begin{center}
\epsfig{figure=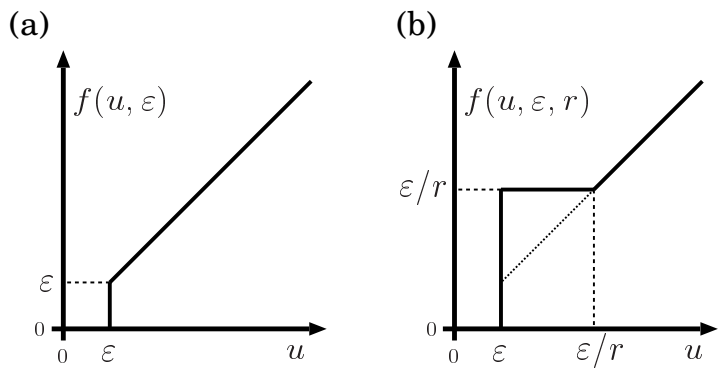,width=0.75\linewidth} 
\end{center}
\caption[]{ (a) The nonlinear function $f(u,\varepsilon)$ used by
Brunet and Derrida \cite{derrida} in the F-KPP equation to study the effect of a
cutoff on the growth function at $\varepsilon=1/N$. Since $\varepsilon 
\ll 1$, the nonlinear behavior of $f$ for values of $u$ of ${\mathcal
  O}(1)$ is not visible on this scale. (b) The nonlinear
function $f(u,\varepsilon,r)$ analyzed by Panja {\em et al.}
\cite{deb4} (thick line). In a small
interval of $u$ of order $\varepsilon$ the growth is enhanced. For
$r < r_{\rm c}= 0.283833 $ the fronts are pushed in the limit
$\varepsilon \to 0$. Stochastic simulations confirm this effect. } \label{figfbrunetandours} 
\end{figure}

To perform an explicit calculation, one of course needs an explicit
model. In their paper \cite{derrida}, Brunet and Derrida did the detailed matching
analysis  for the deterministic F-KPP equation (\ref{fkpp}) with a
cutoff for the nonlinear growth function $f$ of the form sketched in
Fig.~\ref{figfbrunetandours}{\em (a)}, i.e., 
$f(u,\varepsilon)= \Theta(|u|-\varepsilon)u(1-u^n)$, where
$\varepsilon =1/N$.

This  deterministic front model has the advantage that many aspects
can be calculated explicitly, and that it helps one to get  intuitive
insight into the behavior. For instance, since there is no growth for
$|u|<\varepsilon$  {\em arbitrarily small} perturbations about the
state $u=0$ {\em do not grow}:  a Gaussian initial condition with
maximum less than $\varepsilon=1/N$ just spreads out diffusively,
without growing in amplitude. {\em Strictly speaking}, therefore, the state
$u=0$ in this model is {\em not linearly unstable} in the sense we have
defined it. An immediate consequence of this is that for $\varepsilon
\neq 0$ fronts in this equation can not be of the pulled
type\footnote{Keep in mind that this is only true for this particular
  model. If one takes instead a modified models in which $f(u) = \mu u$ 
  for $|u|<\varepsilon$ and with $\mu < 1$, then the state $u=0$ is linearly unstable,
  but strictly speaking the linear spreading speed is then equal to
  $2\sqrt{\mu}$. Since this is  less than the spreading speed to which the
  model converges in the limit $\varepsilon\to 0$, the velocity
  correction is than still given by (\ref{bdresult}). Also from this point of view we conclude that
the dynamically relevant front is a pushed one. The  details of
  the matching analysis and of the calculation of the relaxation  spectrum are different in 
  this case, but the final conclusions are the unchanged.} --- they 
are  pushed!  In line with the fact that the stability spectrum
of a pulled front is generally continuous, while it is discrete for a
pushed front, one can calculate the relaxation times $\tau_m$ of the
slowest relaxation modes of the fronts in the above model to find \cite{kessler,deb2}
\begin{equation}
\frac{1}{\tau_m } \simeq \frac{\pi^2 [(m+1)^2 -1]}{\ln^2 N}\label{tauexp},
~~~~~~~~~~(m\geq 1).
\end{equation}
The modes associated with this slow relaxation extend throughout the
whole leading edge; that they are also relevant for
fluctuating fronts is for  the same reason that for $N\to \infty$ the dominant
velocity correction is determined by the matching of the mean field  
front profile to a finite cross-over region where fluctuations are important.
Even though the relaxation times diverge as $N\to\infty$ the
logarithmic divergence is so slow that for many practical values of
$N$ the relaxation time is still relatively short. If so, convergence
to the asymptotic speed $v_{\rm N}$ is relatively fast \cite{kessler}, and for all
practical purposes the front behaves like a pushed front.

The prediction (\ref{bdresult}) for the asymptotic large-$N$ velocity correction
of fluctuating fronts has been corroborated\footnote{Kessler {\em et
al.} \cite{kessler} have reported a discrepancy of a factor of 2  in the
prefactor of the $1/\ln^2 N$ correction, but the simulations by Panja \cite{deb3}
do not find evidence for such a large deviation. } by simulations of  a
variety of stochastic lattice models
\cite{noise4,derrida,bd2,derrida3,kessler,deb3,vanzon}; also
field-theoretic arguments  are in agreement with  this prediction \cite{levine} for
reasons we shall come to below in the next section.

For any finite $N$, a front in a stochastic model also shows a
stochastic wandering about the average position. This diffusive
wandering, which for the type of models we are interested in
can be studied by tracking, e.g., the position of the point where the particle density
reaches a certain fraction of the average particle density behind the front,\footnote{In models
in which the total number of particles remains fixed, like in the
clock model of section \ref{sectionvanzon}  or the model studied by
Brunet and Derrida \cite{derrida}, one can equivalently track the
center of mass of the particles. However, in models where the particle
density behind the front region is finite, one has to be careful not
to include the fluctuations from this region ---  the total mass fluctuations in this region
will grow  proportionally to $\sqrt{N L}$ where $L$ is the size of
this region. Since $L $ grows linearly in time, this will look like a
diffusive behavior, but it has nothing to do with the {\em diffusive
wandering}  of the {\em front position} that is of interest here and in the
stochastic Langevin models discussed in the next section. }\label{footnotebd}
 can be characterized  by a {\em front diffusion coefficient
} $D_{\rm front}$. The large-$N$ asymptotic scaling of $D_{\rm front}$
is very difficult to study by simulations since one needs reliable data over
several decades  while $D_{\rm front}$ rapidly becomes very
small. Nevertheless,  Brunet and Derrida have
been able to study the scaling in simulations of a stochastic model
which is closely related to the clock model of section
\ref{sectionvanzon}, by using a clever trick which allowed them to go
up to values of $N$ of order $10^{150}$. They obtained a scaling
\begin{equation}
D_{\rm front} \sim 1/ \ln^3 N, ~~~~~~~~(N\to \infty),\label{dscaling}
\end{equation}
and empirically found that this asymptotic scaling originates from
fluctuations in the tip region: In a simplified  version of their model
in which only the very first lattice site of the front shows
stochastic fluctuations, they observed the same asymptotic behavior
for $D_{\rm front}$ as in their full model. It appears that the behavior is associated with the
fluctuation  behavior of the  low-lying  modes  of the stability operator
of the deterministic equation, whose   relaxation time
$\tau_m$ is given in (\ref{tauexp}) above.\footnote{If one follows the
arguments of \cite{rocco2} by expanding the fluctuation behavior of the fronts in
terms of the eigenmodes of the stability operator, one finds that for
large $N$ the coefficients in this expansion obey Langevin-equations
with noise which is exponentially dominated by  fluctuations at the tip. Moreover, the
noise that drives the different coefficients is correlated.  It is easy
to convince oneself that the resulting expressions (and
Eq.~(\ref{dfformula}) below) lead to logarithmic scaling of $D_{\rm
front}$ but whether such arguments lead to the behavior
(\ref{dscaling}) is at present unclear to me. It is also possible
that non-perturbative effects like those discussed in the next section
are important.}

The analysis we have sketched above only identifies the  leading order
finite-$N$ velocity correction. This term is remarkably universal; it
appears that if one wants to go beyond this term, one is forced to
perform a detailed analysis of the fluctuations in the tip region, and 
to match the behavior there to the front profile in the rest of the
leading edge. This is a complicated problem, because the number
densities in the tip region are small so that standard perturbation
methods do not apply. Although a first-principles theory is still
lacking, quite a bit of insight is obtained from  an approximate
analysis \cite{deb3} in the tip region, which focuses on the stochastic behavior
of the foremost occupied lattice site (an idea that has also played a
role in earlier stochastic lattice models for small values of $N$
\cite{ba2,kerstein1,kerstein2,deb5}). This analysis shows that 
corrections beyond the leading velocity correction  term depend on  virtually all
the details  of the microscopic model, including the precise
mean-field behavior in the region behind the front. In fact, in the
tip region the lattice and finite-particle effects are so important
that the profile is not truly a uniformly translating profile anymore \cite{deb2}.

Finally, it is good to stress once more that the analysis summarized
above is based on the assumption that the fluctuation effects in the
tip region {\em suppress} the growth. Obviously, if the growth rate for $u$ of order $1/N$ is 
{\em enhanced}, the very tip region can move faster than $v^*$ of the unperturbed case,
and this will lead to a pushed front with asymptotic velocity {\em larger than} $v^*$. 
It easy to show this explicitly following the lines of \cite{derrida} by using the growth 
function shown in Fig.~\ref{figfbrunetandours}{\em (b)}. The explicit analysis shows that 
for $r< r_{\rm c}= 0.283833$ the asymptotic front speed in the limit $\varepsilon=1/N\to 0$ is indeed
larger than $v^*$. Simulations of a stochastic particle model in which growth rates for particle
occupancies of 3 or less are enhanced in the same way as suggested by
the function $f(u,\varepsilon,r)$  of Fig.~\ref{figfbrunetandours}{\em
(b)} already show this enhancement of the average front speed. Note
that these results clearly illustrate that the limits do not commute:
if we take the limit $\epsilon\to 0$ in the function $f$, we arrive at
the classic F-KPP equation with asymptotic velocity $v^*=2$, while if
we consider fronts with $\varepsilon >0 $ and $r<r_{\rm c}$ and then
take the limit $\varepsilon \to 0$, we get a pushed front with speed
$v^\dagger$ larger than
$v^*$!

Since the main conclusions of this section affect the gross features of fluctuating
fronts in more than one dimension, we summarize them below:

\begin{tabular}{p{0.1cm}|p{13.2cm}}
& {\em Fronts in stochastic particle models  which in the mean field limit
$N\to \infty$ converge to a pulled front solution, are generically
pushed fronts for any finite $N$. Their speed relaxes
exponentially in time to an asymptotic speed $v_{\rm N} $ which for
models in which the growth at small particle occupancies is suppressed
relative to the mean field value, is given for large $N$ by
(\ref{bdresult}). Different growth rules at very small particle
numbers can give pushed fronts for finite $N$ with speed $v^\dagger$ larger than
the spreading speed $v^*$ of the mean field limit. }
\end{tabular}

For  a fixed $N$ one has to consider such fluctuating fronts as being
``pushed''  as far as their long-time large-size scaling properties is
concerned (see section \ref{sectionfluctuatinguniclasses} below). On the other hand, all the special
scaling properties as $N\to \infty$ originate from the fact that in
this limit one approaches a pulled front solution. To remind us of
their special character, we have therefore sometimes
referred to such fronts as fluctuating ``pulled'' fronts
\cite{deb2}. However, as we will see below, even this name does not do
full justice to the fact that in particle models, the fluctuation
properties of the fronts are essentially like those of pushed fronts,
not like those of pulled fronts. 

A remarkable consequence of the effects discussed in this section has
been demonstrated by Kessler and Levine \cite{kessler2}. As we
discussed in section \ref{chemicalbacterial}, there are deterministic
pulled fronts which do now show a long-wavelength instability for any
value of the parameters, but which do so for suitable diffusion ratios
as soon as they are pushed. As we have just seen, finite particle
effects are sufficient to induce this crossover, and as was shown in
\cite{kessler2} in this way finite particle effects can induce a
long-wavelength interfacial instability. 

 \subsection{Related  aspects of fluctuating fronts  in stochastic Langevin equations} \label{langevin}

The previous discussion has already mentally prepared us for another
surprise concerning fluctuating  fronts in field-theoretic
approaches to the study of  fronts propagating into an unstable
state. In a field theoretic formulation, the natural starting point to
study propagating fronts in the presence of noise
\cite{armero1,armero2,doering,doering2,gardiner,garciasancho,lemarchand,lemarchand2,noise1,mikhailov,levine,schimansky,noise2} is to start with a
stochastic version of the F-KPP equation (\ref{fkpp}),
\begin{equation}
\partial_t u(x,t) = \partial^2_x  u(x,t) + f(u(x,t)) + g(u(x,t)) \eta(x,t)\label{stochdiffeq}.
\end{equation}
In this  Langevin-type  equation  $\eta$ is
a stochastic noise term which is delta-correlated in space and time
with strength $2 \mu$, 
\begin{equation}
\langle  \eta(x,t) \eta(x^\prime , t^\prime)\rangle = 2 \mu\,
\delta(x-x^\prime)\delta (t-t^\prime),\label{noisestrength}
\end{equation}
where the brackets denote an average over the noise. Since  $\eta$ is multiplied by some
function $g(u)$ of the fluctuating field $u$ in (\ref{stochdiffeq}), the noise term is
``multiplicative'' rather than additive.\footnote{As is well known, for a stochastic differential
equation with multiplicative noise, one faces the question whether the noise is interpreted in the
It\^o or in the Stratonovich sense. The issues we address here are unaffected by the particular 
choice one makes, so we refer to the literature
\cite{arnoldstochastic,garciasancho,gardiner} for further discussion.}

The form of the function $g$ in (\ref{stochdiffeq}) that gives the
noise strength  as a function of the dynamical variable $u$, depends on
the physical model for the noise. When the origin of the noise lies in
the fluctuations in the number of particles per correlated volume or
per lattice site, then it is customary to take $g(u) \propto
\sqrt{u}$. The rationale behind this is that quite generally  the
fluctuations in the number of stochastic moves of $n$ independent particles in
some correlated volume or at some lattice site is of order
$\sqrt{n}$; after normalization this gives fluctuations of order
$\sqrt{u}$ in the particle density. In simulations, it is often
advantageous to take the function $g(u)$ to vanish too in the
saturated state behind the front, so that fluctuations in the trivial
state behind the front are suppressed (compare footnote \ref{footnotebd}).

If the noise originates from independent random  fluctuations in one of
the parameters of the function $f(u)$, it is natural to take the noise
strength $g(u) \propto u$ for small $u$ (see \cite{rocco3} and
references therein). We shall come back to this
case in the next section.

\begin{figure}[t]
\begin{center}
\epsfig{figure=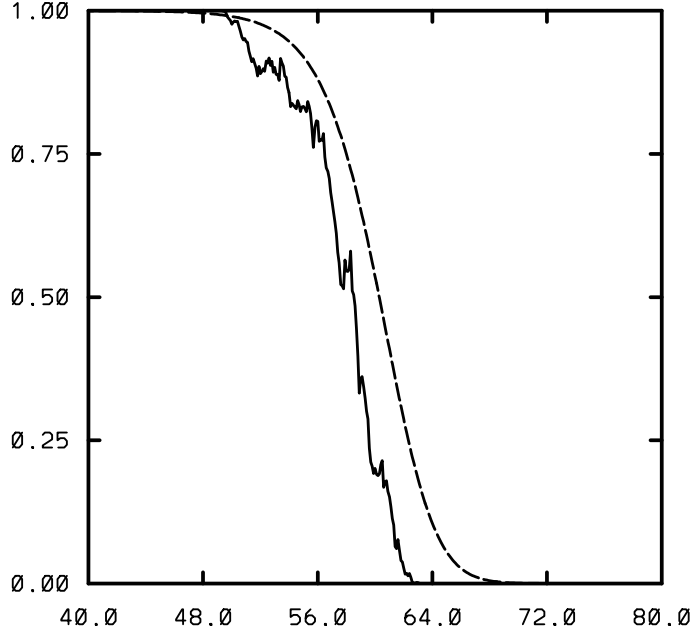,width=0.65\linewidth} 
\end{center}
\caption[]{Snapshot of a stochastic front in a simulation by Doering
{\em et al. } \cite{doering} of the
stochastic F-KPP equation (\ref{stochdiffeq})  with $f(u)=u(1-u)$ and $g(u)=\sqrt{u(1-u)}$
(full line). With this function $g(u)$ there are no fluctuations in
the stable state  $u=1$ behind the front. The dashed line is a plot of
a uniformly translating front profile of the deterministic F-KPP equation (\ref{fkpp}). 
Note that the stochastic front solution has compact support: the dynamical variable is
{\em identically zero} beyond some point. The deterministic front solution, on the other hand, is nonzero
at every position.  } \label{figdoering}
\end{figure}

Let us return to the case in which $g(u)\propto \sqrt{u}$ for small
$u$. The remarkable finding  \cite{doering,doering2,compact} is that
 these stochastic fronts  have {\em compact
support}! In other words, 
these stochastic  fronts  are identically zero beyond some
finite value of $x$ \cite{compact}.  This  is illustrated in
Fig.~\ref{figdoering}. Since for any finite time front
solutions in the deterministic F-KPP equation 
are nonzero  {\em for all} $x$, this implies that this type of noise has a
{\em non-perturbative effect } on fronts --- we can not think of the
noise as being a small stochastic perturbation on top of an underlying
smooth front profile that obeys a mean-field like deterministic
equation!

The fact that  fronts in stochastic front equations with
multiplicative noise with $g\propto \sqrt{u}$ for small $u$ have
compact support has as a remarkable consequence that they show the
same surprising $1/\ln^2 N$ scaling of the correction to the front
velocity in the small noise limit ($\mu \to 0$ in
(\ref{noisestrength}) \cite{levine}. The reason for this is clear
from the discussion of the previous section: as long as the
non-perturbative effects of the noise is limited to a finite region
near the tip of the front, then as we saw the $1/\ln^2N$ scaling of
the velocity correction results from the matching analysis of the part
of the leading edge of the front where the noise effect becomes
arbitrarily small in the limit $N\to \infty$. 

There are many intriguing and open questions regarding such
stochastic fronts, but we refer to the two recent papers by Doering
{\em et al.} \cite{doering,doering2} for further
discussion of such issues. To my knowledge, an open question is
a proper understanding  of the scaling
(\ref{dscaling}) of the front diffusion coefficient $D_{\rm front}$, e.g,  by exploiting the fact that in
some limits the behavior of such Langevin equations can be mapped 
onto finite particle models using a duality transformation
\cite{doering,doering2,levine}. 

We stress once more that these considerations are in practice essentially only 
of importance for fronts
which are pulled in the mean-field limit. As we mentioned earlier, the 
discussion of   section  \ref{sectionmbabreakdown} showed that
singular perturbation theory does apply to pushed fronts, and this
means that in the weak noise limit one can study the effect of noise
on the motion of pushed fronts perturbatively. As an example, we note
that for small noise strength the  diffusion coefficient $D_{\rm front}$ of
pushed fronts  is given by
\begin{equation}
D_{\rm front} = \mu \, \frac{ \int_{-\infty}^{\infty } {\rm d}\zeta\, e^{2v_{\rm as} \zeta} \, 
  ({\rm d} U_0/{\rm d}\zeta )^2  \, g^2(U_0)}{\left[
    \int_{-\infty}^{\infty} {\rm d} \zeta\,
    e^{v_{\rm  as}\zeta} \,
    ({\rm d}U_0/{\rm d}\zeta)^2 \right]^2}. \label{dfformula}
\end{equation}
Here $g$ is the multiplicative noise strength in (\ref{stochdiffeq})
and $U_0(\zeta)= U_0(x-v_{\rm as}t)$ is the  pushed front solution
of the deterministic limit  of this stochastic differential equation
\cite{armero2,mikhailov,rocco2}. This expression has the
typical structure discussed in section \ref{sectionmbabreakdown}, a
ratio of two solvability-type integrals with the perturbing term in
the numerator --- compare, e.g., Eq.~(\ref{solvab1}). For pushed
fronts the integrands vanish exponentially for large positive $\zeta$
and hence pushed fronts are quite insensitive to the non-perturbative
effects in the tip region when the stochastic noise is turned on. But, 
as we already discussed in section \ref{sectionmbabreakdown}, for
pulled deterministic fronts the integrals do not converge; the
anomalous scaling properties of fluctuating  pulled fronts that we will
discuss in the section below are a
result of this. Armero {\em et al.} \cite{armero2} have compared the above expression 
for $D_{\rm front}$ with extensive numerical simulations of pushed
fronts with noise, and found that the formula works quite well, even
at appreciable values of the noise strength $\mu$.

     \subsection{The universality class of the scaling properties of
fluctuating fronts} \label{sectionfluctuatinguniclasses}

Now that we understand some of the important features of the behavior
of stochastic fronts in  one dimension, let us close with considering
the implications for the universality class of the scaling properties
of these propagating fronts in more than one dimension. 

As we noted in the very beginning of this section on stochastic fronts, the scaling
properties of fluctuating front and interfaces is usually studied in
terms of interface models like the KPZ equation
\cite{barabasi,halpin-healy,kpz,krug2}. Now, even though  in
the limit $N\to \infty$ some stochastic fronts can become pulled, for
any fixed $N$ or noise strength $\mu$ the stochastic fronts we have
examined all are  pushed fronts. Therefore, in view of
the conclusion of 
section \ref{sectionmbabreakdown} that pushed fronts do map onto an
effective interface model for their long-wavelength and long-time
dynamics, these models should all effectively map onto one of the
familiar type of interface models that have been studied so
extensively in the
literature\cite{barabasi,halpin-healy,kpz,krug2}. Or, to put it
differently: 

\begin{tabular}{p{0.1cm}|p{13.2cm}}
& {\em the particle lattice models or the stochastic
differential equations with multiplicative noise and  $g(u)\propto
\sqrt{u}$ which in the mean field limit map onto 
dynamical equations with pulled fronts, are in practice in the same
universality classes as the usual interface models that one would
naively write down for them.}
\end{tabular}

As a historical note, it is only  fair to admit that I got this issue
wrong initially \cite{tripathy1}. When we started to
wonder about the question whether fluctuating pulled fronts might be
in a different universality class, we were motivated by several
observations:  {\em (i)} the fact that  deterministic pulled fronts do not map onto an
effective interface model; {\em (ii)} the fact that their dynamics is driven by the
dynamics in the semi-infinite region ahead of the front itself rather
than the dynamics in the front region itself; and {\em (iii)} the fact
that there were simulations of stochastic versions of the F-KPP
equation \cite{noise5} which appeared to  suggest that in two and higher dimensions
such fronts are in a different universality class than the KPZ
equation. This made us argue \cite{tripathy1} that stochastic lattice model
realizations of the F-KPP equation with pulled fronts would exhibit
non-KPZ scaling. The insight summarized above that both lattice model
realizations and Langevin-type versions with multiplicative noise
strength proportional to $\sqrt{u}$ make the fronts always
pushed invalidates this conjecture. In fact, soon after our proposal,
Moro \cite{moro} demonstrated this and presented numerical simulation results in
support of this. 

There is actually one rather special case where fluctuating fronts
deserve to be viewed as genuinely pulled, and for these the scaling
properties seem to be anomalous indeed \cite{rocco,tripathy2}: if one has a
medium in which the parameters entering the nonlinear growth function
$f(u)$ are fluctuating (see, e.g., \cite{rocco3}), a natural model is to take $g(u)\approx  u$ for
small $u$ in (\ref{stochdiffeq}). Let us consider the case that the
growth function $f(u)$ is such that the fronts are pulled, and that $f^\prime(0)=1$. In the
presence of noise the leading edge equation for a front in $d$
bulk dimensions then reads\footnote{We immediately focus on the analysis in
the leading edge, where nonlinear terms in $u$ can be ignored. In
section \ref{sectionmatching2} we showed that the nonlinear terms play
the role of a boundary condition for the leading edge dynamics.}
\begin{equation}
\partial_t u = \nabla^2_x u + (1+\eta) u. \label{kpz+1}
\end{equation}
If  the front propagates on average in the $x$ direction, then upon
transforming to the co-moving frame $\xi=x-v^*t=x-2t$ the equation for
the leading edge variable $\psi=e^\xi u$, which already played an
important role in the analysis of the universal relaxation of pulled
fronts in section \ref{twofeatures}, reads
\begin{equation}
\partial_t \psi = \nabla^2 \psi +  \psi \eta .
\end{equation}
As is well known, with a Cole-Hopf transformation $\psi=e^{h}$ this
equation reduces to the standard KPZ equation
\cite{barabasi,halpin-healy,kpz,krug2} for $h$,
\begin{equation}
\partial_t h = \nabla^2 h + (\vec{\nabla}h)^2 + \eta.
\end{equation}
The point to note about this simple argument is that one arrives at
the KPZ equation {\em  in the dimension of the system in which the front
propagates}: $h$ and the noise $\eta$ are functions of both the
$\xi$ coordinate perpendicular to the front and the coordinates
parallel to the front. This implies that the universality class of
such  genuinely pulled fluctuating fronts is that of the KPZ equation, but in one
dimension higher than one would naively guess if one would think of
the front as a moving interface \cite{rocco,tripathy2}. One 
implication of this is that such fronts in one dimension show
sub-diffusive behavior, an observation that is corroborated by
numerical simulations \cite{rocco}. 

We emphasize again that this anomalous scaling --- KPZ-type scaling
but in one dimension higher than one would naively expect --- is only
expected if the effective equation is of the form (\ref{kpz+1}), i.e.,
if the strength $g(u)$ of the noise is linear in the dynamical field
$u$. As we discussed above, intrinsic fluctuations like those that arise naturally in
any  particle model correspond naturally to fluctuations of
strength proportional to $\sqrt{u}$ and in such models one indeed never
sees signs of anomalous diffusion or scaling behavior. I believe that this is due
to the following. Whatever the precise model, the relative strength
$g(u)/u$ of intrinsic fluctuations is stronger the smaller $u$, i.e.,
the stronger the smaller the number of particles. Now,
it is easily seen from (\ref{dfformula}) that whenever $g(u)/u$ 
behaves as $u^{-\gamma}$ with $\gamma > 0$ for small $u$, the integrand in the numerator is
{\em exponentially dominated}  by the contributions at the tip of the
front. This means that fronts with intrinsic fluctuations behave like
objects with fluctuations in a {\em finite} region and hence with normal
scaling behavior. In other words, only if fluctuations throughout the
whole leading edge contribute equally to the fluctuating behavior ($\gamma=0$), as
they do when they are driven by fluctuations in an external parameter (or
possibly for error propagating fronts --- see the last paragraph of
section \ref{sectionerrorpropagation}), will one
see anomalous (transient) scaling. In realistic particle models, one won't.

These observations illustrate that the behavior of a fluctuating
front is subtle. Because they operate close to a bifurcation point,
whether or not we should think of them as being close to being pulled
depends on the quantity under consideration. Indeed, regarding 
their velocity, we can consider such fronts as being  ``weakly
pulled''  for
large $N$ in the sense that  their velocity approaches $v^*$ for 
$N\to\infty$, but when we consider the fluctuation properties  of a
front with intrinsic fluctuations we should not think of them as
being weakly pushed.\footnote{This was
stressed in particular to me by E. Moro.}. Hopefully future research
will clarify these  subtleties and the scaling behavior
(\ref{dscaling}) of $D_{\rm front}$.

\section{Outlook}
In this article we have aimed to introduce the subject of front
propagation into linearly unstable states from a unifying  perspective, that
allows us to bring together essentially all important developments in
this field. The concept of  linear spreading
speed $v^*$ not only aids in developing an intuitive understanding
and in sharply defining pulled and pushed fronts, but it also lies at
the basis of the formalism that allows one to derive the universal
relaxation towards the pulled front speed, using matched asymptotic
expansions. In addition, ``global modes'', the breakdown of an
effective interface formulation of pulled fronts or the universal large-$N$
velocity corrections of stochastic fronts can all be naturally
approached from this perspective.  The many examples that we have
discussed show the ubiquity of the problem of front propagation into
unstable states, and illustrate how one can understand an enormous
variety of systems with a few simple tools.

There are several open issues. Concerning deterministic fronts, the most pressing one is a sharp
definition, using an appropriate averaging procedure, of an incoherent
pushed front. In what sense are  incoherent fronts which arise from the nonexistence
of coherent pattern forming fronts different from incoherent fronts
which originate from the instability of  coherent fronts? Furthermore,
while our definition of  coherent pattern 
forming pushed fronts is consistent with what is know for the quintic
CGL equation or extensions of the Swift-Hohenberg equation, such
types of fronts remain relatively unexplored. For pulled fronts in
periodic media, we have conjectured that their asymptotic speed can be
calculated using Floquet analysis, but this proposal remains to be
explored. Also our understanding of stochastic fronts, especially
when they are close to the mean field limit, is still
incomplete.  In addition, I have the impression that the difference
between pulled and pushed fronts may be underestimated in the fields of
turbulent combustion and wave propagation  in periodic media.
 Finally, even though we have argued that there are hints of slow
convergence to the asymptotic speed in experiments specifically aimed
at probing pulled fronts, only very accurate new experiments can
settle this issue. 

Of a different nature are the challenges posed by pushed fronts. Even
though the mechanism of pushed front propagation for uniformly
translating and coherent pattern forming fronts can be considered
known, making concrete predictions for a given problem requires the
explicit construction of the nonlinear pushed front solution. Even for
nonlinearly translating fronts, this is a highly nontrivial problem
for any equation beyond the F-KPP equation. Moreover, I do not think there is
much hope that there will ever be a general framework that allows one
to calculate pushed front solutions for large classes of equations, as
every detail counts. Guessing how one can add a term to a dynamical
equation to make the front propagation become pushed is often trivial,
but any serious analysis of pushed fronts is often virtually
impossible --- the fact that the pushed front solutions of the quintic
CGL equation can be obtained analytically is an exceptional miracle.

For those mathematicians who only accept rigorous proofs, almost
everything in this paper can be considered an open issue.
The route we have chosen to follow here is quite different from the
usual one favored in the mathematics literature. I believe the
approach we have advanced here is ready to be put on rigorous footing.
If this will be done, it will undoubtedly allow one to approach
large classes of equations at one fell swoop. 

\section{Acknowledgment} 
I would like to take this opportunity to express my thanks to Ute
Ebert for intense collaborations and discussion on front propagation
issues in the last few years. Much of the presentation in this article
is shaped by the interaction with her. In addition I want to
acknowledge 
 all the other  collaborators, colleagues and friends with whom I have
worked on or discussed front propagation and related issues in the
past decade: 
Mischa Anisimov,
Chiara Baggio,
Markus B\"ar,
David Bensimon,
Henri Berestycki,
Tomas Bohr,
Daniel Bonn,
Haim Brezis,
Christiane Caroli,
Jaume Casademunt,
Hugues Chat\'e,
Pat Cladis,
Elaine Crooks,
Mike Cross,
Greg Dee,
Patrick De Kepper,
Bernard Derrida,
Arjen Doelman,
George Gilmer,
Ray Goldstein,
Henry Greenside,
Tim Halpin-Healy,
Danielle Hilhorst,
Pierre Hohenberg, 
Robert Ho{\l}yst,
Desz\"o Horv\'ath,
Martin Howard,
Joost Hulshof,
Reijer Jochemsen,
Tasso Kaper,
Julien Kock\-el\-ko\-ren,
Lorenz Kramer,
Joachim Krug,
Philip Maini,
Hiroshi Matano,
Mayan Mimura,
Chaouqi Misbah,
Esteban Moro,
Judith M\"uller, 
Alan Newell,
Yasumasa Nishiura,
A\-lex\-an\-der Morozov,
Peter Palffy-Muhoray,
Debabrata Panja, 
Bert Peletier,
Mark Peletier,
Werner Pesch,
Zoltan Racz,
Andrea Rocco,
Vivi Rottsch\"afer,
Bjorn Sandstede,
Boris Shraiman,
Willem Spruijt,
Kees Storm,
Alessandro Torcini,
Goutam Tripathy,
Henk van Beijeren,
Jan-Bouwe van den Berg,
Martin van Hecke, 
Willem van de Water,
Ramses van Zon,
John Weeks
and Rinke Wijn\-gaar\-den.
Finally, I also want to thank Kees Storm for detailed comments on an
earlier version of the manuscript, Deb Panja and Alexander Morozov for help with some of
the figures, and Yolande van der Deijl for her perspective.

\newpage
\begin{appendix}
\section{Comparison of the two ways of evaluating the asymptotic linear 
spreading problem}\label{appendixconnection}

There are two ways to extract   the long-time asymptotics of the general equation (\ref{inverseeq}),  
\begin{equation}
\underline{\hat{\phi}} (k,\omega) =
\underline{\underline{\hat{G}}}(k,\omega)\cdot 
\underline{\underline{\hat{H}}}(k,\omega)\cdot 
\underline{\bar{\phi}}(k,t=0), \label{inverseeqapp}
\end{equation}
depending  on whether one first evaluates  the $\omega $-integral of the inverse
Laplace transform, or  the inverse $k$-integral of the Fourier
transform. The first route most closely parallels our earlier
discussion of section \ref{sectionv*}. Indeed, if we first evaluate
the $\omega$-integral --- which runs 
along a line in the upper half $\omega $-plane, as (\ref{lftransform}) 
converges for sufficiently large $\omega_{\rm i}$, so the functions
are analytic sufficiently far up in the upper half $\omega $-plane --- 
then the poles where   $|   \underline{\underline{\hat S}}|=0$
determine the dispersion relations $\omega_{\alpha}(k)$ of the
various branches which we label by $\alpha$. The remaining spatial
Fourier inversion then has a form similar to (\ref{finverse}),
except that there is an additional sum over the branch index $\alpha$, 
and that associated with each branch $\alpha$ and wavenumber $k$ there 
is an eigenvector $\underline{\hat U}$ of the linear dynamical
matrix (see section 5.5 of \cite{evs2} further details). From there
on, the analysis is essentially similar to the one given for a single
field in section \ref{sectionv*} for each branch. If it so happens
that there is more than one branch where modes are unstable, then
it is possible that there is more than one linear spreading velocity
$v^*_{\alpha}$. {\em Since we are
considering a fully linear problem, it is clear that the largest value 
of $v^*_{\alpha}$ is the relevant linear spreading velocity } for generic
initial conditions which have nonzero overlap with the associated
eigenvector $\underline{\hat U}$. We  refer to the largest
velocity  of these as {\em the} linear associated with each of these branches
spreading velocity $v^*$ and likewise associate $k^*$, $\lambda^*$ and 
$D$ with the values at the corresponding linear spreading
point.\footnote{Since the discussion of section
  \ref{sectionexponential}  implies that  each
  velocity $v^*_\alpha$ corresponds to  the minimum of a curve
  $v_{\alpha ,{\rm env}}$, it is sometimes said that the selected speed 
    is the maximum of the minimum velocities. An example
    where this was checked for fronts can be found in \cite{paquette2}.}

The advantage of this line of analysis (first evaluating the
$\omega$-integral by contour integration) is that the importance of
initial 
conditions is immediately clear. In complete analogy with
(\ref{steepincond}) and the earlier
discussion, so-called steep   initial conditions, for 
which the  amplitudes of the components
$\phi_m(x,t=0)$ decay faster than $\exp(-\lambda^*x)$, lead to profiles 
which asymptotically spread with velocity $v^*$.

The second route to extract the long time asymptotics is the one which 
was developed since the late 1950-ies in plasma physics. We only 
summarize the essentials here and refer to   \cite{bers,briggs,huerre1,ll}
for details. The main idea of the analysis is the following. For each
$\omega$, there are poles of the integrand in the complex $k$-plane,
so when $\omega$ varies along the $\omega$-integration path in the
upper half of the complex $\omega$-plane which runs parallel to the
real axis, the poles trace out curves
in the $k$-plane. Imagine now that we lower the $\omega $-integration 
path; then these  lines of poles in the $k$-plane will shift. If one
of these lines of poles  tends to cross the $k$-axis, we just deform the
$k$-integration path, so as to avoid the lines traced out by the
poles. Now, we can continue to deform the $k$-integration path so as
to avoid the lines of poles, until  at some point, upon lowering the
$\omega $ -path, two  lines of poles in the $k$-plane come together
from opposite 
sides of the integration path and ``pinch off'' this path
\cite{bers,briggs,huerre1,ll}. Where will this ``pinch point'' be? Not 
surprisingly, it is the same as the saddle point in the previous
formulation. Indeed, since close to the the point $\omega_{\alpha}^*$, $k_{\alpha}^*$
\begin{equation}
\omega - \omega_{\alpha}^* \approx \half \left. \frac{\ds ^2
    \omega}{\ds k^2}\right|_{k_{\alpha}^*} (k-k_{\alpha}^*)^2 = -i {\mathcal D} (k-k_{\alpha}^*)^2,
\end{equation}
we have
\begin{equation}
k-k_{\alpha}^* \approx \pm \sqrt{ i (\omega -\omega_{\alpha}^*)/{\mathcal D}} .  \label{doubleroot}
\end{equation}
Hence as the $\omega$-integration path is lowered towards $\omega_\alpha^*$,
two poles in the $k$-plane merge together to ``pinch off''  the
$k$-contour. In this analysis, which is often referred to as the
``pinch 
point analysis'', one thus evaluates the $k$-integral first and then
performs the  $\omega$-integration. In view of (\ref{doubleroot}),
after performing the $k$-integration  there is a branch cut in the
$\omega$-plane. Evaluation of the $\omega $-integration around the
branch cut then gives the same expressions as those we found before. 

The above discussion of the pinch point analysis gives  the long-time
asymptotic behavior of the Green's function. This implicitly 
means that delta-function like initial conditions are assumed; the
possibility of exponentially decaying initial conditions is usually
not discussed in the literature, but these do lead to the same
conclusions as before: steep  initial conditions lead to profiles
spreading asymptotically with velocity $v^*$. In fact, since the
$k$-integral is closed first in the pinch-point analysis,  poles
in the complex $k$-plane arising from 
exponentially decaying  initial conditions  give rise to
additional terms in the $\omega$-integral which compete with the
contribution from the pinch point in a way similar to the one we
discussed before in section \ref{sectionexponential}.

One final caveat is important. While the linear spreading point $k^*$ is determined by a local 
condition, the relevance of this point for the dynamics is subject to
some (mild) conditions on the  analytic structure in the $k$-plane. In
the first formulation, where we focus on the $k$-integral evaluation
given the dispersion relation $\omega(k)$, the 
underlying assumption is that the $k$-integration path can be deformed 
continuously to go through the saddle point. In other words, there
should be no branch cuts or non-analyticities to prevent us from
reaching this point. Furthermore, the condition that at the spreading
point  $D>0$ is important. For, if this condition is not fulfilled, then
if we would write $\phi(\xi,t)=e^{-\lambda^*\xi}\psi$, the equation
for $\psi$ is governed by a negative diffusion coefficient [see
section \ref{twofeatures}, Eq.~(\ref{leadingedge1})]. This means 
that convergence to a smooth asymptotic exponential behavior is not
possible for $D<0$. 

Likewise, there are conditions on the analytic behavior in the pinch
point formulation: the ``pinching off'' of the  deformed
$k$-integration path by the two lines of poles in the complex $k$-plane
means that these two lines have to be analytical continuations of
branches which are below and above the $k$-integration path when it
was running along the real $k$-axis.  It actually appears to me that the
precise status of such conditions and of the relation between the
conditions in the two formulations is not fully understood. E.g., the
condition $D>0$ is not generally found in the pinch point
literature.\footnote{In view of (\ref{doubleroot}) this condition
  implies a statement about the local orientation of the two lines of poles
  in the complex $k$-plane: if the $\omega$-contour grazes just
  over $\omega^*$ in the horizontal direction the condition $D>0$
  implies that one of the lines of poles 
  always lies in the upper half $\Delta k$-plane and the other one in
  the lower $\Delta k$-plane. The example give in \cite{bers} of two poles which merge but 
  which do not form a pinch point, is actually an example of a point
  where $D<0$, but whether this is accidental I do not
  know. See also appendix M of \cite{evs2} for further discussion of
  the difference between the two methods.  } Nevertheless, our
pragmatic point of view in this paper is that once can proceed with
the general analysis laid out in this paper, keeping in mind  that 
for any given problem, one can explicitly check whether the conditions 
underlying the general derivation are satisfied.

\section{Additional observations and conjectures concerning front selection}\label{otherconjectures}

It turns out that the exact pushed front solutions of the F-KPP
equation obtained by the
``reduction of order method'' discussed briefly at the end of section
\ref{selectionutfs}, do have  some special properties in the complex
$\zeta$-plane \cite{goriely}. This has led   to the speculation that
these properties hold more generally\cite{goriely}  --- if true, this would allow one
to obtain the pushed front speed $v^{\dagger}$ without explicitly
constructing the strongly heteroclinic front profile itself. However,
the fact that pushed front solutions are sensitive to every change in
the equations indicate that such a remarkable property is very
unlikely; detailed investigations and the construction of an explicit
counterexample have indeed confirmed that the idea is untenable \cite{cisternas}.

Another observation which has been proposed as a possible road to
understanding   the selection problem of uniformly
translating fronts   is to look for front solutions which
monotonically connect  the  {\em exact} unstable state $\phi=0$ and the stable state
in a {\em finite} but large interval $[0,L]$  of the co-moving coordinate
$\zeta$ \cite{theodorakis1,theodorakis2}. It is clear that this requirement does reproduce the selected
speed for uniformly translating fronts. After all, this requirement
amounts to searching for front solutions which go trough zero at $L$
when a larger interval is considered. Now,  in the uniformly translating
regime the selected front is always the only front solution which is
close to a front solution $\Phi_v(\zeta)$ which will just go through
zero when the velocity is lowered. Indeed,  the
pushed front solution $\Phi_{v^{\dagger}}(\zeta)$ has $a_1=0$ and
so in its neighborhood are the first solutions (\ref{decayformula}) which have $a\neq0$ and
which go through zero, while in the pulled case the first solutions
which go through zero appear just below $v=v^*$, because the
 $k_{\rm r}\neq 0$. Thus this approach is consistent with known
results for uniformly translating fronts, but it gives no insight into
the underlying dynamical mechanism and it does not apply to
pattern forming fronts. The convergence of the velocity obtained this
way to the selected one can be analyzed using the methods discussed in
section \ref{sectionfixedboundary}.

Recently, there has been interest in reformulating the front
propagation problem in terms of Hamilton-Jacobi theory (see, e.g.,
\cite{fedotov2}). At present such methods appear to be limited to the
obtaining the asymptotic front speed, not the convergence to it. Since
these results are already contained within those discussed here, we
will not discuss it further.

We finally note that the renormalization group approach  developed for
partial differential 
equations  \cite{goldenfeld} has also been used in the
context of front propagation
\cite{bricmont2,paquette3,paquette1,paquette2}. Viewed as a
reformulation of  singular
perturbation theory, the renormalization group method is very useful
in reinterpreting the basis of amplitude equations and similar methods \cite{bricmont,goldenfeld};
however, as  we  discuss in section \ref{sectionmba} singular perturbation
methods do not apply when one wants to go beyond  the asymptotic speed
of  pulled fronts, so here its  use for the front
propagation problem appear to be  limited. In addition, 
renormalization group methods  can be used to study contraction to the
 asymptotic speed or front stability \cite{bricmont2}. Whether the
universal convergence towards the asymptotic speed can be 
analyzed or proved with this method, is to my knowledge an open question.

\newpage
\section{Index}
Several colleagues have asked me to supply an index for this
paper. Unfortunately, the paper format  of Physics Reports does not
allow for an index with references to the page numbers. We therefore
provide an index to section numbers.

{\footnotesize
\begin{tabbing}
\hspace*{6.8cm} \= \kill
absolute instability \> \ref{sectionconvversusabs}, \ref{sectionnonlinearconv2abs}, \ref{secfrontsnoise1},  \ref{secfrontsnoise2}, \ref{sectionmullinssekerka},  \ref{sectionglobalmodes}         \\
absorbing wall \>   \ref{sectionmatching2}    \\
adiabatic relaxation \>  \ref{relaxtoutpulled}, \ref{relaxtodpfpulled}   \\
amplitude equation \>  \ref{sectiontcrb}, \ref{sectionkupperslorz}     \\
autocalytic  reaction \>  \ref{chemicalbacterial}   \\
bacterial front  \>  \ref{secfrontsnoise2}   \\
ballistic deposition  \>   \ref{meanfieldgrowth}     \\
bending regidity  \>   \ref{secpropRayleigh}    \\
Benjamin-Feir instability \> \ref{sectioncglcubic}, \ref{secfrontsnoise1}    \\
biological invasion/growth model  \>  \ref{chemicalbacterial}, \ref{sectionbiologicalinvasion}, \ref{sectionotherbio}       \\
Bloch theorem \>  \ref{sectioncombustion}   \\
bluff body  \>  \ref{sectionvonkarman}  \\
boundary condition \>  \ref{sectionmatching2}    \\
Boussinesq effect  \>  \ref{sectionpropagatingrt}   \\
branch cut \> \ref{sectionmoregeneral} \\
branches \> \ref{sectionmoregeneral}, \ref{sectiontwospreadingpoints}  \\
Brillouin zone \> \ref{sectionmoregeneral} \\
bunching  \>  \ref{sectiondebunching}   \\
Cahn-Hilliard equation \> \ref{sectionwhenpushed}, \ref{sectionch}, \ref{sectionpropagatingrt}, \ref{secspinodal}       \\
cascade model  \>  \ref{sectioncascade}  \\
Cayley tree  \>  \ref{sectionphasetrans}    \\
chaotic domain/system  \>  \ref{sectionkupperslorz}, \ref{sectionerrorpropagation}     \\
chemical front  \>   \ref{chemicalbacterial}   \\
clock model  \>   \ref{sectionvanzon}   \\
coarsening \> \ref{sectionch}, \ref{sectiondynamicaltest}, \ref{sectionmba}       \\
coherent pattern forming front \> \ref{scopeaim}, \ref{sectiontypesoffronts}, \ref{selectioncpfs}, \ref{structuralstability}, \ref{relaxtodpfpulled}, \ref{sectionsh}, \ref{sectioncglcubic}, \\
  \>  \hspace*{3mm}  \ref{sectioncglquintic}, \ref{sectiontcrb},  \ref{sectionmatching1}       \\
coherent structure solution  \>  \ref{sectioncglcubic}, \ref{sectioncglquintic}   \\
combustion  \>  \ref{sectioncombustion}      \\
comparison argument/theorem \> \ref{sectionexponential} \\
Complex Ginzburg Landau equation \> \ref{section4thorder}, \ref{sectioncglcubic},  \ref{sectionparallelshear}, \ref{secfrontsnoise2}   \\
conservation of steepness \>   \ref{sectiontransients}  \\
conserved dynamics \>  \ref{sectionch}, \ref{secspinodal}      \\
convective instability \> \ref{perspective}, \ref{sectionconvversusabs}, \ref{sectionnonlinearconv2abs}, \ref{sectioncglcubic}, \ref{sectionexamples},  \ref{sectiondebunching},  \ref{sectionparallelshear},  \ref{sectionvonkarman},  \ref{secfrontsnoise1}, \\
  \>  \hspace*{3mm}   \ref{secfrontsnoise2}, \ref{sectionmullinssekerka}, \ref{sectionglobalmodes}          \\
convective versus absolute instability \> \ref{sectionconvversusabs}, \ref{sectionnonlinearconv2abs}, \ref{sectionvonkarman}, \ref{secfrontsnoise1}      \\
Couette flow   \>  \ref{sectionparallelshear}, \ref{secfrontsnoise1}     \\
counting argument \> \ref{selectionutfs}, \ref{stabvsel}, \ref{sectioncglcubic}     \\
coupled map lattice  \>  \ref{sectionerrorpropagation}  \\
Crank-Nicholson \> \ref{sectionmoregeneral}  \\
crossover region \> \ref{sectiontransients} \\
cubic CGL equation \>  \ref{sectioncglcubic}, \ref{secfrontsnoise1}     \\
curvature correction \>   \ref{sectionmbabreakdown}         \\
cutoff effect  \>  \ref{finiteparticles}  \\
debunching \> \ref{sectiondebunching} \\
delay equation  \>  \ref{sectionbiologicalinvasion}    \\
dendrite \>  \ref{sectionstreamers}, \ref{sectionmullinssekerka}, \ref{sectionmba}     \\
deposition model  \>   \ref{meanfieldgrowth}  \\
difference equation  \>  \ref{sectionmoregeneral}, \ref{stabvsel}, \ref{relaxtoutpulled}, \ref{sectionvanzon}       \\
diffusion limited aggregation \>   \ref{chemicalbacterial}, \ref{sectionmullinssekerka}    \\
director fluctuations \>  \ref{sectiondynamicaltest}    \\
discharge \>  \ref{sectionstreamers}, \ref{sectionmullinssekerka}    \\
discrete set  \>  \ref{sectioncglcubic} \\
disordered $XY$ model  \>  \ref{sectionphasetrans}  \\
dispersion relation \>  \ref{sectionv*}, \ref{section4thorder}                \\
effective diffusion coefficient $D$ \>  \ref{sectionv*}               \\
effective  interface approximation \>  \ref{sectionkupperslorz}, \ref{sectionstreamers}, \ref{chemicalbacterial}, \ref{sectionmba}, \ref{secspherical}, \ref{sectionstochasticfronts}         \\
effects of stability \> \ref{sectioneffectsofstab}   \\
entire function \>  \ref{sectionmoregeneral}  \\
Euler approximation \> \ref{sectionmoregeneral} \\
exponential solutions \> \ref{sectionexponential}       \\
Extended Fisher-Komogorov equation \>  \ref{sectionefk}   \\
error propagation  \>  \ref{sectionerrorpropagation}   \\
facet \>  \ref{sectiondebunching}  \\
ferromagnet  \>  \ref{ferromagnet}   \\
finite difference equation \> \ref{sectionmoregeneral}, \ref{relaxtoutpulled}  \\
first order phase transition \>  \ref{sectiondynamicaltest}   \\
F-KPP equation \> \ref{perspective}, \ref{secoverview},
\ref{sectionv*}, \ref{sectionexponential}, \ref{sectiontransients},
\ref{sectionmoregeneral}, \ref{sectionconvversusabs}, \ref{selectionutfs}, \ref{stabilitygeneral}, \\
 \>   \hspace*{3mm}  \ref{relaxtoutpulled}, \ref{sectionefk}, \ref{sectionsmectic}, \ref{sectioncombustion},  \ref{sectionbiologicalinvasion}, \ref{sectionvanzon}, \ref{ferromagnet},  \\
\>   \hspace*{3mm}  \ref{sectionphasetrans}, \ref{twofeatures}, \ref{sectionmatching1}, \ref{secspherical},  \ref{sectionmbabreakdown},  \ref{sectionglobalmodes}, \ref{finiteparticles},  \ref{langevin}                           
        \\
flame  \>  \ref{sectioncombustion}, \ref{sectionmba}  \\
Floquet theory \>  \ref{sectioncombustion}   \\
fluctuations \>  \ref{sectiontcrb}, \ref{sectionstochasticfronts},  \ref{finiteparticles}, \ref{sectionfluctuatinguniclasses}     \\
fluid dynamics \>  \ref{sectionexamples}  \\
fourth order equations \>  \ref{section4thorder}   \\
free energy   \>  \ref{sectionch}, \ref{secpropRayleigh}, \ref{sectionkupperslorz},  \ref{sectionsupercond},  \ref{sectiondynamicaltest}, \ref{sectionsmectic}    \\
front diffusion coefficient \> \ref{finiteparticles}    \\
front propagation/selection problem \>  \ref{scopeaim}, \ref{sectiontypesoffronts}\\
global mode \> \ref{perspective}, \ref{stellar}    \\
generating function  \>   \ref{sectionphasetrans}  \\
Ginzburg-Landau theory  \>  \ref{sectionsupercond}  \\
Green's function \>  \ref{sectionmoregeneral}, \ref{epilogue} \\
group velocity \>  \ref{stabvsel}, \ref{secfrontsnoise2}     \\
Halperin-Lubensky-Ma effect  \>  \ref{sectiondynamicaltest} \\
Hamilton-Jacobi theory  \> App. \ref{otherconjectures}    \\
hard sphere gas  \>   \ref{sectionvanzon}     \\
heated wire experiment  \>  \ref{secfrontsnoise2}     \\
hexagonal pattern  \>  \ref{sectionpropagatingrt}  \\
Hopf bifurcation \>  \ref{secfrontsnoise1}, \ref{secfrontsnoise2}   \\
homoclon solution \>  \ref{secfrontsnoise2} \\
imperfections \> \ref{sectionconvversusabs}   \\
incoherent pattern forming front \>  \ref{scopeaim}, \ref{sectiontypesoffronts}, \ref{sectionincoherentpffs}, \ref{relaxtodincohpfpulled}, \ref{sectionks}, \ref{sectioncglcubic},  \ref{sectioncglquintic},\\
  \>  \hspace*{3mm} \ref{secfrontsnoise2}, \ref{sectionerrorpropagation}, \ref{sectionmatching2}            \\
initial conditions \>  \ref{sectiontransients}    \\
instantaneous front velocity \>  \ref{relaxtoutpulled}, \ref{relaxtodpfpulled} \\
integration contour \> \ref{sectionmoregeneral} \\
integro-differential equation \>  \ref{sectionmoregeneral},  \ref{stabvsel}  \\
intermediate asymptoticd \>  \ref{stabvsel}  \\
invasion model  \>  \ref{sectionbiologicalinvasion}, \ref{sectionotherbio}     \\
kernel \>  \ref{sectionmoregeneral}, \ref{sectionbiologicalinvasion}   \\
kink \> \ref{sectionefk}, \ref{sectionsmectic}      \\
KPZ universality class \>  \ref{sectionks}, \ref{sectionfluctuatinguniclasses}  \\
Kuramoto-Sivashinsky equation \> \ref{scopeaim}, \ref{sectionwhenpushed}, \ref{sectionks}, \ref{sectioncglquintic}, \ref{sectionpropagatingrt}, \ref{chemicalbacterial}     \\
Landau-Lifshitz equation  \>  \ref{ferromagnet}    \\
large-time asymptotics \>  \ref{sectiontransients}, \ref{sectionmoregeneral}, \ref{sectioncglcubic}   \\
langevin equation \>  \ref{langevin}   \\
leading edge \>  \ref{scopeaim}, \ref{selectioncpfs}, \ref{twofeatures}     \\
leading edge dominated dynamics \> \ref{sectionleadingedgedominated}, \ref{sectionmullinssekerka},  \ref{sectionphasetrans}      \\
leading edge transformation  \>  \ref{twofeatures}    \\
leading edge variable  \>  \ref{twofeatures}    \\
level line \> \ref{relaxtoutpulled}  \\
Levy process  \>  \ref{sectioncombustion}   \\
liquid crystal  \>  \ref{sectiondynamicaltest}, \ref{sectionsmectic}     \\
linearly unstable \>  \ref{sectionv*}       \\
linear marginal stability \>  \ref{perspective}\\
linear spreadin point \>  \ref{sectionv*}, \ref{sectiontwospreadingpoints}         \\
linear spreading speed \>  \ref{sectionv*},   \\
lipid \>  \ref{secpropRayleigh}    \\
locality \> \ref{sectiontwofold}   \\
localized initial condition \> \ref{sectiontwofold}, \ref{sectionleadingedgedominated}   \\
locking \>   \ref{sectionmoregeneral} \\
logarithmic shift   \>  \ref{relaxtoutpulled}, \ref{twofeatures}    \\
long wavelength instability \>  \ref{chemicalbacterial}  \\
Lyapunov exponent  \>  \ref{sectionerrorpropagation}, \ref{sectionvanzon}    \\
Lyapunov functional \> \ref{sectionwhenpushed}, \ref{sectionch}, \ref{sectionkupperslorz}, \ref{sectiondynamicaltest}     \\
marginal stability \>  \ref{perspective}, \ref{stabvsel}, \ref{secpropRayleigh}    \\
matching analysis  \>  \ref{relaxtoutpulled}, \ref{sectionmatching1}, \ref{sectionmatching2}, \ref{finiteparticles}            \\
Mathieu equation  \>  \ref{sectioncombustion}   \\
maximum growth condition \>  \ref{sectionexponential}  \\
mean field growth model  \>  \ref{meanfieldgrowth}     \\
membrane  \>  \ref{secpropRayleigh}  \\
memory kernel \> \ref{sectionmoregeneral}, \ref{sectionbiologicalinvasion}     \\
monotonic front profile \>  \ref{sectionefk}, \ref{sectionstreamers}    \\
motion by mean curvature  \>   \ref{sectionmbabreakdown}         \\
moving boundary approximation \> \ref{sectionkupperslorz}, \ref{sectionstreamers}, \ref{chemicalbacterial}, \ref{sectionmba}, \ref{secspherical}, \ref{sectionmbabreakdown}, \ref{sectionstochasticfronts}        \\
Mullins-Sekerka instability  \>   \ref{sectionsupercond}, \ref{chemicalbacterial}, \ref{sectionmullinssekerka}      \\
multiplicative noise  \>   \ref{langevin}   \\
nematic  \>  \ref{sectiondynamicaltest}    \\
node conservation \>  \ref{sectionefk}, \ref{sectionsh}, \ref{sectionch}, \ref{sectioncglcubic}   \\
noise-sustained structures  \>  \ref{secfrontsnoise1},  \ref{secfrontsnoise2}, \ref{sectionmullinssekerka}        \\
non-causal  \>   \ref{stabvsel}     \\
non-conserved dynamics \>  \ref{secspinodal}   \\
non-monotonic front profile \>  \ref{sectionefk}   \\
non-perturbative effect  \>    \ref{langevin} \\
nucleation \>  \ref{sectiontcrb},  \ref{sectionsupercond}   \\
one-parameter family \> \ref{selectionutfs}, \ref{stabilitygeneral}, \ref{sectioncglcubic}    \\
ordinary differential equation \> \ref{selectionutfs}   \\
Orr-Sommerfeld equation  \>  \ref{sectionparallelshear}, \ref{sectionvonkarman}   \\
Painleve  analysis \> \ref{selectionutfs}  \\
partition function  \>   \ref{sectionphasetrans}  \\
pattern selection \>  \ref{perspective}\\
pearling  \>   \ref{secpropRayleigh}  \\
periodic media  \>   \ref{sectioncombustion}      \\
periodic front solution \> \ref{sectiontypesoffronts}  \\

phase field model  \>  \ref{sectionmullinssekerka}, \ref{sectionmba},  \ref{sectionmbabreakdown}              \\
phase separation \>  \ref{sectionch}   \\
phase slip  \>  \ref{sectioncglquintic},  \ref{sectionsupercond}, \ref{secfrontsnoise2}     \\
phase winding solution \> \ref{sectioncglcubic}, \ref{sectiontcrb}    \\
pinching  \>   \ref{secpropRayleigh}    \\
pinch point \>  \ref{sectionv*}, \ref{sectionmoregeneral}, App. \ref{appendixconnection}               \\
pipe flow  \>  \ref{sectionparallelshear}     \\
pitch  \>   \ref{sectionsmectic}   \\
plasma  \> \ref{sectionmoregeneral}, \ref{sectionstreamers}  \\
Poiseuille flow  \> \ref{secpropRayleigh},  \ref{sectionparallelshear}    \\
pole \>  \ref{sectionmoregeneral} \\
polymers  \>  \ref{secspinodal}   \\
population dynamics  \>  \ref{sectionotherbio}  \\
power law initial conditions  \>  \ref{sectionexponential}\\
pulled front \>  \ref{scopeaim}, \ref{sectiontwofold}, \ref{selectionutfs}, \ref{selectioncpfs}, \ref{sectionincoherentpffs}, \ref{sectionwhenpushed},  \\
  \> \hspace*{3mm}  \ref{relaxtoutpulled}, \ref{relaxtodincohpfpulled}, \ref{sectionefk}, \ref{sectionsh}, \ref{sectionch}, \ref{sectionks},   \\  \>  \hspace*{3mm}  \ref{sectioncglquintic}, \ref{sectionkupperslorz}, \ref{sectionstreamers}, \ref{sectiondebunching}, \ref{secspinodal}, \ref{secfrontsnoise2}, \ref{chemicalbacterial}, \ref{sectioncombustion}, \\ 
\>  \hspace*{3mm}   \ref{meanfieldgrowth}, \ref{sectionerrorpropagation}, \ref{sectionmba}, \ref{sectionmbabreakdown},  \ref{sectionglobalmodes}                           \\
``pulled front'' \> \ref{finiteparticles},  \ref{langevin}, \ref{sectionfluctuatinguniclasses}  \\
pulled versus pushed  \>  \ref{sectiontwofold}, \ref{selectionutfs}, \ref{selectioncpfs}, \ref{stabvsel}, \ref{sectioncglquintic}, \ref{epilogue}, \ref{sectionkupperslorz}, \ref{sectionstreamers},   \\  \>  \hspace*{3mm}   \ref{chemicalbacterial}, \ref{meanfieldgrowth}, \ref{sectionerrorpropagation},  \ref{sectionmbabreakdown}                         \\
pulse-type solution \>  \ref{sectionks}, \ref{sectioncglcubic}, \ref{sectionwounds}   \\
pushed front \>  \ref{scopeaim}, \ref{sectiontwofold}, \ref{selectionutfs}, \ref{selectioncpfs}, \ref{sectionincoherentpffs}, \ref{sectionwhenpushed},  \ref{sectionefk},    \\      \>      \hspace*{3mm}  \ref{sectionsh}, \ref{sectionks}, \ref{sectioncglquintic}, \ref{epilogue}, \ref{sectionkupperslorz}, \ref{sectionstreamers}, \ref{secfrontsnoise2}, \\ \>  \hspace*{3mm}   \ref{chemicalbacterial}, \ref{sectioncombustion}, \ref{sectionerrorpropagation}, \ref{sectionmba},  \ref{sectionmbabreakdown},  \ref{sectionglobalmodes}                               \\
quintic CGL equation \> \ref{selectionutfs}, \ref{selectioncpfs}, \ref{sectionwhenpushed}, \ref{structuralstability}, \ref{sectioncglquintic},   \ref{sectionsupercond},  \ref{sectionmatching2},  \ref{sectionglobalmodes}              \\
ramp  \>   \ref{globalmodeseps}   \\
Rayleigh instability \>   \ref{secpropRayleigh}, \ref{sectionsupercond} \\
Rayleigh-Benard instability  \>  \ref{sectiontcrb}, \ref{sectionkupperslorz}, \ref{sectionwounds}, \ref{sectionmba}     \\
Rayleigh-Taylor instability \>  \ref{sectionpropagatingrt}  \\
reaction-diffusion model  \>  \ref{chemicalbacterial}    \\
reduction of order methods \> \ref{selectionutfs}, \ref{sectionwhenpushed}, \ref{sectioncglquintic}, App. \ref{otherconjectures}  \\
relaxation mode  \>  \ref{finiteparticles}  \\
renormalization group  \>  \ref{sectionwounds}, \ref{sectioncarpentier}, App. \ref{otherconjectures} \\
retracting fronts  \>  \ref{sectioncglquintic},  \ref{sectionvanzon}    \\
Reynolds number  \>  \ref{sectionparallelshear}   \\
saddle point \>   \ref{sectionv*}, \ref{stabvsel}, \ref{twofeatures}                 \\
second order phase transition \>  \ref{sectiondynamicaltest}    \\
selected wavenumber  \> \ref{sectionefk},  \ref{sectioncglcubic}, \ref{sectiontcrb}    \\
separation of scales  \>  \ref{sectionmba},  \ref{sectionmbabreakdown}           \\
sidebranches \> \ref{section4thorder}, \ref{sectionmullinssekerka}     \\
shape mode  \>  \ref{sectionmatching1}  \\
similarity analysis  \>   \ref{sectioncascade}, \ref{twofeatures}   \\
similarity variable  \>  \ref{sectionuniversalrel}    \\
singular perturbation theory  \> App. \ref{otherconjectures}   \\
sink \>  \ref{sectioncglcubic}  \\
slowly varying control parameter  \>    \ref{globalmodeseps} \\
smectic  \>  \ref{sectiondynamicaltest}, \ref{sectionsmectic}    \\
solar cycle  \>   \ref{stellar}   \\
solvability condition \>  \ref{sectionmoregeneral},  \ref{sectionmbabreakdown}          \\
source  \> \ref{sectioncglcubic}, \ref{secfrontsnoise2}    \\
spark  \>  \ref{sectionstreamers}  \\
spherically expanding front  \>  \ref{secspherical}     \\
spinodal decomposition \>  \ref{secspinodal},  \ref{sectionsupercond}    \\
spots   \>  \ref{sectionparallelshear}   \\
stability  \> \ref{stabvsel}, \ref{relaxtoutpulled},  \ref{sectionmbabreakdown}            \\
steep initial conditions \>  \ref{sectiontransients}, \ref{sectiontwofold}, \ref{twofeatures}    \\
steepness \>  \ref{sectionexponential}, \ref{stabvsel}                     \\
stellar activity cycle  \>   \ref{stellar}   \\
step instability \>  \ref{sectiondebunching}    \\
stochastic fronts \>  \ref{sectionstochasticfronts}    \\
stratified fluid \>  \ref{sectionpropagatingrt}   \\
strongly heteroclinic orbit \> \ref{selectionutfs}, \ref{stabvsel}, \ref{epilogue}     \\
structural phase transition \>  \ref{secsalje}     \\
structural stability \> \ref{structuralstability}, \ref{sectioncglquintic}      \\
subcritical bifurcation/transition  \>  \ref{sectioncglquintic}, \ref{secpropRayleigh},  \ref{sectionparallelshear}, \ref{secfrontsnoise2}      \\
sufficiently localized perturbation \>   \ref{sectionv*}           \\
superconductor  \>  \ref{sectionsupercond}   \\
supercritical bifurcation/transition \>   \ref{sectionexamples}, \ref{sectiontcrb}, \ref{secfrontsnoise1},  \ref{secfrontsnoise2}      \\
surface tension  \>  \ref{secpropRayleigh}   \\
surface step  \>  \ref{sectiondebunching}   \\
Swift-Hohenberg equation \> \ref{scopeaim}, \ref{selectioncpfs}, \ref{sectionsh}, \ref{sectionch}, \ref{sectiontcrb},  \ref{sectionmatching1}        \\
switching  \>  \ref{sectionsmectic}    \\
Taylor-Couette cell/flow \> \ref{sectiontcrb},  \ref{secfrontsnoise1}, \ref{sectionwounds}     \\
Taylor vortex  \>  \ref{sectiontcrb},  \ref{secfrontsnoise1}  \\
thermal plume  \>  \ref{sectionmba}  \\
thin film (equations) \>  \ref{sectionpropagatingrt}, \ref{secpropRayleigh} \\
throughflow \> \ref{sectionconvversusabs}, \ref{secfrontsnoise1}     \\
tip of a front  \>  \ref{finiteparticles}  \\
transients \>  \ref{sectiontransients}, \ref{secsalje},  \ref{sectionwounds}     \\
traveling wave  \>  \ref{secfrontsnoise1}  \\
tricritical point   \>   \ref{sectiondynamicaltest}     \\
tweed pattern  \>  \ref{secsalje}    \\
two-parameter family \> \ref{selectioncpfs}, \ref{sectiontwospreadingpoints}, \ref{stabilitygeneral}, \ref{sectioncglcubic}     \\
turbulence   \>  \ref{sectionparallelshear}, \ref{sectioncombustion}, \ref{sectioncascade}      \\
Turing instability  \> \ref{chemicalbacterial}  \\
uniformly translating front \>  \ref{scopeaim}, \ref{sectionexponential}, \ref{sectiontypesoffronts}, \ref{selectionutfs},  \ref{relaxtoutpulled}, \ref{sectionmatching1}      \\
universality classes \> \ref{sectiontwofold}, \ref{meanfieldgrowth},  \ref{sectionfluctuatinguniclasses}   \\
universal relaxation \> \ref{sectionunirelsimple}, \ref{relaxtoutpulled}, \ref{sectionsh}, \ref{sectionvanzon}, \ref{sectionphasetrans}, \ref{sectionuniversalrel}, \ref{twofeatures}, \ref{sectionmatching1},  \\
\>  \hspace*{3mm}    \ref{sectionmatching2}            \\
$v_{\rm co}$ \>  \ref{sectiontransients}, \ref{epilogue}\ref{epilogue} \\
$v_{\rm env}$ \> \ref{sectionexponential}, \ref{sectionmoregeneral}, \ref{sectionleadingedgedominated}, \ref{sectiontwospreadingpoints}, \ref{meanfieldgrowth},  \ref{sectionvanzon}          \\
von Karman vortex street  \>  \ref{sectionparallelshear}, \ref{sectionvonkarman}   \\
vortex   \>  \ref{sectionsupercond},  \ref{sectionparallelshear}, \ref{sectionvonkarman}    \\
WKB-analysis  \>  \ref{globalmodeseps}   \\
wound healing  \>  \ref{sectionwounds}   \\
zero mode \> \ref{stabilitygeneral}    \\
\end{tabbing}
}

\end{appendix}

\end{document}